%% file: main.tex
\pdfoutput=1
\documentclass[10pt,a4paper,oneside]{book}
\usepackage{graphicx}
\usepackage{amsmath}
\usepackage{amsfonts}
\usepackage{amsthm}
\usepackage{amssymb}
\usepackage{fancyhdr}
\usepackage{psfrag}
\usepackage{indentfirst}
\usepackage{enumerate}
\usepackage{floatflt}
\usepackage{cite}
\usepackage[small]{caption}
\usepackage{wrapfig}
\usepackage{float}
\usepackage{hyperref}
\usepackage{algorithm} 
\usepackage{algorithmic}
\usepackage[]{tocbibind} 
\usepackage{appendix}
\usepackage[french,magyar,USenglish]{babel}
\usepackage[utf8]{inputenc}
\usepackage{setspace}
\newcommand{\be}{\begin{equation}}
\newcommand{\ee}{\end{equation}}
\newcommand{\bea}{\begin{eqnarray}}
\newcommand{\eea}{\end{eqnarray}}

\newcommand{\emptypage}
{
\thispagestyle{empty}
}

\fancypagestyle{plain}
{
\fancyhead[RO]{\slshape\footnotesize\leftmark}
\fancyhead[RE]{}
\fancyhead[LO]{}
\fancyhead[LE]{\slshape\footnotesize\leftmark}
\fancyfoot[C]{}
\fancyfoot[LE, RO]{\thepage}
}

\newtheorem{theorem}{Theorem}[section]

\newtheorem{proposition}[theorem]{Proposition}

\newtheorem{definition}[theorem]{Definition}

\begin{document} 

\begin{spacing}{1.0}
\frontmatter
\input{frontpage}

\pagenumbering{roman}
\tableofcontents
\input{Notations}

\mainmatter
\pagenumbering{arabic}
\end{spacing}

\begin{spacing}{1.0}
\input{Intro}
\input{GeneralOverview}

\input{complexnetw}
\input{NumericalMethods}

\input{NetwContactProc}
\input{NetwRBPM}
\input{PottsPerco}
\input{TAFIM}
\input{Discussion}

\end{spacing}
\begin{spacing}{1.0}
\pagenumbering{roman}
\setcounter{page}{0}
\bibliography{db_thesis}
\bibliographystyle{thsty}
\end{spacing}


\input{Appendix}

\backmatter
\input{Aknowledgement}
\input{Declaration}

\input{SummaryEng}

\input{SummaryHun}
\input{SummaryFr}

\end{document}

%% file: frontpage.tex
\emptypage
\begin{center}
  \vspace{5mm}
  \LARGE
  \textbf{COOPERATIVE BEHAVIOUR IN COMPLEX SYSTEMS} \\
  \Large
  \vspace{5mm}
  \textbf{by} \\
  \vspace{5mm}
  \LARGE
  \textbf{M\'arton Karsai} \\
  \vspace{20mm}
  \Large
  {\bf{\textsl{THESIS}}} \\
  \textsl{presented for the degree of} \\
  \vspace{2mm}
  {\bf{\textsl{Doctor of Philosophy}}} \\
  \vspace{5mm}
\begin{center}
\normalsize
\begin{tabular}{ccc}
\includegraphics[width=4cm]{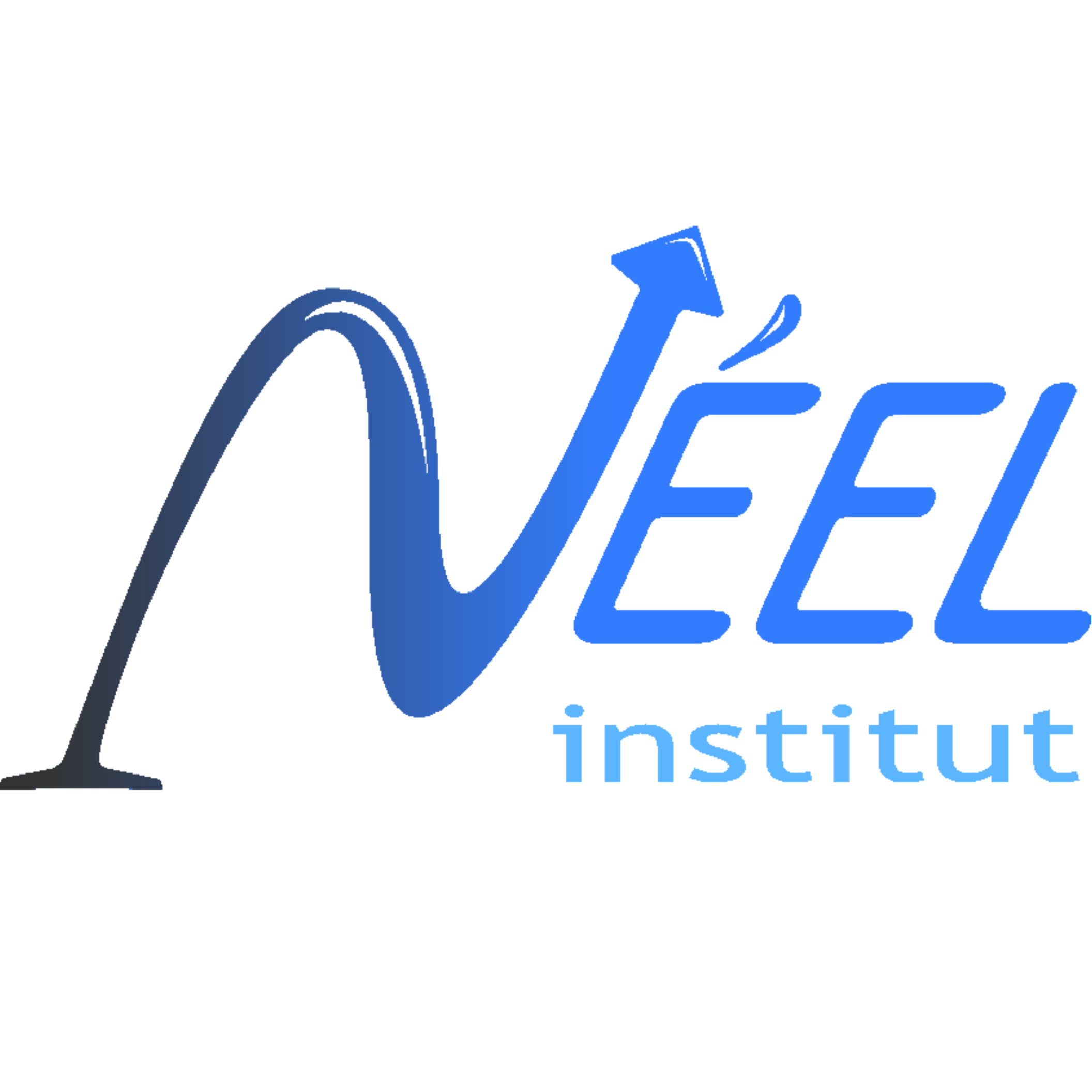} \hspace{2mm} & \includegraphics[height=3.5cm]{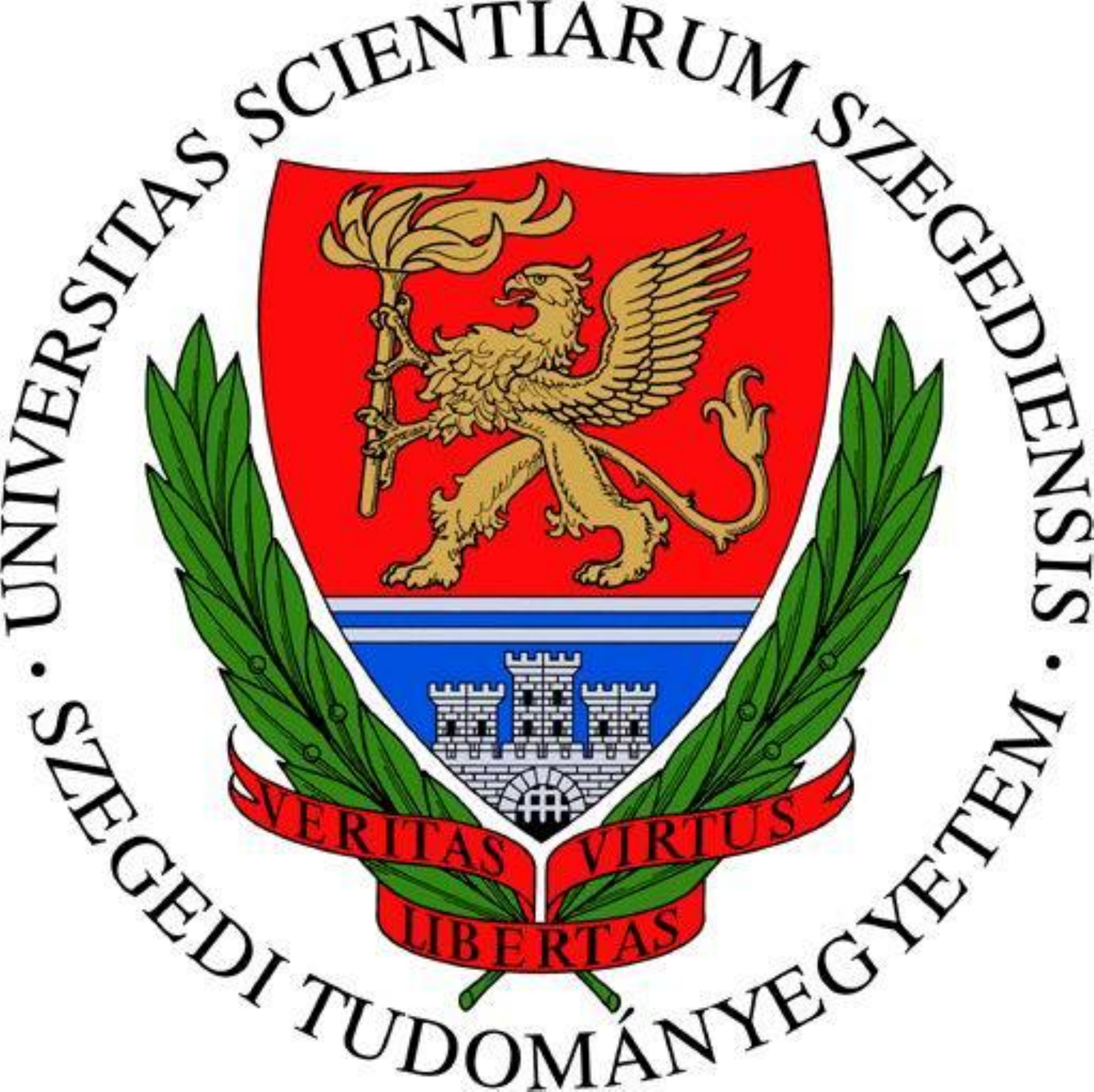} \vspace{3mm} \hspace{2mm}& \includegraphics[height=4cm]{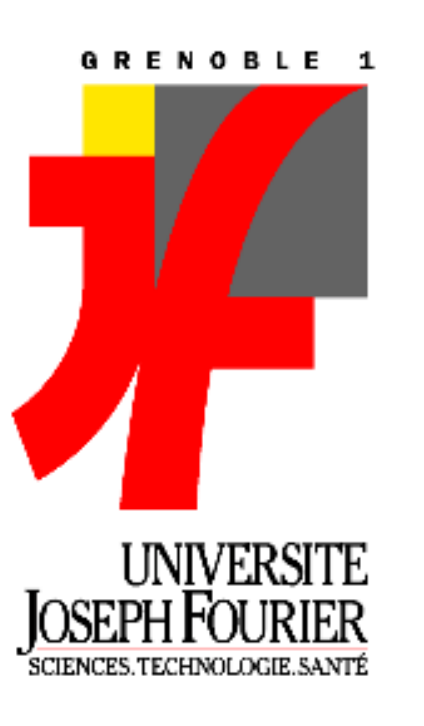}  \\ 
\textsl{CNRS - Institut N\'eel} \hspace{2mm}& \textsl{Department of Theoretical Physics}  \hspace{2mm}& \textsl{Universit\'e Joseph Fourier} \\
\textsl{Grenoble}\hspace{2mm} & \textsl{Faculty of Science and Informatics} \hspace{2mm}& \textsl{Grenoble} \\
 & \textsl{University of Szeged} &
\end{tabular}
  \vspace{13mm}
\end{center}
\begin{center}
\begin{tabular}{ll}
 \textsl{Supervisors:} & \textsl{Prof. Dr. Ferenc Igl\'oi} \\
& \textsl{Dr. Jean-Christian Angl\`es d'Auriac}
\end{tabular}
\end{center}
  \vspace{8mm}
\Large
  \textsl{Doctoral School of Physics} \\
  \vspace{9mm}
  \textsl{Szeged - Grenoble} \\
  \textsl{2009}\\
  \normalsize
\end{center}

\newpage
\emptypage
 \vspace*{2cm}
 \hspace{8cm} \textit{\LARGE {Dedicated to my parents}}\\

%% file: Notations.tex
\chapter{Selected notations}
\begin{tabular}{lll|lll}
$\langle...\rangle$ & \hspace{.1in} & sample average 	&\hspace{.05in}	$v$ & \hspace{.1in} & vertex 		\\
$A(s,t)$ & \hspace{.1in} & autocorrelation function 	&\hspace{.05in}	$V$ & \hspace{.1in} & node set				\\
$b$ & \hspace{.1in} & rescaling factor			&\hspace{.05in}	$x_b$ & \hspace{.1in} & bulk magnetization scaling dimension\\
$c(...)$ & \hspace{.1in} & number of connected components&\hspace{.05in}$x_s$ & \hspace{.1in} & surface magnetization scaling dimension\\
$C$ & \hspace{.1in} & specific heat			&\hspace{.05in}	$z$ & \hspace{.1in} & dynamical exponent\\
$d$ & \hspace{.1in} & dimension				&\hspace{.05in}	$\mathcal{Z}$ & \hspace{.1in} & partition function\\
$d_f$ & \hspace{.1in} & fractal dimension		&\hspace{.05in}	$\alpha$ & \hspace{.1in} & critical exponent ($C_H(t)\sim|t|^{-\alpha}$)\\
$e$ & \hspace{.1in} & edge				&\hspace{.05in}	$\beta$ & \hspace{.1in} & inverse temperature 		\\
$E$ & \hspace{.1in} & energy				&\hspace{.05in}	 & \hspace{.1in} & critical exponent ($M(t)\sim(-t)^{\beta}$)		\\
    & \hspace{.1in} & edge set				&\hspace{.05in}	$\gamma$ & \hspace{.1in} & degree exponent ($P(k)\sim k^{-\gamma}$)\\
$f$ & \hspace{.1in} & free energy density 		&\hspace{.05in}	 & \hspace{.1in} & critical exponent ($\chi_H(t)\sim|t|^{-\gamma}$)\\
$\mathcal{F}$ & \hspace{.1in} & free energy 		&\hspace{.05in}	$\Gamma(...)$ & \hspace{.1in} & correlation function	\\
$G$ & \hspace{.1in} & graph				&\hspace{.05in}	$\Delta$ & \hspace{.1in} & strength of disorder\\
$h$ & \hspace{.1in} & reduced magnetic field 		&\hspace{.05in}	$\delta(...)$ & \hspace{.1in} & Kronecker delta function\\
$H$ & \hspace{.1in} & external magnetic field 		&\hspace{.05in} 	$\delta$ & \hspace{.1in} & critical exponent ($M(h)\sim|h|^{-1/\delta}$)\\
$\mathcal{H}$ & \hspace{.1in} & Hamiltonian		&\hspace{.05in}	$\eta$ & \hspace{.1in} & critical exponent ($\Gamma(r)\sim r^{-(d-2+\eta)}$)\\
$J$ & \hspace{.1in} & exchange energy between spins	&\hspace{.05in}	$\lambda$ & \hspace{.1in} & strength of infection\\
$k$ & \hspace{.1in} & degree				&\hspace{.05in}	$\mu$ & \hspace{.1in} & degradation exponent\\
$k_B$ & \hspace{.1in} & Boltzmann constant 		&\hspace{.05in}	$\nu$ & \hspace{.1in} & critical exponent ($\xi(t)\sim|t|^{-\nu}$)\\
$\bar{\ell}$ & \hspace{.1in} & average path length	&\hspace{.05in}	$\nu_h$ & \hspace{.1in} & critical exponent ($\xi(h)\sim|h|^{-\nu_h}$)\\
$L$ & \hspace{.1in} & linear size			&\hspace{.05in}	$\nu_\shortparallel$ & \hspace{.1in} & time correlation exponent \\
$\mathcal{L}$ & \hspace{.1in} & Liouville operator	&\hspace{.05in}	$\xi$ & \hspace{.1in} & correlation length		\\
$m$ & \hspace{.1in} & order parameter density 		&\hspace{.05in}	$\rho$ & \hspace{.1in} & density\\
    & \hspace{.1in} & initial connectivity		&\hspace{.05in}	$\tau$ & \hspace{.1in} & finite size relaxation time\\
$M$ & \hspace{.1in} & order parameter 			&\hspace{.05in}	$\tau(L)$ & \hspace{.1in} & characteristic time\\
$N$ & \hspace{.1in} & system size			&\hspace{.05in}	$\chi$ & \hspace{.1in} & susceptibility 		\\
$\mathcal{N}$ & \hspace{.1in} & largest correlated volume & & & \\
$\mathcal{O}(...)$ & \hspace{.1in} & ordo function	& & & \\
$p$ & \hspace{.1in} & probability (value)		& & & \\
$P(...)$ & \hspace{.1in} & probability (function)	& & & \\
$q$ & \hspace{.1in} & number of Potts state		& & & \\
$r$ & \hspace{.1in} & geometrycal distance		& & & \\
$s$ & \hspace{.1in} & waiting time			& & & \\
 & \hspace{.1in} & source				& & & \\
$S$ & \hspace{.1in} & entropy				& & & \\
$t$ & \hspace{.1in} & reduced temperature		& & & \\
 & \hspace{.1in} & time					& & & \\
 & \hspace{.1in} & sink					& & & \\
$T$ & \hspace{.1in} & temperature 			&  & & \\
\end{tabular}

%% file: Intro.tex
\chapter{Introduction}
\label{Intro}

Cooperative phenomena is an ordinary pattern in various part of science where interacting entities attempt to satisfy some optimal conditions. Such kind of behaviour is observable in equilibrium physics where the degree of cooperation can be characterized by macroscopic functions but cooperation could emerge in out of equilibrium systems also, like in sociology, economy, biology or physics etc., where the correlation evolves in the system when an entity aspire to satisfy personal optimal conditions and it is content to cooperate with other participants for the sake of its cause. In this case we cannot characterize the system with macroscopic functions, but we can describe it on microscopic level. Even in different systems the correlative agents are defined differently but their behaviour presents universal attributes. In this sense the complexity of a system can be defined by its interconnected features arising from properties of its individual parts. Such parts could be either which govern the interactions or the dynamical rules to control the time evolution or the backgrounding structure which also can influence the system behaviour. This kind of combined effects leads to a complex system with non-trivial cooperative behaviour and brings on many interesting questions which give the motivation for advanced studies. 

Correlated systems can be examined efficiently within the frame of statistical physics. The subject of this area is to study systems which depend on random variable, and describe their behaviour obtained from large number of observation using physical terminology. An interesting subject of this segment is the one of phase transitions which has been intensively studied since the beginning of the 20th century. Phase transitions occur in many part of Nature where counteractive processes compete to determine the state of the system controlled by an external parameter. The best-known examples in physics are the liquid-vapor or the ferromagnetic-paramagnetic phase transitions but similar behaviour can be observed in model systems like in different spin models, in epidemic spreading problems or in critical percolation. 

The first relevant approach was the \textit{mean field theory} which gave a phenomenological description of phase transitions and critical phenomena and is capable to describe a wide range of model systems. It was first defined by Pierre Weiss in 1907 for ferromagnetic systems where he assumed that the spins interact with another through a molecular field, proportional to the average magnetization \cite{Weiss1907}. This \textit{molecular field approximation} method neglects the interactions between particles and replaces them with an effective average field which enables to simplify the solution of the problem. At the same time the disadvantage of the method originates from the average molecular field also, since it neglects any kind of fluctuations in the system. Therefore the mean field approximation is valid only in higher dimensional systems or in case of models where the fluctuations are not important.

One of the simplest and most elegant mean field speculations concerning the possible general form of a thermodynamical potential near to the critical point was introduced later by Lev. D. Landau \cite{Landau1936,Landau1980} in 1936. The \textit{Landau theory} allows a phenomenological reproduction of continuous phase transitions based on the symmetry of the order parameter. He assumed that the free energy can be expanded as a power series of the order parameter where the only terms which contribute are the ones compatible with the symmetry of the system.

The observation that the correlation length diverges at the phase transition point and the fluctuations in the system are self-similar in every length scale, led to the recognition of scale invariant behaviour of critical systems. The first comprehensive mathematical approach of such phenomena was given by Leo Kadanoff in the 1960s \cite{RevModPhys.39.395,Kadanoff1966}. A few years later in the beginning of the 1970s based on his results, Kenneth G. Wilson gave the relevant discussion of the subject in his celebrated papers \cite{Wilson1971,Wilson1971b}. He introduced the \textit{renormalization group theory} which serves predictions about critical behaviour in agreement with experimental results and give a possibility to categorize critical systems into universality classes due to them singular behaviour. His investigations were awarded with a Nobel prize in 1982.

In the beginning of the 1970s another theory emerged which completed the knowledge about the universal behaviour around the criticality. The \textit{conformal field theory} was firstly investigated by Polyakov \cite{Polyakov1972}, who capitalized that the group of conformal mappings is equivalent to the group of complex analytical functions in two dimensions \cite{Belavin1984,Henkel1999}. Exploiting this property and the scale invariant behaviour, the partition function of a critical system is derivable which leads to exact calculations of critical exponents. In the case of conform transformations the order parameter correlation functions can be calculated also which then enables to deduce the order parameter profiles along the system boundaries.

As a first approximation generally the investigated system is assumed to be homogeneous, which condition simplifies its study and the physical description. However, in Nature a substance is often characterized by certain degree of inhomogeneity which could perturb its critical behaviour \cite{Ziman1979}. Frequently stated examples of this abnormality, are the lattice defects and impurities in crystals. In theoretical description this kind of feature is introduced by the concept of disorder which is defined as random distributed values of certain properties of the investigated \textit{disordered model}. These random properties can be the strength of interaction, a random external field or can arises by random dilution. These properties are able to modify the critical features of the system as changing the order of phase transition or shift the critical point and exponents, thus transforming the model into a new universality class.

However, inhomogeneity can arise from the geometrical structure of the backgrounding framework of materials also. Beyond solid-state physics where crystals have a regular geometry, many self-organizing media in the Nature can form random structures. Two Hungarian mathematicians, P\'al Erd\H{o}s and Alfr\'ed R\'enyi defined the first related \textit{random graph model} in their pioneering papers \cite{Erdos1959,Erdos1960,Erdos1961} in the beginning of the 1960s which was the origin of the new science of networks. However, as the informations of real world networks became available by the emergence of large databases, a deeper view into the underlying organizing principles suggested a more complex structure. In the beginning of the 1990s Barab\'asi, Albert and Jeong after they examined the structure of the World Wide Web found an algebraic decay of the network degree distribution. This observation made them realize the real rules which govern the evolution of such kind of \textit{complex networks}. They defined a dynamical growing network model, \textit{the Barab\'asi-Albert model} \cite{Barabasi1999}, where the sites are not connected homogeneously but follow a so-called preferential attachment rule leading to a \textit{scale-free graph} with a power-law degree distribution. This model gave a very good approximation of real complex networks and became popular after many complex systems in science and technology were found to present the same structure \cite{albert-2002-74}. This interdisciplinary behaviour which suggests universal rules behind self-organizing networks keeps this discipline to the frontline of science up to this day.

Analyzing a complex many body system was difficult earlier since generally these systems have large degrees of freedom and those could be in many possible states. However, by the improvement of computational engines new possibilities appeared and high precision numerical methods were developed to examine the relevant models. It was the base of the new discipline of numerical physics which then became the third pillar of the science beyond the experiments and theory. By using numerical calculations, the statistical description of many body systems became available which then led to a renaissance of the statistical physics. This relation induced the evaluation of \textit{Monte Carlo methods} which provided special computational techniques for statistical physics simulations.

One of the first and most frequently used method which is capable to simulate interacting spin systems is the \textit{Metropolis algorithm}, introduced in 1953 \cite{metropolis:1087}. This so-called single spin flip algorithm evaluate the system through a Markov chain where the system energy change depends on local configurations. However, some other algorithms of \textit{Swendsen-Wang} \cite{PhysRevLett.58.86} and \textit{Wolff} \cite{PhysRevLett.62.361} provided more efficient methods where instead of one-by-one spin flips, complete domains are turned in one step and evaluate the system faster toward its equilibrium.

Beyond that, many other kind of statistical methods were developed in order to calculate some thermodynamical function using mathematical considerations. A recent algorithm which capitalize \textit{combinatorial optimization} is capable to calculate exactly the free energy for models where the free energy function can be recognized as a \textit{submodular function} \cite{dAuriac2002,dAuriac2004}. This iterative method which provided results in strongly polynomial time was applied frequently in course of my work.

\vspace{.3in}

My motivation during my PhD studies was to examine cooperative behaviour in complex systems using the methods of statistical and computational physics. The aim of my work was to study the critical behaviour of interacting many-body systems during their phase transitions and describe their universal features analytically and by means of numerical calculations. In order to do so I completed studies in different subjects which are presented in the thesis in the following order:

After this introduction in the second chapter we summarize the capital points of the related theoretical results. We shortly discuss the subjects of phase transitions and critical phenomena and briefly write about the theory of universality classes and critical exponents. Then we introduce the important statistical models which will be examining later in the thesis. We close this part with a description of disordered models which will be helpful during later discussions.

In the third chapter first, we point out the definitions of graph theory that we need to introduce the applied geometrical structures. Then we review the main properties of regular lattices and define their general used boundary conditions. We close this chapter with a short introduction to complex networks.

The fourth chapter contains the applied numerical methods that we used in the course of numerical studies. We write a few words about Monte Carlo methods and introduce a combinatorial optimization algorithm and its mathematical background. As a last point we describe our own technique to generate scale-free networks.

In the fifth chapter we consider nonequilibrium phase transitions, such as epidemic spreading, in weighted scale-free networks, in which highly connected nodes have a relatively smaller ability to transfer infection. We solve the dynamical mean-field equations and discuss finite-size scaling theory. The theoretical predictions are confronted with the results of large scale Monte Carlo simulations on the weighted Barab\'asi-Albert network.

In the sixth chapter we examine the ferromagnetic large-q state Potts model in complex evolving networks, which is equivalent to an optimal cooperation problem, where the agents try to optimize the total sum of pair cooperation benefits and the supports of independent projects. We study the critical behaviour on Barab\'asi-Albert networks in case of homogeneous couplings and consider the effects of bond disorder by the magnetization behaviour. We locate the phase transition point and some critical exponents using the shift of the distribution of finite-size transition points. We also examine the structural evolution of the largest cluster.

The seventh chapter contains our consideration of the classical random bond Potts model in the large-q limit with disorder-dominated critical points and study the distribution of clusters that are confined in strips and touch one or both boundaries. We study optimal Fortuin-Kasteleyn clusters using the combinatorial optimization algorithm and obtain accurate numerical estimates for the critical exponents and demonstrate that the density profiles are well described by conformal formulas.

In the eighth chapter we study the antiferromagnetic Ising model on triangular lattice at zero temperature. We consider its critical dynamical behaviour in the aging regime and search for proof to decide that its non-equilibrium behavior is governed by a diffusive growth with logarithmic correction or if it follows a subdiffusive behaviour with an effective dynamical exponent. In order to do so, we define a quantity which depends exclusively on the dynamical exponent and that is capable to recognize the true dynamical behaviour. We also examine the scaling of the two-time autocorrelation function in equilibrium and in the aging regime.

During my PhD studies I investigated another subject where I examined the mapping between the surface behaviour of the random bond Potts model and random field Ising model in two dimensions. The surface mapping between the two systems was expected to be exact following from some theoretical consideration \cite{PhysRevLett.79.4063}, which was partly confirmed by the numerical calculations, however these results are not included into this current paper.

The new results presented in this work are published in \cite{Karsai2006,Karsai2007,Karsai2008} or their publication is in progress \cite{Karsai2009}.

%% file: GeneralOverview.tex
\chapter{General Overview}
\section{Phase transitions and critical phenomena}
\label{PhTr_CritPh}
A material pass through a phase transition when it is transformed between its different macroscopic phases and its properties change controlled by an external parameter. The type of the phase transition is related to the order of the transition which was defined by Ehrenfest as the singularity of the free energy or one of its derivatives. This singularity takes place at a critical value of the external control parameter where the phase transition arises and separates the macroscopic phases in the phase space. In completely different systems, thermodynamic properties show the same critical behavior at the vicinity of their phase transition point, which kind of behaviour is related to the subject of universality and critical phenomena. In the following chapter we are going to discuss briefly the theory of phase transitions and critical phenomena, define some models which are of general importance or which are related to the following problems.

\subsection{Ferromagnetic phase transition}
In ferromagnetic materials the equilibrium phase depends on the external field $H$ and the temperature $T$. These quantities control the macroscopic behavior of the system \cite{Stanley1971}. The magnetization $M$ plays the order parameter role and enables to identify the different phases of the system. In ferromagnets varying $H$ or $T$, two kind of phase transition can be identified following different orders.

The first phase transition happens at low temperature between two separated macroscopic phases, where the sign of $M$ is either positive or negative determined by the dominant spin direction. The external field controls the phase transition. Changing $H$ at a critical point $H_c=0$, the ferromagnet jumps from one of its macroscopic ground state to the other, the magnetization immediately changes sign and the system goes through a sharp phase transition. Since the magnetization, which is the first derivative of the free energy (see in Section.\ref{Thermodynamical quantities}), is discontinuous at the critical point this phase transition is of first-order. 

\begin{figure}[htb]
\begin{center}
\includegraphics*[ width=10.0cm]{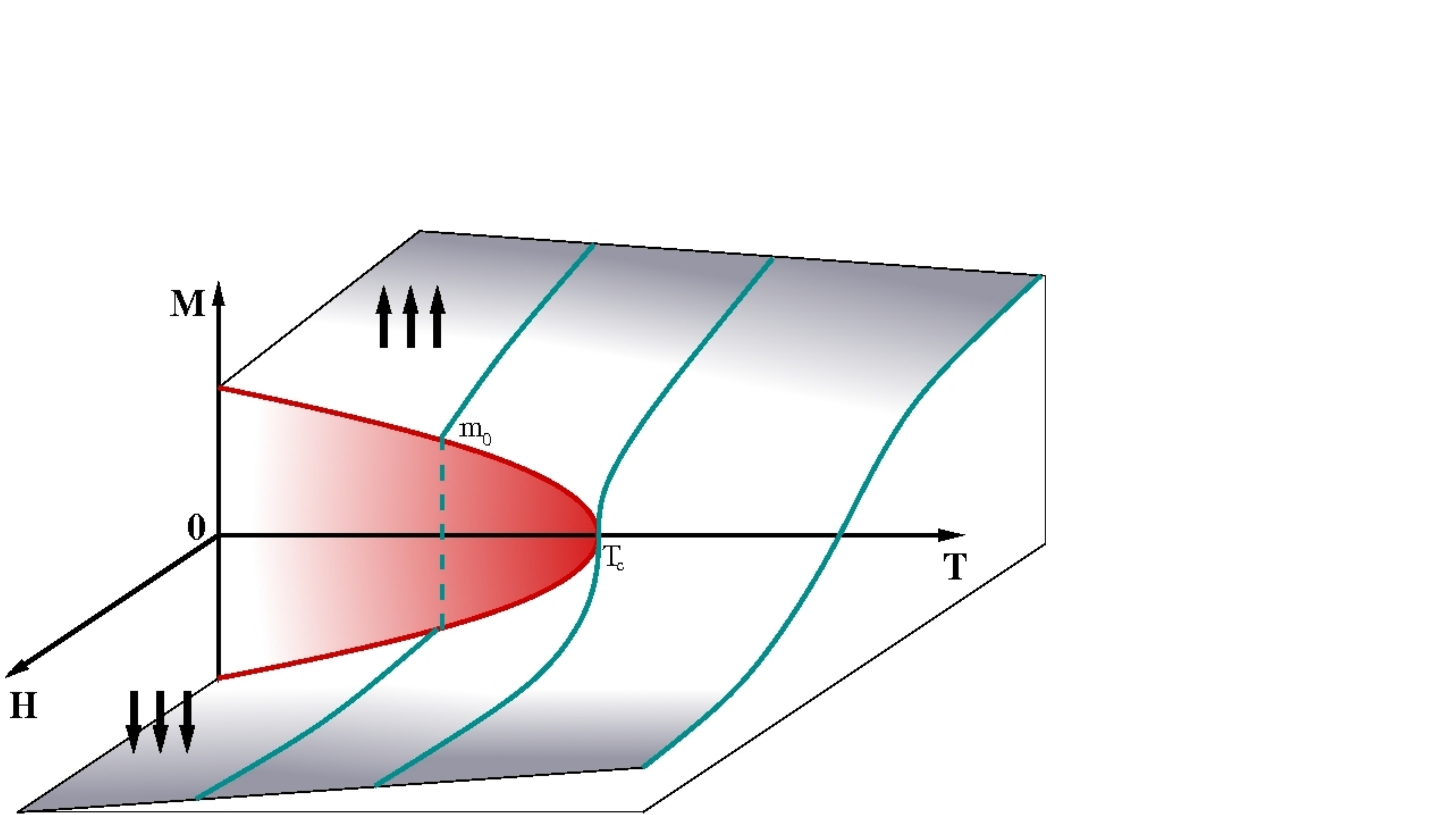}
\caption{Phase diagram of the two dimensional Ising ferromagnet. Below $T_c$ the system paths through a first order phase transition driven by the external field $H$. Here at $H=0$ the magnetization $m_0(H)$ is discontinuous. On the critical line $H=0$ another phase transition of second order occurs at $T_c$ which above the magnetization is zero.}
{
\label{ferromagnet}
}
\end{center}
\end{figure}

The second phase transition happens if we keep $H=0$ and increase the temperature from zero and following the critical phase coexisting line (Fig.\ref{ferromagnet}). The spontaneous magnetization decreases until a critical temperature $T_c$ is reached where it smoothly vanishes out and goes to zero. The system at this critical temperature (also called Curie-temperature), looses its ferromagnetic properties and enters the paramagnetic phase through a continuous phase-transition (Fig.\ref{ferromagnet}). If $T>T_c$ the system can be transformed from the field driven up to driven down phase and vica versa without any phase transition since the magnetization is zero and there is no fundamental difference between the two region (Fig.\ref{ferromagnet}).

\subsection{A microscopic model}
\label{IsingSimple}
To understand better the magnetic behaviour of the system at different phases we need to study it on a microscopic level. To do so we define the general Ising model which enables us to illustrate the behavior of magnetic spins at different macroscopic phases and during the phase transition.

Consider a $d$-dimensional hypercubic lattice, where at each site $i$ a spin $\sigma_i$ is present. All spins can take a value of $\sigma_i=\pm 1$ and interact only with their nearest neighbours by an exchange interaction $J$. The Hamiltonian of the Ising model is
\begin{equation}
	\mathcal{H}=-J\sum_{\langle i,j \rangle} \sigma_i \sigma_j
\end{equation}
where the summation runs over all spin pairs $\langle i,j \rangle$ of nearest neighbours. If the exchange interaction $J>0$, neighbouring spins favor to align in the same direction, driving ferromagnetic order into the system. The order parameter is the average magnetization per site defined as $m=1/L^d \sum_i \sigma_i$ where $L$ is the linear size of the system and where the sum goes over all spins. This model has been defined and solved in $1925$ by Ising \cite{Ising1925} in one dimension, where the system does not show any phase transition for $T>0$ since it does not have any stable ordered phase. However an exact solution exists in two dimension by Onsager \cite{PhysRev.65.117} in absence of external field, where the phase diagram is the same as for ferromagnets. In three or higher dimensions the Ising model is still not solved analytically however many precise numerical result exist about the critical point and exponents.

\subsection{Critical phenomena}
Critical phenomena takes place at the critical point of a continuous phase transition, when the correlation length diverges. Different thermodynamical quantities follow singular behaviors and are connected by scaling relations \cite{Simons2001}. To discuss such phenomena, we study the ferromagnetic-paramagnetic continuous phase transition from the point of view of the two dimensional Ising model on a microscopic scale. To do so we analyze an ``Ising system'' in the thermodynamical limit, when the system size tends to infinity. We follow the zero field line (Fig.\ref{ferromagnet}), vary the temperature and measure the spin-spin correlation length. The correlation length $\xi$ (for exact definition see Section \ref{Thermodynamical quantities}) represents the degree of correlation at a given temperature and it can be measured like the linear size of the largest magnetic domain.

We start from $T=\infty$ limit, where the system is completely disordered and $\xi=0$, since the magnetic order is destroyed by thermal fluctuations. Decreasing $T$ though, but keeping it above $T_c$, the exchange interactions become more dominant and the neighbouring spins try to align parallel. Even thermal fluctuations influence the system down to the scale of lattice-spacing, local magnetic domains evolve and correlation appears in short range order. Accordingly the correlated spin clusters grow and the correlation length becomes finite $\xi>0$ and increases as the temperature approaches $T_c$.

The critical temperature $T_c$ is located at the point where the correlation length becomes infinite. At the same time some thermodynamical properties of the system denote singular power-law behaviors characterized by a set of exponents which enables us to classify critical systems. This critical phenomena will be the subject of the next section.

At the vicinity of the critical point, fluctuations become more relevant up to the correlation length scale. These fluctuations are correlated and self-similar, so the system shows the same structure from the lattice constant scale up to $\xi$, independently of the scaling resolution. Since the correlation length turns to be infinite at $T_c$ the critical system becomes scale-invariant. This causes a fractal like structure for the dominant ordered cluster which now spreads all over the system on every length scale.

After the system tries to leave its critical phase as we decrease the temperature, a spontaneous symmetry breaking appears at $T_c$ since the system must choose to evaluate into the up or down spin state. This explains the symmetrical behavior of the magnetization below $T_c$ presented in the (Fig.\ref{ferromagnet}). Below $T_c$ the system is in the long-range order ferromagnetic phase, and the correlation contribution to energy becomes more significant than the entropic one. The magnetization moves off from $M=0$ and increases until its saturation value at $T=0$ where only the exchange interaction dominates and the systems becomes completely ordered.

\section{Universality classes and critical exponents}

At the vicinity of the critical point, various thermodynamical properties follow singular behaviors. This non-analytical attribute is characterized by a set of critical exponents, allows us to compare and classify phase transitions in different physical systems. These critical exponents appear only at the critical point of second order phase transition and fall into a limited number of universality classes, featured fairly different systems driven by the same fundamental symmetries \cite{Yeomans1994,Bagamery2006}. In the following we are going to define thermodynamical quantities and the related critical exponents to analyze universal behaviors.

\subsection{Thermodynamical quantities}
\label{Thermodynamical quantities}

The partition function of a canonical statistical is system given by:
\begin{equation}
	\mathcal{Z}(T,H)=\sum_{\mu} e^{-\beta E_{\mu}}
\label{Z}
\end{equation}
where the sum runs over all the possible microscopic states. Each microscopic states with energy $E_{\mu}$ appears with the probability $P({\mu})=e^{-\beta E_{\mu}}/\mathcal{Z}$ where the thermal term is introduced as $\beta=1/k_B T$, also called inverse temperature and $k_B$ denotes the Boltzmann constant. Then we define the free energy to be proportional to the logarithm of the partition function:
\begin{equation}
	\mathcal{F}(T,H)=-k_B T \mbox{ln}(\mathcal{Z}) \qquad \mbox{and} \qquad f(T,H)=-k_BT\lim_{L\to \infty} L^{-d} \mbox{ln}(\mathcal{Z})
\label{FreeE}
\end{equation}
where $f(T,H)$ denotes the free energy density which gives the energy of one particle in the system since it is normalized with the system size. All the thermodynamical quantities could be defined then by the partition function and the free energy or one of its derivative. The magnetization is given at constant temperature by:
\begin{equation}
 	M=-\left(\dfrac{\partial\mathcal{F}}{\partial H} \right)_T \qquad \mbox{,} \qquad \chi_T=\left( \dfrac{\partial M}{\partial H} \right)_T 
\label{M}
\end{equation}
where $\chi_T$ is the  magnetic response function or magnetic susceptibility at constant temperature. The internal energy arises as the derivative of the logarithm of the partition function:
\begin{equation}
 	U=-\left( \dfrac{\partial \mbox{ln}\mathcal{Z}}{\partial \beta}\right) \qquad \mbox{,} \qquad C_H= \left( \dfrac{\partial U}{\partial T}\right)_H
\end{equation}
where $C_H$ is the thermal response function or the specific heat at constant $H$ field. The entropy of the all system is given also at constant field by:
\begin{equation}
 	\qquad S= -\left( \dfrac{\partial \mathcal{F}}{\partial T}\right)_H \qquad \mbox{,} \qquad C_{H,M}= \left( T \dfrac{\partial S}{\partial T}\right)_{H,M}
\end{equation}
where $C_{H,M}$ gives the specific heat at constant field and magnetization.

The above presented thermodynamical parameters give an image of the system on a macroscopic scale. However to understand better the behavior of phase transitions it is crucial to follow the evolution on a microscopic level. The connected correlation function gives a chance for that, since it measures how local fluctuations affect different parts of the system. It is introduced as a measure of the spin-spin correlation between two spins $\sigma_i$ and $\sigma_j$ at the position $\vec{r}_i$ and $\vec{r}_j$ separated by distance $r=|\vec{r}_i-\vec{r}_j|$. 
\begin{equation}
	\Gamma(\vec{r}_i,\vec{r}_j) = \Gamma_{i,j} =\langle (\sigma_i-\langle \sigma_i\rangle)(\sigma_j-\langle \sigma_j\rangle)\rangle
\end{equation}
where $\langle ... \rangle$ denotes the thermal average. If the system is in the short-range order the correlation function decays rapidly and spin fluctuations at large distance do not show any correlation. However if the decay of the correlation function is slow, fluctuations are correlated through a long distance and the system is in the long-range order. If the system is symmetric under spatial translations, then $\langle \sigma_i\rangle=\langle \sigma_j\rangle$ and the correlation function depends only on the spatial distance $r$, so
\begin{equation}
	\Gamma_{i,j}=\Gamma(r)=\langle \sigma_i \sigma_j \rangle-\langle \sigma \rangle^2
\label{CFDef}
\end{equation}
Above the critical temperature $T>T_c$, spins are correlated only on short distance and the correlation decays exponentially as:
\begin{equation}
	\Gamma(r)=r^{-(d-2+\eta)}e^{-r/\xi}
	\label{CF}
\end{equation}
here $\xi$ defines the correlation length which is supposed to be rotational invariant near the critical point for large $r$ distance. In the power of $r$, $d$ denotes the spatial dimension and the exponent $\eta$ features critical singularity of the correlation function which will be treated in the following part.

All these thermodynamical parameters that we defined above are calculated by their expected value in statistical thermodynamics which for an arbitrary $Y$ are quantity defined as:
\begin{equation}
	\langle Y \rangle =\dfrac{1}{\mathcal{Z}}\sum_{\mu}Y_{\mu} e^{-\beta E_{\mu}}
\end{equation}
where $Y_{\mu}$ denotes the value of the quantity $Y$ for the spin configuration ${\mu}$ with energy $E_{\mu}$.

\subsection{Critical point exponents and Universality}
\label{Critical point exponents and Universality}

In continuous phase transitions close to the critical point, the correlation function becomes infinite and the second derivatives of the thermodynamical potential diverge \cite{Kadanoff1976}. In magnetic systems the free energy plays the potential role and the divergent parameters are the specific heat and the zero field susceptibility. This behavior strongly depends on the distance to the criticality controlled by some scaling fields like the reduced temperature $t$ and the reduced magnetic field $h$ defined as:
\begin{equation}
 t=\dfrac{T-T_c}{T_c} \qquad \mbox{and} \qquad h=\dfrac{H}{k_BT_c}
\label{ReducedT}
\end{equation}
Then the critical behavior of an arbitrary continuous and positive $Y(t)$ thermodynamical function can be characterized with its critical exponent
\begin{equation}
 \epsilon=\lim_{t\to 0}\dfrac{\mbox{ln}(Y(t))}{\mbox{ln}(t)} \qquad \mbox{where} \qquad Y(t)\sim |t|^{\epsilon}
\end{equation}
This pure power-law behavior however does not appear in this clear form, but more generally
\begin{equation}
  Y(t)=A|t|^{\epsilon}(1+Bt^{\epsilon_1}+\cdots) \qquad \mbox{,} \qquad \epsilon_1>0
\end{equation}
where, close to the critical point, only the leading term dominates. This exponent is measurable and suitable to describe the behavior of the thermodynamical parameters at the vicinity of the phase transition point, while the whole function might not be. Since in ferromagnets the system passes through a continuous phase transition if $h=0$ and $t=0$, the singular behavior can be recognized as keeping one of the scaling field at zero and studying the thermodynamical behavior varying the other field around the critical point. If $h=0$ and the temperature approach $T_c$, the zero-field magnetization, the isothermal susceptibility, the specific heat and the zero-field correlation length diverge as:
\begin{equation}
 M(t)\sim (-t)^{\beta} \qquad \mbox{,} \qquad \chi_H(t) \sim |t|^{-\gamma} \qquad \mbox{,} \qquad C_H(t) \sim |t|^{-\alpha} \qquad \mbox{,} \qquad \xi \sim |t|^{-\nu}
\label{critExps}
\end{equation}
However following the $t=0$ critical isothermal curve on the $H-M$ plane (Fig.\ref{ferromagnet}), the magnetization and the correlation length decays as a power law around $h\sim 0$ introducing critical exponents:
\begin{equation}
M(h) \sim |h|^{1/\delta} \qquad \mbox{,} \qquad \xi(h)\sim |h|^{-\nu_h}
\end{equation}
Finally the above described pair correlation function (Eq.\ref{CF}) can be characterized with the exponent $\eta$ since at $T_c$ the exponential part vanishes so its asymptotic behavior is:
\begin{equation}
 \Gamma(r)\sim r^{-(d-2+\eta)}
\end{equation}

Following from experiments the short range interacting systems, which have the same dimensions and order parameter symmetries, show the same critical exponents. This universal behavior enables us to compare drastically different and complex physical systems and define simple models with the same fundamental properties to analyze their behavior in a simpler way. These systems with similar critical exponents are conjugated to universality classes, which are labeled by the simplest theoretical model belonging to the same universality class. 

\subsection{Scaling Hypothesis}
\label{ScalingHyp}

At the critical point the correlation length is infinite and the system is self-similar on every length scale. This scale invariant character and the homogeneous behavior of the divergent thermodynamical functions establishes the scaling hypothesis.

Changing the length scale of a lattice system and increasing the lattice spacing $\mathbf{r}$ with a factor $b>1$, the number $N$ of spins is decreasing to $N'$, the degree of freedom is reducing and all the distance in the lattice are rescaled as $\mathbf{r}'=\mathbf{r}/b$ where $b^d=\frac{N}{N'}$. The spins which gather at the same block in the scaled system are replaced by their average value and the free energy density which was distributed on N sites initially, now belongs to $N'$ sites and scales as $\widetilde{f}(\mathbf{r}/b)=b^d f(\mathbf{r})$.

During this rescaling process the two scaling field $t$ and $h$ change by the factor $b^{d-x}$ where $x$ is defined as the scaling dimension \cite{Igloi1993,Juhasz2002}. This scaling dimension governs the displacement of the system in the parameter space. If $d-x<0$, then the corresponding field is increased under rescaling and moves away the system from a fixed point which determines the universal behaviour of the system in the parameter space. This type of fields are called to be relevant since these are capable to change the critical behaviour of the system. However if $d-x>0$ the system approaches the fixed point, the critical exponents remain unchanged and the field is irrelevant. If $d=x_i$, called marginal field, then only higher order fields are important. The system becomes invariant under rescaling only when the relevant fields vanish which corresponds to a certain set of critical properties. Since irrelevant variables finally vanish under rescaling, only the relevant and marginal parameters influence the critical behavior and exponents.

In second order phase transitions near to the fixed point the free energy density can be decomposed to a regular and a singular part as:
\begin{equation}
f(t,h)=f_{reg}(t,h)+f_{sing}(t,h)
\end{equation}
The regular term is analytical and can be neglected, however according to the scaling hypothesis \cite{Widom1965,Domb1965,PhysRev.158.176,Kadanoff1966} the singular part of the free energy density can be written as a homogeneous function of the linear relevant scaling fields. For magnetic systems with short range interactions the two relevant scaling fields are the reduced temperature $t$ and reduced external field $h$ so the singular part of the free energy density inside a bulk behaves as:
\begin{equation}
 f_b(t,h,1/L) \sim b^{-d}\widetilde{f}(b^{1/\nu}t,b^{d-x_b}h,b/L)
\label{scaling}
\end{equation}
where $\nu$ and $x_b$ define the scaling dimensions of the scaling fields of reduced temperature $t$ and bulk magnetization field $h$ respectively. In Eq.\ref{scaling} we included the inverse of the linear size of the system as a new relevant scaling field, since at its bulk critical point a system is truly critical only in the thermodynamical limit, i.e. when $1/L = 0$.

Differentiating Eq.\ref{scaling} in both sides with respect to the field, following the definition Eq.\ref{M} and by setting $h=b/L=0$ and $b=t^{-\nu}$, we get the scaling form for the magnetization:
\begin{equation}
 m(t,h,1/L) \sim t^{\nu x_b}\widetilde{m}(b^{1/\nu}t,b^{d-x_b}h,b/L)
\end{equation}
However if we set $t=b/L=0$ and $b=h^{-\frac{1}{d-x_b}}$ then the scaling of the magnetization with field is:
\begin{equation}
 m(t,h,1/L) \sim h^{\frac{x_b}{d-x_b}}\widetilde{m}(b^{1/\nu}t,b^{d-x_b}h,b/L)
\end{equation}
Following the same method, it is possible to derive the scaling forms of thermodynamical functions and write all critical exponents in terms of the scaling dimensions $x_b$ and $\nu$ as:
\begin{equation}
\beta=\nu x_b \quad \mbox{,}\quad  \delta=\dfrac{d}{x_b}-1 \quad \mbox{,}\quad  \alpha=2-d\nu \quad  \mbox{,}\quad  \gamma=\nu(d-2x_b) \quad \mbox{,}\quad  \eta=2x_b
\label{scaleexps}
\end{equation}

When the system is limited in space through a surface, besides the bulk term $f_b$ in Eq.\ref{scaling}, new local contributions (surface, corner,...) appear to the free energy density, with singular parts behaving as above for the bulk. For example, the contribution of a surface with dimension $d-1$ transforms as \cite{Binder1984}:
\begin{equation}
 f_s(t,h_s,1/L) \sim b^{-(d-1)}\widetilde{f}_s(b^{1/\nu}t,b^{d-1-x_s}h_s,b/L)
\end{equation}
where a surface magnetic field $h_s$ has been included and $x_s$ denotes the scaling dimension of the surface magnetization. Following the previous method it is possible to recognize relations between surface critical exponents also.

The consequence of homogeneous behavior of critical quantities is that they have the same exponents below and above the transition point. Many exact equalities can be identified between these quantities, since we have relations (Eq.\ref{scaleexps}) between them which can be all defined by the same two independent parameter. Some of these, the so-called scaling laws, are collected in Table \ref{table1}.

\begin{table}[h]
\centering
\begin{tabular}{ l l }
\hline
\\[-.09in]
\vspace{.05in}		
  $\alpha + 2\beta + \gamma = 2$ & Rushbooke's identity \\
\vspace{.05in}
  $\gamma = \beta(\delta-1)$ & Widom's identity \\
\vspace{.05in}
  $\gamma = \nu(2-\eta)$ & Fisher's identity \\
\vspace{.05in}
  $\alpha = 2-\nu d$ & Josephson's identity \\
\hline  
\end{tabular}
\caption{Scaling laws}
\label{table1}
\end{table}

These equalities can be examined with three independent exponents and they are widely supported by experimental works. Here the Josephson's identity used to call the hyperscaling relation since it includes the space dimension, but it is relevant only below an upper critical dimension $d_u$, defined in Mean Field Theory. Upper that marginal dimension $d_u$, mean field theories provide exact values of the critical exponents and determine the scaling functions. It has been shown that the failure of usual scaling functions within the mean field regime is related to variables (scaling fields) which become dangerously irrelevant for $d>d_u$. These variables affect the scaling behaviour and breakdown that hyperscaling relations which connect critical exponents to the spatial dimension $d$.

\subsection{Finite size scaling}
\label{FiniteSC}
Theoretical considerations are true if we always assume that the system size is infinite. However in real experiments and Monte Carlo simulations the size of the examined system is necessary finite which modify the critical quantities of the system. To extract values of critical exponents and critical temperature by observing the variation of measured quantities with the system size $L$ is the subject of \textit{finite-size scaling} methods.

If we measure on a finite system at the vicinity of the critical point of a continuous phase transition, the correlation length $\xi$ cuts off as it approaches the system size $\xi>L$. As long as $\xi\ll L$ the system behaves the same as for the infinite size limit and the thermodynamical functions have homogeneous form as we have seen in Sec.\ref{ScalingHyp}. Then the free energy can be written at the form of Eq.\ref{scaling}:
\begin{equation}
f(t,h,L)\sim L^{-(2-\alpha)/\nu}\widetilde{f}(L^{1/\nu}t,L^{d-x_b}h,1)
\end{equation}
where we used the substitution $b=L$ using the same fashion like in scaling hypothesis. Note that here $\nu$ and $\alpha$ assume their infinite lattice values. The choice of the scaling variable $L^{1/\nu}t$ was motivated by observation that the correlation length diverges as $\xi\sim t^{-\nu}$ at the critical point but limited by the system size $L$. So $L/\xi$ scales as $L/\xi\sim Lt^{\nu}\sim L^{1/\nu}t$ which is true in the asymptotic regime $L/\xi \gg 1$ \cite{Landau2000,Binder1986}. Now if we differentiate it with respect to the scaling fields $h$ or $t$, we can get the finite-size scaling form of the thermodynamical functions. For example after differentiate it with respect to the magnetic field the magnetization:
\begin{equation}
m(t,h,L)\sim L^{-\beta/\nu}\widetilde{m}(L^{1/\nu}t,L^{d-x_b}h,1)
\label{OrdPScale}
\end{equation}

Since this scaling form is supposed to be the same for all system sizes, they should fall together if we plot them on the same graph \cite{Newman2001d}. This collapse is expected only with the true (infinite size limit) values of critical exponents and $T_c$. This leads us to the crucial idea of finite-size scaling, since if we do not know these exact values but we measured the thermodynamical functions for different sizes, then vary the exponents and $T_c$, the best collapse of the function curves will determine the critical temperature and exponents.

\vspace{.2in}

At the critical point of a second order phase transition an infinite correlated cluster takes place in the system which indicates a fractal like structure. The mass of this percolating cluster scales with the linear system size as:
\begin{equation}
M(L)\sim L^{d_f}
\end{equation}
where $d_f$ is defined as the fractal dimension of the system, and it is allowed to be a non-integer number. The magnetization density for finite systems is $m(L)=L^{-d}M(L)$ by definition where $d$ denotes the euclidean dimension of the system. Thus following from Eq.\ref{OrdPScale} the fractal dimension is related to the scaling dimension of magnetization as
\begin{equation}
d=d_f+x_b
\label{fractdimxb}
\end{equation}

\subsection{Conformal invariance}
\label{ConfInv}

Covariance under conformal transformations is expected to hold at the critical point of systems with short range interactions, which possess translational and rotational symmetry and are invariant under uniform scaling \cite{Igloi1993}. A conformal transformation $\textit{\textbf{r}}\rightarrow\textit{\textbf{r}}'(\textit{\textbf{r}})$ can be seen as a generalization of uniform scaling, where the structure of the lattice locally preserved, but the rescaling factor $b(\textit{\textbf{r}})$ becomes a smooth function of the position. Geometrically those mappings are conformal invariant where the angle $\mbox{cos}(\theta)= \textit{\textbf{r}}\cdot\textit{\textbf{r}}'/(r^2r'^2)^{1/2}$ between $\textit{\textbf{r}}$ and $\textit{\textbf{r}}'$ holds unchanged under the transformation.

Conformal transformations compose a group. This conformal group is finite-dimensional for systems with dimension higher than two and contains rotations, uniform dilatation, translations and inversions. A special conformal transformation, which is a composition of the previous ones is:
\begin{equation}
\dfrac{\textit{\textbf{r}}'}{r'^2}=\dfrac{\textit{\textbf{r}}}{r^2}+\textit{\textbf{a}}
\label{invtrinv}
\end{equation}
contains an inversion, a translation and an inversion again. It is especially useful, since a semi-infinite plane with a flat surface which contains the origin, is invariant under Eq.\ref{invtrinv} if $\textit{\textbf{a}}$ is parallel with the surface. The covariance under such an infinitesimal transformation determines the form of critical two-point functions.

The method of conformal invariance is especially powerful in two dimensions where the conformal group, being isomorphic with the group of complex analytic functions, becomes infinite-dimensional and strongly restricts the possible values of critical exponents for a broad class of systems. Then it is natural to use complex coordinates $z=x+iy$ (and $\overline{z}=x-iy$) and use complex mapping $w(z)$ to go from one geometry to another. If some critical correlation is determined in the first geometry it can be transformed into the second geometry. It is known \cite{Cardy1999a,Henkel1999} that under conformal mapping $z\rightarrow w(z)$ in two dimension, the correlation functions of given operators $\phi_i(z_i,\overline{z}_i)$ transforms as:
\begin{multline}
\langle \phi_1(z_1,\overline{z}_1)\phi_2(z_2,\overline{z}_2)... \rangle=\Big[ \prod_i |w'(z_i)|^{\Delta_i}|w'(\overline{z}_i)|^{\overline{\Delta}_i}\Big]\\
\times \langle \phi_1(w(z_1),\overline{w}(\overline{z}_1))\phi_2(w(z_2),\overline{w}(\overline{z}_2)) ... \rangle
\label{CICorrF}
\end{multline}
where the local dilatation factor is then $b(z_i)=|\mbox{d}w(z_i)/\mbox{d}z_i|^{-1}$ and $x_i=\Delta_i+\overline{\Delta}_i$ is the scaling dimension of the operator $\phi_i(z_i,\overline{z}_i)$.

A generally used conformal mapping in two dimensions is a logarithmic transformation
\begin{equation}
w(z) = u+iv = \dfrac{L}{2\pi}\mbox{ln}z
\end{equation}
which maps an infinite $z$ plane onto a periodic strip of $(u,v)\in (-\infty,\infty)\times \left[ 0;L \right]$ with $L$ width and infinite length. The two-point correlation function in the new cylindrical system can be deduced using the transformation in Eq.\ref{CICorrF}:
\begin{equation}
\langle \phi(u_1,v_1)\phi(u_2,v_2) \rangle=\dfrac{(2\pi/L)^{2x_{\phi}}}{\left[ 2\mbox{cosh}\big(\frac{2\pi}{L}(u_1-u_2)\big)-2\mbox{cos}\big(\frac{2\pi}{L}(v_1-v_2)\big)\right]^{x_{\phi}} }
\end{equation}
which for large distance $|u_1-u_2|\gg L$ in longitudinal directions is reduced to a decreasing exponential form:
\begin{equation}
\langle \phi(u_1,v_1)\phi(u_2,v_2=v_1) \rangle \underset{|u_1-u_2|\gg L}{\sim} \bigg( \dfrac{2\pi}{L}\bigg)^{2x_{\phi}} e^{-\frac{2\pi x_{\phi}}{L}(u_1-u_2)}
\end{equation}
Consequently the correlation length $\xi$ is related to the $L$ width of the strip as \cite{Cardy1984}:
\begin{equation}
\xi=\dfrac{L}{2\pi x_{\phi}}
\end{equation}

Another logarithmic transformation is used to map a half-plane ($v>0$) onto a strip with free boundaries. This conformal transformation is:
\begin{equation}
w=\dfrac{L}{\pi}\mbox{ln}z
\label{CISemiInf}
\end{equation}
which is capable to study surface correlations.

\subsection{Critical dynamics}
\label{CritDyn}

As far as here we discussed only the static aspects of critical phenomenas, where we left out of consideration the time evolution of the system. However the dynamical behaviour of a ferromagnetic system becomes critical also at the vicinity of a continuous phase transition point where the $\xi_{\bot}$ spatial and $\xi_{\parallel}$ time correlation lengths become infinite. The characteristic of such phenomena depends on the initial state whereby the system was started \cite{Li1994,PhysRevE.53.2940,Hohenberg1977,Kornyei2008}. If we initiate our infinite system at a temperature $T<T_c$ and heat it up to $T=T_c$ then the system pass through a \textit{equilibrium relaxation}. The magnetization which is not zero initially behaves under a scaling transformation $\mathbf{r}'=\mathbf{r}/b$ (see in Section \ref{ScalingHyp}) as:
\begin{equation}
m(t)\sim b^{-x}\widetilde{m}(t/b^z) \qquad \mbox{which goes as} \qquad m(t)\sim t^{-\beta/\nu z}
\label{eqDyn}
\end{equation}
if we set the time $t=b^z$. Here $\beta$ and $\nu$ are the static critical exponents defined in \ref{Critical point exponents and Universality} and $z$ is a new dynamical exponent.

Another rules control the system during \textit{non-equilibrium relaxations}. Here we start the system from a random initial state ($T=\infty$ limit) where the correlation length is $\xi_{\bot}\ll L$, and quench it to $T=T_c$. Then the magnetization grows as:
\begin{equation}
m(t)\sim t^{\theta}
\end{equation}
for a macroscopic time before the actually expected decay towards equilibrium takes over \cite{Li1994}. The $\theta$ exponent cannot be expressed in terms of scaling relations of static exponents. It is due to the behaviour of the magnetization in this short-time limit which is controlled by a different scaling dimension $x_0$ than in equilibrium where the scaling dimension is $x_m=\beta/\nu$ (see Eq.\ref{scaleexps}). This new scaling dimension can be expressed of the form $x_0=\theta z+\beta/\nu$. After the magnetization is reached its maximum, then it starts to decrease toward its equilibrium value $m(t\rightarrow \infty)=0$ and following the same behaviour as it was defined in Eq.\ref{eqDyn}.

In non-equilibrium phase transitions the system is not invariant under time translation and the time correlation lengths behaves as:
\begin{equation}
\xi_{\parallel}\sim|t|^{-\nu_{\parallel}}
\label{relaxTdef}
\end{equation}
where $\nu_{\parallel}(=\tau)$ is the time correlation exponent also called relaxation time. In the scaling regime the relation $\xi_{\parallel}\sim \xi^z$ obtains between the spatial (for definition see Eq.\ref{critExps}) and the time correlation lengths which define the critical dynamical exponent as $z=\nu_{\parallel}/\nu$, where $\nu$ is the spatial critical exponent defined in Eq.\ref{critExps}.

Eventually, for a finite system in the long-time regime the decay of the magnetization becomes exponential
\begin{equation}
m(t)\sim e^{-t/\tau'}
\label{mDynScExp}
\end{equation}
where $\tau'$ is the finite size relaxation time. This is scaling with the system size as $\tau'\sim L^z$ and cause a \textit{critical slowdown} for larger sizes. However if we use sufficiently large lattice and short time a good estimate of $z$ can be determined from Eq.\ref{eqDyn} \cite{Landau2000}.

Another way to characterize dynamical features of a system is to measure the two-times autocorrelation function which is defined as:
\begin{equation}
A(s,t)=\left\langle \sigma_i(s)\sigma_i(t) \right\rangle-\left\langle \sigma(t) \right\rangle^2 
\label{autocorrDef}
\end{equation}
where the second term on the right hand side is the square magnetization which is zero at the critical point. Here $s$ and $t$ denotes two time moments and $\sigma_i(t)$ is the value of the i\textit{th} spin at time $t$. From the two moments the smaller $s$ is called waiting time, used to be constant during the measure while $t$ is the measuring time which evaluate the system. 
To measure the autocorrelation function we fix the waiting time $s$ and measure the time-correlation as a function of $t$ when $t\geq s$. At the critical point the autocorrelation is a homogeneous function and scales as \cite{PhysRevB.71.094424}:
\begin{equation}
A(s,t)\sim b^{-2\beta/\nu}\widetilde{A}(s/b^z,t/b^z)
\label{AutCorrSF}
\end{equation}
In \textit{equilibrium} relaxations by definition this function depends only on $t-s$. Thus if $s\rightarrow 0$ and $t\rightarrow \infty$, and we use the substitution $t=b^z$, this scaling function can be rewritten as:
\begin{equation}
A_{eq}(s,t)\sim (t-s)^{-2\beta/\nu z}
\label{AutCorrEq}
\end{equation}

In \textit{non-equilibrium} case when the system quenched below its critical temperature, the characteristic domain size grows as
\begin{equation}
\xi(t)\sim t^{1/z}
\label{dynCorrL}
\end{equation}
and the autocorrelation function can be decomposed into two parts \cite{Hohenberg1977,Walter2008,Calabrese2005}:
\begin{equation}
A(s,t)=A_{st}(t-s)+A_{ag}(t/s)
\label{AutCorrDecomp}
\end{equation}
where the first short-time component $A_{st}(t-s)$ is induced by reversible processes inside domains and vanishes out rapidly. Then the second aging term $A_{ag}(t/s)$ becomes relevant which is driven by the irreversible motions and annihilation of domain walls. At the critical point we neglect the first term so the autocorrelation function becomes a homogeneous function of $t/s$. Then it can be written in a scaling form following from Eq.\ref{AutCorrSF} if we set $s=b^z$ as:
\begin{equation}
A_{ag}(s,t)\sim s^{-2\beta/\nu z}\widetilde{A}(t/s)
\label{AutCorrAgscaling}
\end{equation}
If the system size is finite, after a long time process the autocorrelation function decays as an exponential function
\begin{equation}
A(s,t)\sim e^{-(t-s)/\tau}
\end{equation}
presents the same fashion of finite size effect as we saw for the magnetization above in Eq.\ref{mDynScExp}.

\section{The Potts model}
\subsection{The general Potts model}
\label{GeneralPotts}
The Potts model is a generalization of the Ising model with arbitrary number of spin state $q$. It was introduced by Renfrey B. Potts in his PhD thesis \cite{Potts1951} suggested by Cyril Domb and defined in general as follows. Take a graph $G=(V,E)$ where at each vertex a spin variable $\sigma_i$ is located, taking one state from $\sigma_i=0...q-1$ with equal probability. Consider nearest neighbour interactions only and include external field in the system, the Hamiltonian of the Potts model forms into:
\begin{equation}
 \mathcal{H}=-\sum_{\langle i,j\rangle}J_{ij}\delta(\sigma_i,\sigma_j)-\sum_i h_i\sigma_i
\label{PottsH}
\end{equation}
where the first sum runs over all nearest neighbours $\langle i,j\rangle$. $J_{ij}$ denotes the strength of exchange interaction between neighbouring spins $\sigma_i$ and $\sigma_j$ and $h_i$ is the external field influence on the spin located at the $i$th vertex. If $J_{ij}>0$ the system is ferromagnetic however if $J_{ij}<0$ then antiferromagnetic rules govern the interactions. If $J=0$ then the system behaves as a paramagnet. The Kronecker delta functions is $\delta(\sigma_i,\sigma_j)=1$ if $\sigma_i=\sigma_j$ otherwise it is $\delta(\sigma_i,\sigma_j)=0$. The partition function of the Potts model can be written as
\begin{equation}
	\mathcal{Z}(q,J,h,\beta)=\sum_{\{ \sigma\}} e^{-\beta \mathcal{H}(\sigma_i)}
\label{PottsZ}
\end{equation}
where the sum runs over all possible $q^{|V|}$ spin configuration and $|V|$ denotes the number of vertices in $V$. From this partition function $\mathcal{Z}$ one can calculate the different thermodynamical properties and critical exponents of the system (see Section \ref{Thermodynamical quantities} and Section \ref{Critical point exponents and Universality}). 

To present the high temperature expansion of the partition function at zero external field $h_i=0$, we may use the linearization $e^{\alpha \delta}=1+(e^{\alpha}-1)\delta$ which is valid only if $\delta \in \left\lbrace 0,1\right\rbrace $. Now introducing $v_{ij}=e^{\beta J_{ij}}-1$ then the partition function can be evaluated as
\begin{equation}
\mathcal{Z}(q,J,\beta)=\sum_{\{ \sigma\}} e^{-\beta\sum_{\langle i,j \rangle} J_{ij}\delta(\sigma_i,\sigma_j)}=\sum_{\{ \sigma\}} \prod_{\langle i,j \rangle} e^{ -\beta J_{ij}\delta(\sigma_i,\sigma_j)}=\sum_{\{ \sigma\}} \prod_{\langle i,j \rangle}\Big( 1+v_{ij}\delta(\sigma_i,\sigma_j) \Big)
\label{dedHTZPotts}
\end{equation}
Then we rearrange the sum and give it in terms of any (possibly not connected) subgraph. The number of possible subgraphs is $2^{|E|}$, where $|E|$ is the number of edge in $G$. The partition function can be then rewritten as
\begin{equation}
 \mathcal{Z}(q,J,\beta)=\sum_{G'\subseteq G} q^{c(G')} \prod_{e \in G'} v_e
\label{HTZPotts}
\end{equation}
where $c(G')$ denotes the number of connected components of $G'$. Within a connected component all spins possess the same state, distributed independently from the possible $q$ state. The product runs over all edges included in the subgraph $G'$ with the convention that if $G'$ is completely isolated i.e. there is no edge in the subgraph, the sum is equal to one. In the high temperature limit ($\beta\ll 1$), only the subgraphs with few edges contribute to the sum, so a possible approximation is to take only these $G'$ subgraphs with small number of edges. This approximation is a high temperature expansion since it is getting better as the temperature is increasing. Note that Eq.\ref{HTZPotts} gives an extension for Potts model, when the previous defined $\mathcal{Z}$ (in Eq.\ref{PottsZ}) is not valid, since we can introduce it with non-integer values of $q$ relating this model to other problems.

For ferromagnetic $q=2$ Potts model on two dimensional square lattice in absence of external field, the critical temperature is known from the exact solution of Onsager \cite{PhysRev.65.117}, which gives the partition function for arbitrary $T$ as well. If $q>3$ the critical temperature is exactly determined, however the partition function is given only at $T_c$. When $q=3$ the critical point is only known exactly in isotropic case. In anisotropic case $T_c$ is predicted by numerical calculations only, however it is strongly believed to be true \cite{Wu1982}. The phase transition of the Potts model in two dimensions depends on the value of $q$. An exact solution exists to establish the phase transition first order if $q>4$ and higher order if $q\leq 4$ \cite{Baxter1973}. When $q=2$ the Mean Field solution of the model is valid only above the upper critical dimension $d_c=4$. In the presence of external field or in higher dimensions, the critical behaviour of the Potts model is known by numerical predictions and series expansion calculations only, and no exact result exists.

A few more specifications of the Potts model exist, transform the general model to another fundamental problems such like the six-vertex ice rule model, the Potts lattice gas \cite{Temperley1971,PhysRevB.17.3650} or the colouring problem of antiferromagnetic Potts model. Some of them will be introduced later here, however all of them demonstrate the expedience of the original model, and explain the prominent attention of scientists during the last few decades. In the following we are going to outline the Random-Cluster model which is capable to describe the partition function of the $q$-state Potts model. Some special cases of Potts model will be introduced which we will use later in the thesis.

\subsection{Random-cluster measure}
\label{RandomClust}
In the 1970s Fortuin and Kasteleyn realized, that the series-parallel laws of Kirchoff \cite{Kirchhoff1847}, which was defined in 1847 on electrical networks, were adaptable for Potts model with different values of $q$. This observation led them to define the Random Cluster model (RC model) \cite{Kasteleyn1969,Fortuin1972,Fortuin1972a,Fortuin1972b} and obtain a high temperature expansion of Potts model. This model is capable to find macroscopic properties in terms of local structures and define the partition function of the $q$-state Potts model using only combinatorial considerations. Here we are going to discuss the essence of their method.

Lets take a finite graph $G(V,E)$ where $V$ denotes the set of vertices and $E$ the set of edges of the graph. The configuration space of RC model is the set of all subset of the edge-set $E$, present like $\Omega=\left\lbrace 0,1 \right\rbrace^E$, where an edge $e\in E$ is called open if $\omega \in \Omega$ and $\omega(e)=1$ otherwise it is closed when $\omega(e)=0$. For a $\omega \in \Omega$, let $\eta(\omega)=\left\lbrace e\in E:\omega(e)=1\right\rbrace$ denote the set of open edges and $c(\omega)$ the number of connected components of the subgraph $G'=(V,\eta(\omega))$ including the separated vertices. 

The model has two parameters, the probability density of open edges $p\in \left\lbrace 0,1\right\rbrace $ and the cluster weight factor $q\in \left\lbrace 0,\infty \right\rbrace $, which gives the number of cluster types. Now a $\phi_{p,q}(\omega)$ measure on the configuration space $\Omega$ is given by: 
\begin{equation}
\phi_{p,q}(\omega)=\dfrac{1}{Z_{RC}}\left\lbrace \prod_{e\in E} p^{\omega(e)}(1-p)^{1-\omega(e)} \right\rbrace q^{c(\omega)} 
\end{equation}
where the partition function is defined as:
\begin{equation}
 \mathcal{Z}_{RC}=\sum_{\omega \in \Omega}\left\lbrace \prod_{e \in E} p^{\omega(e)}(1-p)^{1-\omega(e)} \right\rbrace q^{c(\omega)}
\label{Z_RC}
\end{equation}

When $q\in \mathbb{Z}$ is an integer, and $q \geq 2$ the previous defined partition function $\mathcal{Z}_{CL}$ corresponds to the Potts partition function defined in Eq.\ref{HTZPotts} since it can be written as:
\begin{equation}
\mathcal{Z}_{RC}=(1-p)^{|E|}\sum_{\omega \in \Omega}\left\lbrace \prod_{e \in E} \left( \dfrac{p}{1-p} \right) ^{\omega(e)} \right\rbrace q^{c(\omega)}
\end{equation}
where $|E|$ denotes the number of the edges in $G=(V,E)$. Substituting $v_e=p/(1-p)$, and choosing $p=1-e^{-\beta}$ we find that the partition function of RC model and Potts model are related:
\begin{equation}
\mathcal{Z}_{RC}=e^{-\beta |E|}\mathcal{Z}_{Potts}
\label{ZeqZ}
\end{equation}

We discuss this relation between the random cluster model and Potts model in a different way the in Appenddix \ref{RCCondMeas}. 

\subsection{$q=1$ - Percolation}
\label{percolation}
The percolation model as a stochastic mathematical theory was firstly introduced by Broadbent and Hammersley in 1957 \cite{Broadbent1957} for modeling fluid or gas flow through a porous medium. This model turned to be equivalent to the $q$-state Potts model when $q=1$ and gives one of the simplest example of phase-transition and critical phenomena.

To define the model we proceed from a graph $G=(E,V)$ and declare each edge $e \in E$ independently to be open with a probability $p$ or set it closed otherwise with probability $(1-p)$. The open edges forms connected clusters of neighbouring vertices in the graph. This is the definition of the Bond Percolation model (BP). It is equivalent with a probability measure on a configuration set $\Omega_p=\prod_{e\in E}\{0,1\}$, where each configuration represented by a subgraph $G'=(E',V)\subseteq G$ whose edge set $E'$ contains precisely the open edges. The probability $p$ plays the control parameter role in the system. For $p$ above some percolation threshold $p_c$ an infinite cluster exists besides the many finite clusters.

\begin{figure}[htb]
\begin{center}
\includegraphics*[ width=13.0cm]{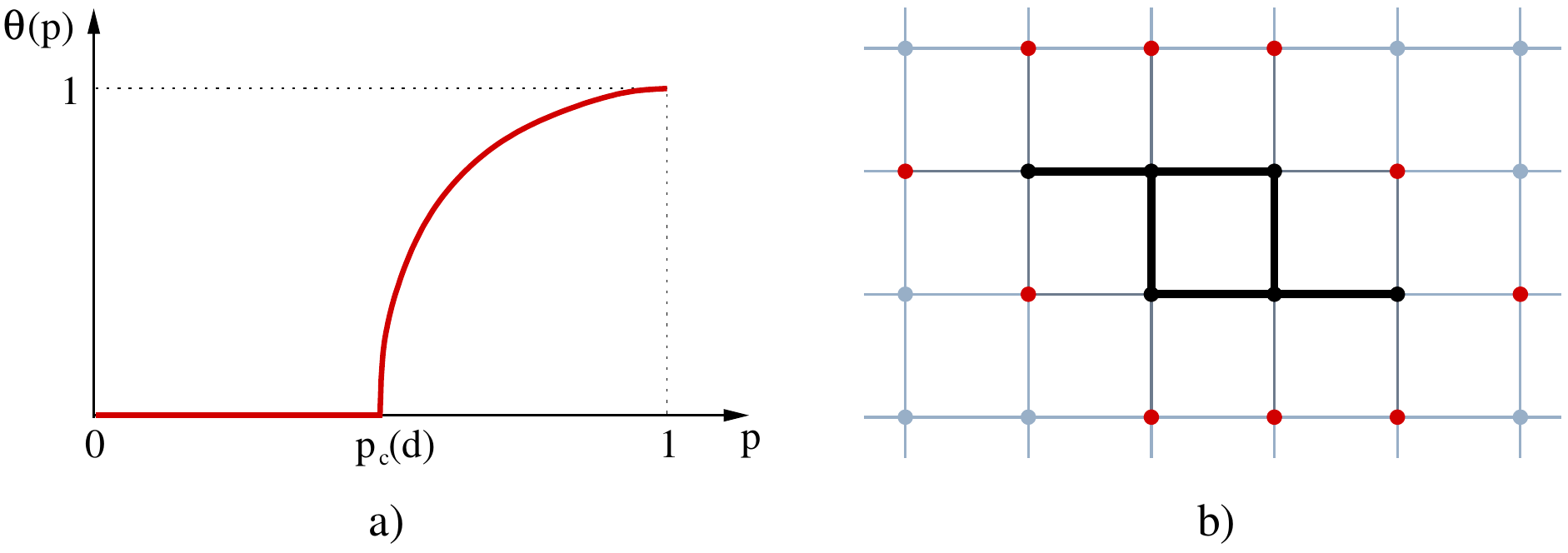}
\caption{a: Phase diagram of percolation model. Here the percolation probability plays the order parameter role and the control parameter is $p$. b: Graphical representation of an animal with size 6 and perimeter 10.}
{
\label{percoMulti}
}
\end{center}
\vspace{-10pt}
\end{figure}

The order parameter like percolation probability $\Theta(p)$ is defined as the probability that a randomly chosen site of a graph belongs to an infinite percolating cluster in the thermodynamical limit. This quantity corresponds to magnetization of spin systems and $\Theta(p)=0$ at the sub-critical phase $p<p_c$, and becomes $\Theta(p)>0$ above the critical point $p\geq p_c$ when the system is at the supercritical phase. It is straightforward that $\Theta(p)=1$ when $p=1$ and all edges are open in the graph $G$ (see Fig.\ref{percoMulti}.a). For finite lattice the percolation cluster can evolve below $p_c$, thus the system presents the exact behavior only at the thermodynamical limit when $N\rightarrow \infty$.

To define further macroscopic quantities we need to calculate $c_n$, the number of clusters of a given size $n$ per a site. It can be given by:
\begin{equation}
 c_n(p)=\sum_s g_{ns}p^{n}(1-p)^{s}
\end{equation}
where $g_{ns}$ denotes the number of finite cluster configurations (animals) with size $n$ and perimeter $s$. Here the perimeter is defined as the number of nearest neighbours of the sites of a finite cluster excluded from the cluster (see Fig.\ref{percoMulti}.b). The average of $\langle c_n(p) \rangle$ corresponds to the free energy of the system. Then the average cluster size can be written as
\begin{equation}
 S(p)=\sum_n\dfrac{c_n n^2}{\sum_n c_n n}
\end{equation} 
which becomes $S(p)=\sum_n c_n n^2$ when $p\geq p_c$, since the denominator turns to be finite. $S$ is behaving as the susceptibility of the system and divergent around the critical point.

All these quantities have critical behaviors around $p_c$, and are described by a set of critical exponents which accomplish the scaling laws. The critical point $p_c(G)$ and exponents depend on the geometry of graph $G$ and lead the systems to different universality classes. 

\begin{figure}[htb]
\begin{center}
\includegraphics*[ width=15.0cm]{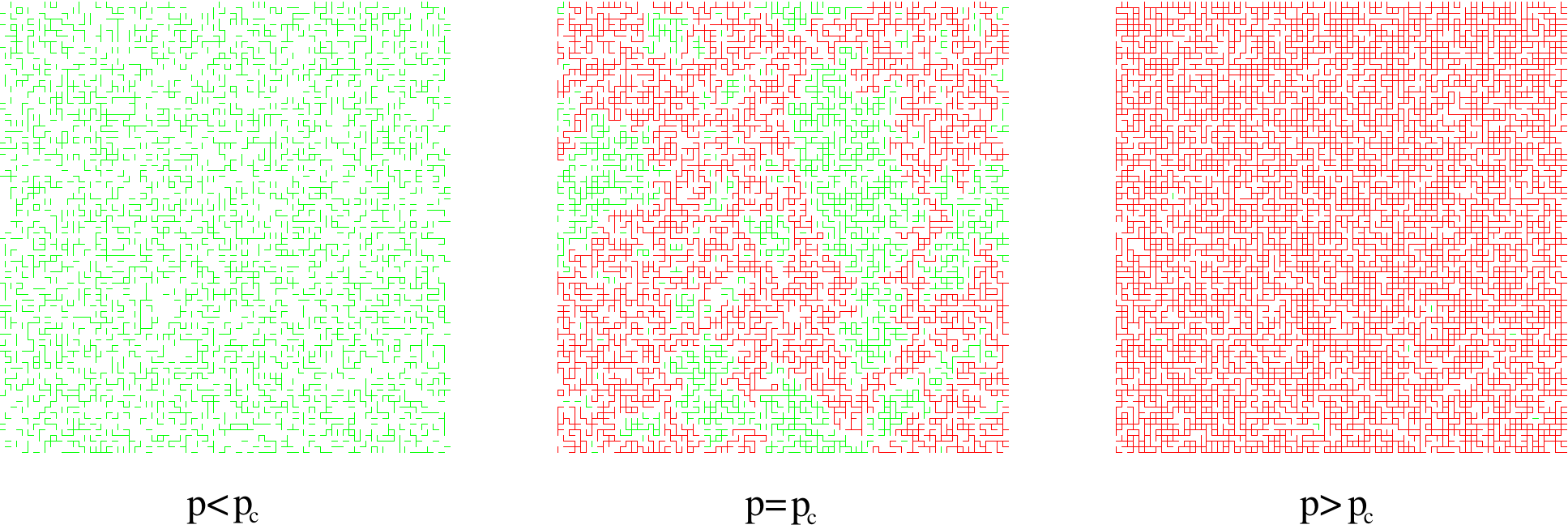}
\caption{Illustration of different phases of percolation. On the left plaquette the system is below the critical probability $p<p_c$ where only finite clusters (green) occurs. At the critical point $p=p_c$ (middle plaquette) an infinite cluster (red) takes place all over the lattice. Above the critical probability $p>p_c$ (left plaquette) the infinite cluster contains most of sites and only finite number of finite clusters are present.}
{
\label{percoCL}
}
\end{center}
\vspace{-10pt}
\end{figure}

Another analogous model exists, called the site percolation (SP), where sites are chosen to be open with probability $p$ and to be closed otherwise. Here the neighbouring open sites shape clusters and follow the same behaviour presented above for BP. The critical points of the two models are relating as $p_c^{BP}(G) \leq p_c^{SP}$ and depends on the choice of the graph. It is equal for trees, and not equal for finite dimensional lattices with $d \geq 2$. For example for two dimensional square lattice $p_c^{BP}=1/2$ since the lattice is self dual, while $p_c^{SP}\simeq 0.59746(5)$ follows from numerical results. Even if the critical points of the two percolation problem could be different, the critical behavior of the two models can be characterized with the same critical exponents.

Since the Bond Percolation is equivalent to the Random Cluster Model with $q=1$ it gives the possibility to consider bond percolation as a special case of the $q$-state Potts model with only one possible state. This enables us to study a percolation problem in the terminology of spin systems. Many other modification of the percolation problem exist, with different critical behaviors e.g. directed percolation (where the edges are directed in the graph), and was specified in many part of research giving expedient models from epidemic spreading to microelectronics.


\subsection{$q=2$ - Ising model}
\label{IsingModel}

As we have seen before, the Ising model was the first capable model to describe analytically a phase transition. Since the Potts model is a generalization of this model, we can define the Ising model as a specific Potts system with $q=2$ number of states. In general we are taking account only nearest neighbours interactions which gives the Hamiltonian:
\begin{equation}
	\mathcal{H}_{\mathit{Ising}}=-J\sum_{\langle i,j \rangle}\sigma_i\sigma_j-h\sum_{i}\sigma_i
\label{IsingHGen}
\end{equation}
where $\sigma_i=\pm 1$ and the first summation runs over all first neighbouring spins $\langle i,j \rangle$ of the system. The critical behaviour of the Ising model in absence of external field have been shortly discussed in Section.\ref{IsingSimple}.

Beyond the Potts model, a few other generalizations of the Ising model do exist. Since in the Ising model the spins are restricted to align parallel to the magnetic field, this model is capable to describe strongly anisotropic systems only. However if the spins $\sigma_i$ are $n$ dimensional normal vectors, then if $n=1$ we get back the original Ising problem, but when $n>1$ we get more realistic models with $O(n)$ continuous rotational symmetries. If $n=2$ the spins can align to any direction continuously around a unit circle and the system energy can be written like:
\begin{equation}
	\mathcal{H}_{XY}=-J\sum_{\langle i,j \rangle}(\sigma_i^x\sigma_j^x+\sigma_i^y\sigma_j^y)-h\sum_{i}\sigma_i^x
\end{equation}
This is the classical XY model. Then if we choose $n=3$ we arrive to the classical Heisenberg model defined \cite{Heisenberg1928,Bloch1930,Bloch1932} with the Hamiltonian:
\begin{equation}
	\mathcal{H}_{H}=-J\sum_{\langle i,j \rangle}(\sigma_i^x\sigma_j^x+\sigma_i^y\sigma_j^y+\sigma_i^z\sigma_j^z)-h\sum_{i}\sigma_i^z
\end{equation}
Finally if $n\rightarrow \infty$ the system is transformed to the so-called Spherical model \cite{PhysRev.176.718}. The classical Heisenberg and XY model are more capable to describe the behaviour of ferromagnets, however it is possible to define the related quantum models where the spin operators are determined by Pauli matrices. Even quantum models have greater difficulty in analytic and numeric treatments \cite{Yeomans1994} in many cases they give a better approximation then the classical description. 

Following from the Mermin-Wagner-Hohenberg theorem \cite{PhysRevLett.17.1133,PhysRev.158.383} in two dimensional systems if $n\geq 2$ the continuous symmetries cannot be broken spontaneously at any finite temperatures $T\neq 0$. There is no magnetic order at $T>0$ since the thermal fluctuations destroy any long range order even at low temperatures. It is not true for the Ising model which has discrete symmetry, and indicate spontaneous symmetry breaking below the critical point $T_c>0$. 

For the XY model another type of phase transition occurs which does not affect any spontaneously broken symmetry. At high temperature the system is disordered and the correlation length is falling exponentially, however another phase exists at low temperature when the correlation length is following a power law and a quasi-long range order appears in the system. Between the two phases at a critical temperature, a \textit{Kosterlitz-Thouless phase transition} of infinite order occurs.

Other possible generalization of the Ising model if we are not taking into account only the first-neighbour couplings during the calculation of the energy, but define interactions between second or further neighbours as well. In the limit when every spin is interacting with all spins of the system we arrive to the mean field presentation of the model. This can be interpreted in different ways, like studying the Ising model on a complete graph (where every site connected to every site) or defining an average molecular field which includes the influence of all spins but neglect any fluctuation. This two definition is proved to be equivalent and gives the Hamiltonian:
\begin{equation}
	\mathcal{H}_{MF}=-\dfrac{zJ}{N-1}\sum_i\sum_{i \neq j}\sigma_i \sigma_j - h \sum_i \sigma_i
\end{equation}
where $z$ denotes the coordination number of the graph \cite{Baxter1982}. Following this Hamiltonian the magnetization per site can be deduced analytically in a self-consistent form \cite{Bragg1934}:
\begin{equation}
	m=\mbox{tanh}\left[\beta (z J m+h)\right] 
\end{equation}
which is capable to calculate exactly the free energy and the critical properties of the system. The mean field problem can be solved analytically on Bethe lattice and it gives the same critical exponents as the former solution. The mean field approximation is relevant only above the upper critical dimension $d_c=4$ and far from the critical point since at the criticality the fluctuations become important, so the calculated critical exponents usually deviate from the measured ones.

\subsection{$q=\infty$ - Potts model with infinite number of states}
\label{Potssqinf}

In two dimensions for the pure Potts model (when $J_{ij}=J$) an exact result of Baxter \cite{Baxter1973} ensures a first order phase transition when $q>4$. However in the large $q$-limit the pure Potts model is solvable in any dimension since thermal fluctuations are reduced. As we have seen before, in absence of an external field in general the Hamiltonian of the Potts model can be defined as:
\begin{equation}
	\mathcal{H}=-\sum_{\langle i,j\rangle} J_{i,j} \delta(\sigma_i \sigma_j)
\label{PottsHinfq}
\end{equation}
where $\sigma_i=0...q-1$ and the sum runs over all edges. Then proceeding from the definition in Eq.\ref{PottsZ} the partition function evolves as:
\begin{equation}
\mathcal{Z}=\sum_{\{ \sigma\}} e^{-\beta \mathcal{H}(\sigma)}=\sum_{G'\subseteq G}q^{c(G')}\prod_{e\in G'}(e^{\beta J_e}-1)
\end{equation}
where we applied the Fortuin-Kasteleyn high temperature development (see Section \ref{RandomClust}). Now we introduce a reduced temperature $T'=T \mathrm{ln} q$ or its inverse as $\beta '=\beta/ \mathrm{ln} q$. This reduction is needed to keep the critical temperature finite, since if $q \rightarrow \infty$ then $T_c \rightarrow 0$ however $T_c \mathrm{ln} q \rightarrow 2$ \cite{PhysRevE.64.056122,PhysRevLett.90.190601,PhysRevE.69.056112,Mercaldo2005}. If we apply now the substitution $e^{\beta J_e}=e^{\mathrm{ln}q \beta ' J_e}=q^{\beta ' J_e}$ then the partition function arises of the form:
\begin{equation}
	\mathcal{Z}=\sum_{G'\subseteq G}q^{c(G')}\prod_{e \in G'} \left(q^{\beta ' J_e}-1\right)
\label{PottsZHTL2}
\end{equation}
which is equivalent to Eq.\ref{HTZPotts} with the parameterization $v_e=q^{\beta ' J_e}-1$. Since all $J_e\geq 0$ we can rearrange the product on the right hand side and write the partition function until the second term as:
\begin{eqnarray}
	\mathcal{Z} = \sum_{G'\subseteq G}q^{c(G')}(q^{\beta '\sum_{e} J_e} & - &  q^{\beta ' \sum_{e \neq e_{min}} J_e}+...) =
\nonumber\\
=\sum_{G'\subseteq G} q^{c(G')+\beta '\sum_e J_e} & - & q^{c(G')+\beta '\sum_{e\neq e_{min}} J_e} + ...
\end{eqnarray}
Here the first term is always the largest one for a given subgraph, since it contains all the ferromagnetic edges $e\in G'$, then the second term is the second largest, because we exclude only the weakest edge $e_{min}$ from the summation, etc. Following this method we are able to write all the terms in order to choose the maximum from the set of the first terms to find the subgraph $G'$ which contributes to the dominant term of the partition function. Then the second largest term would be the maximum of the remaining first terms of $G''\neq G'$ and the second term of the previous chosen $G'$.
\begin{equation}
\mathcal{Z}=\mbox{max}_{G'\subseteq G} \hspace{.04in} q^{c(G')+\beta '\sum_{e\in G'} J_e}  + \mbox{max}_{G''\neq G'}\hspace{.04in} ( q^{c(G'')+\beta ' \sum_{e\in G''} J_e}, q^{c(G')+\beta ' \sum_{e\neq e_{min}} J_e}) +...
\label{PottsZterms}
\end{equation}
The first term in the right hand side of Eq.\ref{PottsZterms} has the largest exponent of $q$, so when $q$ goes to infinity the partition function can be rewritten as
\begin{equation}
\mathcal{Z} \simeq \sum_{G'\subseteq G} q^{\phi(G')} \qquad \mbox{where} \qquad \phi(G')=c(G')+\beta '\sum_{e \in G'}J_e
\label{qinfZ}
\end{equation}
which is dominated by the largest term $\phi^* = \mbox{max}_G \phi(G')$ only which defines an \textit{optimal set} of edges $G^*$ which degeneracy is likely to be one. Thus the study of the large $q$-state model is reduced to an optimization problems which solution depends on the temperature and the given realization. The free energy of the system, following from the definition given in Eq.\ref{FreeE}, can be calculated as
\begin{equation}
	\mathcal{Z}=e^{-\beta \mathcal{F}}=q^{-\beta ' \mathcal{F}}=q^{\phi^*}
\end{equation} 
thus the free energy per site is given by $-\beta 'f=\phi^*/N$ where $N$ denotes the number of sites of the system.

In the pure ferromagnetic Potts model when $J_{ij}=J$ the structure of the optimal set in the different thermodynamical phases are trivial which follows from the next proposition.

\begin{definition}
A type $\mathrm{I}$ graph is a graph for which exists $w_c=J\beta_c$ such that the optimal set $G^*$ with edges all having the same weight $w=J\beta$ is
\begin{center}
$G^* =
\begin{cases}
 \emptyset & \mbox{if} \quad w \leq w_c \\
E & \mbox{if} \quad w \geq w_c \\
\end{cases}$
\end{center}
\end{definition}
\begin{definition}
Let $G=(V,E)$ be a graph and $\Pi$ the set of permutations of $E$ (automorphy group) such that $\forall$ $(i,j)\in E$, $(\pi(i),\pi(j)) \in E$, where $\pi \in \Pi$. 
\end{definition}
\begin{proposition}
If for any $e\in E$ and $f\in E$ there exists $\pi\in\Pi$ such that $f=\pi(e)$, then $G$ is a type $\mathrm{I}$ graph.
\label{PropIgraph}
\end{proposition}
\begin{proof}
If an edge $e\in E$ belongs to an optimal set, then every edge belongs to an optimal set since there exists $f=\pi(e)$ for any $f,e \in E$. The union of optimal sets forms an optimal set as well. Consequently if one edge is belonging to an optimal set, all edges $e\in E$ belong to the same optimal set, so $E$ is an optimal set. On the contrary, if even one edge is not belonging to any optimal set, then either of the edges belong to any optimal set, so the optimal set is $\emptyset$ and we proved that $G$ is a type $\mathrm{I}$ graph. We remark that all the regular lattices of statistical mechanics are type $\mathrm{I}$ graph.
\end{proof}
\begin{figure}[htb]
  \begin{center}
    \begin{minipage}[b]{0.48\linewidth}
\includegraphics*[ width=7cm]{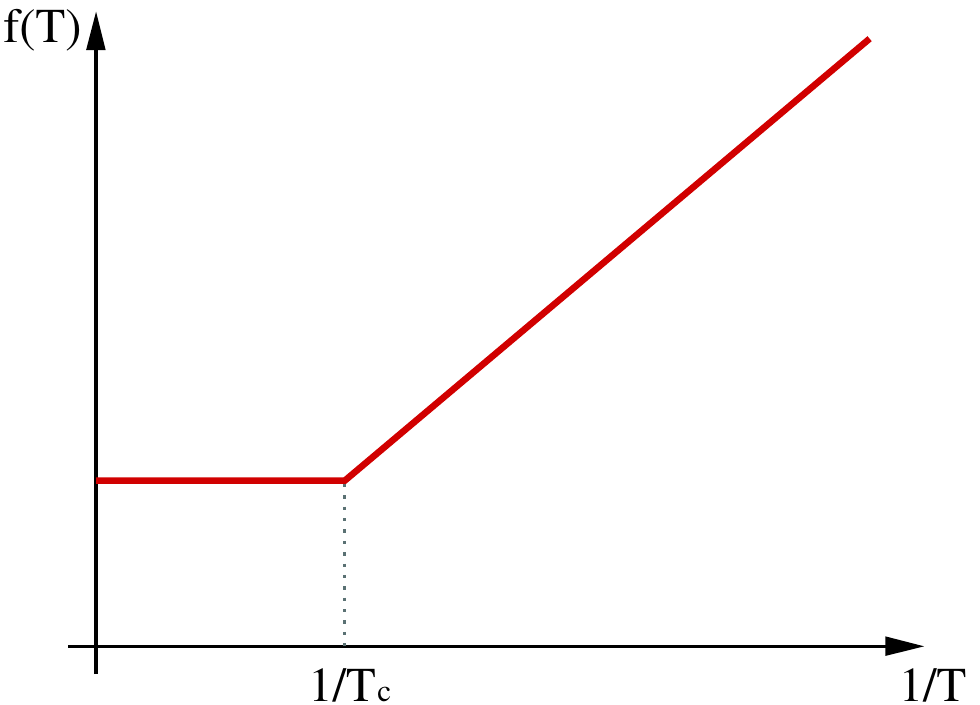}
    \end{minipage}\hspace{.2in}
    \begin{minipage}[b]{0.38\linewidth}
      \caption{Free energy of the pure $q$-state Potts model to the inverse temperature. Below $T_c$ (above $1/T_c$) the optimal set contains all the edges and $f(T)$ is linear. Above $T_c$ (below $1/T_c$) the optimal set is empty and $f(T)$ is a constant function.\qquad\qquad\qquad\qquad\qquad}
\label{PottsFE}
    \end{minipage}
  \end{center}
\end{figure}

The consequent of the proposition in Prop.\ref{PropIgraph} is that in the pure ferromagnetic case the optimal set is either fully connected or empty. A phase transition of first order takes place between the low temperature and the high temperature phase with non-analytical behaviour of the free energy at the critical temperature when the two phases coexist. This critical temperature $T_c$ can be calculated easily since below the phase transition point $T<T_c$ the optimal set contains all the edges $e\in E$ and the free energy is $-\beta ' f N=1+N\beta' Jd$, while above $T_c$ the optimal set is an empty diagram so the free energy is $-\beta ' f N=N$. Consequently at the phase coexistence point $1+N\beta ' Jd=N$ so the critical temperature is $T_c=Jd/(1-1/N)$. 

Thus all the information about the Potts model in the large $q$ limit is contained in the optimal set $G^*$, since the partition function, the free energy and the other thermal properties can be deduced from that. Moreover the magnetization and correlation function can be obtained from the geometrical structure of $G^*$ what will be discussed in details later for the disordered case when the dominant graph has a nontrivial structure.

\section{Disordered models}

Disorder in physical systems can be characterized by the deviation from the completely ordered phase \cite{Ziman1979}. For example in ferromagnets at $T=0$ and $h=0$ the system is completely ordered since the energy minima determines all the coupling of magnetic moments and the system is in its ground state. By increasing the temperature the order is weakening as certain magnetic moments deviate from their ground states and disorder arises in the system. The origin of such spontaneous disorder is twofold: it can rise from thermal fluctuations or quantum fluctuations. However it is possible to introduce non-spontaneous disorder which affects the critical behavior of the system. This additional disorder used to be introduced in two ways. If the system is conducted by random variables which do not evolve with time, the system is called to be \textit{quenched disordered}. It is opposite to \textit{annealed disorder}, where random variables are allowed to evolve. In the following section we discuss the effect of quenched randomness on phase transitions and define some related models.

\subsection{Effects of disorder}
\label{EffectsOfDisorder}
To understand the effects of disorder in statistical systems we need to use the terminology of renormalization group theory \cite{Kadanoff1966,Wilson1971,Wilson1971b,RevModPhys.45.589}. The idea of renormalization is to change the scale of the system, which decreases its degree of freedom and transform its parameter into a point, where the system becomes scale invariant as we have discussed in Section \ref{ScalingHyp}. In real space renormalization it can be described as an energy scale transformation, where the Hamiltonian is replaced by another Hamiltonian with the same structure, but with different parameters and reduced degree of freedom. This parameter change is equivalent to moving at the parameter phase during the renormalization iteration. The critical point corresponds to a fixed point at parameter phase, where the correlation length becomes infinite and the system is scale invariant.

\subsubsection{Relevance-irrelevance criterion}
\label{RelevanceIrrelevanceCriterion}
For homogeneous systems with continuous phase transition, perturbation can change the system's criticality. Using renormalization methods we can see how a suitably large disorder can move the fixed point at the parameter phase, and change the critical exponents of the pure system, transform the model into a new universality class. The stability of the fixed point against disorder were firstly studied in diluted models (see later) by Harris, but his results can be generalized for any kind of bond disorder. 

The Harris-criterion predicts the perturbation relevant if at the pure model
\begin{equation}
d\nu/2-1>0
\label{HarrisCrit}
\end{equation} 
where $d$ is the dimension of the system and $\nu$ is the correlation exponent. In other words it is valid if the specific heat exponent is positive $\alpha>0$ (see Eq.\ref{critExps}). In that case the system moves towards the parameter phase into a new fixed point with new exponents, different from the original pure system. When  $\alpha<0$ the disorder is irrelevant and the system keeps the original pure exponents during the renormalization. At the marginal case $\alpha=0$ further considerations are needed. However the above defined criterion is valid only at weak disorder, close to the pure system fixed point, since sufficiently strong disorder can transform the system into a new fixed point even though it was resistance against weak randomness.

Another consideration exists by Imri and Ma \cite{PhysRevLett.35.1399} about the effect of random field on systems with continuous phase transitions. To present their idea we proceed from the Ising model at low temperature and suppose that there are two domains of spins in the system with linear size $L$ which are orientating to the opposite direction. Now the energy loss by the surface between the two domains is $E_{surf}\sim 2JL^{d-1}$. However, applying random field, the system earns $E_{rf} \sim hL^{d/2}$ energy by the spin fluctuation inside the domains. Now if we compare the two energies:
\begin{equation}
	\dfrac{E_{surf}}{E_{rf}} \sim \dfrac{J}{h}L^{d/2-1}
\end{equation}
This means that if $L \rightarrow \infty$ then the ferromagnetic phase disappears at any finite temperature if the dimension is smaller than two.

When the pure system has first order phase transition, the effect of disorder is more complicated. We do not have any kind of criterion of relevance or perturbation expansion to describe the behavior around the discontinuous transition point. However it is known rigorously \cite{PhysRevLett.62.2503} that in two dimensions a first order phase transition softens into a continuous one if any amount of disorder is applied. Such behaviour is true in three dimensions as well, but there exists a crossover region where for weak disorder the transition keeps first order and presents continuous behaviour only for sufficiently strong disorder. 

If the system stays inhomogeneous on macroscopic and microscopic scale during the coarse-graining renormalization iteration and relative magnitude of inhomogeneities (i.e. ratios of parameters in the Hamiltonian) do not remain finite in the long-wavelength limit then the system is controlled by an \textit{infinite randomness fixed point}. In such systems the distributions of logarithmic magnitudes of the terms in the Hamiltonian become broaden as the energy scale tends to zero. This means that in finite quenched system the thermal quantities become sample-dependent. Consequently if we are interested in average quantities in a finite system with size $N$, it is necessary to perform an additional quenched disordered average over the disordered configurations. The fluctuations over samples of an arbitrary thermodynamical quantity $X$ then can be described by the following relative variance:
\begin{equation}
	V_N(X)=\dfrac{[X^2]_{av}-[X]^2_{av}}{[X]^2_{av}}
\label{VarianceDef}
\end{equation}
where $[...]_{av}$ denotes the sample-to-sample average of quantity $X$. If $V_N(X)$ tends to zero as the system size $N\rightarrow\infty$ then the quantity $X$ is self-averaging and a sufficiently large system is a good representation  of the whole ensemble. However if $V_N(X)$ tends to a constant as $N\rightarrow\infty$, and even the critical point is known exactly, statistics cannot be improved by going over larger size. In this case $X$ is called to be non-self-averaging and in order to characterize $X$, its whole distribution is needed. 

\subsection{Models with disorder}
\subsubsection{Diluted magnets}
For classical systems beyond thermal fluctuations another type of disorder can be introduced. One of the simplest realization of such systems are the diluted magnets. In this case the Hamilton operator of the Ising model can be written as
\begin{equation}
	\mathcal{H}_{DM}=-\sum_{\langle i,j\rangle}J_{ij}(\eta_i\sigma_i)(\eta_j\sigma_j)
\end{equation}
which is capable to describe the two main types of dilution. For site diluted models the exchange energy is constant $J_{ij}=J$ and for each site we choose $\eta_i=0$ with probability $p$ and $\eta_i=1$ otherwise with probability $1-p$. This model can represent for example a binary alloy where magnetic and non-magnetic particles occupy the lattice sites randomly with a given concentration $p$. However we can introduce another type of dilution (called bond dilution) where each site is occupied ($\eta_i=1$) but the exchange energy is $J_{ij}=0$ with probability $p$ and $J_{ij}=J>0$ with probability $(1-p)$ independently for each couplings.
\begin{figure}[htb]
  \begin{center}
    \begin{minipage}[b]{0.48\linewidth}
\includegraphics*[width=6.5cm]{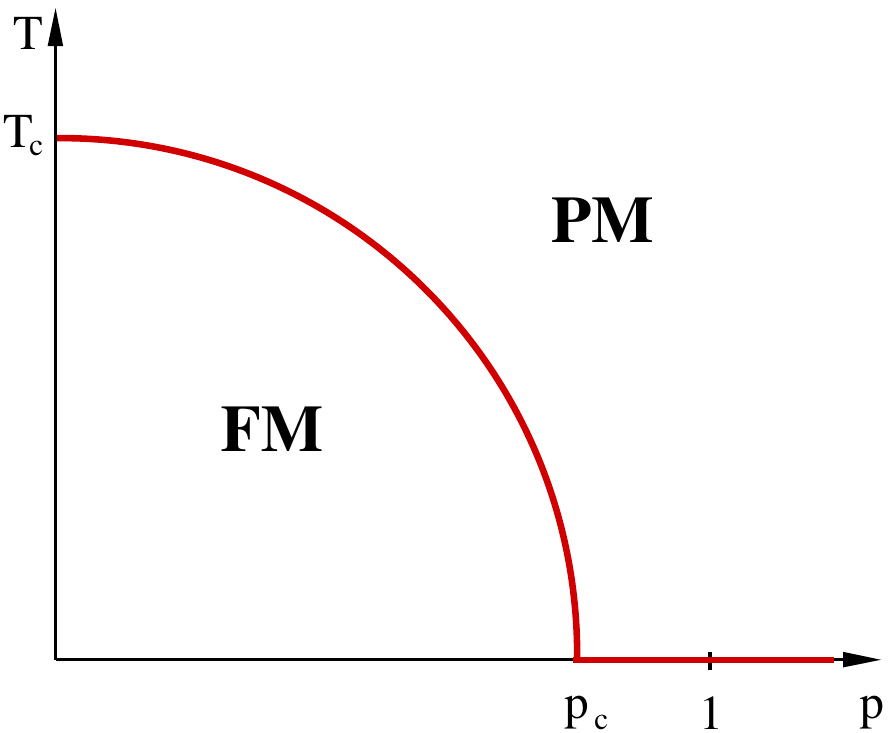}
    \end{minipage}\hspace{.2in}
\hspace{-.2in}
    \begin{minipage}[b]{0.38\linewidth}
      \caption{Phase diagram of diluted magnets. The ferromagnetic (FM) and paramagnetic (PM) phase are separated by a critical line. \qquad}
\label{DiluteM}
    \end{minipage}
  \end{center}
\end{figure}
Both models have two relevant parameters, the temperature $T$ and the $p$ concentration of non-magnetic components. At $T=0$, decreasing $p$ from one towards zero at a given critical $p_c$, an infinite cluster of connected sites arises all over the system and we get back the problem of percolation that we have already discussed before in Section \ref{percolation}. At the phase diagram if we cross the separatrix, a continuous phase transition occurs belonging to different universality classes. When $T=0$ it is the universality class of percolation, while at $p=0$ it is belonging to the university class of pure ferromagnetic systems. If $p$ is between $0<p<p_c$ then the exponents are constant. If the dilution is irrelevant the critical exponents are the same like in the pure case, however if the disorder is relevant then a new fixed point determines the phase transition in the hole region $0<p<p_c$ with new critical exponents.

\subsubsection{Random bond models}
\label{RandomBondModels}
Another type of bond dilution can be defined on random spin systems. Here the couplings $J_{i,j}\geq 0$ are independently distributed random variables with positive mean and $\Delta$ deviation, which define the third relevant parameter of such random systems. This kind of dilution can change the order of phase transition, typically softening it from first order to continuous as we increasing $\Delta$ away from zero \cite{PhysRevB.61.3215,PhysRevE.64.036120,mercaldo:026126,Mercaldo2005}. 

The ferromagnetic \textit{random bond Potts model} in the limit $q=\infty$ corresponds to this case where the couplings $J_{ij}$ are independently distributed positive random variables sampled from a probability distribution $P(J_{ij})$. The Hamiltonian of the model is defined in Eq.\ref{PottsHinfq}. Since the system is ferromagnetic ($J_{ij}>0$), the ground state must be one of the $q$ states where all the spin have the same value. On a two dimensional square lattice if $0<\beta J_{ij}<1$ and the distribution $P(\overline{J}+\Delta)=P(\overline{J}-\Delta)$ is symmetric, the critical point can be derived from self-dual relations \cite{PhysRevB.23.3421}:
\begin{equation}
T_c=\dfrac{1}{\beta_c}=2\overline{J}
\label{RBPMTc}
\end{equation}
where $\overline{J}>0$ denotes the mean of the coupling values. This critical point separates the ferromagnetic and paramagnetic phases. For the $\Delta=0$ pure case, Eq.\ref{RBPMTc} equivalent of the form we have seen in Section.\ref{Potssqinf} when the size $N\rightarrow\infty$ and the phase transition is of first order. As a quenched disorder is switched on, the nature of the phase transition changes. This question was studied by Cardy anf Jacobsen \cite{Cardy1997}, who construct an interface Hamiltonian which was mapped onto that of the random field Ising model. They found that for $d=2$ for any small amount of disorder destroys the phase coexistence at $T_c$ and the phase transition softens to second order \cite{PhysRevE.69.056112,PhysRevLett.62.2503}. In higher dimensions $d>2$ for weak disorder the correlation length stays finite at the critical point and a phase coexistence persists define the phase transition first order. However for strong enough disorder the correlation becomes divergent at a tricritical point and the phase transition is of second order, the phase transition line is broken and an additional phase arises \cite{Mercaldo2005}.

Some kind of bond dilution is capable to change the critical behaviour not only at the critical point, but at the vicinity of the criticality too. If we set the couplings in the ferromagnetic Hamiltonian to $J_1$ with probability $p$ and $J_2$ otherwise $(J_1>J_2>0)$ then some quantities present singular behavior in an extended phase around the critical point \cite{PhysRevLett.23.17}. This phase is located between the critical temperature for the onset of magnetic long-range order and the critical temperature of the pure model\cite{PhysRevLett.59.586}. This so-called \textit{Griffiths phase} \cite{PhysRevLett.54.1321} is caused by rare local domains which are in the contrary phase that the whole system itself.

\subsubsection{Random field models}
\label{RFIMDef}
If we introduce disorder into the field term $h_i$ instead of the couplings $J_{i,j}$ in the Hamiltonian Eq.\ref{IsingHGen} we arrive to another important class of disordered systems called \textit{random field models} which were defined first by Imry and Ma \cite{PhysRevLett.35.1399}. Suppose that the couplings are $J_{i,j}=J$ homogeneous and define $h_i$ as an independent random variable with a given distribution $P(h_i)$ with zero mean. In such systems there is competition between the long range order and the random ordering field, since the neighbouring spins tend to align parallel while the applied external field tries to pin each spin according to the sign of the local field.

In the \textit{random field Ising model} (RFIM) the system has two relevant parameters, the temperature $T$ and the variance of the field distribution $V(h_i)$ defined in Eq.\ref{VarianceDef}. For sufficiently small $V(h_i)$ and $T$, the exchange interactions are dominant and the system is in a ferromagnetic phase with non-vanishing magnetization. If we increase the field over a critical value $V_c(h_0)$ the system passes through a phase transition, which after each spin earns more if those align in the direction of the local random field, thus any long-range order is destroyed. The mean field approximation of the model on complete graphs is always solvable and it determines the order of the phase transition depending on the sign of the second derivative of the $P(h_i)$ distribution respect to the field at $h=0$. If it is positive then the phase transition is first order while if it is negative a phase transition of second order takes place.

For example if $P(h_i)$ is a Gaussian distribution then $P''(h=0)<0$ and the phase transition is second order and the system has the same phase diagram than diluted models if we replace $p$ with $h_0$. However if $P(h_i)$ is bimodal so thus $P''(h=0)>0$, there exists a tricritical point, i.e. the phase transition becomes first order for larger critical fields \cite{PhysRevB.18.3318}. 

No exact solution exists for any finite dimensional system except in one dimension \cite{Nieuwenhuizen1986,PhysRevLett.50.1494} however it is following from numerical treatments that in three dimension the universal behaviour of the RFIM depends on the choice of $P(h_i)$ distribution also \cite{dAuriac1997}. The ground state of the model is not degenerated if the distribution $P(h_i)$ is continuous and it can be weakly degenerated only if $P(h_i)$ is discrete. To find the ground state of the system we need to solve an optimization problem which can be calculated in strongly polynomial time (see later in Section \ref{CalcofRFIM}) \cite{Hartmann2002}.

\subsubsection{Spin glasses}

Spin glasses are disordered materials exhibiting high magnetic frustration. The origin of disorder can arise from disordered structure or the disordered magnetic behaviour of the material. In addition the presence of competing interactions and disorder cause a high frustration to the system, since it is not able to reach its lowest energy scale.

The convenient statistical model to study such materials is the Ising spin glass model \cite{RevModPhys.58.801}. The Hamiltonian can be deduced from Eq.\ref{IsingHGen} if we define $h_i=h$ homogeneous and choose the exchange couplings $J_{ij}= \pm J_{ij}$ with probability $p$ and $1-p$ allowing antiferromagnetic couplings between neighbouring spins.
\begin{figure}[htb]
  \begin{center}
    \begin{minipage}[b]{0.48\linewidth}
      \includegraphics*[ width=7.0cm]{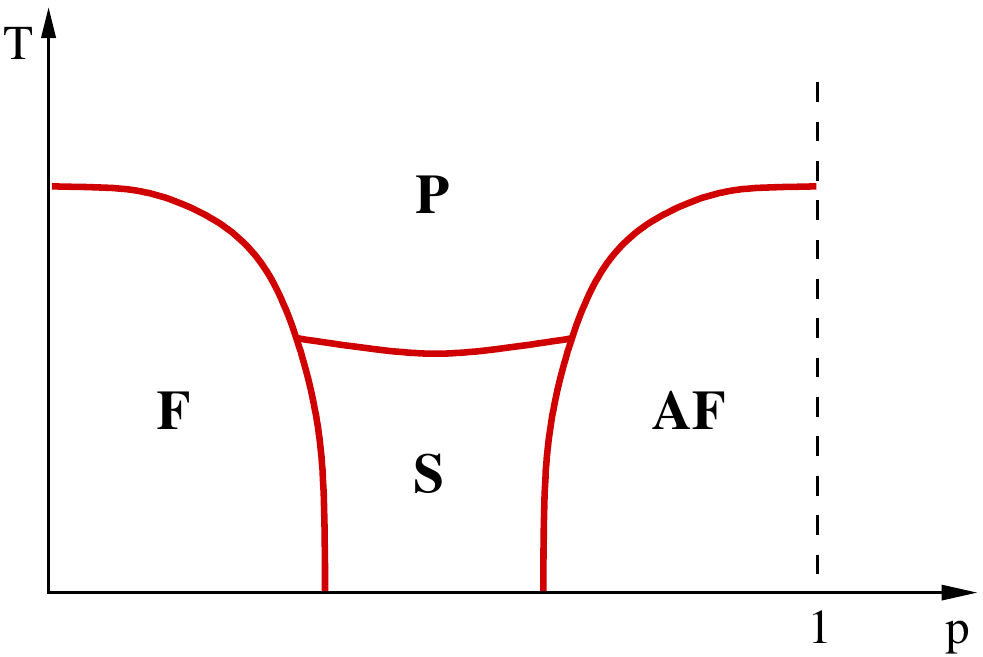}
    \end{minipage}\hspace{.4in}
    \begin{minipage}[b]{0.38\linewidth}
      \caption{Phase diagram of three dimensional Ising spin glass on cubic lattice. Beyond the ferromagnetic (F) and paramagnetic (P) phase an antiferromagnetic (AF) and a spin glass (S) phase takes place at the phase diagram.\qquad}
\label{iSpin}
    \end{minipage}
  \end{center}
\end{figure}

The phase diagram of the model is more complex than for diluted ferromagnets, since here beyond the ferromagnetic and paramagnetic phase, two other phases occur (see Fig.\ref{iSpin}). When $p$ is sufficiently large, the antiferromagnetic couplings dominate all over the system, the neighbouring spins favor to tend antiparallel and an antiferromagnetic phase evolve. 

However between the ferromagnetic and antiferromagnetic phase at low enough temperature with intermediate density of antiferromagnetic couplings, a so-called spin glass phase is located. If an external field is applied and the system is cooled under a critical temperature, after the field is removed, the magnetization of the system has a rapid exponential decay to a certain value (remnant magnetization). Then the magnetization follows a slow non-exponential relaxation to zero. This slow decay is identical to spin glasses and makes large difficulties for experimental studies.

The system has non-degenerated ground state but it has exponentially large number of states having approximately the same energy. To find the minimal energy state is an optimization problem which is solvable in polynomial time in two dimension however it is an NP complete problem for higher dimensions. At the energy minima state any long range order is present since only local ferromagnetic domains evolve in the system with randomly directed magnetic moments. That is why the average magnetization is zero, however its second moment does not vanish so it is suitable for order parameter.

%% file: complexnetw.tex
\chapter{Complex Networks}
\label{ComplexNetworks}
\section{Introduction to Graph Theory}

In Nature, many cases the structure of a complex system can be featured by entities and some type of links between them. The subject of graph theory is to mathematically describe this kind of structural features and characterize the main properties and laws of the underlying geometries. Historically Leonard Euler was the mathematician, who firstly illustrated a structure like this, later called graph, in his demonstration of the K\"{o}nigsberg bridge problem in $1736$. However his idea was forgotten for more than $100$ years later until 1847, when Kirchoff used again graphs in the proof of his two laws. Following them in the 20th century the graph theory became a powerful model of many parts of science. 

Such a description were found to be relevant to describe structures of complex systems in biology, sociology, economy and in many other subjects. Enormous information and technological networks grew up like the World Wide Web or the Internet, and parallel many facilities appeared to analyze existing networks using new methods and computational tools. Although these distinct segment of science were different and they used their own terminologies, many of them were found to ruled by the same principles and presented similar structures. We will focus on this structures in the following chapter.

\subsection{Definitions}

Consider a graph $G=(V,E)$ described by a set of vertices (nodes) $V$ and a set of edges (links) $E$, where the edges link pairs of nodes $(p,q)\in E$ (with $p,q \in V$). If the starting node $p$ and the ending node $q$ of an edge is ordered, we talk about directed graphs, otherwise it is undirected. If $p=q$ we call it loops, which could be directed and undirected as well. Hereafter typically we are going to use simple undirected graphs which do not contain any loops and multiple edges.

There are two general ways to represent a graph in mathematics and computer science. If the number of vertices $N$ denotes the size of the graph, the $N \times N$ \textit{adjacency matrix} $\mathcal{A}$ may describe the graph with elements $a_{pq}=1$ if a directed edge is presented between vertices $p$ and $q$, otherwise $a_{pq}=0$. In weighted graphs the $a_{pq}$ elements of the adjacency matrix are equal to the weights of existing edges or being zero. If the graph is undirected, the adjacency matrix is symmetrical so it may be converted into an upper triangular matrix. This representation is useful if $|E| \simeq |V|^2$ so the graph is dense. 

However if $|E| \ll |V|^2$ thus the graph is sparse, the so called \textit{adjacency list} representation is usually preferred because it provides a compact way to represent such structures. This representation based on a point array $Adj[p]$ of every $p \in V$, where each $Adj[p]$ elements own an adjacency list of all the nodes $q \in V$ which are the neighbors of the given $p$. If the graph is weighted it is easy to include the weight values for each element into the adjacency lists.

The advantage of the adjacency matrix representation is the direct access of the edges in the matrix however this description is fairly robust. The disadvantage of the adjacency list description is that we cannot access the edges in one step, but this description does not use more memory.

\subsection{Graph properties}

To characterize different graphs we need to examine some relevant properties which are capable to discriminate the graphs with different structures. Here we introduce the main properties which are frequently used in the literature.
\vspace{.1in}

\textbf{Degree distribution:} It is possible to describe a graph with given structural feature geometrically or we can analyze its statistical properties if its size is sufficiently large. The first and most frequently studied statistical parameter is the degree distribution of vertices. The $k$ degree of a node is defined as the number of neighbours it has if the graph is undirected. However for directed graphs we make difference between $k_{in}$ in-degree and $k_{out}$ out-degree, where $k_{in}$ and $k_{out}$ are the numbers of incoming and outgoing edges respectively for a given node. The degree distribution shows the fraction of nodes that own a given number of connections with an average, defined as the number of connections per site: $\langle k\rangle=2N_{edge}/N$. Here $N_{edges}$ denote the number of edges.

\textbf{Average path length:} Average path length $\bar{\ell}$ is another important graph property, which measures the average shortest path length between two connected nodes. Here the shortest path length $\ell_{pq}$ between vertex $p$ and $q$ is defined as the shortest series of edges with unit length, called geodesic distance as well. If the graph is directed, $\ell_{pq}$ and $\ell_{qp}$ are probably different. To calculate it we need to average the distance over all pair of nodes of a given graph. Between separated vertices the average path length is $\bar \ell = \infty$ by definition, and in a fully connected graph $\bar{\ell}=1$, however for a $d$ dimensional lattice with system size $N$, it scales as $\bar{\ell} \sim N^{1/d}$.

\textbf{Clustering coefficient:} The clustering coefficient $C$ is a local parameter of a graph and gives the cliquishness of the neighborhood of a node. Denote $n_1^{(p)}$ the number of first neighbors of site $q$, and $n_{edge}^{(q)}$ the number of edge presented between them. The clustering coefficient of vertex $q$ defined as $C_q=n_{edge}^{(q)}/(n_1^{(p)}(n_1^{(p)}-1)/2)$, where $n_1^{(p)}(n_1^{(p)}-1)/2$ is the maximum number of possible edges between the $n_1^{(p)}$ neighbours if those would form a complete graph. Averaging $C_q$ over all vertices $q \in V$, it yields to the clustering coefficient $C$ of the graph. Practically, $C$ gives the probability that if three nodes are connected with two edges then a third edge is presented. $C$ was first introduced in sociology \cite{Wasserman1994}, but Watts and Strogatz defined it in network science in the previous form \cite{Watts1998}.

\textbf{Connected components:} By definition a connected component of an undirected graph is a maximally connected subgraph, where each node can be reached by paths along edges. Generally a graph is likely to contain more than one connected component with size $1$ to $N$.
In directed graphs the maximally connected subgraphs or strongly connected components, where there are directed paths between each vertex. The largest connected component or also called giant component or giant cluster has a primary importance in percolation theory and network science.

\textbf{Graph spectra:}
The spectrum of a graph is the set of $\lambda_i$ eigenvalues of its adjacency matrix $\mathcal{A}$. It is graph invariant, and it gives a better tool to compare graphs than $\mathcal{A}$ itself, which depends on the graph labeling. It is also convenient to define the graph spectral density, coming from the random matrix theory \cite{Mehta2004}, as $\rho(\lambda)=\sum_{i=1}^{N}\frac{\delta(\lambda-\lambda_i)}{N}$, where $\lambda_i$ the i$th$ biggest eigenvalue of $\mathcal{A}$ \cite{PhysRevE.64.026704}. This density becomes continuous if $N \rightarrow \infty$ and is related directly to the topology of the network.

\section{Geometrically ordered graphs}

\subsection{Regular lattices}

Since the atoms of a crystal are arranged in a fix periodical structure, in solid state physics a special type of graph is used to describe such systems which is called lattice. A lattice is defined as a symmetry group with translational symmetry in $n$ direction, or in other words, it is a space ordered graph with translational invariance. It is arranged by unit cells which fill periodically the $d$-dimensional space. In theoretical physics many models defined on lattices (lattice models) are exactly solvable and also easy to simulate using computational methods.
\begin{figure}[htb]
\begin{center}
\includegraphics*[ width=11.0cm]{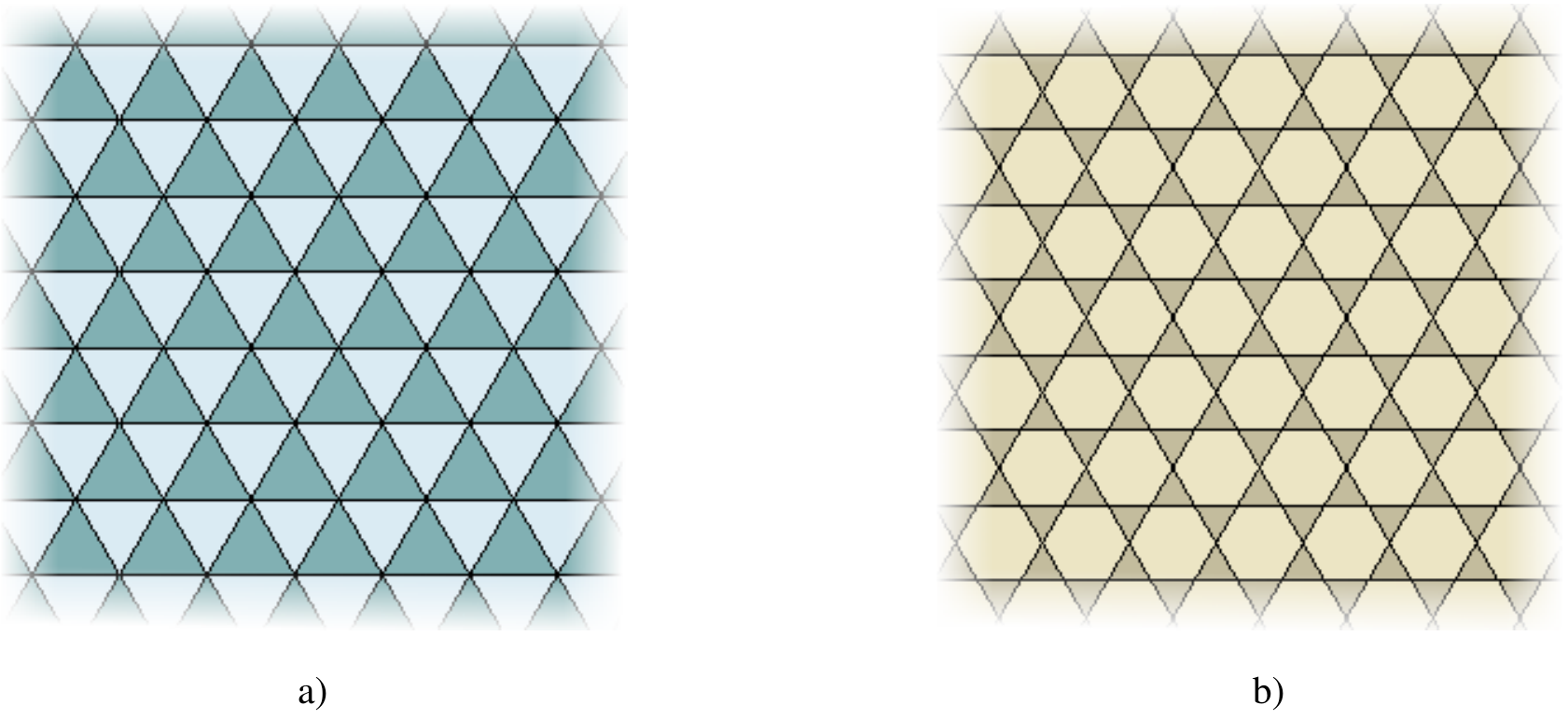}
\caption{The triangular lattice (a) and the Kagom\'{e} lattice (b) are the most studied regular structure which can induce geometrical frustration in antiferromagnetic lattice models.}
{
\label{kagometriang}
}
\end{center}
\end{figure}

Another usually required main property of a lattice is the regularity. In graph theory a graph is called regular if its each vertex $p \in V$ has the same number of neighbors, thus they have the same degree $k$. We called $k$-regular graphs those graphs which contain vertices with degree $k$ only.

The geometrical properties of a crystal lattice can induce frustration in condensed matters like in antiferromagnetic systems. The simplest regular lattices which cause such frustration in two dimension are the triangular lattice and the Kagom\'{e} lattice (Figure \ref{kagometriang}), which were intensively studied from the early 50s \cite{PhysRev.79.357}. In these lattices a geometrical constrain arises from the structure of the lattice which does not let the system relax to its ground state and induce residual entropy at zero temperature. The water ice was the first example which presented such behaviour, found in 1936 \cite{Giauque1936}, but later other matters showed similar features.

\subsection{Boundary conditions}

Since it is possible to study only finite lattice systems via computer simulations, an important question arises about the influence of the lattice boundaries. Beyond the finite size effects, on the margin of a finite lattice, all edges which link to the last nodes are hanging and change the local free energy. However, by applying special boundary conditions we can eliminate these boundary and corner effects. Here we are going to outline the main types of the frequently used boundary conditions.

\begin{figure}[htb]
\begin{center}
\vspace{-0.25in}
\includegraphics*[ width=16.0cm]{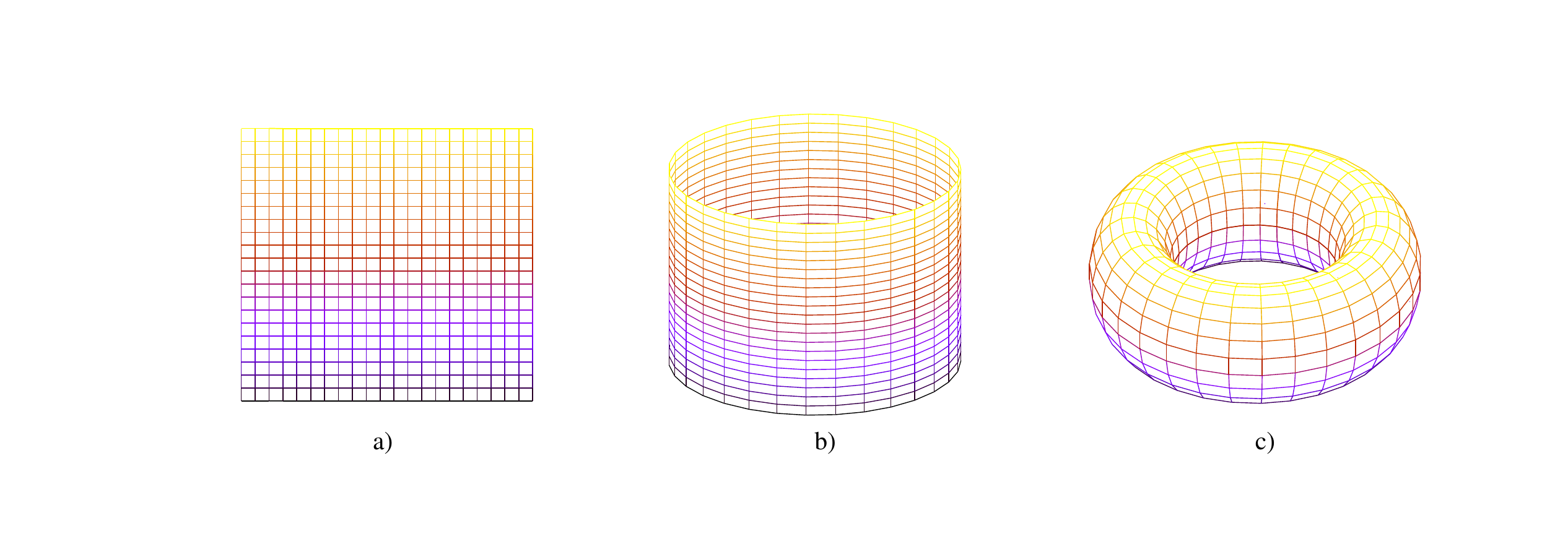}
\vspace{-0.4in}
\caption{Two dimensional square lattice with free boundary condition (a), half periodical boundary condition (b) and periodical boundary condition (c).}
{
\label{bouncond}
}
\vspace{-0.1in}
\end{center}
\end{figure}

Proceed from a simple two dimensional square lattice with linear size $L$ (Figure \ref{bouncond}.a), the simplest conventional choice is the free boundary condition, where we do not involve any kind of link between different boundaries, but let dangling bonds on the circumference. This case is suitable to study such problems where free edge boundaries are more realistic, like modeling local surface effects which arise on the boundaries of different matters.

If we link two opposite boundaries of a square lattice, as we connect each last node to the first node of a given row, it forms a cylinder (Figure \ref{bouncond}.b). Here parallel with the periodical condition, the system is homogeneous and the correlation length is maximally $L/2$. The name of such a geometry is the half periodical condition.  Another cylindrical structure, which is the easiest to implement, is the helical (also called screw periodical) boundary condition, where the nodes of the lattice sit on a one dimensional chain and wrapped around the system. Practically in the end of the chain the very last node connected to the very first node and cause an inhomogeneity which is only negligible in the limit of infinite size.

The $d$ dimensional lattice which has periodical boundary condition in each $d$ direction covers a $d+1$ dimensional torus (Figure \ref{bouncond}.c). Here the system is translational invariant in each directions, and eliminates every boundary effects, however the correlation length is still reduced by the system size so finite size effects arise in the system.

Another boundary condition which reduces finite size effects, and which does not need any additional edges is the mean-field boundary condition. Here an additional external field is introduced on the boundaries, which induces the mean bulk conditions on the borders, eliminates boundary and corner effects.

\section{Geometrically disordered graphs}
A spontaneously evolving network in the real world usually follows rules which are controlled by random properties. Such self-organized systems exclude regularity and show a fairly different structure which finds its origin in special features like dynamical growth or randomness. In the following section we are going to overview the brief history of random networks and define general models which belong to this segment.

\subsection{Random graphs - Erd\H{o}s-R\'{e}nyi model}
\label{ERmodel}

In the 50s Solomonoff and Rapoport \cite{Solomonoff1951} and independently Erd\H{o}s and R\'{e}nyi \cite{Erdos1959,Erdos1960,Erdos1961} were the first to define a random graph model in order to describe real world nets. They supposed that the networks in nature are driven by uncorrelated random rules, and they defined a simple model, later called Erd\H{o}s-R\'{e}nyi model, which produces a graph (called random graph, Poisson random graph or Bernoulli graph) where each pair of nodes connected randomly with a given probability $p$.

There are two equivalent definitions for random graphs. In the one was given by Erd\H{o}s and R\'{e}nyi \cite{Erdos1959}, they proceed from $N$ labeled nodes connected with $n$ edges, which are randomly chosen from the possible $\frac{N(N-1)}{2}$. Such graphs $G_{N,n}$ with $N$ nodes and $n$ edges form a probability space, where each realization appears with equal chance. The other definition called binomial model start from $N$ separated nodes, where every pair of nodes is connected with a given probability $p$. Then the expectation value of total number of edges in a $G_{N,p}$ graph is $E(n)=p\frac{N(N-1)}{2}$. In other words $G_{N,p}$ is an ensemble of all such graphs where a graph with $n$ edges is realized with probability $p^n(1-p)^{\frac{N(N-1)}{2}-n}$.

\subsubsection*{Degree distribution}
The $P(k)$ degree distribution of the Erd\H{o}s-R\'{e}nyi graph follows a binomial distribution:
\begin{equation}
 P(k)=\binom{N-1}{k} p^k (1-p)^{N-1-k}
\label{ERbinom}
\end{equation}
where $p^k$ is the probability that $k$ edges are present, $(1-p)^{N-1-k}$ is the probability of absence of additional edges and $\binom{N-1}{k}$ is the number of ways to select endpoints for the present $k$ edges. If the size $N$ tends to infinity, this binomial form transforms to a Poisson distribution:
\begin{equation}
 P(k)=e^{-p(N-1)}\dfrac{p(N-1)^k}{k!}=e^{-\langle k \rangle}\dfrac{\langle k \rangle^k}{k!}
\label{ERPoisson}
\end{equation}
where we denote the average degree as $\langle k \rangle=p(N-1)\sim pN$. The Poisson distribution implies that a random graph can be characterized by an average degree without a significant deviation of any other degree.

\subsubsection*{Topological phase transition}
Varying probability $p$, the topology of a random graph passes through a percolation type phase transition, where the following phases are located:
\begin{itemize}
 \item If $\langle k \rangle \simeq pN < 1$ the edge density is low, the graph contains isolated trees, and the cluster size distribution is exponential. The diameter (here the longest path length) of the graph commensurate with the diameter of a tree.
 \item If $\langle k \rangle > 1$ a giant component appears which holds most of the vertices. The diameter of the graph equal to the linear size of the giant cluster.
 \item If $\langle k \rangle > \mbox{ln}(N)$ the graph is fully connected.
\end{itemize}
The critical point is located where $\langle k \rangle \simeq pN=1$ consequently at $p_c=1/N$.

\begin{figure}[htb]
  \begin{center}
    \begin{minipage}[b]{0.58\linewidth}
\includegraphics*[ width=8.5cm]{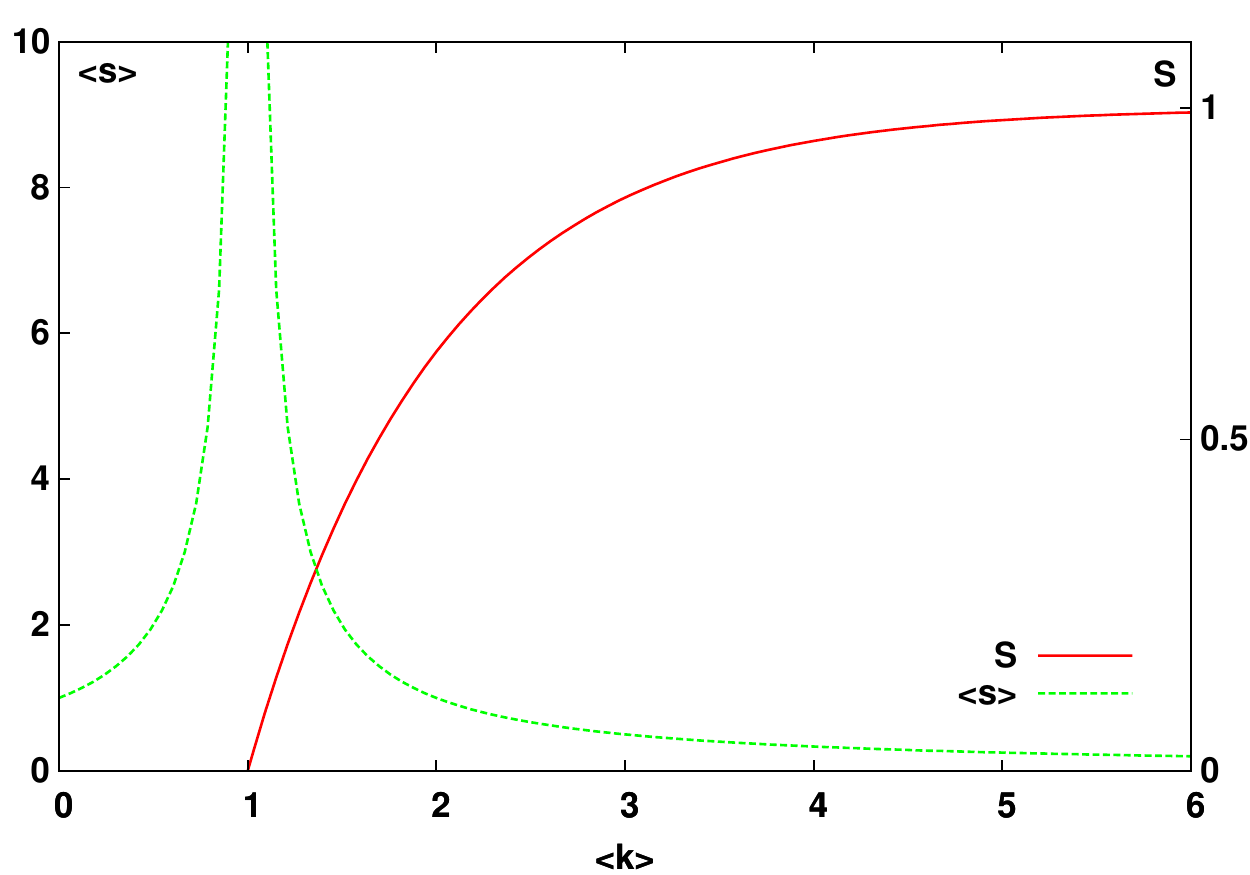}
    \end{minipage}\hspace{.2in}
    \begin{minipage}[b]{0.30\linewidth}
      \caption{Demonstration of the topological phase transition in random graphs. Increasing the mean degree per site, the $S$ size of giant component grows at the critical point and the mean component size $\langle s \rangle$ diverges. \qquad\qquad\qquad\qquad\qquad\qquad\qquad}
\label{ERPhaseTr}
    \end{minipage}
  \end{center}
\end{figure}

It is possible to calculate the size $S$ of the giant cluster which plays the order parameter role in the phase transition \cite{Molloy1995,Molloy1998,newman-2003-45}. Denote $w$ the probability that a randomly chosen point does not belong to the giant cluster, in other words $P_w(k)=w^k$ is the probability that the chosen point with degree $k$ does not have any neighbours in the giant cluster. Averaging it over $P(k)$, $w$ goes as
\begin{equation}
 w=\sum_{k=0}^{\infty} p_k w^k = e^{-\langle k \rangle} \sum_{k=0}^{\infty}\dfrac{(\langle k \rangle w)^k}{k!} = e^{\langle k \rangle (w-1)}
\end{equation}
consequently the fraction of nodes belonging to the giant component is given in a self consistent form as $S=1-w=1-e^{- \langle k \rangle S}$.

We can calculate also the specific heat like parameter $\langle s \rangle$, which is defined as the mean size of the components where a randomly chosen node belongs. It goes as $\langle s \rangle=\frac{1}{1-\langle k \rangle - \langle k \rangle S}$, and at $p_c$ it has a singularity. Even $S$ and $\langle s \rangle$ do not have a solution in a closed-form, it is easy to see that if $\langle k \rangle <1$, the only non-negative solution of $S$ is $S=0$ and if $\langle k \rangle >1$ the finite solution is rational to the size of the giant cluster. The phase transition point is located at $\langle k \rangle =1$ where $\langle s \rangle$ diverges and the giant component appears in the graph. Figure \ref{ERPhaseTr} demonstrates such kind of transition.

\subsubsection*{Graph properties}
The \textit{spectral density} also presents different behavior in the two separated phases. When a giant cluster occurs in the graph, the density shows a semicircle distribution \cite{PhysRevE.64.026704}, where the largest eigenvalue dominates. In the other phase, when trees are present typically, the spectrum deviates from this semicircle because the odd moments are missing since it is impossible to have any closed walk with odd number of step in the graph.

The \textit{average path length} $\bar{\ell}$ of a random graph can be studied by the mean number of neighbors at distance $\ell$ separated from a given vertex \cite{Bollobas1984,Chung2001,Bollobas1981}. The mean number of the nearest neighbors is $\langle k \rangle$ so approximately $\langle k \rangle^{\ell}$ number of vertices are located at a distance $\ell$ or closer to a chosen point. Hence $\langle k \rangle^{\bar{\ell}} \sim N$ so the average path length $\bar{\ell} \simeq \frac{log N}{log \langle k \rangle} \simeq \frac{log N}{log (pN)}$ is small even for large networks.

Reconsidering the \textit{clustering coefficient} for random graph, it is possible to give a different but an equivalent definition as before. In random graphs, the probability that two neighbors of a chosen vertex are connected is equal to the probability that two randomly chosen nodes are connected \cite{albert-2002-74}, which is $P=C_{rand}=\frac{\langle k \rangle}{N}$. Following this definition, the clustering coefficient is size dependent.

A generalization of the random graph model which is suitable to produce a graph with arbitrary degree distribution is the \textit{configuration model}. Here first we define the $P(k)$ degree distribution, and then following the chosen statistic, we assign a degree to each node in the graph. After we randomly pick up pairs of stubs and connect the chosen vertices together, we can produce every possible topology of a random graph with an optional degree distribution. 

\subsubsection*{Adaptability}
Even though the random graphs have short average path length, most of the other properties do not correspond to real world networks. It shows a Poisson degree distribution with exponential tail for large $k$ and the clustering coefficient depends on the system size which features are usually uncharacteristic for empirical networks. These three major properties characterize the random graphs and any deviation from these must be explained by non-random processes, which drive us to real world behavior. However the importance of the Erd\H{o}s-R\'{e}nyi model and random graphs are indisputable since they induced the base of a new science and gave many tools which became crucial to characterize many real and theoretical networks later.

\subsection{Small-world networks - Watts-Strogatz model}
The small-world phenomena first appeared in literature in 1929 in the novel Chain \cite{Karinthy1929} published by a Hungarian writer Frigyes Karinthy. In his short story he wrote about six steps of separation between any randomly chosen people around the world. This 'six degrees of separation' was first proven by Stanley Milgram in his famous experiment in 1967 \cite{Milgram1967}, where he examined the average path length in social network of people in the United States. In his research he sent letters to randomly selected citizens in the US, who had then to forward those to their friends. Milgram counted the number of steps after the letters arrived back to him and found the average path length $\bar \ell=5.5$. After that in the following decades the small-world effects had been found in various type of networks and worked around in many scientific papers.

\begin{figure}[htb]
\begin{center}
\includegraphics*[ width=13.0cm]{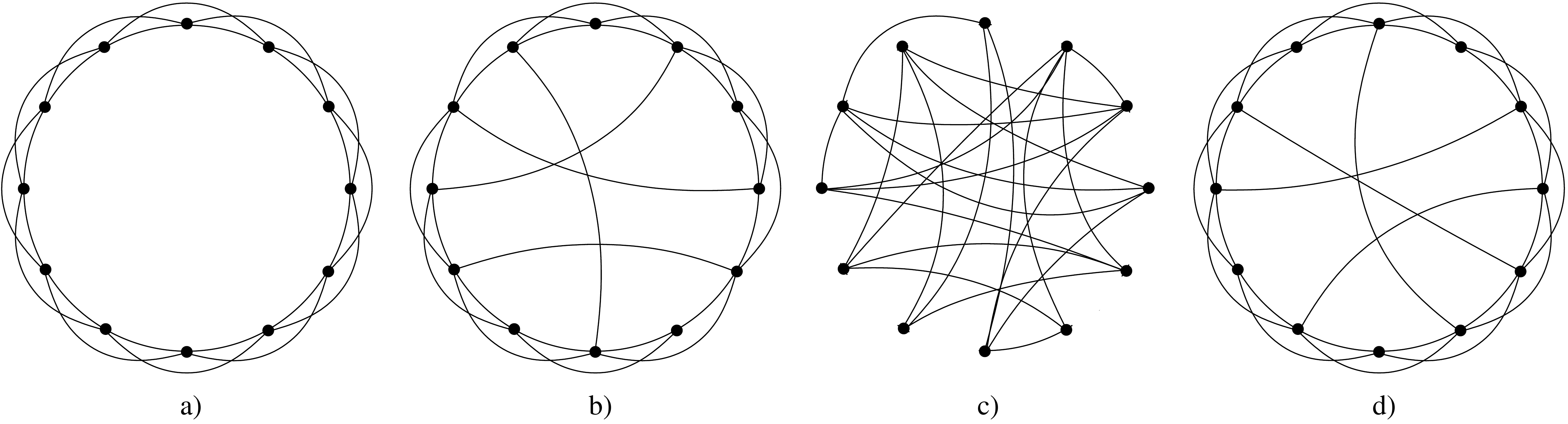}
\caption{The evolution of the Strogatz-Watts network. Starting from a regular ring (a), following the rewiring process (b), it arrives to a random graph structure (c). A alternative definition (d), where only shortcuts are added to remaining original ring.}
\label{SWLattice}
\end{center}
\vspace{-15pt}
\end{figure}

An approach of the previously described real world networks was defined by Watts and Strogatz \cite{WattsBook1999,Watts1998,Watts1999} in the so-called small-world model, which makes transition between regular and random graphs. Their definition proceed from an initial one dimensional regular graph with periodic boundary condition - a chain - where each vertex is linked to its first $K$ neighbors in both direction. Then, going through all the vertices, each edge is rewired with a probability $p$, and an opposite ending point is chosen randomly. When $p=0$ (Figure \ref{SWLattice}.a) the network keeps its initial regular structure, but when $p=1$ (Figure \ref{SWLattice}.c), the network transforms to a random graph. The clustering coefficient at $p=0$ is large, $C=(3K-3)/(4K-2)$ which tends to $3/4$ for large $K$. However the average path length is long $\bar \ell = L/4K$ for large $L$ since a regular lattice does not show a small-world behavior and the average distance between vertices goes as $N^{1/d}$. When $p \rightarrow 1$ the system relaxes to random network, where $\bar \ell = (\mbox{ln} L)/(\mbox{ln} K)$ and the clustering coefficient becomes lower: $C\simeq 2K/L$.

A modified definition was given by Monasson \cite{Monasson1999} and independently by Newman and Watts \cite{Newman1999a,Newman1999c} where they kept the original regular chain but gave additional shortcuts between randomly chosen vertices (Figure \ref{SWLattice}.d). The probability $p$ here gives the density of these additional edges. This model for sufficiently large $L$ and small $p$ is equivalent to the original Watts-Strogatz model.
\subsubsection*{Average path length}

\begin{figure}[htb]
  \begin{center}
    \begin{minipage}[b]{0.58\linewidth}
\includegraphics*[ width=8.5cm]{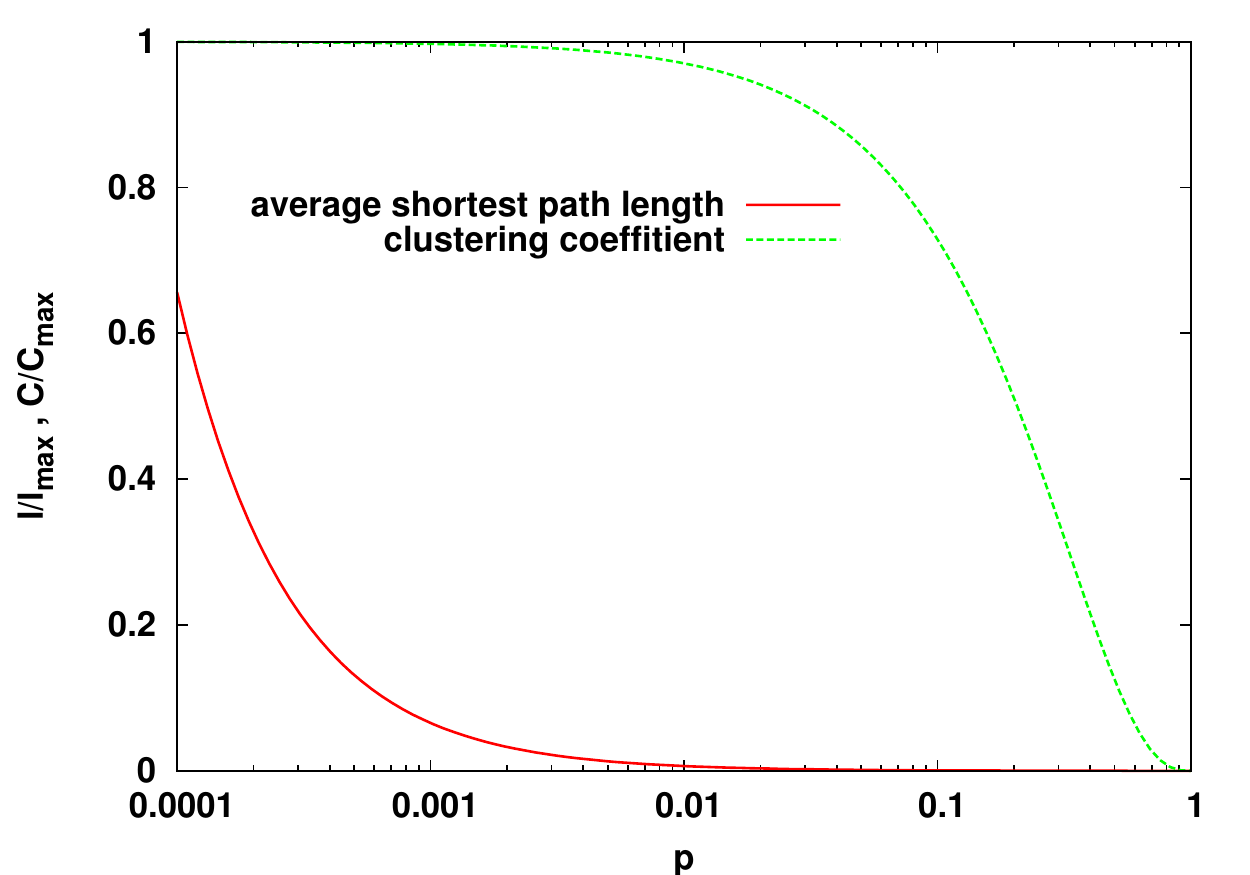}
    \end{minipage}\hspace{.2in}
    \begin{minipage}[b]{0.30\linewidth}
      \caption{The clustering coefficient $C(p)$ and average path length $\bar \ell(p)$ of the Watts-Strogatz network in the function of rewiring probability $p$. For convenience $C(p)$ and $\bar \ell(p)$ were divided with their maximal value $C(p=0)$ and $\bar \ell(p=0)$. \qquad\qquad\qquad\qquad\qquad\qquad\qquad\qquad\qquad\qquad\qquad\qquad\qquad\qquad}
\label{SWPathLC}
    \end{minipage}
  \end{center}
\end{figure}

Proceeding from a regular 'large-world' lattice, where $\bar \ell= L/4K$, and varying $p$ from $0$ to $1$ the graph passes through a structural change and arrives to a random 'small-world' graph where $\bar \ell = (\mbox{ln} L)/(\mbox{ln} K)$. In between this two phases there is a crossover regime, where $\bar \ell$ begins to decrease drastically (Figure \ref{SWPathLC}). The reason of this falling is that the additional shortcuts which appears first at $p \geq 1/NK$, decrease suddenly the geodesic distance between opposite parts of the graph. Many attention have been focused to determine the behavior of $\bar \ell$ in this range, but it is still not exactly solved. A widely accepted explanation that $\bar \ell$ satisfies a scaling relation \cite{PhysRevLett.82.3180}:
$$
\bar \ell \sim \xi g(L/ \xi) \qquad ~~\mbox{and}~~ \qquad
g(x) = \left\{ 
\begin{array}{l l}
  x & \quad \mbox{if $x \ll 1$}\\
  \mbox{ln}(x) & \quad \mbox{if $x \gg 1$}\\
\end{array} \right.
$$
where the correlation length scales as $\xi \sim p^{-1/d}$ \cite{barrat-1999}. Using renormalization group treatments \cite{Newman1999b} an equivalent scaling form has been found:
\begin{equation}
\bar \ell \sim \frac{L}{K}f(L/\xi)
\label{eqno1_4}
\end{equation}
which differs only by a factor $K$ and where $\xi=1/pK$. According to the scaling form in Eq.\ref{eqno1_4} the graph can pass through the transition controlled by $p$ or $L$ as well, since $LKp$ is equal to the mean number of shortcuts, which induce the topological change. Finally a mean-field treatment \cite{Newman2000} of the model in one dimension shows approximately that:
\begin{equation}
 f(x)=\dfrac{1}{2 \sqrt{x^2+2x}}\mbox{tanh}^{-1}\dfrac{x}{\sqrt{x^2+2x}} \qquad
\mbox{and so} \qquad
\bar \ell = \dfrac{\xi}{2 K \sqrt{1+2\xi /L}}\mbox{tanh}^{-1}\dfrac{1}{1+2\xi /L}
\end{equation}

\subsubsection*{Clustering coefficient}
The clustering coefficient, which is large at the initial regular graph, is invariant of the system size $L$ at $p=0$ since it depends only on the coordination number $z=2K$ of the lattice. If disorder is introduced into the system by rewired edges, it remains close to $C(p=0)$, as long as a large fraction of original neighbors keep connected. The probability that three vertices which were connected at $p=0$ still construct a triangle when $p>0$ is $(1-p)^3$, since there are three edges which need to keep intact. It follows that the clustering coefficient changes as \cite{Barrat2000}:
\begin{equation}
C(p)=C(0)(1-p)^3=\frac{3K-3}{4K-2}(1-p)^3
\end{equation}
while for the alternative definition, where instead of rewiring edges, only shortcuts are added to the system, $C(p)$ behaves as \cite{Newman2000a}
\begin{equation}
C(p)=\dfrac{(3K-3)}{(4K-2)+4Kp(p+2)}
\end{equation}

\subsubsection*{Degree distribution}

Since in the Watts-Strogatz model the initial network is regular and the degree of each vertex is $k=2K$, the degree distribution $P_{WS}(k,p)$ at $p=0$ is a delta function at $2K$. On the other hand, a non-zero $p$ introduces disorder into the network, which broadens the degree distribution, but maintains the average degree around $\bar k = 2K$. The degree distribution for $p>0$ follows a Poisson distribution, since the network is homogeneous and relaxes into a random graph.

During the rewiring process only one end of a chosen edge is reconnected, so each node has at least $K$ neighbors in the end. It means the network remains connected for each $p$ if $K\geq 1$. The real connectivity of a node can be written as $c_i=K+n_i$, with $n_i \geq 0$, where $n_i$ is the number of additional connections that a node has beyond $K$ \cite{Barrat2000}. These $n_i$ number of edges can be of two kinds also: it contains $n_i^1$ edges which the rewiring process left unchanged with probability $(1-p)$ and $n_i^2=n_i-n_i^1$ number of edges which became connected (each with probability $p/(N-1)$) to node $i$ during the process. The probability distribution of these two kinds of edges can be written as:
\begin{equation}
P_1(n_i^1)=\binom{K}{n_i^1}(1-p)^{n_i^1}p^{K-n_i^1} \qquad \mbox{and} \qquad P_2(n_i^2)=\dfrac{(Kp)^{n_i^2}}{n_i^2!}e^{-pK}
\end{equation}
if $N \gg 1$, the complete degree distribution turns into:
\begin{equation}
P_{WS}(k)=\sum_{n=0}^{min(k-K,K)}\binom{K}{n}(1-p)^{n}p^{K-n} \times \dfrac{(Kp)^{k-K-n}}{(k-K-n)!}e^{-pK} \quad \mbox{if} \quad k \geq K
\end{equation}
However this $P_{WS}(k)$ degree distribution can never be identical to the degree distribution of a random graph, since in the latter case the graph can contain isolated components, while a small-world network is always connected.

\subsubsection*{Real small-world networks}

Following Milgram's experiment the 'six degrees of separation' phenomena were characterized in many other real world networks \cite{Amaral2000}. For example the same behavior was found in e-mail networks too, with an average path length between five and seven \cite{Dodds2003}. A few experiments are in progress to characterize small-world behavior in nowadays popular community sites and instant messaging services.

In collaboration networks, there are two well studied examples of small-world behavior. One experiment was defined on movie-actor collaboration network in Hollywood, also called 'The Kevin Bacon Game' \cite{WattsBook1999,Watts1998}, and another collaboration network can be defined in science between researchers who have common publications \cite{newman-2001-98,PhysRevE.64.016131,PhysRevE.64.016132}. Both networks show small-world behaviors with smaller average path length than six, but these networks have an additional property. Namely, their structure is non-homogeneous and their degree distribution follows a power-law, what we are going to discuss in the following section.

\subsection{Scale-free networks - Barab\'{a}si-Albert model}
\label{BAmodel}
Many self-organized networks in the Nature show such kind of dynamical behaviors, which were not included in the earlier network models. These dynamics induce a sufficiently different structure for these systems. The dynamical growth seems to be an actual request for real-world network models, however an other important role influences the evolution which drives strong inhomogeneity into the evolving network and produces degree-distribution with a power-law tail.

The citation network of scientific paper was the first example which was found and showed scale-free behavior studied by Price in 1965 \cite{PRICE1965}. He counted the directed citations between papers and found the in- and also the out-degree distributions following power-law degree distribution as:
\begin{equation}
P(k) \sim A k^{-\gamma}
\label{PowerLawDegrDistr}
\end{equation}
He then developed a model \cite{Price1976} where he gave an alternative description of inhomogeneity. His idea was based on the "the rich get richer" theory of Simon \cite{Simon1955}, namely that a paper which was published earlier and became popular has larger probability to obtain new citations than a paper without many references. He called this phenomenon "cumulative advantage". However these kind of systems had difficulties to be studied since strong computational resources were not accessible at this age, so Price's idea was forgotten almost for three decades. In the late 90s this phenomenon was re-found in some other type of networks, using computational methods, and has been enduring popularity up to date in many segments of research.

\subsubsection*{Barab\'{a}si-Albert model}
\label{BAmodelSec}
Barab\'{a}si, Albert and Jeong were the first who studied the structure of the World Wide Web (WWW) where the nodes of the network were defined as the websites and the edges were the directed hyperlinks between them. It turned out that a few extremely connected website hold most of the links in the network, and beyond these sites the largest part of the webpages were slightly connected. They found a power-law behavior for the in- and out-degree distributions, and gave a relevant explanation for the growing dynamics of the network. Their model of evolving networks, today called scale-free model or Barab\'{a}si-Albert model, is excellent to simulate systems which show scale-free features. Such properties were found later in biology, sociology, technology and in many fields of the science.

In their model they included the dynamical growth and the above described inhomogeneity role, which they called "preferential attachment". The model contains the following steps:

\begin{algorithm}
\caption{Barab\'{a}siAlbertModel($m_0,m,t$)}
\begin{algorithmic}[1]
\STATE Starting from a fully connected graph with $m_0$ vertices
\FORALL{timestep $t$}
	\STATE add a new node with $m(\le m_0)$ edges
	\FORALL{$m$ number of edges proceed from the new node}
		\STATE choose a node $i$ to which the new node connects with a probability rational to its degree such that: $\Pi(k_i)=\frac{k_i}{\sum_j k_j}$.
	\ENDFOR
\ENDFOR
\end{algorithmic}
\label{BAmodelAlg}
\end{algorithm}

\begin{figure}[htb]
\begin{center}
\includegraphics*[ width=10.0cm]{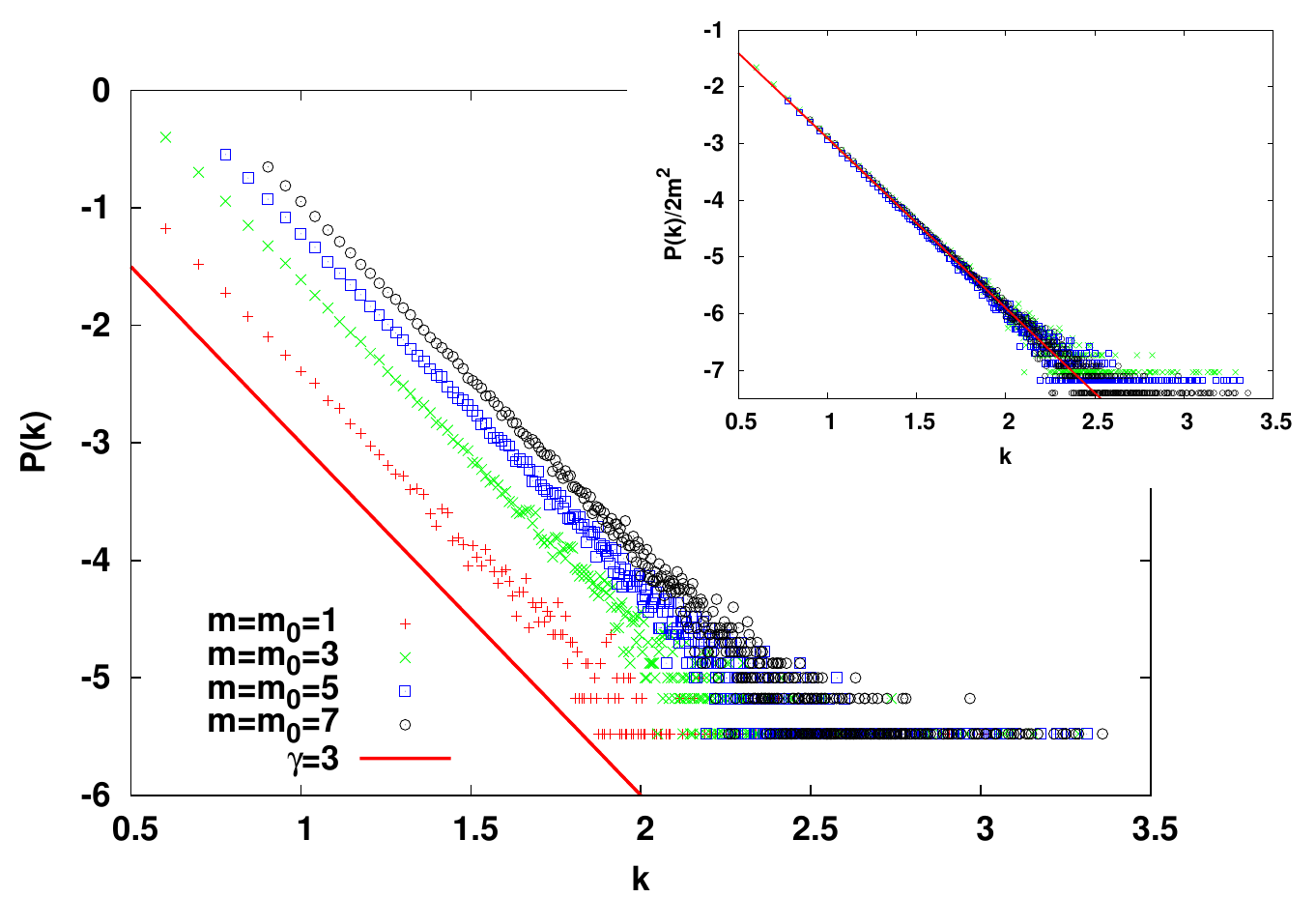}
\caption{Degree distribution of Barab\'{a}si-Albert network for various $m$ with system size $N=300000$. The slope of the skew line is $\gamma=3$ and gives the power-law degree exponent. In the inset we demonstrate, that the degree distributions are independent of $m$ after rescaling as $k^{-\gamma}\sim P(k)/m^2$. The fitted line gives the expected exponent $\gamma=3$.
\label{BA_DegrDist}
}
\end{center}
\end{figure}

This model was defined on undirected graphs  and gave a degree exponent $\gamma=3$ (Figure \ref{BA_DegrDist}), which is a good approach for the exponents of real world networks.
It can be solved in the large size limit with an approximate solution given by Barab\'{a}si and Albert \cite{Barabasi1999,barabasi-1999}, and it can be proved exactly with two another equivalent methods: the master equation approach published by Dorogovtsev \textit{et.al.} \cite{Dorogovtsev2000b,379835} and rate-equation approach introduced by Krapivsky \textit{et.al.} \cite{PhysRevLett.85.4629}. For these solutions see Appendix \ref{BASolutions}.

\subsubsection*{Types of correlation}
There are two types of correlation in evolving Barab\'asi-Albert networks found by analytical studies \cite{PhysRevE.63.066123}. The first nontrivial correlation is between the age and the degree of a node. The vertices, which were added earlier to the network had better chance to acquire edges, so these nodes have higher degrees. For the case $m=1$, when the evolving graph is a tree, the probability distribution of the degree of vertex $i$, with age $a$ (the elapsed time since $i$ was given to the network) is:
\begin{equation}
P_k(a)=\sqrt{1-\dfrac{a}{N}}\left( 1-\sqrt{1-\dfrac{a}{N}} \right)^k
\end{equation}
so the degree distribution for nodes with fixed age decays exponentially, with a characteristic degree scale which diverges as $\langle k \rangle \sim (1-(a/N))^{-1/2}$ if $a\rightarrow N$. It means that young nodes (where $a/t \rightarrow 0$) have typically small degrees, while for the older ones the average degree is large. This age-degree correlation was not found in real networks \cite{Adamic2000}, however using an extension, called fitness parameter \cite{Bianconi2001,Bianconi2001a}, this correlation can be reduced in current model too.

The other type of correlation was found between the degrees of adjacent vertices \cite{PhysRevE.63.066123}. If $N_{jk}$ denotes the number of edges which are connecting two type of nodes with degree $j$ and $k$, in the case when $m=1$:
\begin{align*}
N_{jk}=\frac{4j}{(k+1)(k+2)(j+k+2)(j+k+3)(j+k+4)}+ \qquad \qquad \qquad \qquad \qquad \qquad \qquad \\
+ \frac{12j}{(k+1)(j+k+1)(j+k+2)(j+k+3)(j+k+4)}
\end{align*}
It is independent of the time and depends on $k$ and $j$ only, suggesting that the correlation evolves spontaneously between the degrees of adjacent vertices, unlike in classical random graphs, where the degrees of connected nodes are uncorrelated \cite{Newman2000b}.

\subsubsection*{Graph Properties}

The most important property of the Barab\'asi-Albert network \cite{albert-2002-74} is its \textit{degree distribution} that follows a power-law. This feature comes from the growth and preferential attachment roles which take place together in the model. To find, which property is responsible for the spontaneous evaluational scale-free feature, we can analyze these independently. If we study a growing network with equal probability attachment, it evolves to a graph with exponential degree distribution. However if we keep the network size constant and introduce edges following the role of preferential attachment, the network evolves into a state where all nodes have the same degree. Consequently both properties need to evaluate a network with a power-law degree distribution.

The \textit{average path length} $\bar \ell$ of a Barab\'asi-Albert network is smaller than a random graph having the same system size. It means that through the few well connected nodes, all the vertices are located closer to each other in the network than in a homogeneous topology. That is why the scale-free network is called ultra small-world network in this sense. $\bar \ell$ is growing logarithmically with the system size, but nevertheless there is no exact theoretical expression for the average path length.

There is no analytical prediction also of the \textit{clustering coefficient} for scale-free graphs. Even this property in Watts-Strogatz networks is independent of the system size, here it has a strong size dependence. It decreases faster with the system size than in random graphs, and follows a power-law $C\sim N^{-\alpha}$ with an exponent $\alpha=0.75$.

The \textit{spectral density} for the Barab\'asi-Albert networks shows a fairly different shape than we saw for random graphs \cite{PhysRevE.64.026704,Goh2001}. It has a continuous distribution and form a triangle where the edges have power-law tails. The first eigenvalue $\lambda_1$ is separated from the main bulk of the spectrum like for Erd\H{o}s-R\'enyi graphs, and it increases approximately as $\lambda_1\sim N^{1/4}$.

The Barab\'{a}si-Albert model has exceptional popularity, since it was first published, and many variation appeared to make it more realistic e.g. introducing fitting parameter we mentioned before. Another modification when we let the attachment probability $\Pi(k+k_0)$ to depends on an additional constant $k_0$ too, instead of $k$ only. Modify $k_0$ between $-m < k_0 < \infty$, it scales the degree distribution \cite{Dorogovtsev2000b,PhysRevE.63.066123}, which then goes as $P(k)\sim k^{-\epsilon}$, with $\epsilon=\gamma+k_0/m$, where $\gamma=3$ the original degree exponent.

If we scale the attachment probability as a power-law $\Pi(k^{\varepsilon})$ and not linearly as before, when $\varepsilon=1$ we get back the original graph, however if $\varepsilon < 1$ or $\varepsilon > 1$, different network behaviors appear \cite{PhysRevE.63.066123,PhysRevLett.85.4629}.

Another variation of the model is given if we increase the average degree over time \cite{PhysRevE.63.025101}. It seems to be realistic, since in the WWW the average degree increasing by time as well. This modification varies the degree exponent too, and approximates more the original Barab\'{a}si-Albert network to real systems.

\subsubsection*{Real scale-free networks}

Such kind of systems which shows a power-law degree distribution can be found in various self-organized systems. In social science the first found example was the citation network \cite{Price1976,PRICE1965} of scientific papers, what we have already discussed. However human sexual networks \cite{Liljeros2001}, the word web of human languages \cite{Ferrer2001} or the above-mentioned collaboration networks following scale-free behavior too \cite{WattsBook1999,Watts1998,newman-2001-98,PhysRevE.64.016131,PhysRevE.64.016132}. In technology, electric circuits \cite{Ferrer2001a}, phone-call nets \cite{Aiello2000} or airplane networks has scale-free topology, and after the WWW \cite{albert-1999-401,Adamic1999}, some other information network, like the Internet (defined as the network of wired rooters) indicates the same behaviour also \cite{Faloutsos1999,Govindan2000}. We can find various type scale independent biological system too, like cellular networks \cite{Jeong2000,Jeong2001,Fell2000}, protein folding networks \cite{Jeong2001,Wagner2001}, ecological and food networks \cite{Montoya2002,Sole2001} or the net of metabolic interactions \cite{Jeong2000}.

These various examples prove that self-organized networks in Nature are not driven by only random rules, but follow such universal laws which introduce a type of order into the system. 

%% file: NumericalMethods.tex
\chapter{Numerical methods}
\label{NumMethds}

The numerical physics became a sterling part of the science of physics since high-performance computational engines and methods became available from the middle of the 20th century. The main expediency of numerical simulations is two fold, related to the theoretical and experimental part of physics and other segments of natural science. In many cases theoretical treatments are available for models for which there is no perfect physical realization (at least at the present time). In this situation the only possible test for an approximate theoretical solution is to compare with numerical results generated from a computer simulation. There are also real physical systems which are sufficiently complex that they are not presently amenable to theoretical treatment. An example is the problem of understanding the specific behavior of a system with many competing interactions and which is undergoing a phase transition. Following this two directions many numerical methods have been developed in the last few decades what we cannot discuss here \cite{Landau2000}. In this chapter we confine oneself to the methods only we applied during our work or strongly related to the subject of statistical physics and Monte Carlo simulations. After a brief introduction into the base idea of general Monte Carlo methods we discuss shortly the subject of combinatorial optimization and its application what we used intensively in the course of our work. We close this chapter with a short discussion of an effective algorithm to simulate scale-free networks.

\section{Principles of Monte Carlo methods}
The subject of Monte Carlo methods in physics is to calculate thermodynamical averages of observable quantities in interacting many body systems. These numerical methods are based on stochastic techniques and use random numbers and probability statistics to study and simulate problems on computers. In this chapter we are going to discuss the general ideas of Monte Carlo methods and review algorithms which are related to our purpose.

\subsection{Importance sampling}
\label{ImportanceSampling}
Lets take a lattice system at a temperature $T$ which contains $N$ particles \cite{Binder1986,Landau2000,Newman2001d}. The time evolution of the system can be described as a serial of states which the system goes through at given state space. Suppose that the system is at the state $\mu$ at time $t$ and define a transition probability $R(\mu \rightarrow \nu)$ that it is in the state $\nu$ a $dt$ time later. We can determine the dynamics of the system thus we define such a rate for all possible states $\nu$ that the system can reach. If we introduce a set of weights $w_{\mu}(t)$ which gives the probability that the system is in the state $\mu$ at time $t$ then the time evolution can be described by the master equation:
\begin{equation}
	\dfrac{dw_{\mu}(t)}{dt}=\sum_{\nu}\big[ w_{\nu} R(\nu \rightarrow \mu) - w_{\mu} R(\mu \rightarrow \nu)\big]
\label{ME}
\end{equation}
where the first term on the right hand side represents the probability that the system gets into the state $\mu$ while the second terms is the probability that it leaves that. Since the system at any time $t$ must be at one of the points of the state space then the weights must obey $\sum_{\mu}w_{\mu}(t)=1$. The expected value of some quantity $Y$ at time $t$ then can be written as:
\begin{equation}
	\left\langle Y \right\rangle = \sum_{\mu} Y_{\mu}w_{\mu}(t)
\end{equation}
In equilibrium when the transition probabilities between two states are equal
\begin{equation}
\sum_{\nu}w_{\mu}R(\mu \rightarrow \nu)=\sum_{\nu}w_{\nu}R(\nu \rightarrow \mu)
\label{equilibr}
\end{equation}
the right hand side of the master equation is canceled out and all weights $w_{\mu}$ become constant since $dw_{\mu}(t)/dt=0$. For thermodynamical systems the equilibrium values of $w_{\mu}$ are known \textit{a priori} exactly and define the equilibrium occupation probability in the infinite time limit as $p_{\mu}=\lim_{t\rightarrow\infty}w_{\mu}(t)$. For canonical systems in equilibrium this probability (as it was already described in Section.\ref{Thermodynamical quantities}) is following a Boltzmann distribution \cite{Gibbs1902}:
\begin{equation}
	p_{\mu}=\dfrac{1}{\mathcal{Z}}e^{-\beta\mathcal{H}_{\mu}} \qquad \mbox{thus the average of quantity } Y \mbox{ is } \qquad \left\langle Y \right\rangle=\dfrac{1}{\mathcal{Z}}\sum_{\mu}Y_{\mu}e^{-\beta \mathcal{H}_{\mu}}
\end{equation}
where $\mathcal{Z}$ is defined in Eq.\ref{Z}.

This average can be calculated only for small systems which have sufficiently enough number of states. However the energy of an equilibrated system changes only on a small scale, which means that the system occupies only a limited number of state with a large probability and has small but not zero chance to reach the rest of the state space. It is straightforward from the quantitative image that the number of states is reduced as the temperature approaches zero since the system does not have energy to reach excited energy levels. If we follow this consideration thus instead of measuring on all states but we make a representative selection $M$ of the most proper states, then average quantities can be calculated over this state set $M$ with acceptable accuracy. Now if we define a distribution $p_{\mu}$ to select randomly the important states $\mu_1,...,\mu_M \in M$ from the complete state set then an estimator of a quantity can be calculated as:
\begin{equation}
Y_M=\dfrac{\sum_{i=1}^M Y_{\mu_i}p^{-1}_{\mu_i}e^{-\beta \mathcal{H}_{\mu_i}}}{\sum_{i=1}^M p^{-1}_{\mu_i}e^{-\beta \mathcal{H}_{\mu_i}}}
\label{AvrQuant2}
\end{equation}
which becomes a more accurate estimate of $\left\langle Y \right\rangle$ as the number of states in $M$ increases and when $M\rightarrow \infty$ then $Y_M=\left\langle Y \right\rangle$.

As a first approximation if we choose $p_{\mu}$ equal for each state then we get back to Eq.\ref{AvrQuant2}. However we have seen that the range of energies of the states sampled by a typical system is very small compared with the total energy. So the system does not sample all states with equal probability but samples them according to a Boltzmann distribution. Then a better strategy would be not to pick up the representative $M$ state with equal probability but proportional to its Boltzmann weight. Now we can define the probability that a particular state $\mu$ gets chosen as $p_{\mu}=\mathcal{Z}^{-1}e^{-\beta \mathcal{H}_{\mu}}$ so the estimator is simplified to:
\begin{equation}
Y_M=\dfrac{1}{M}\sum_{i=1}^{M}Y_{\mu_i}
\end{equation}
where the Boltzmann factors have canceled. This approximation works much better since we pick up most properly the states where the system spends the majority of time, with a relative frequency corresponding to the time the system spends there. This method of \textit{importance sampling} was one of the base idea of Monte Carlo methods and commonly used in many algorithms. 

Now the question arises how to choose these proper states from the state space since choosing randomly and then accept them with probability $e^{-\beta \mathcal{H}_{\mu}}$ is still not effective. Almost all Monte Carlo schemes rely on \textit{Markov processes} to generate the set of these states. It is a mechanism which transforms the system into a new state $\nu$, with a transition probability $R(\mu \rightarrow \nu)$, if the system was in the state $\mu$ before. It chooses the new state $\nu$ in a random manner from the set of possible transitions corresponding to the transition probabilities, which probabilities need to satisfy the constrain
\begin{equation}
\sum_{\nu}R(\mu\rightarrow\nu)=1
\label{stateNorm} 
\end{equation}
By definition the transition probabilities must be time independent and the choice of the new state $\nu$ depends only on the properties of the former state $\mu$ (in other words the system has no memory).

The serial of states which was generated by a repeatedly applied Markov process is called \textit{Markov chain}. We can specify the Markov process thus when we start the system from an arbitrary state, after enough iterations the states appear with a probability given by the Boltzmann distribution. Then the system reaches its equilibrium. To ensure this specialty for the Markov process the system needs to satisfy two conditions.

The \textit{ergodicity} is the first required but not sufficient condition which means that the system, started from any initial state $\mu$, should be able to reach any other state $\nu$ in the state space, via the Markov process if we run it long enough. This condition is required since each state appears with a non-zero probability in the Boltzmann distribution. However in an ergodic system if a transition probability between two states is $R(\mu \rightarrow \nu)=0$ it does not mean that the system cannot reach the state $\nu$ initiated from $\mu$. In real systems most of the state transitions are forbidden, but there exists a sequence of transition between each state where the intermediate transitions are allowed.

The other condition we place on our Markov process is the condition which ensures that it is the Boltzmann distribution that we generate after the system reaches its thermal equilibrium. A system is in equilibrium if it is true for each states that the probabilities to get in a state and leave that are equal to each other, as it was drawn in Eq.\ref{equilibr}. We can simplify this expression apply Eq.\ref{stateNorm} and write to the form:
\begin{equation}
p_{\mu}=\sum_{\nu}p_{\nu}R(\nu\rightarrow\mu)
\end{equation}
However this condition still does not guarantee that we reach the equilibrium from an arbitrary initial state and we need one more consideration related to the Markov process:
\begin{equation}
	p_{\mu}R(\mu\rightarrow\nu)=p_{\nu}R(\nu\rightarrow\mu)
\label{detailedBal}
\end{equation}
which is called the condition of \textit{detailed balance}. It tells us that the average rate of transformation between two states is equal and independent of the direction which is always the case for real systems also.

Now if in the Markov process we suppose that the $p_{\mu}$ distribution of states satisfies the condition of detailed balance and we wish the equilibrium distribution to be the Boltzmann distribution, rearranging Eq.\ref{detailedBal}, it tells us that the transition probabilities should relate as:
\begin{equation}
\dfrac{R(\mu\rightarrow\nu)}{R(\nu\rightarrow\mu)}=\dfrac{p_{\nu}}{p_{\mu}}=e^{-\beta(\mathcal{H}_{\nu}-\mathcal{H}_{\mu})}
\label{Transratio}
\end{equation}

This constrain with the one defined in Eq.\ref{stateNorm} limits the transition probabilities, but if the Markov process obey these and our system is ergodic then the equilibrium distribution of states will follow a Boltzmann probability distribution.

If we choose the transition probabilities of the form of Eq.\ref{Transratio}, during the simulation the system can be trapped in a intermediate sate for a longer period or even locked for the whole processing time. It happens if the chosen transition probabilities are even correct Boltzmann probabilities but those are too small to change the current state. A solution is to brake the transition probabilities into two parts as $R(\mu\rightarrow\nu)=g(\mu\rightarrow\nu)A(\mu\rightarrow\nu)$ where $g(\mu\rightarrow\nu)$ defines a selection probability and gives the chance that the system in sate $\mu$ will choose the state $\nu$ as a next state during the Markov process. $A(\mu\rightarrow\nu)$ is the acceptance ratio which gives the ratio of the acceptance of the transition, chosen by $g(\mu\rightarrow\nu)$. Then the Eq.\ref{Transratio} can be rewritten of the form
\begin{equation}
\dfrac{g(\mu\rightarrow\nu)A(\mu\rightarrow\nu)}{g(\nu\rightarrow\mu)A(\nu\rightarrow\mu)}=\dfrac{p_{\nu}}{p_{\mu}}=e^{-\beta(\mathcal{H}_{\nu}-\mathcal{H}_{\mu})}
\label{DetailedBal2}
\end{equation}
which equation still satisfies the detailed balance but the ratio $A(\mu\rightarrow\nu)/A(\nu\rightarrow\mu)$ can be arbitrary chosen between zero and $\infty$ which means that the selection probabilities can take any values we wish. The best choice if the acceptance ratios are close to one. If we fix their ratio for every transitions then the selection probabilities can be scaled arbitrary if we multiply the counter and denominator with the same number. An average choice to set the larger selection probability to one and gear the other probability to that. This modification can significantly fasten up our simulation while the system still satisfies all the required constrains.

\section{General Monte Carlo algorithms}

In the following section we are going to introduce general Monte Carlo algorithms of statistical physics which are capable to simulate phase transition at finite temperatures. All of our considerations will be drawn with the Ising model paradigm, however all of our discussions are relevant for other models too. The nearest neighbour Ising model here is defined on a finite size lattice with N sites and periodical boundary condition to eliminate boundary effects. The initial state of the simulated system usually is a completely ordered ground state belonging to $T=0$ or a random state defined as the $T=\infty$ limit. 

\subsection{Metropolis algorithm}

The most frequently used method to simulate spin lattice models is the \textit{Metropolis algorithm} which was firstly defined on simulation of hard-square liquids in 1953 \cite{metropolis:1087}. Here we use a single spin-flip techniques to generate a new state from the current state of the system. Since the system size is $N$ and each spin can choose between two directions $\sigma_i=\pm 1$, the total number of the states is $2^N$. But if we flip only one of the $N$ spins at each iteration, the system has $N$ different $\nu$ states which can be reached from its current state $\mu$. Then if we choose the selection probability $g(\mu\rightarrow\nu)$ equal for each possible state transition $\mu\rightarrow\nu$ thus there are $N$ non-zero selection probabilities, that take the same value $g(\mu\rightarrow\nu)=1/N$, otherwise they are zero \cite{Newman2001d}. Substitute it into the Eq.\ref{DetailedBal2} we get the form:
\begin{equation}
\dfrac{R(\mu\rightarrow\nu)}{R(\nu\rightarrow\mu)}=\dfrac{A(\mu\rightarrow\nu)}{A(\nu\rightarrow\mu)}=e^{-\beta(\mathcal{H}_{\nu}-\mathcal{H}_{\mu})}
\label{MetropDetB}
\end{equation}
This ratio depends only on the energy difference $\Delta E=\mathcal{H}_{\nu}-\mathcal{H}_{\mu}$ between two states at a given temperature. Now we can choose any acception rate which satisfies the detailed balance condition. Historically the first choice was the form of Metropolis \cite{metropolis:1087}:
\begin{center}
\begin{equation}
A(\mu\rightarrow\nu) =
\begin{cases}
 e^{-\beta\Delta E} & \mbox{if} \quad \Delta E > 0 \\
1 & \mbox{otherwise}\\
\end{cases}
\label{MetropEq}
\end{equation}
\end{center}
where we set the larger acceptance probability to one and scale the smaller pro rata as we proposed to do in Section.\ref{ImportanceSampling}. Eq.\ref{MetropEq} means that we accept every transition which does not increase the energy of the system, otherwise we take it with a Boltzmann probability depending on the energy difference. The implementation of this method is simply defined in Alg.\ref{alg1} where one Monte Carlo step (MCS) is recognized as $N$ complete \textit{for} iteration between the lines $3-8$, thus that each spin could be chosen at least once per a MCS. 
\begin{algorithm}
\caption{Metropolis()}
\begin{algorithmic}[1]
\STATE Choose an initial state
\FOR{1 to $N_{MCS}$}
\FOR{$1$ to $N$}
	\STATE Choose a spin $\sigma_i$
	\STATE Calculate the energy change $\Delta E$ which results if the spin at site $i$ is overturned
	\STATE Generate a random number such that $0<r<1$
	\STATE Flip the spin if $r<e^{-\beta\Delta E}$
\ENDFOR
\ENDFOR
\end{algorithmic}
\label{alg1}
\end{algorithm}
The choice of spins in the 4\textit{th} line can be systematic or done by random fashion.

One of the slowest part of the algorithm is to calculate the energy difference of the system between the current state and after we flip a spin. However the energy of the initial and final state in one iteration depends only on the relation of the chosen spin and its direct neighbours since the Hamiltonian contains interactions for only first neighbours. So we need to calculate the energy difference on the first neighbour terms only and the energy of the rest of the system stays invariant during the iteration. Another trick may be applied if the lattice is regular and the number of neighbours is the same for each spin. Then the variation of the states of the chosen spin and its first neighbours is finite. So, if we calculate initially the energy difference for all these substates then during the iterations we do not need to recalculate always the energy but only pick up the related pattern. 

After the system has reached the equilibrium, to calculate an average quantity we need to measure in different MCS which are related to independent states. Such kind of time average and the multitude average over samples in equilibrium are equal in ergodic systems. 

An alternative method exists which defines the single spin-flip transition probability as \cite{glauber:294}:
\begin{equation}
R(\mu\rightarrow\nu)=1+\sigma_i \mbox{tanh} (\beta \mathcal{H}_i)
\end{equation}
where $\sigma_i \mathcal{H}_i$ is the energy of the $i$th spin in state $\mu$ \cite{Landau2000}. This method commonly called the \textit{Glauber dynamics} gives the same results as the Metropolis dynamics, however in the high temperature limit it stays ergodic unlike the Metropolis method \cite{Muller-Krumbhaar1973}.

\subsection{Cluster algorithms}

A more efficient class of algorithms to simulate lattice systems are the cluster-flip algorithms. Here instead of flip spins one by one we flip whole clusters of neighbouring spins which are at the same spin state. However the cluster selection techniques is not trivial since the size of clusters should depends on the temperature. Swendsen and Wang were the first who defined a relevant method, using the Fortuin-Kasteleyn transformation (see Section.\ref{RandomClust}) in their Monte Carlo algorithm \cite{PhysRevLett.58.86}. Following their idea Wolff found a different algorithm which was proven to be faster in higher dimensions \cite{PhysRevLett.62.361}. 

The common idea was to evaluate clusters of neighbouring spins which are at the same state, by adding bonds between them with a probability $p_{add}=1-e^{-2\beta J}$ (following the same method like in Fig.\ref{RCmodel}.b). In the Swendsen-Wang algorithm at every MCS we divide the whole lattice into clusters and then turn each cluster with a probability $1/2$. However in the algorithm of Wolff we proceed from a randomly selected spin and clustering (with $p_{add}$) only that domain where it belongs to. Then we turn the selected sub-domain with probability $1$.

The probability $p_{add}$ scales with temperature and shapes small clusters at high temperatures, but it assembles large amount of spins at the low temperature phase, where $p_{add}$ is large and a system wide domain is present. The cluster methods have a special advantage at the criticality where they are much faster than any single spin-flip algorithm. 

At the vicinity of the phase transition point, where the correlation length $\xi$ becomes infinite we need to flip many spins to produce uncorrelated patterns. The correlation time then diverges as $\tau\sim\xi^{z}$, where $z$ is the earlier defined dynamical exponent. Since in Monte Carlo simulations the system size is always finite, than at the critical point the correlation length becomes commensurable to the linear size $\xi\sim L$, which means that the relaxation time $\tau$ is increasing by the system size and causes the symptom of \textit{critical slowing down}. The cluster flip algorithms reduces the effect of this slowing and indicates various dynamical exponents (see Table.\ref{DynamExp}) which identify different dynamical universality classes.
\begin{table}[h]
\centering
\begin{tabular}{cccc}
\hline
dimension & Metropolis & Wolff & Swendsen-Wang \\ 
\hline \hline
$2$ & $2.167\pm 0.001$ & $0.25\pm 0.01$ & $0.25\pm 0.01$ \\ 
$3$ & $2.02\pm 0.02$   & $0.33\pm 0.01$ & $0.54\pm 0.02$ \\ 
$4$ & -                & $0.25\pm 0.02$ & $0.86\pm 0.02$ \\
\hline
\end{tabular}
\caption{Different values of dynamical exponents for the Metropolis, Wolff and Swendsen-Wang algorithm at different dimensions \cite{Newman2001d}. As we see in Table.\ref{DynamExp}, the Wolff algorithm is faster in higher dimensions, but in two dimensions in case of any applied external field the Swendsen-Wang method provides a faster solution.}
\label{DynamExp}
\end{table}

\section{Combinatorial optimization}
\label{CombinOpt}

Combinatorial optimization is a method to find extremity of different quantities and calculate their correct values numerically. Instead of examining all states of a finite system which can be done only in $t\sim e^L$ non-polynomial time (NP-complete problems), here we exploit some special feature which lets us apply algorithmic methods and find a solution in $t\sim L^k$ polynomial time (P-type problems). In case of a problem that cannot be solved analytically or algorithmically we need to apply Monte Carlo methods to calculate average quantities as we discussed before.

For example to calculate the average free energy of the ferromagnetic random bond Potts model, it is necessary to calculate the average over the whole probability distribution $P(J_{ij})$ as:
\begin{equation}
f(\beta)=\int dJ_1 \cdots \int dJ_m P(J_1,...,J_m)f_{J_1...J_m}(\beta)
\end{equation}
which is an NP-complete problem. Here we show how to compute exactly $f_{J_1...J_m}(\beta)$ for a particular choice of the couplings $J_1,...,J_m$ in the $q$ infinite limit \cite{dAuriac2004}. As we have seen in Section.\ref{GeneralPotts}, the partition function of the Potts model on a graph $G$ can be expressed in terms of its subgraphs $G'$ in the high temperature limit (see Eq.\ref{HTZPotts}). When $q=\infty$, the partition function can be rewritten in the form of Eq.\ref{qinfZ}, where only the subgraph $G^*$ contributes which maximizes $\phi(G')=c(G')+\sum_{e\in G'} J_e$. Then to calculate the partition function and the free energy, the problem is reduced to find the optimal subgraph $G^*$ which maximizes $\phi(G')$ or equivalently minimize:
\begin{equation}
\phi_P(G')=-\Big(c(G')+\sum_{e\in G'} J_e\Big)
\label{PottsFreeMin}
\end{equation}
This function has a special property which allows us to minimize it efficiently using optimization methods, which will be introduced in the following.

\subsection{Minimization of Submodular Functions}

Lets consider now $\phi_P(G')$ as a set function $f(A)$ which are defined as $f:2^E\rightarrow \mathbb{R}$ where $2^E$ denotes the set of all subsets of a finite set $E$. If the cardinality of set $2^E$ is large, it is impossible to calculate all its values in order to find the smallest one, however if $f(A)$ is recognized as a submodular function then to find its minimum becomes a optimization problem.
\begin{definition}
A set function is submodular if for all subsets $A\subseteq V$ and $B \subseteq V$:
\begin{equation}
	f(A)+f(B)\geqslant f(A\cap B) + f(A\cup B)
\end{equation}
\end{definition}
To find a subset $A^*$, called the optimal set, which minimizes $f(A)$ such that $f(A^*)\leq f(A)$ for any $A\subseteq E$ is possible in strongly polynomial time. To introduce a method we need to consider some other characters of submodular functions. Lets denote $f_E$ the function which we need to minimize on all subsets of $E$. Then for any subset $F\subseteq E$ we can define a set function thus $f_F(A)=f_E(A)$ for any $A\subseteq F$. This means if $f_E$ is a submodular function so thus $f_F$ is also a submodular function and satisfies the following proposition \cite{dAuriac2004}.
\begin{proposition}
$F\subseteq E$ is a subset and $e \in E$ is an element of $E$. If $A_F$ is an optimal set of the set function $f_F$ defined on $F$, then there will be an optimal set $A_{F\cup \{ e \}}$ of the function $f_{F\cup \{ e \}}$ defined on $F \cup \{ e \}$ such that $A_F \subseteq A_{F\cup \{ e \}}$.
\end{proposition}
This property has an important consequence. If we have found the optimal set $A^*_F$ for a subset $F$ of $E$ then all elements in $A^*_F$ will still belong to one optimal set $A^*_G$ of all the sets which $G\supseteq F$. It means that we can find the optimal set of $E$ , finding successively the optimal set for its subgraphs. In the following section we use this property to construct a method to find the optimal set to minimize a submodular function.

\subsection{The Method to calculate the Potts free energy}
\label{OCMethodFE}

It is simple to show that free energy $\phi_P(G')$ of the $q=\infty$ Potts model, which was defined in Eq.\ref{PottsFreeMin} on a graph $G=(V,E)$ is submodular \cite{dAuriac2002}. Using this property we can calculate this free energy in strongly polynomial time using combinatorial optimization predictions as follows. 

Let us suppose that the optimal set $A^*_k$ of the subgraph $G_k\subseteq G$ has been found and contains the vertices $V_k=\{v_0,v_1,...,v_k\}$ with the edges which have both extremities in $V_k$. Then the edges of the optimal set can be grouped into clusters (connected subsets of vertices), where the clusters are connected with edges, which are not belonging to the optimal set. The weight $J_{ij}$ of an edge between two clusters is defined as the sum of weights of all edges having one extremity in each cluster. The isolated vertices are considered as clusters of size one. We know from the previous proposition that the current cluster structure will be embedded into the optimal set of $G_{k+1}$. However when we add a new site to $G_k$, then some clusters of $A^*_k$ (possibly 0) are merged together with the new site $v_k$ while the other clusters keep unchanged. So the next step is to find the clusters which merge together with the new vertex $v_{k+1}$.
\begin{figure}[htb]
\begin{center}
\includegraphics*[ width=11cm]{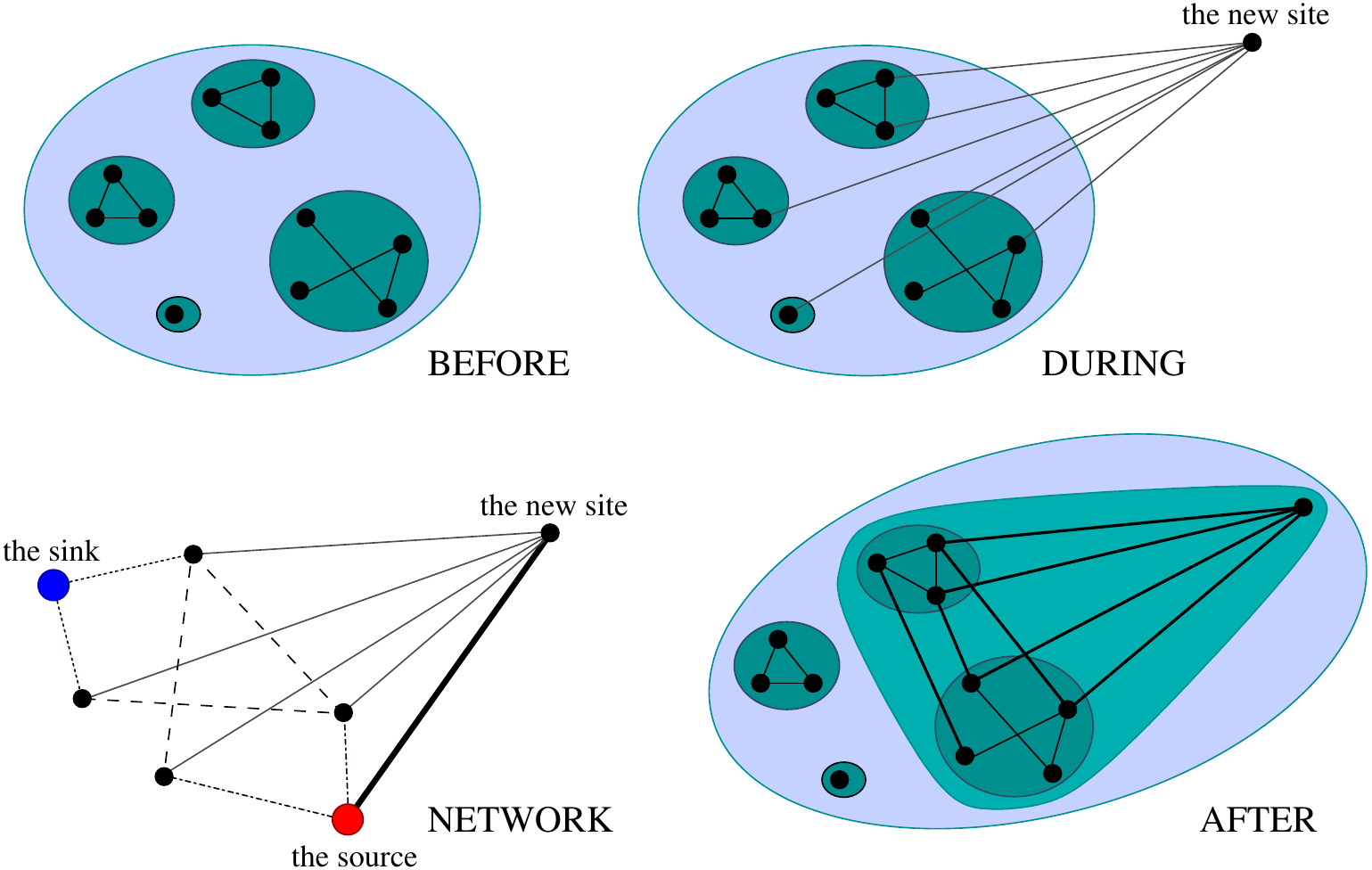}
\caption{Add a new vertex during the combinatorial optimization algorithm of the random bond Potts model. On ``BEFORE'' and ``AFTER'' only the edges of the optimal set are presented. Proceeding from the subgraph ``BEFORE'', we add the new site, which connects to the presented sites (``DURING''). Then the ``NETWORK'' is generated with the additional source and sink sites as we were discussed above with the capacities defined in \ref{OSweights1}-\ref{OSweights3}. At the ``NETWORK'' the thick line carries an infinite weight and the dashed edges are not belonging to the optimal set. After we calculated the minimum cut and some new edges included in the optimal set, signed with thick line at the state ``AFTER''.}
{
\label{CombOpt}
}
\end{center}
\vspace{-10pt}
\end{figure}

To solve this problem we introduce another weighted and directed graph, called network (see Fig.\ref{CombOpt}). The vertices in this network are of three kinds: (i) the clusters defined by the optimal set $A^*_k$ in $G_k$, (ii) the new vertex $v_{k+1}$, (iii) and two additional sites, which we called source $s$ and sink $t$. Between that vertices which corresponds to such clusters which are connected in the original graph $G_k$, we set two directed edges, one in each direction. Moreover we introduce additional edges from the source $s$ to each vertex and from every vertex to the sink $t$. After we defined the network, a convenient selection of edge capacities will reduce the optimization problem to finding the minimum cut between $s$ and $t$ which can be solved in strongly polynomial time. Here a \textit{cut} $S$ by definition is a partition of all vertices into two sets $S$ and $T$ such that $s\in S$ and $t\in T \equiv \overline{S}$. The weight $C(S)$ of a cut is the sum of edge weights $c(u,v)$, where the sites $u$ and $v$ are in different sets of $S$ and $T$. Then the suitable choice for the capacities:
\begin{eqnarray}
c(u,v)&=&c(v,u) =  \dfrac{1}{2}J_{uv} \label{OSweights1} \\
c(s,v) & = & \mbox{max}\Bigg( 0, \frac{1}{2}\sum_{u\in V(u)}J_{uv} -1 \Bigg) \label{OSweights2} \\
c(v,t) & = & \mbox{max}\Bigg( 0, 1-\frac{1}{2}\sum_{u\in V(u)}J_{uv} \Bigg)
\label{OSweights3}
\end{eqnarray}
where $V(u)$ denotes the set of vertices, different from $s$ and $t$, linked by an edge to site $v$. Then the minimum cut $S$ which separates $s\in S$ and $t\notin S$ in the network, gives the optimal set $A^*_k$ as the set of such edges which have both extremities in $S$.

\subsection{Implementation}
\label{OSImplement}

Following the previous considerations we can define an algorithm to calculate the free energy of the system as finding the optimal set of the graph $G$. The algorithm proceeds by computing the optimal set on $|V|-1$ sublattice with an increasing number of sites and arbitrary choice of adds order. To represent the actual data structure in the computer memory we can use two main options. We can construct the whole data structure and rebuild the network after each step, or one can define the whole structure in the beginning and increase the edge weights between vertices, which are at the same cluster, as large as that those cannot be in other clusters during the calculation. The latter case allocates less memory, however the former representation is faster since the extra time for building the new structure in every step is largely compensated by the smaller network to process. 

\begin{algorithm}
\caption{OptimalCooperation()}
\begin{algorithmic}[1]
\STATE $Network \Leftarrow$ \O
\FORALL{vertex $v$}
	\STATE \textbf{call} AddVertexToNetwork($Network,v$)
	\STATE $S$ $\Leftarrow$ MinCut($Network$)
	\STATE \textbf{call} ContractNetwork($Network$)
\ENDFOR
\STATE Uncontract all vertices
\STATE $OptimalSet \Leftarrow$ the edge set induced by the cut $S$
\end{algorithmic}
\label{algOC}
\end{algorithm}

The optimal cooperation algorithm (see Alg.\ref{algOC} and also Fig.\ref{CombOpt}) at each iteration includes three steps \cite{dAuriac2004}: (i) build the network (Alg.\ref{algAddVert}), (ii) find the minimum cut, (iii) contract all the connected vertices on the source side into a single vertex in a way that we can uncontract them at the end of the algorithm (Alg.\ref{algContrNetw}).

\begin{algorithm}
\caption{AddVertexToNetwork(Network,v)}
\begin{algorithmic}[1]
\STATE \textbf{comment} \textit{connect vertex v to its neighbours sites as in the original graph}
\FORALL{neighbours $u$ of $v$}
	\FORALL{vertex $x$ in the network to which $u$ belongs}
		\STATE $c(v,x)\Leftarrow c(v,x)+J_{uv}/2$
	\ENDFOR
	\STATE $c(x,v)\Leftarrow c(v,x)$
\ENDFOR
\STATE connect the source $s$ and the sink $t$ using weights defined in Eq.\ref{OSweights3} and Eq.\ref{OSweights2}
\end{algorithmic}
\label{algAddVert}
\end{algorithm}

The crucial point of this method is to find the minimum cut in a series of $|V|-1$ networks. Many algorithms exist to calculate the minimum cut (or equivalently the maximum flow) of a weighted directed graph in polynomial time. We applied the Goldberg\&Tarjan and the Ford-Fulkerson algorithm in the course of our work which are explained in details in \cite{Cormen2001}.

\begin{algorithm}
\caption{ContractNetwork(Network)}
\begin{algorithmic}[1]
\STATE Contract the vertices in $S$ into a single vertex $v^*$
\STATE \textbf{comment} \textit{modify the weights accordingly}
\FORALL{vertex $u$ not in $S$}
	\STATE $c(u,v^*) \Leftarrow 0$
	\FORALL{vertex $v$ neighbour of $u$ in $S$}
		\STATE $c(u,v^*)\Leftarrow c(u,v^*)+c(u,v)$
	\ENDFOR
	\STATE $c(v^*,u)\Leftarrow c(u,v^*)$
\ENDFOR
\end{algorithmic}
\label{algContrNetw}
\end{algorithm}

The optimal cooperation algorithm is running in strongly polynomial time. However the running time depends on the distance from the critical point and the different realizations of the random bonds. An estimation can be calculated as an average of the execution time of a system with a given size but different disordered realizations. This evaluation gives $\mathcal{O}(AN^{2.4})$ if the size $N$ of the graph is larger than a thousand.

The previous algorithm does not give any information about the number of optimal sets. However we can make some statements about this cardinality if we use the property that the union and the intersection of the optimal sets are optimal sets as well. If we calculate now the union and intersection of all optimal sets and those are equal, then there is only one unique optimal set \cite{dAuriac2004}.

\subsection{Exact calculation for random field Ising model at $T=0$}
\label{CalcofRFIM}
The minimum cut - maximum flow method is also applicable to find the exact ground state at $T=0$ of the random field Ising model (defined in Section \ref{RFIMDef}) on an arbitrary graph without thermalization problems \cite{dAuriac1985}. 

Allow the random field to admit positive and negative values. To define the network we need to introduce again two additional points, the sink and the source, and connect all sites $v$ in the original graph to the source $s$ if the local field $h_v>0$ and to the sink $t$ otherwise, if $h_v<0$. Then we define the capacities in the network as follows:
\begin{eqnarray}
c(u,v)&=&c(v,u) =  J \\
c(s,v) & = & h_v \\
c(v,t) & = & -h_v
\label{OSweightsRFIM}
\end{eqnarray}
After we calculate the minimum cut of the network, to get the corresponding spin configuration we set to $+1$ those spins which are belonging to the source $s$ and to $-1$ those which are in the set of sink. If the random field is discrete, maybe more than one minimum cut exists with the same value so the ground state can be degenerated. Following this method we can find the ground state of the random field Ising model with one minimum cut calculation in strongly polynomial time.

\section{Simulation of scale-free networks}

Many methods exist to simulate a scale-free network with power law degree distribution. One of the first defined model was the Barab\'{a}si-Albert model that we discussed in detail in the Section \ref{BAmodel}. This model describes dynamically growing network where the sites, added at each time step to the network, choose their neighbours from the already presented sites proportionally to their degree. In the corresponding algorithm (Alg.\ref{BAmodelAlg}) this rule of \textit{preferential attachment} means to recalculate the attaching probabilities $\Pi(k_i)$ (defined in Alg.\ref{BAmodelAlg}) of all sites at each time step which causes an $\mathcal{O}(AN^{2})$ execution time.

In the course of our work we applied another but equivalent algorithm to simulate scale-free networks. Instead of recalculating all the attaching probabilities at each step we introduced a degree map. This data structure is a container of data elements containing unique data pairs: a key number which represents a degree $k$ and a list of nodes $list(n_i(k_i=k))$ having this given degree. At each timestep $t$ we need to update this map only, in a way that the degree elements with zero list size are forbidden to be present. The size of this map is smaller than the system size if the $N$ is sufficiently large which can be explained with the special structure of the scale-free network. Such networks are non-homogeneous thus many sites have the same degree, consequently the number of different degrees in the network is much smaller than the number of sites. Following this consideration a possible algorithm can be defined as follows:

\begin{algorithm}
\caption{ScaleFreeNetwork($m_0,m,t$)}
\label{SFAlg}
\begin{algorithmic}[1]
\STATE Start from a fully connected graph with $m_0$ vertices
\STATE Create a $degreemap(k,list(n))$ initially with one element $degreemap(m_0-1,list(n))$
\FORALL{timestep $t$}
	\STATE Add a new node with $m(\le m_0)$ edges
		\FORALL{$m$ number of edges proceed from the new node}
			\STATE Choose an element $j$ from the $degreemap(k,list(n))$ with the probability $\Pi(k)=\frac{kN_k}{\sum_k kN_k}$
			\STATE Choose a node $n_l$ randomly from the list of $degreemap(j,list(n))$ to which the new node connects
			\STATE Erase $n_l$ from the list of $degreemap(j,list(n))$ and delete element $j$ if the list empty
			\STATE Create $degreemap(j+1,list(n))$ if does not exists and insert $n_l$ into its node list
		\ENDFOR
	\STATE Create element $degreemap(m,list(n))$ if does not exists and insert the new point into its node list
\ENDFOR
\end{algorithmic}
\end{algorithm}
Double edges are not allowed in the network and $N_k$ denotes the number of sites with degree $k$. Here the time to calculate the probability $\Pi(k)$ at each time step is in the order of $\mathcal{O}_{\Pi}(AN^{1/3})$ following from numerical averages. Consequently the execution time of the whole algorithm for large enough size is $\mathcal{O}(AN^{4/3})$ leads to a faster simulation than the original method.

It is easy to see that the Alg.\ref{BAmodelAlg} and the Alg.\ref{SFAlg} are equivalent. In the Alg.\ref{BAmodelAlg} at each step the nodes with the same degree have the same probability to receive a new edge and the sum of their probabilities must be $\sum_i\Pi(k_i)=1$. In the Alg.\ref{SFAlg} the situation is the same, but here the nodes with same degree and probability are collected into bulks, which is just a special case of the Alg.\ref{BAmodelAlg}. Then we define a probability $\Pi(k)=\frac{kN_k}{\sum_k kN_k}$ for each bulk, relational to the $N_k$ number of nodes in the bulk and to the common $k$ degree they have. These probabilities must be normalized $\sum_k\Pi(k)=1$ too. After all to choose an ending point of a new edge is equivalent in the two methods since to choose one node randomly with probability $\Pi(k_i)$ where nodes with the same probability are allowed or to choose randomly a bulk with probability $\Pi(k)$ belonging to the degree $k$ and then randomly select one node having this chosen degree is the same since $\sum_i\Pi(k_i=k)=\Pi(k)$ for each different degree $k$.

Consequently the two methods indicate the same type of network following a $P(k)\sim k^{-\gamma}$ power law degree distribution with an exponent $\gamma=3$ as we have proved in Section \ref{BAmodel}.

%% file: NetwContactProc.tex
\chapter{Non-equilibrium phase transitions and finite-size scaling in weighted scale-free networks} 

\section{Introduction}
\label{CPIntro}

Complex networks, which have a more complicated topology than regular lattices have been observed in a large class of systems in different fields of science, techniques, transport, social and political life, etc, as we have seen in Chapter \ref{ComplexNetworks}.
We described previously that such networks have connections between remote sites, and are ruled by small world effects. Moreover a non-democratic way of the distribution of the links characterizes the system which induces a degree distribution with a power-law tail (see Eq.\ref{PowerLawDegrDistr}). Hence these networks are called scale-free. In real networks the $\gamma$ degree exponent is generally: $2<\gamma<3$ which is usually the result of growth and preferential attachment as it was shown by Barab\'asi and Albert \cite{Barabasi1999}. They defined a scale-free network model which is capable to generate a similar kind of graph with an exponent $\gamma=3$, what we have discussed in detail in Section \ref{BAmodel}.

Since the agents of a network interact with one another, it is natural to ask about the cooperative behavior of the system. In particular if there exists some kind of thermodynamical phase and if there are singular points when the strength of the interaction or other suitable parameter is varied (such as the strength of disordering field, temperature, etc.). In this respect, static models \cite{Aleksiejuk2002,Leone2002,PhysRevE.66.016104,Bianconi2002,igloi-2002-66,Dorogovtsev2004} (Ising, Potts models, etc.), as well as non-equilibrium processes \cite{PhysRevLett.85.5468,PhysRevE.66.036113,Pastor-Satorras2001a,PhysRevE.66.016128} (percolation, spread of epidemics, etc.) have been investigated. However, generally non-weighted networks were considered, in which the strength of the interaction at each bond is constant. 

Studying cooperative processes on scale-free networks, due to long-range interactions a conventional mean-field behavior is expected to hold. However it turned out that it is true only if the network degree exponent is large enough, i.e. $\gamma>\gamma_u$, when the behavior of the system is similar to a high dimensional regular lattice. For the Ising model this upper critical exponent is $\gamma_u=5$, whereas for the percolation and epidemic spreading $\gamma_u=4$. As we decrease $\gamma$, the most connected nodes start to play an important role since they induce strong correlations in their neighbourhoods at short distance. At $\gamma=\gamma_u$ there are logarithmic corrections to the mean-field singularities, whereas for $\gamma_u>\gamma>\gamma_c$, where $\gamma_c$ is a lower threshold value, we arrive to the unconventional mean-field regime in which the critical exponents are $\gamma$ dependent. The effect of topology of scale-free networks becomes dramatic if $\gamma$ is lowered below the lower threshold value, $\gamma \le \gamma_c$, and the average of $k^2$, defined by $\langle k^2 \rangle=\int P(k) k^2 dk$, as well as the strength of the average interaction, becomes divergent. Then for any finite value of the interaction, the scale-free networks are in the ordered state, consequently there is no threshold value of any phase transition. Since $\gamma_c=3$, in realistic networks with homogeneous interactions, this type of phenomena should always occur.

Recently, much attention has been paid on weighted networks, in which the interactions are not homogeneous. Generally the strength of the interactions of highly connected sites are comparatively weaker than the average, that can be explained by technical or geographical limitations.  To model epidemic spreading, one should keep in mind that sites with a large coordination number are generally earlier connected to the network and in the long period of existence they have larger chance to be immunized\cite{PhysRevE.65.055103}.

An interesting class of degraded networks has been introduced recently by Giuraniuc et al\cite{PhysRevLett.95.098701} in which the strength of interaction in a link between sites $i$ and $j$ is rescaled as:
\be
\lambda_{i,j}=\lambda \frac{(k_i k_j)^{-\mu}}{\langle k^{-\mu}\rangle ^2}\;,
\label{lambda}
\ee
where $k_i$ and $k_j$ are the connectivities of the given sites. Here $0<\mu<1$ is the degradation exponent with which, if $\mu=0$, we recover the nondegraded network with homogeneous interactions $\lambda$. The properties of this type of network in equilibrium critical phenomena, in particular for the Ising model have been studied in detail in Ref\cite{PhysRevLett.95.098701}. Most interestingly the equilibrium critical behavior is found to depend only on one parameter, the effective degree exponent,
\be
\gamma'=(\gamma-\mu)/(1-\mu)\;,
\label{gammaeff}
\ee
thus topology and interaction seem to be converted. One important aspect of the degraded network in Eq.\ref{lambda} is that phase transitions in realistic networks with $\gamma \le 3$ are also possible, if the degradation exponent is sufficiently large , $\mu>(3-\gamma)/2$. Therefore theoretical predictions about critical singularities can be confronted with the results of numerical calculations.

In this chapter we study nonequilibrium phase transitions in weighted networks. Our aim with these investigations is twofold. First, we want to check if the simple reparametrization rule in Eq.\ref{gammaeff} stays valid for nonequilibrium phase transitions, too. For this purpose we make dynamical mean-field calculations and perform large scale Monte Carlo simulations. Our second aim is to study the form of finite-size scaling in nonequilibrium phase transitions in weighted scale-free networks. To analyze our numerical results we use recent field-theoretical calculations\cite{PhysRevE.72.016119} in which finite-size scaling in Euclidean lattices above the upper critical dimension has been studied. In the conventional mean-field regime of scale-free networks, analogous scaling relations are expected to apply.

The structure of this chapter is the following. First we introduce a reaction-diffusion model, the contact process, which is the prototype of a non-equilibrium phase transition belonging to the universality class of directed percolation. Then we give the dynamical mean-field solution of the recent model, and in the following section finite-size scaling theory is shown. Finally results of Monte Carlo simulations of the contact process on weighted Barab\'asi-Albert networks are presented.

\section{Reaction-diffusion models - The Contact Process}

Reaction-diffusion systems are mathematical models to describe how the concentration of one substance or more  distributed in space, changes under the influence of two processes: local reactions in which the substances are converted into each other, and diffusion which causes the substances to spread out in space. These kind of descriptions are naturally applied in chemistry but also capable to describe non-chemical dynamical processes like in biology, physics or information science etc.

These non-equilibrium processes are described by stochastic models where the reactions and diffusions are defined as probabilistic transition rules. The state of the system is characterized by $w_{\mu}(t)$ which gives the probability distribution that it is in state $\mu$ at time $t$. It can be shown that in the limit of very large system sizes, the temporal evolution of the probability distribution $w_{\mu}(t)$ evolves deterministically according to a master equation (see Eq.\ref{ME}). Since the state of the system at time $t$ depends only on the previous state, the master equation describes a Markov process. Using a matrix notation the master equation may be written of the form
\begin{equation}
\partial_t |w_{\mu}(t) \rangle=-\mathcal{L}|w_{\mu}(t) \rangle
\end{equation}
where $|w_{\mu}(t) \rangle$ denotes a vector with the components of probabilities in $w_{\mu}(t)$ \cite{Hinrichsen2000}. The temporal evolution is governed by the Liouville operator defined through the matrix elements as:
\begin{equation}
\langle \mu|\mathcal{L}|\nu\rangle=-R_{\mu\rightarrow \nu}+\delta(\mu,\nu)\sum_{\omega}R_{\mu\rightarrow \omega}
\end{equation}
where the coefficient $R_{\mu\rightarrow \nu}$ denotes the state transition rate from state $\mu$ to $\nu$ and $\delta(\mu,\nu)=1$ if only $\mu=\nu$ otherwise it is zero. A formal solution of the master equation is given by $|w_\mu(t)\rangle=\mbox{exp}(-\mathcal{L}t|w_{\mu}(0)\rangle)$ where $|w_{\mu}(0)\rangle$ denotes the initial probability distribution at $t=0$. Therefore, in order to determine $|w_\mu(t)\rangle$, the Liouville operator has to be diagonalized which is usually a nontrivial task.

In the following we will focus on simple reaction-diffusion lattice model with only one type of particles. It can be recognized as simple two-state model since each site can either be occupied (infected) by a particle $A$ or empty (healthy) $\varnothing$. The reaction in the system is defined in terms of transition rates of two types: ($i$) an occupied site can infect its empty neighbours with a rate $\lambda$, ($ii$) any infected sites may recover with a rate $\kappa$ decreasing the total amount of spreading agent. The competition of this two processes characterizes the behaviour of such spreading processes. Depending on the relative rates $\Delta=\lambda/\kappa$ which controls the balance between infection and recovery, the system can evolve into two final states. If $\Delta>\Delta_c$ the infection process dominates, the epidemic disease will spread over all the systems, reaching a stationary state in which the infection and recovery balance one another. However, if $\Delta<\Delta_c$ the recovery dominates and the total amount of spreading agents continues to decrease and eventually vanishes. Then the system is trapped to a completely inactive state, the so-called absorbing state. Since this absorbing phase can be reached but not be left, the system cannot satisfy the condition of detailed balance so it is driven by a non-equilibrium process. The transition between the stationary infected and absorbing phase is continuous and characterized by universal behaviour \cite{Odor2003}. The simplest model exhibiting such transition is the directed case of percolation model (see Section \ref{percolation}) which denominate the related universality class of directed percolation (for a review see \cite{Odor2004,Odor2008}).

In directed percolation an orientation is defined for each percolating edge of the lattice which determines the direction that the spreading agent can only move along (like water in porous medium in gravitation field). A phase transition occurs for directed percolation which is similar in many respects to the transition of isotropic case. The two phases are defined by the existence of an infinite percolating cluster and controlled by a probability $p$. It has a critical value $p_c$ when the percolating cluster appears and the order parameter like percolation probability becomes $\Theta(p\geq p_c)>0$ as we have discussed in Section \ref{percolation}. However here $p_c$ is larger than for the isotropic case and the critical behavior is also different. Instead of having rotational invariant long-range correlation, duality symmetry and rational critical exponents, here the critical properties present anisotropy in space leading to different critical exponents. The numerical values of exponents are not known analytically but seem to be irrational numbers.

\subsection{The Contact Process}

\begin{figure}[htb]
\begin{center}
\includegraphics*[ width=13cm]{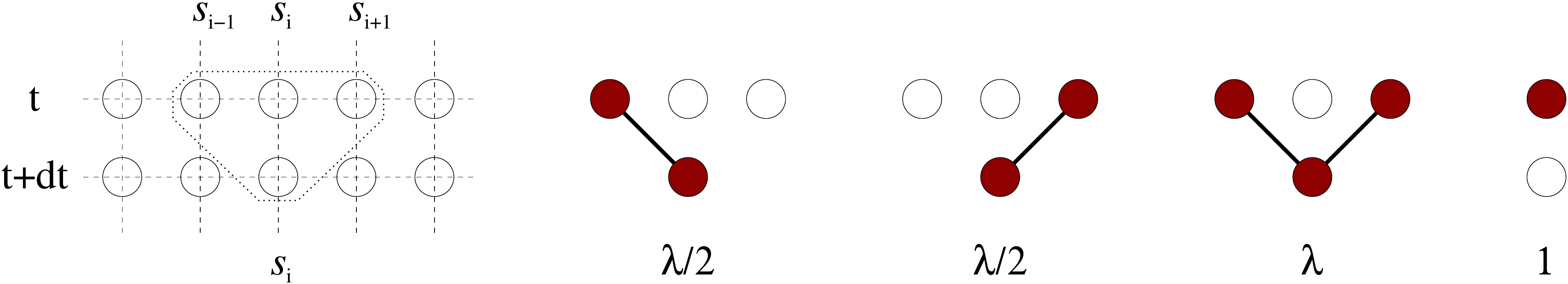}
\caption{Contact process on 1+1 dimensional lattice. The empty point at $s_i$ denote the inactive point which becomes infected by its active neighbours $s_{i-1}$ or $s_{i+1}$ with a rate $\lambda/2$ per site. If $s_i$ is infected (fullfield point) then it recovers with a probability 1 in one time iteration \cite{Hinrichsen2000}.}
{
\label{CPfig}
}
\end{center}
\vspace{-10pt}
\end{figure}

The contact process is a lattice model belonging to the universality class of directed percolation, first introduced by Harris \cite{Harris1974,Marro1999} as an infection spreading model without immunization. It is defined on a graph, whose site can be either active ($s_i(t)=1$) or inactive ($s_i(t)=0$). During the time evolution we use a random sequential update and pick up a site $i$ at each step randomly. At the next $t+dt$ iteration step the state of the chosen site $s_i(d+dt)=0,1$ depends on its previous state $s_i(t)$ at time $t$ and the number of its active first neighbours $n_i(t)=\sum_{\left\langle i,j\right\rangle }s_j(t)$. The new state is assigned according to certain transition rates $R(s_i(t)\rightarrow s_i(t+dt),n_i(t))$ which are defined for the standard contact process as:
\begin{eqnarray}
R(0\rightarrow 1,n_i) &=& \lambda n/k_i \\
R(1\rightarrow 0,n_i) &=& \kappa=1
\end{eqnarray}
where $k_i$ denotes the degree of the $i$th site (see Fig.\ref{CPfig}). We set the recovery probability $\kappa=1$ so the infection probability $\lambda$ controls the spreading process. 

We get the simplest representation of the model at $1+1$ dimension, when the underlying graph is a chain. Then the critical point is located at $\lambda_c=3.29785(8)$ suggested by Monte Carlo simulations and series expansions \cite{PhysRevE.58.4266,Jensen1993,PhysRevLett.67.2391}. The master equation of the $1+1$ dimensional contact process with periodical boundary condition and size $N$ can be written as \cite{Hinrichsen2000}:
{\setlength\arraycolsep{0.1em}
\begin{eqnarray}
\partial_t w_{t}(s_1,...,s_N)=\sum_i^N(2s_i-1)&\big\{&\lambda_{s_{i-1}}w_t(s_1,...,s_{i-2},1,0,s_{i+1},...,s_N) \nonumber\\[-8pt] 
					      &&+\lambda_{s_{i+1}}w_t(s_1,...,s_{i-1},0,1,s_{i+2},...,s_N) \nonumber\\
					      &&-w_t(s_1,...,s_{i-1},1,s_{i+1},...,s_N) \big\}
\end{eqnarray}
}
where $w_t(s_1,...,s_N)$ denotes the probability that the system is in the configuration $\{s_1,...,s_N\}$ at time $t$. Here we involved only two-site interactions. Then the corresponding Liouville operator $\mathcal{L}_{CP}=\sum_i\mathcal{L}_i$ is given by:
\[ \mathcal{L}_i =  \dfrac{1}{2} \left( \begin{array}{cccc}
0 & -1 & -1 & 0 \\
0 & 1+\lambda & 0 & -1\\
0 & 0 & 1+\lambda & -1 \\
0 & -\lambda & -\lambda & 2 \end{array} \right)\] 

In higher dimensional lattice systems or in complex networks the master equation becomes as complicated as it is not solvable analytically. Then approximation methods become appropriate to describe the critical behavior of the model. A corresponding method arises from mean field theory what we will apply to give a dynamical solution of the contact process in scale-free networks.

\section{Dynamical mean field solution of contact process in scale-free networks}
\label{DynMF}
In many cases the macroscopic properties of reaction-diffusion processes can be predicted by solving the corresponding mean field theory. If the system contains a large number of particles and the diffusion is dominating over the reaction then this solution gives a relevant approximation and becomes exact in infinite dimensions. Here we present a dynamical mean field solution of the contact process \cite{Harris1974,Hinrichsen2000} as a dynamical adaptation of the Weiss molecular field theory. We apply this model on a scale-free network with power-law degree distribution characterized by a degree exponent $\gamma$ as we defined in Section \ref{BAmodel}. This solution is expected to be exact, due to long-range interactions in the system. We initially assume that the infected agents are homogeneously distributed all over the network and ignore any spatial correlation as well as instabilities with respect to inhomogeneous perturbations.

We start with a set of equations of the time derivative of the mean active site density, $\rho_i$, at site, $i=1,2,\dots,N$:
\be 
\frac{\partial \rho_i}{\partial t}= \sum_j \lambda_{i,j}(1-\rho_i) \rho_j - \rho_i\;,
\label{mf}
\ee
where the correlations between densities of different sites are omitted. Here $\lambda_{i,j}$ gives the strength of infection, the $(1-\rho_i)$ term is the probability that the $i$th site is empty and the $\rho_j$ gives the probability that the $j$th site is infected. The last $-\rho_i$ term represents the recovery in the equation. In the next step following the spirit of the Weiss molecular mean-field approximation we replace the interactions using Eq.\ref{lambda}:
\be
\lambda_{i,j}=\lambda \frac{k_i k_j}{\sum_j k_j}\frac{(k_i k_j)^{-\mu}}{\langle k^{-\mu}\rangle^2}\;,
\label{lambda_mf}
\ee
i.e. there is an interaction between each site having the (mean) value, which is proportional to the probability of the existence of the current bond ($P(k_ik_j)=k_ik_j/\sum_j k_j$).  Now in terms of an average density:
\be
\rho=\frac{\sum_j k_j^{1-\mu} \rho_j}{\sum_j k_j^{1-\mu}}\;
\label{rho}
\ee
we can rewrite the dynamical equation as:
\begin{eqnarray}
\frac{\partial \rho_i}{\partial t} &=& \sum_j \lambda \frac{k_i k_j}{\sum_j k_j} \frac{(k_ik_j)^{-\mu}}{\langle k^{-\mu}\rangle^2} (1-\rho_i) \rho_j - \rho_i \\ 
&=& \lambda k_i^{1-\mu}\sum_j \frac{k_j^{1-\mu}}{\sum_j k_j\langle k^{-\mu}\rangle^2} (1-\rho_i) \rho_j - \rho_i
\end{eqnarray}
Then we multiply the first term on the right hand side with $\sum_j k_j^{1-\mu}/\sum_j k_j^{1-\mu}$ and rewrite the equation of the form:
\begin{eqnarray}
\frac{\partial \rho_i}{\partial t} &=& \lambda k_i^{1-\mu} \sum_j \frac{k_j^{1-\mu}}{\sum_j k_j\langle k^{-\mu}\rangle^2} (1-\rho_i) \dfrac{k_j^{1-\mu}\rho_j}{k_j^{1-\mu}} - \rho_i \\
&=& \lambda k_i^{1-\mu} \sum_j \frac{k_j^{1-\mu}}{\sum_j k_j\langle k^{-\mu}\rangle^2} (1-\rho_i) \rho - \rho_i
\end{eqnarray}
where we used a substitution with $\rho$ defined in Eq.\ref{rho}. If we multiply again the first term with $N/N$, we arrive to the form:
\begin{eqnarray}
\frac{\partial \rho_i}{\partial t} \hspace{.1in} = \hspace{.1in}  k_i^{1-\mu} \lambda \frac{ \langle k^{1-\mu} \rangle}{\langle k \rangle \langle k^{-\mu}\rangle^2} (1-\rho_i) \rho - \rho_i
\hspace{.1in} = \hspace{.1in} k_i^{-\mu} \widetilde{\lambda}(1-\rho_i) \rho - \rho_i
\end{eqnarray}
with $\tilde{\lambda}=\lambda \langle k^{1-\mu}\rangle /\langle k\rangle \langle k^{-\mu}\rangle^2$.

In the stationary state, $\partial \rho_i/\partial t=0$, the local densities are given by:
\begin{equation}
\rho_i=\frac{\widetilde{\lambda} k_i^{1-\mu}\rho}{1+\widetilde{\lambda} k_i^{1-\mu} \rho}
\label{rho_i} 
\end{equation}
i.e. they are proportional to $k_i^{1-\mu}$. Putting $\rho_i$ from Eq.\ref{rho_i} into Eq.\ref{rho} we obtain an equation for $\rho$:
\begin{eqnarray}
\rho=\frac{1}{\sum_i k_i^{1-\mu}} \sum_i k_i^{1-\mu}\frac{\widetilde{\lambda} k_i^{1-\mu}\rho}{1+\widetilde{\lambda} k_i^{1-\mu} \rho}
\end{eqnarray}
which can be rearranged as:
\begin{eqnarray}
\sum_i k_i^{1-\mu} \hspace{.1in} = \hspace{.1in} \frac{1}{\rho}\sum_i k_i^{1-\mu}\frac{\widetilde{\lambda} k_i^{1-\mu}\rho}{1+\widetilde{\lambda} k_i^{1-\mu} \rho}
\hspace{.1in} = \hspace{.1in} \widetilde{\lambda}\sum_i\frac{k_i^{2(1-\mu)}}{1+\widetilde{\lambda}k_i^{1-\mu}\rho}
\end{eqnarray}
Now if we take the continuous limit thus the summation over $i$ is replaced by an integration over the degree distribution, $P_D(k)$, and choose the upper limit of the integration for $k_{max} \to \infty$ (thermodynamic limit) then:
\begin{equation}
\langle k^{1-\mu}\rangle =\widetilde{\lambda} \int_{k_{min}}^{k_{max}} P_D(k) \frac{k^{2(1-\mu)}}
{1+\widetilde{\lambda} k_i^{1-\mu} \rho} {\rm d} k
\label{rho1}
\end{equation}
The solution of Eq.\ref{rho1} in the vicinity of the transition point, $\rho \ll 1$, depends on the large-$k$ limit of the degree distribution, defined in Eq.\ref{PowerLawDegrDistr}. This solution is given in terms of the integration variable, $k'=k^{1-\mu}$, as:
\begin{equation}
\langle k' \rangle=\widetilde{\lambda} A \int_{k'_{min}}^{k'_{max}} k'^{-\gamma'} \frac{k'^{2}}
{1+\widetilde{\lambda} k' \rho} {\rm d} k' \equiv Q(\rho,\gamma'),\quad \rho \ll 1 \;,
\label{rho2}
\end{equation}
where $\gamma'$ was defined in Eq.\ref{gammaeff}. Note, that the functional form of the Eq.\ref{rho2} is identical to that for standard scale-free networks, just with an effective degree exponent, $\gamma'$.

To analyze the solution of Eq.\ref{rho2} we apply the method in \cite{igloi-2002-66}, which is somewhat different from the original method in \cite{Pastor-Satorras2001}. For small $\rho$, $Q(\rho,\gamma')$ in Eq.\ref{rho2} can be expanded in a Taylor series at least up to the term with $\sim \rho^2$ as:
\begin{equation}
Q(\rho,\gamma')=\widetilde{\lambda} A \int_{k'_{min}}^{k'_{max}} k'^{-\gamma'} k'^{2} (1-\widetilde{\lambda}k'\rho + (\widetilde{\lambda}k'\rho)^2-...) {\rm d} k'
\end{equation}
where the first three terms can be written of the form:
\begin{eqnarray}
a_1(\rho) &=& \widetilde{\lambda}A\int_{k'_{min}}^{k'_{max}} k'^{-\gamma'} \dfrac{ k'^{2}}{1+\widetilde{\lambda} k' \rho} {\rm d} k' \Bigg\rvert_{\rho=0}\\
a_2(\rho) &=& -\widetilde{\lambda}^2 \rho A\int_{k'_{min}}^{k'_{max}} k'^{-\gamma'} \dfrac{ k'^{3}}{1+\widetilde{\lambda} k' \rho} {\rm d} k' \Bigg\rvert_{\rho=0}\\
a_3(\rho) &=& \widetilde{\lambda}^3 \rho^2 A\int_{k'_{min}}^{k'_{max}} k'^{-\gamma'} \dfrac{ k'^{4}}{1+\widetilde{\lambda} k' \rho} {\rm d} k' \Bigg\rvert_{\rho=0}
\end{eqnarray}

Following from these terms we can detect three range of $\gamma'$ where the critical behavior is different:
\begin{itemize}
\item $\gamma' >4$

The first term of the series expansion is $\langle k'\rangle=\widetilde{\lambda} A \langle k'^2\rangle$ consequently there is a finite transition point,   $\tilde{\lambda}_c=\langle k'\rangle /A\langle k'^2\rangle $, and the density in the vicinity of the transition point behaves as: $\rho(\lambda) \sim (\lambda-\lambda_c)$. This is the conventional mean-field regime since the degree exponent is larger than the upper critical value $\gamma'>\gamma'_u(=4)$ and the critical behavior on the scale-free network is the same as on a complete graph. At the borderline case, $\gamma'=4$, there are logarithmic corrections to the mean-field singularities.

\item $3 < \gamma' < 4$
  
Here the effect of the connectivity of the scale-free network is relevant, so the singularities of the thermodynamical quantities of the system are different from the conventional mean field behavior. For small $\rho$ only the linear term in the Taylor expansion of $Q(\rho,\gamma')$ exists and the $\rho$-dependence of the next term becomes singular. The critical behaviour in this regime is due the interplay between the regular linear first term (which exists since $\gamma'>3$) and the negative singular next-to-leading term in the expansion.

The $\rho$-dependence can be estimated by noting that for a small, but finite $\rho$ there is a cut-off value, $\widetilde{k'} \sim 1/\rho$, so that
\be
a_2(\rho) \sim -\widetilde{\lambda}^2 \rho  A \int_{k'_{min}}^{\widetilde{k'}} k'^{-\gamma'} k'^{3} {\rm d} k'
\sim \rho^{\gamma'-3}\;.
\label{a22}
\ee
Consequently the density at the transition point behaves anomalously,
\be
\rho \sim (\lambda-\lambda_c)^{\beta},\quad \beta=1/(\gamma'-3)\;.
\label{unconv}
\ee
This is the unconventional mean-field region where the critical behaviour depends on $\gamma'$.

\item $\gamma' < 3$ 

In this case $Q(\rho,\gamma)$ is divergent for small $\rho$. Its behavior can be estimated as in Eq.\ref{a22} leading to $Q(\rho,\gamma') \sim \lambda^{\gamma'-2} \rho^{\gamma'-3}$. Consequently the system for any non-zero value of $\lambda$ is in the active phase. As $\lambda$ goes to zero the density vanishes as:
\begin{equation}
\rho(\lambda) \sim \lambda^{(\gamma'-2)/(3-\gamma')}\;.
\label{ord}
\end{equation}
Here at the border, $\gamma'=3$, the system is still in the active phase, but the density is related to a small $\lambda$ as: $|\ln (\rho\lambda)| \sim 1/\lambda$.
\end{itemize}
Before we confront these analytical predictions with the results of numerical simulations we discuss the form of finite-size scaling in scale-free networks. 

\section{Finite-size scaling for $d>d_c$}
\label{sec:fss}

In a numerical calculation, such as in Monte Carlo (MC) simulations, one generally considers systems of finite extent and the properties of the critical singularities are often deduced via finite-size scaling. As we have seen in Section \ref{FiniteSC} it is known in the phenomenological theory of equilibrium critical phenomena that due to the finite size of the system, $L$, critical singularities are rounded and their position is shifted\cite{Barber1984}. As it is elaborated for Euclidean lattices finite-size scaling theory has different forms below and above the so-called upper critical dimension, $d_c$. For $d<d_c$ in the scaling regime the singularities are expected to depend on the ratio, $L/\xi$, where $\xi$ is the spatial correlation length in the infinite system\cite{PhysRevLett.28.1516}. This case has been discussed in Section.\ref{FiniteSC}. On the other hand for $d > d_c$, when mean-field theory provides exact values of the critical exponents, following from renormalization group treatments, the finite-size scaling theory involves dangerous irrelevant scaling variables\cite{Fisher1974}, which results in the breakdown of hyperscaling relations. For equilibrium critical phenomena predictions of finite-size scaling theory\cite{Brezin1982,Brezin1985} above $d_c$ are checked numerically, but the agreement is still not satisfactory\cite{PhysRevLett.76.1557,Chen1998,Luijten1999}.

For non-equilibrium critical phenomena finite-size scaling above $d_c$ is considered first in the frame of scaling theory \cite{PhysRevLett.49.478} and for the percolation process \cite{Aharony1984} and applied also for a coupled directed percolation system \cite{PhysRevLett.94.145702}. A field-theoretical derivation of the results has been made only very recently \cite{PhysRevE.72.016119,Janssen2007} and here we recapitulate the main findings of the analysis. For directed percolation, which represents a broad class of universality \cite{Hinrichsen2000}, dangerous irrelevant scaling variables are identified in the fixed point.  As a consequence scaling of the order-parameter is anomalous:
\begin{equation}
\rho=L^{-\beta/\nu^*} \widetilde{\rho}(\delta L^{1/\nu^*}, h L^{\Delta/\nu^*})\;,
\label{fss}
\end{equation}
Here, $\delta$, is the reduced control parameter, with the notations of Section \ref{DynMF} $\delta=(\lambda-\lambda_c)/\lambda_c$ and $h$ is the strength of an ordering field. The critical exponents, $\beta=1$ and $\Delta=2$, are the same as in conventional mean-field theory. However, the finite-size scaling exponent is given by, $\nu^*=2/d$, and thus depends on the spatial dimension, $d$. Note, that below $d_c=4$ it is the correlation length exponent, $\nu$, which enters into the scaling expression in Eq.\ref{fss}, but above $d_c$, due to dangerous irrelevant scaling variables it should be replaced by $\nu^*$. At the critical point, $\delta=0$, the scaling function, $\tilde{\rho}(0,x)$, has been analytically calculated and checked by numerical calculations.

In the following we translate the previous results for complex network, in which finite-size scaling is naturally related to the volume of the network via the relation $N \leftrightarrow L^d$, where $N$ is the number of sites. We examine two different cases in the conventional mean-field regime with respect to a site with average or maximum degree. For a typically connected site, i.e. with a coordination number , $k \sim \langle k \rangle$, we arrive from Eq.\ref{fss} to the finite-size scaling prediction:
\begin{equation}
\rho_{typ}=N^{-\beta/\omega} \widetilde{\rho}_{typ}(\delta N^{1/2}, h N^{\Delta/2})\;,
\label{fss1}
\end{equation}
where $\omega$ is defined as the correlation volume critical exponent $\omega=d\nu^*$. On the other hand for the maximally connected site with $k_{max} \sim N^{1/(\gamma-1)}$ \cite{newman-2003-45} according to Eq.\ref{rho_i} the finite-size scaling form is modified by:
\begin{equation}
\rho_{max}=N^{-\beta/\omega+(1-\mu)/(\gamma-1)} \widetilde{\rho}_{max}(\delta N^{1/2}, h N^{\Delta/2})\;.
\label{fss_max}
\end{equation}
Since in the derivation of the relation in Eq.\ref{fss} the actual value of $\beta$ has not been used, we conjecture that the results in Eqs.\ref{fss1} and \ref{fss_max} remain valid in the unconventional mean-field region, too. The suggestions above expected to hold for scale-free networks, however those cannot be exact even in the conventional mean-field regime since the system size is finite.

\section{Monte Carlo simulation}
\label{sec:mc}

Next we are going to review some algorithmic methods and results of Monte Carlo simulations in order to study numerically the previously presented theoretical predictions. First we represent the contact process in an algorithmic manner, then we describe our results of numerical simulations about its critical behaviour and confront it with the theory. 

\subsection{Implementation of Contact Process}

As we have seen before the contact process is a lattice model which is governed by two processes, the infection what is spreading site-to-site and the recovery which retard the infestation during the dynamic. Following this property the related algorithm can be decomposed in two independent parts in one time step, which gives a possibility to parallelize the algorithm. 

As a first step we read the network, calculate the weight degradations $\lambda_{i,j}$ for each edge as it was defined in \ref{lambda} and choose an initial site to infect (see Alg.\ref{CPAlg}). During the method we contain the sick nodes in a list \textit{InfectedList()} that we update at every iteration. At each time step we check all the nodes $i$ in \textit{InfectedList()} and try to infect their neighbours $j$ with a rate $\lambda_{i,j}$. The newly infected sites are stored in a separated list \textit{NewInfectedList()}. At the same time step we recover all the occupied nodes in \textit{InfectedList()} with a rate $\kappa$ and delete them from the list if they become healthy again. Finally we merge the \textit{NewInfectedList()} and the rest of \textit{InfectedList()} to produce the new set of infected nodes for the next time step.
\begin{algorithm}
\caption{ContactProcess()}
\begin{algorithmic}[1]
\STATE Initialization
\STATE Choose an initial point and infect it
\FORALL{timestep $t$}
	\FORALL{nodes $i$ in \textit{InfectedList()}}
		\FORALL{neighbours $j$ of $i$}
			\IF{Infection($\lambda_{i,j}$,$i$,$j$)}
				\STATE \textit{NewInfectedList.insert(j)}
			\ENDIF
		\ENDFOR		
	\ENDFOR
	\FORALL{nodes $i$ in \textit{InfectedList()}}
		\IF{Recover($\kappa$,$i$)}
			\STATE \textit{InfectedList.delete($i$)}
		\ENDIF
	\ENDFOR
	\STATE \textit{InfectedList()}$\Leftarrow$ Merge(\textit{InfectedList()},\textit{NewInfectedList()})
\ENDFOR
\end{algorithmic}
\label{CPAlg}
\end{algorithm}

The running time of the algorithm can be approximate with the function $f(N)\simeq t(N\langle k\rangle +N)$ which is strongly influenced by system properties like the degree distribution. During our calculation we applied the process on scale-free networks with average degree $\langle k\rangle\simeq 1$ but we let the system running for $t\simeq 2N$ time steps, consequently the execution time was $\mathcal{O}(4N^2)$.

\subsection{Numerical results}

In the actual calculation we considered the contact process on the Barab\'asi-Albert scale-free network\cite{Barabasi1999}, which has a degree exponent, $\gamma=3$, and used a degradation exponent, $\mu=1/2$. Consequently from Eq.\ref{gammaeff} the effective degree exponent is $\gamma'=5$, thus conventional mean-field behavior is expected to hold. (We note that the same system is used to study the equilibrium phase transition of the Ising model in Ref.\cite{PhysRevLett.95.098701}.) Networks of sites up to $N=4096$ are generated using the Alg.\ref{SFAlg} by starting with $m_0=1$ node and having an average degree: $\langle k \rangle=2$.  Results are averaged over typically $10000$ independent realizations of the networks.

In the calculation we started with a single particle at site $i$, (which was either a typical site or the maximally connected site) and let the process evolve until a stationary state is reached in which averages become time independent. In a finite system in particular in the vicinity of the critical point this state is only quasi-stationary and has a finite life-time before in the system all particles are annihilated. In order to obtain a true (quasi-)stationary average value we have considered only the samples which have survived up to the time where the average is done \cite{PhysRevA.41.5294}. In the calculation we monitored the average value of the occupation number, $\rho_i$, as introduced in mean-field theory in Eq.\ref{mf}, and the fraction of occupied sites, $m_i$, (order parameter) as a function of the control parameter $\lambda=\lambda/\kappa$, whereas $\kappa=1$ was set to be unity.  In a regular lattice in the stationary state and in the thermodynamic limit $\rho_i$ and $m_i$ are the same quantities. In a complex network, however, in which the $\rho_i$ are position dependent they are not strictly equivalent. Their singular  behavior in the vicinity of the transition point, however, is expected to be the same. Indeed at the critical point the fraction of active sites is vanishing with $N$, see in Eqs.\ref{fss1} and \ref{fss_max}, therefore starting from a typical site the probability of the infection of a largely connected site is vanishing.

\subsection{The $\lambda_c$ phase transition point and $x$ finite-size scaling exponent}

\begin{figure}[htb]
\begin{center}
\includegraphics*[width=9.0cm]{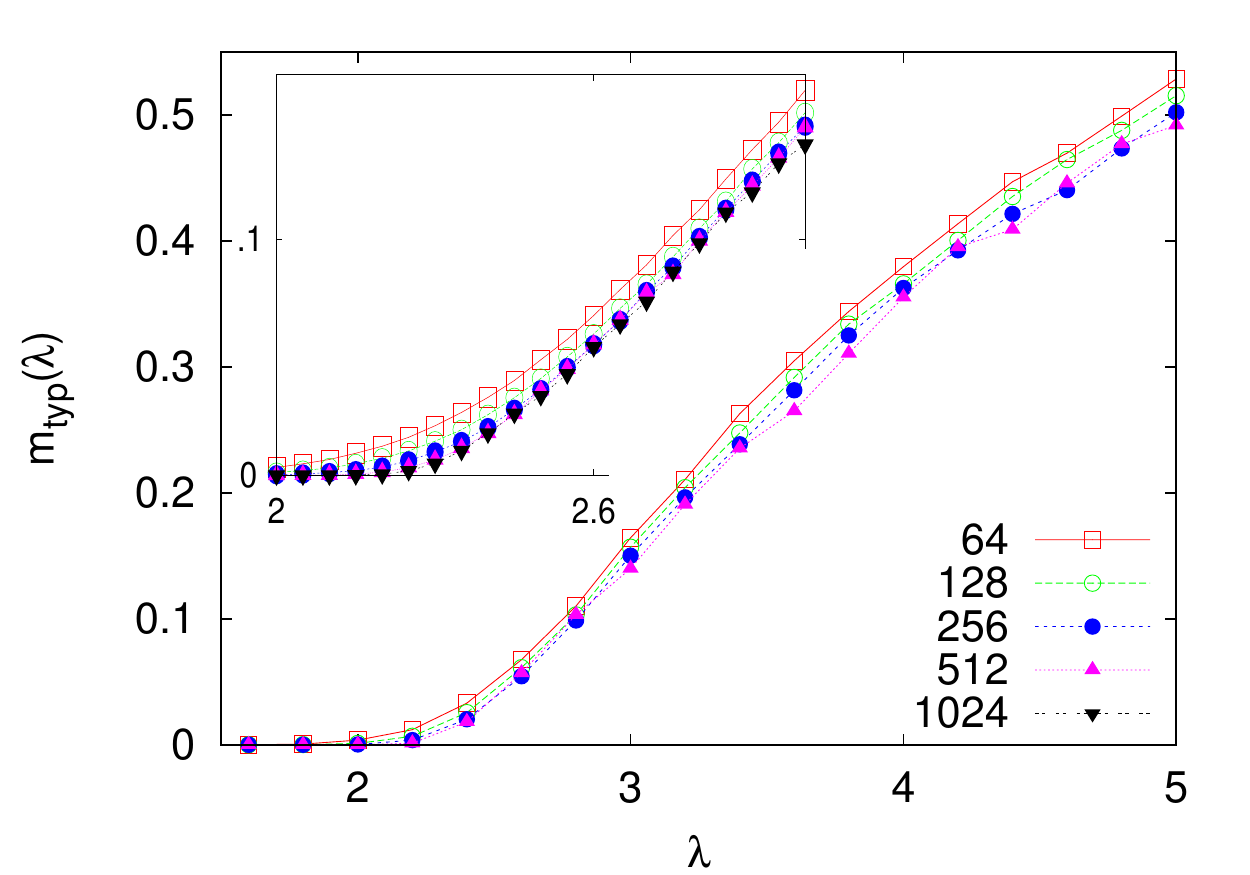}
\caption{Variation of the order parameter at a typical site as a function of the creation rate, $\lambda$. Inset: Enlargement in the critical region.}
\label{CPfig1}
\end{center}
\vspace{-15pt}
\end{figure}

\begin{figure}[htb]
\begin{center}
\includegraphics*[width=9.0cm]{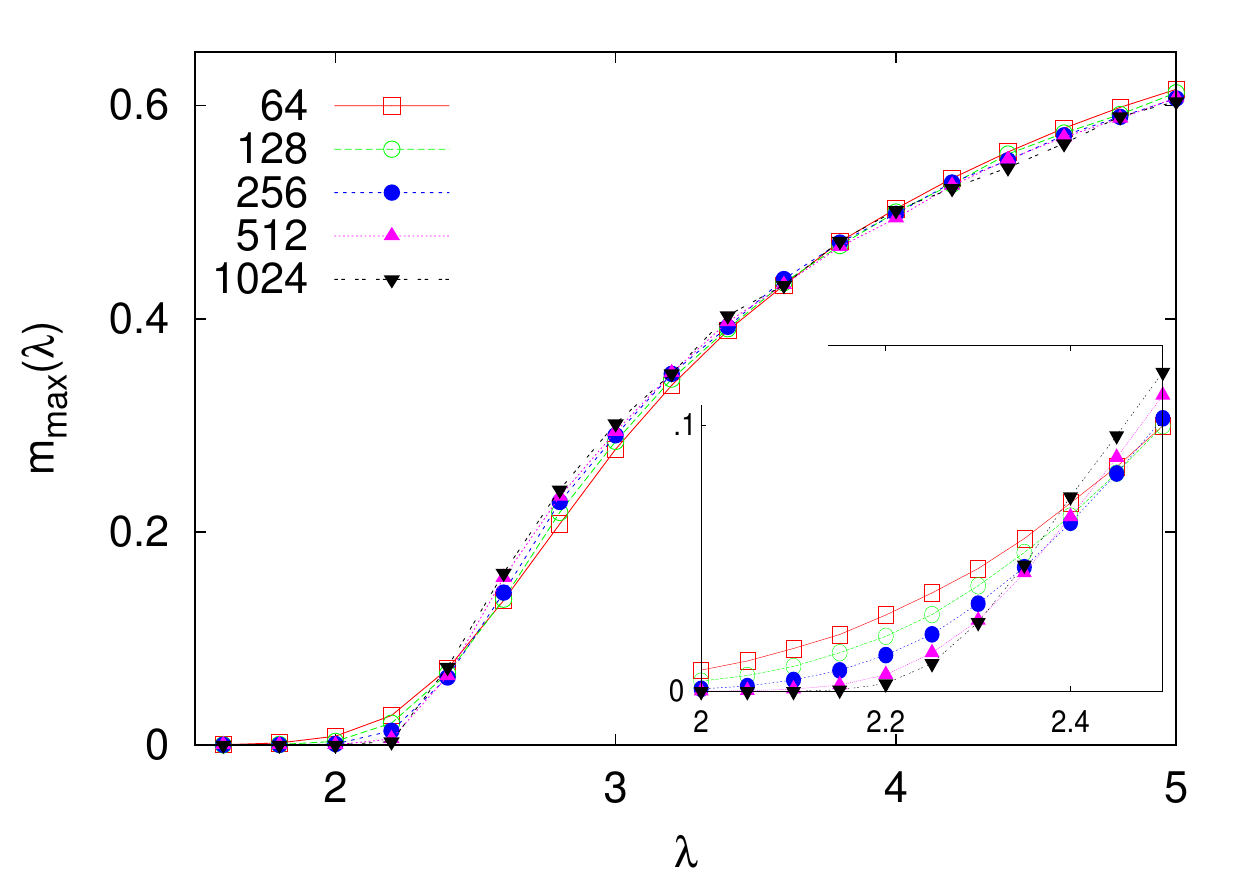}
\caption{As in Fig.\ref{CPfig1} for the maximally connected site.}
\label{CPfig2}
\end{center}
\vspace{-15pt}
\end{figure}

The $\lambda$ dependence of the order parameter is shown in Fig.\ref{CPfig1} for a typical site and in Fig.\ref{CPfig2} for the maximally connected site. Evidently there is a phase transition in the system in the thermodynamic limit between the absorbing state and a stationary infected state, which is rounded by finite size effects as shown in the insets of Figs. \ref{CPfig1} and \ref{CPfig2}.

\begin{figure}[htb]
\begin{center}
\includegraphics*[width=14.5cm]{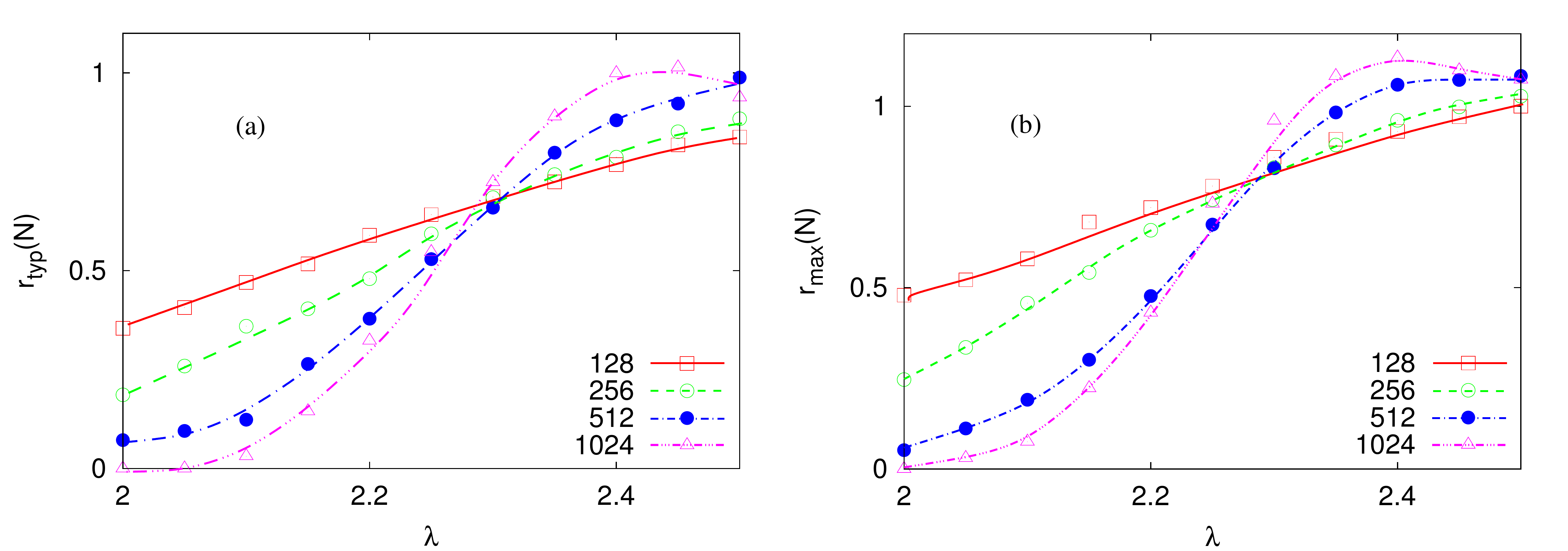}
\caption{The ratio $r(N)=m(N)/m(N/2)$ a) for a typical site, b) for the maximally connected site. Note that the location of the crossing points, which defines $\lambda_c$ is the same in the two cases, whereas the value at crossing, which is related to the finite-size exponent, $x$, through $r(N,\lambda_c)=2^{-x}$ is different.} 
\label{CPfig3}
\end{center}
\vspace{-15pt}
\end{figure}

As we have seen in Eq.\ref{OrdPScale} the order parameter at the vicinity of the critical point of continuous phase transitions is scaling as a power law which arises of the form following from \ref{sec:fss} as:
\begin{equation}
m(t,N)\sim N^{-x}\widetilde{m}(N^{1/\omega}t)
\label{MFmScale}
\end{equation}
where $x=\beta/\omega=\beta/d\nu^*$ denotes the finite-size exponent as given in scaling theory in Eqs.\ref{fss1} and \ref{fss_max}. At the critical point the magnetization can be written as $m(N,t=0)\sim A N^{-x}$ where $A$ is a constant. Consequently the ration of magnetization of the systems with size $N$ and  $N/2$ can be written in the form:
\begin{equation}
r(N)=\dfrac{m(t=0,N)}{m(t=0,N/2)}=\dfrac{N^{-x}}{(N/2)^{-x}}=2^{-x}
\end{equation}
where the order parameter curves for different $N$ cross each other and the crossing point can be used to identify $\lambda_c$ and exponent $x$ through extrapolation. To locate the phase transition point we form this ratios for different finite sizes. As shown in Fig.\ref{CPfig3} $r(N)$ tends to zero in the inactive phase, $\lambda < \lambda_c$, and tends to a value of one in the active phase, $\lambda > \lambda_c$. 

The transition point is found to be the same within the error of the calculation both for a typical site and for the maximally connected site and given by: $\lambda_c=2.30(1)$. The finite-size scaling exponent, however, calculated from $r(N,\lambda_c)$ is different in the two cases. For a typical site we estimate: $x_{typ}=0.54(7)$, which should be compared with the field-theoretical prediction in Eq.\ref{fss1}, which is $x_{typ}=\beta/2=1/2$. In the maximally connected site the finite-size scaling exponent is measured as $x_{max}=0.27(4)$, which again agrees well with the field-theoretical prediction in Eq.\ref{fss_max}: $x_{max}=\beta/2-(1-\mu)/(\gamma-1)=\beta/2-1/(\gamma'-1)=1/4$.

\subsection{The $\omega$ correlation volume exponent}

\begin{figure}[htb]
\begin{center}
\includegraphics*[width=9.0cm]{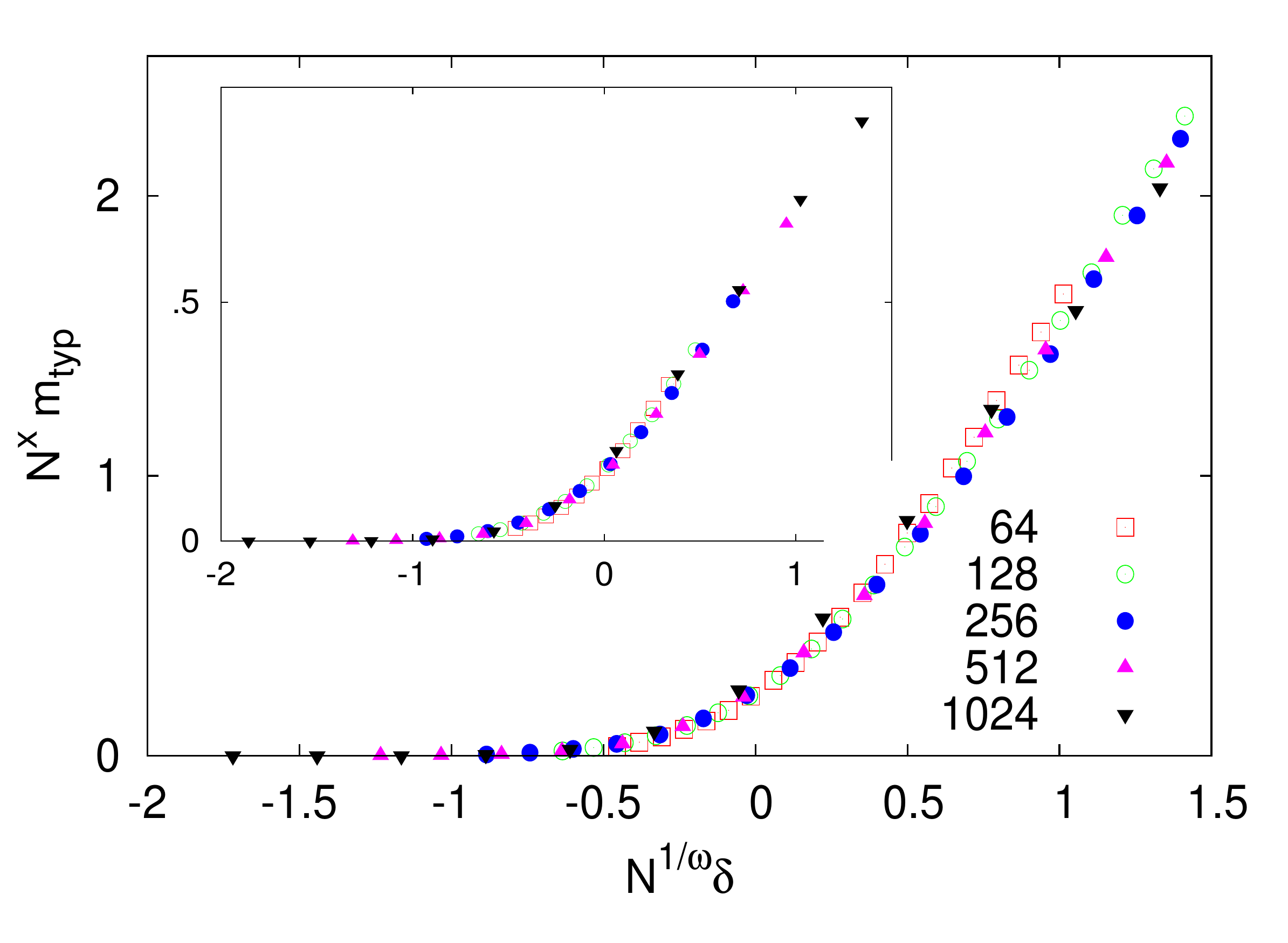}
\caption{Scaling collapse of the order-parameter near the transition point using the functional forms in Eq.\ref{fss1} for a typical site and in Eq.\ref{fss_max} for the maximally connected site (inset). The finite-size scaling exponents, $x_{typ}$ and $x_{max}$ are fixed by the previous analysis and the correlation exponent, $\omega$, is used to have an optimal collapse of the data, see text.} \label{CPfig4}
\end{center}
\vspace{-15pt}
\end{figure}

Next, we consider correlations in the vicinity of the transition point and calculate the relation between the correlated volume $\mathcal{V}\sim\xi^d$, and the distance from the critical point, $\delta$. Following the consideration in Eq.\ref{critExps} it is expected that the correlation volume arises in a power-law form, $\cal{V} \sim |\delta|^{-\omega}$, where $\omega=d\nu^*$ is the correlation volume exponent (see Eq.\ref{fss1}). According to field-theoretical results in Eqs.\ref{fss1} and \ref{fss_max} this exponent is $\omega=2$ in the mean field regime since there $\nu^*=2/d$ (see Section \ref{sec:fss}), both at a typical site and at the maximally connected site. Now in the limiting case, ${\cal V} \sim N$, the scaled order parameter, $\widetilde m= m N^{x}$, is expected to depend on the scaling combination, $N^{1/\omega} \delta$ (see Eq.\ref{MFmScale}), which is demonstrated in Fig.\ref{CPfig4}, both in the typical site and in the maximally connected site (inset). In both cases, $x$ and $\lambda_c$ are fixed by the previous analysis and $\omega$ is obtained from the optimal scaling collapse, as $\omega_{typ}=2.1(2)$ and $\omega_{max}=2.0(1)$. Thus, once more we have agreement with the field-theoretical results.

\subsection{Dynamical scaling}

\begin{figure}[htb]
\begin{center}
\includegraphics*[width=9.0cm]{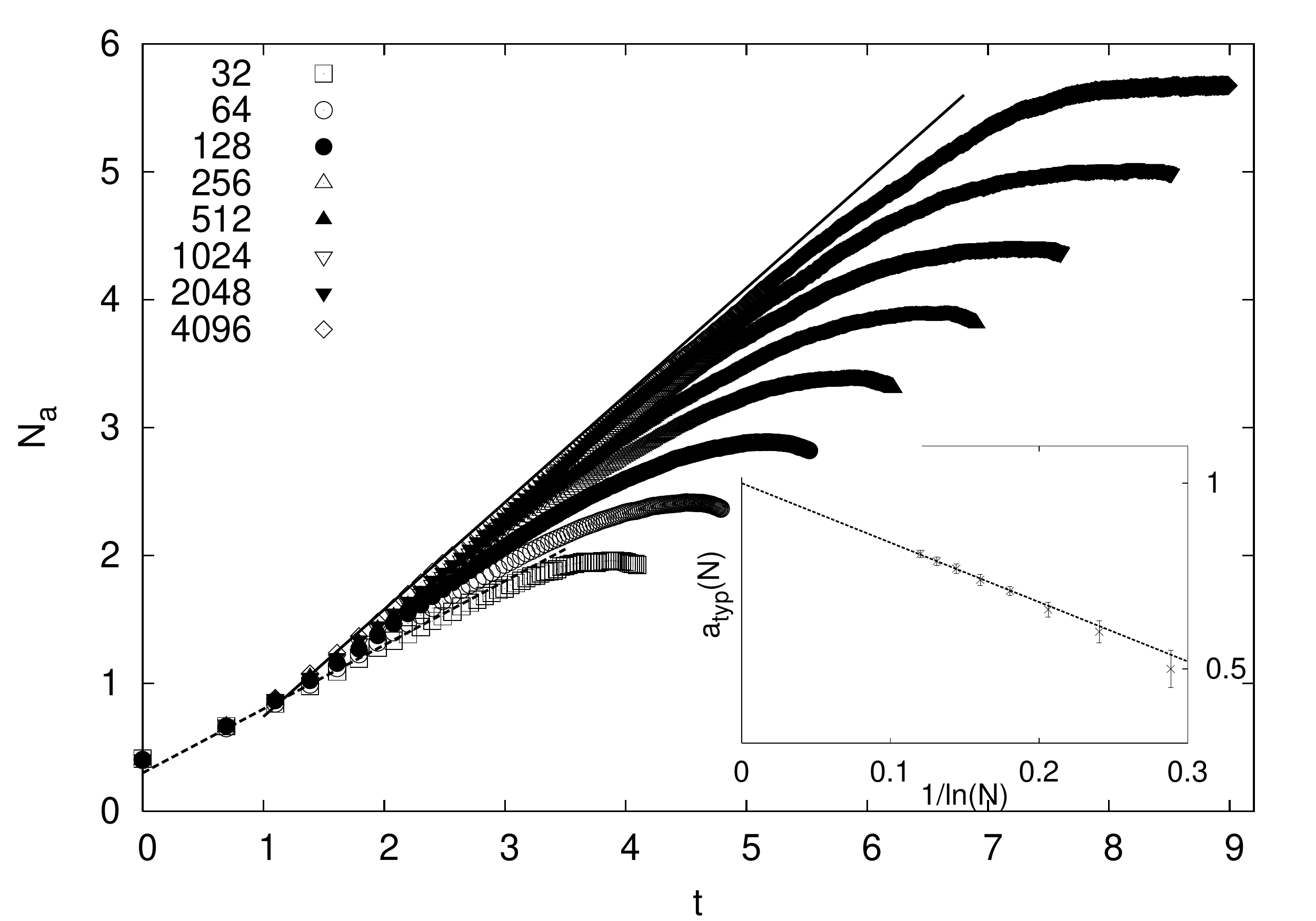}
\caption{Time dependence of the number of active sites starting from a typical site at the critical point in a log-log scale. For $t > N^{\zeta}$, when the infection has reached the border of the system the curves are saturated. The slope of the limiting curve for $N  \to \infty$  defines $a_{typ}$. The full straight line with slope $0.84$ fits the largest finite-size data. For small sizes there is a cross-over and the local slope of the curves is $a_{max} \approx 0.5$, see text. Here the broken straight line has a slope of $1/2$. Extrapolation of the finite-size slopes, $a_{typ}(N)$, as a function of $1/\ln N$ is shown in the inset.}
\label{CPfig5}
\end{center}
\vspace{-15pt}
\end{figure}

\begin{figure}[htb]
\begin{center}
\includegraphics*[width=9.0cm]{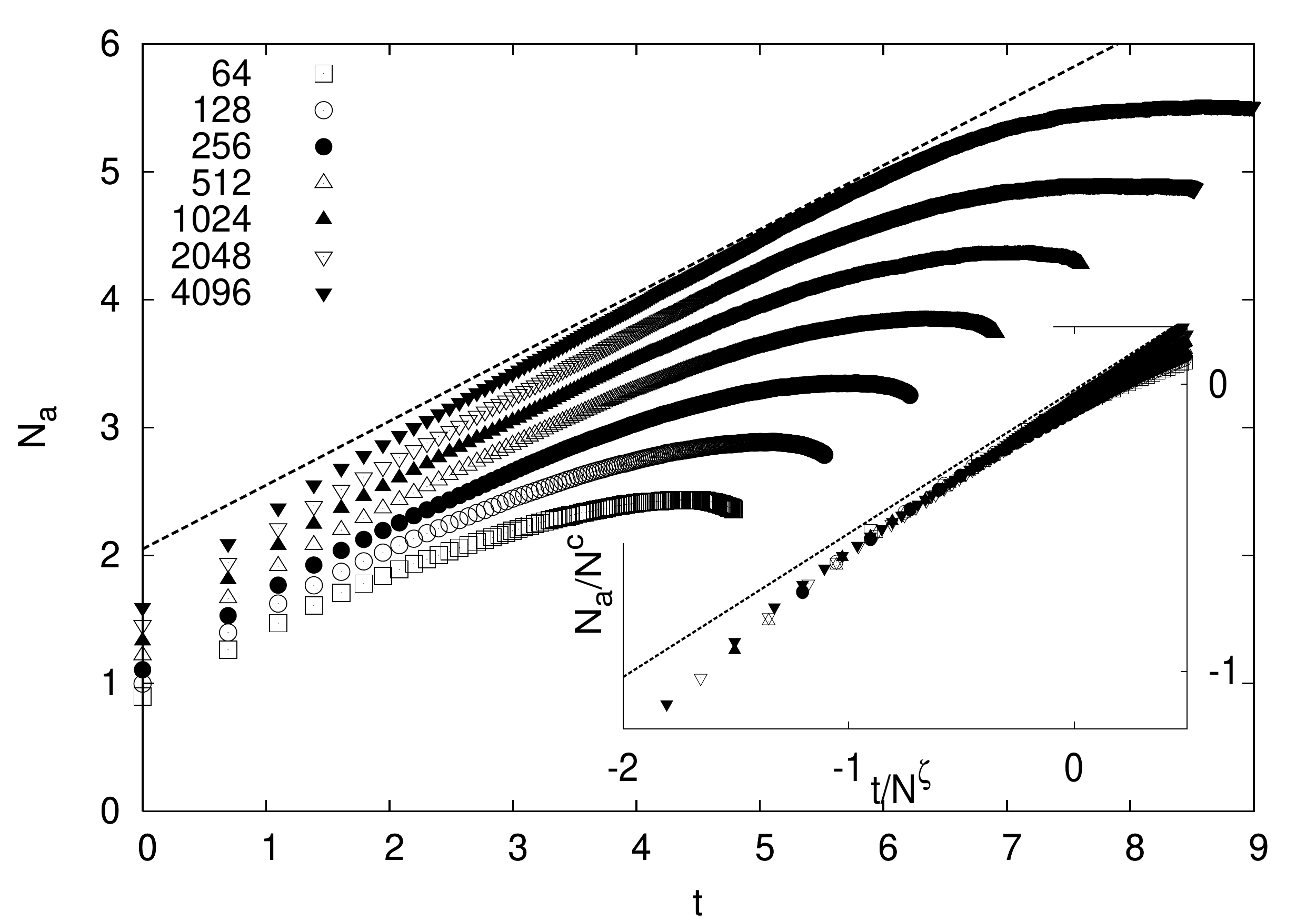}
\caption{The same as in Fig.\ref{CPfig5} for the maximally connected site.
Inset: scaling collapse of $N_a/N^c$ versus $t/N^{\zeta}$ in a log-log plot
with $c=0.5$ and $\zeta_{max}=0.5$. The slope of the straight lines both
in the main panel and in the inset indicates the prediction $a_{max}=1/2$.}
\label{CPfig6}
\end{center}
\vspace{-15pt}
\end{figure}

Finally, we turn to analyze dynamical scaling in the system. At the critical point, $\lambda=\lambda_c$ we have measured the number of active sites, $N_a$, as a function of time, $t$, which is shown in Fig.\ref{CPfig5} when the starting point is a typical site and in Fig.\ref{CPfig6} for the maximally connected site. As seen in Figs.\ref{CPfig5} and \ref{CPfig6} after a starting period (which is discussed below) there is an asymptotic region with a power-law dependence, $N_a \sim t^a$, which turns to a saturation regime around $t \sim N^{\zeta}$, when the active sites reach the boundary of the system.

Here the networks were generated with a parameter $m_0=m=1$ (Alg. \ref{BAmodelAlg}) thus the initial network was always a single node and during the network evaluation. At each iteration step we added new a site with $m=1$ number of edge, consequently the generated graph became a tree which boundary could be defined as the set of nodes with degree $k=1$. The sites of this boundary set are those which were added later to the network and which are less connected. 

At the critical point, when the infection is initiated from a single site, the disease spreads all over the system until the infection becomes homogeneously distributed all over the network. It means that the infection reached all the boundary sites, the correlation length became commensurable to the diameter of the network and the correlation volume turned to be $\mathcal{V}\sim N$. Following this limiting point the system arrives to its saturation region, where the infection evaluates homogeneously in the whole graph and the number of infected sites fluctuating around a finite value.

It is seen in Figs.\ref{CPfig5} and \ref{CPfig6} that the exponents $a_{typ}$ and $a_{max}$, which refer to a typical site and the maximally connected site, respectively, are different. However the border of the saturation regions, i.e. the exponent $\zeta$ are approximately the same for the two cases. In a regular lattice with dimension, $d>d_c$, (which contains only typical sites) the dynamical exponent, $\zeta$, is conjectured \cite{PhysRevLett.49.478} to be, $\zeta=1/2$, which is derived recently by field-theoretical methods\cite{PhysRevE.72.016119} and checked numerically \cite{Lubeck2005,Janssen2007}. This result for $\zeta$ is expected to hold for networks, too.

In order to analyze the results in Figs.\ref{CPfig5} and \ref{CPfig6} we discuss the scaling behavior of $N_a(t,N)$. In the stationary state the density is independent of the initial conditions and it is proportional to the density at a typical site, $\rho_{typ} \sim N^{-x_{typ}}$. Since $\rho\sim N_a/N$ therefore the scaling form of the active site number is given by:
\begin{equation}
N_a(t,N)=N^{1-x_{typ}} \widetilde{N}_a(t/N^{\zeta})\;,
\label{N_a}
\end{equation}
in which the scaling function, $\widetilde{N}_a(y)$, is different for a typical site and for the maximally connected site, respectively. In the small $y$ limit before the saturation time, we obtain $\widetilde{N}_a(y) \sim y^{a_{typ}}$ and $\widetilde{N}_a(y) \sim y^{a_{max}}$, for a typical site and for the maximally connected site, respectively in this way we recover the previously announced $t$-dependences. For small $t$, $N_a$ is of $\mathcal{O}(1)$ relational the average number of first neighbours, if we start with a particle at a typical site.

Next we analyze the size dependence of $N_a(t,N)$ at a fixed $t$. For a typical site $lim_{N \to \infty} N_a(t,N)$ is expected to have a finite limiting value, which is in accordance with the numerical results in Fig.\ref{CPfig5}. Indeed for typical sites the size of the network should not influence the local growth in the system. Using the small $y$ behavior of the scaling function we obtain the form:
\begin{equation}
N_a (t,N) \sim N^{1-x_{typ}}(t/N^{\zeta_{typ}})^{a_{typ}}
\end{equation}
where for small $t$, $N_a\sim 1$ and we obtain the relation at a fixed $t$:
\begin{equation}
1-x_{typ}-\zeta_{typ} a_{typ}=0\;.
\label{zeta_typ}
\end{equation}
For not too large finite systems, however, there is a starting regime and thus a cross-over phenomenon what can be seen in Fig.\ref{CPfig5}. This cross-over is due to the fact that in a small system the infection can reach the maximally connected site in a short time and afterwards the growth is characterized with an exponent $a_{max} < a_{typ}$. This starting region is also indicated in Fig.\ref{CPfig5}.

If we proceed from the maximally connected site, for small t, $N_a$ has the same order as the number of links at the hub. However the number of neighbors and thus the local density depend on the size of the network. As a consequence $N_a(t,N)$ at a fixed $t$ has a size dependence, which can be seen in Fig.\ref{CPfig6}. Using the result in Eq.\ref{rho_i} we obtain  $\lim_{N \to \infty} N_a(t,N) \sim k_{max}^{1-\mu} \sim N^{(1-\mu)/(\gamma-1)}\sim N^{1/(\gamma'-1)} \sim N^{x_{typ} -x_{max}}$, where we used the approximation $k_{max}\sim N^{1/(\gamma-1)}$ which is true for scale-free networks with degree exponent $\gamma$ \cite{newman-2003-45}. Now comparing the $N$ dependence for small $t$ leads to the equation:
\begin{equation}
N^{x_{typ}-x_{max}}\sim N^{1-x_{typ}}(t/N^{\zeta_{max}})^{a_{max}}
\end{equation}
which leads the relation at a fixed $t$:
\begin{equation}
1-x_{typ}-\zeta_{max} a_{max}=x_{typ} -x_{max}\;.
\label{zeta_max}
\end{equation}

The exponent, $a_{typ}$, can be obtained from the asymptotic slope of the master curve in Fig.\ref{CPfig5}, which has a strong size dependence. Extrapolation as shown in the inset of Fig.\ref{CPfig5} gives $a_{typ}=0.95(10)$ which is compatible with the mean field and finite-size scaling prediction, $a_{typ}=1$ derived from Eq.\ref{zeta_typ}. 

For the maximally connected site in Fig.\ref{CPfig6} the slope of the curves have a weaker size-dependence and can be estimated as: $a_{max}=0.52(4)$. Consequently from the relations in Eqs.\ref{zeta_typ} and \ref{zeta_max} we obtain that both $\zeta_{typ}$ and $\zeta_{max}$ are compatible with the mean field prediction, $\zeta=1/2$. The scaling prediction for $N_a(t,N)$ for small $t$ is checked for the maximally connected site in the inset of Fig.\ref{CPfig6}. Here $N_a/N^c$ is plotted as a function of $t/N^{\zeta}$. There is an appropriate scaling collapse with the expected values: $c=1-x_{typ}=0.5$ and $\zeta_{max}=0.5$ derived from Eq.\ref{zeta_typ} and \ref{zeta_max}. After the infection reached its saturation value, we get back the mean-field value $a=0$ \cite{Odor2008} in both cases since the active sites are homogeneously distributed all over the network and the their number is fluctuating around a constant value.

From the previous results, using $N \sim \delta^{-\omega}$, we obtain for the scaling behavior of the relaxation time $\tau$ (defined in Eq.\ref{relaxTdef}), in the vicinity of the transition point, $\tau \sim \delta^{-\nu_{\parallel}}$, with $\nu_{\parallel}=\zeta \omega$, so that $\nu_{\parallel}=0.95(10)$ both for the typical and the maximally connected sites, in complete agreement with the
mean field and finite-size scaling result, $\nu_{\parallel}=1$ \cite{Lubeck2005}.

\section{Summary}
\label{sec:disc}

In this section we considered non-equilibrium phase transitions in weighted scale-free networks, in which the creation rate of particles at given sites is rescaled with a power of the connectivity number. In this way non-equilibrium phase transitions are realized even in realistic networks having a degree exponent, $\gamma \le 3$. Mean field theory, which is generally believed to be exact in these lattices, is solved and the previously known three regimes of criticality (conventional and unconventional mean-field behavior, as well as only active phase) are identified. The theoretical predictions in the conventional mean-field regime are confronted with the results of Monte Carlo simulations of the contact process on the weighted Barab\'asi-Albert network.

To analyze the simulation results we have applied and generalized recent field-theoretical considerations \cite{PhysRevE.72.016119} about finite-size scaling of non-equilibrium phase transitions above the upper critical dimension $d_c=4$, i.e. in the mean-field regime. For a network the natural variable is the volume $N$ (mass) of the system which enters in a simple way into the scaling combinations.

We investigated two different cases, when the contact process was started from a site with average degree or the infection spreading was initialized from the best connected site with the maximum degree. In the first case we have obtained overall agreement with finite-size scaling theory in which the critical exponents are simple rational numbers. We have also numerically demonstrated that at sites with very large connectivity there are new local scaling exponents, which differ from the values measured at a typical site.

As an extension of our results we could check the change of critical behaviour as the system approaches the border of the conventional mean-field regime. To do that we could perform a similar study using different degradation exponent leading to the borderline value $\gamma'=4$.

Some parts of this chapter were based on a recent paper which contains also all the previously discussed results. It was published in a frequented journal of the field by M\'arton Karsai, R\'obert Juh\'asz and Ferenc Igl\'oi in the Physical Review E \textbf{73} 036116 in 2006 \cite{Karsai2006}.

%% file: NetwRBPM.tex
\chapter{Rounding of first-order phase transitions and optimal cooperation in scale-free networks}

\section{Introduction}

Cooperative behaviour in real systems appears when entities get more advantage from collective acts than going ahead alone. This phenomena depends on the interaction between the agents and external conditions also. A few examples come from sociology where the entities are humans and the external condition can be defined as a political status, or in economy where the interactions are defined between firms with external subsidy, etc. It is straight forward if one participant earns more alone then collective morals are reduced, individual acts become likely and the entities are uncorrelated. However, if the agents cannot prefer a claim lonely then cooperative behaviour becomes favorable and a correlated phase appears. Such kind of cooperative phenomena arise in many parts of physics also. The most obvious example is the appearance of magnetic domains in ferromagnets. In this chapter we examine such kind of cooperative model in scale-free networks. Here we investigate a simple magnetic model in which the agents are represented by classical spin variables: the interactions are described by ferromagnetic couplings, whereas the temperature plays the external role of a disordering field.

\begin{figure}[htbp]
    \begin{minipage}[b]{0.5\linewidth}
      \includegraphics*[ width=7.0cm]{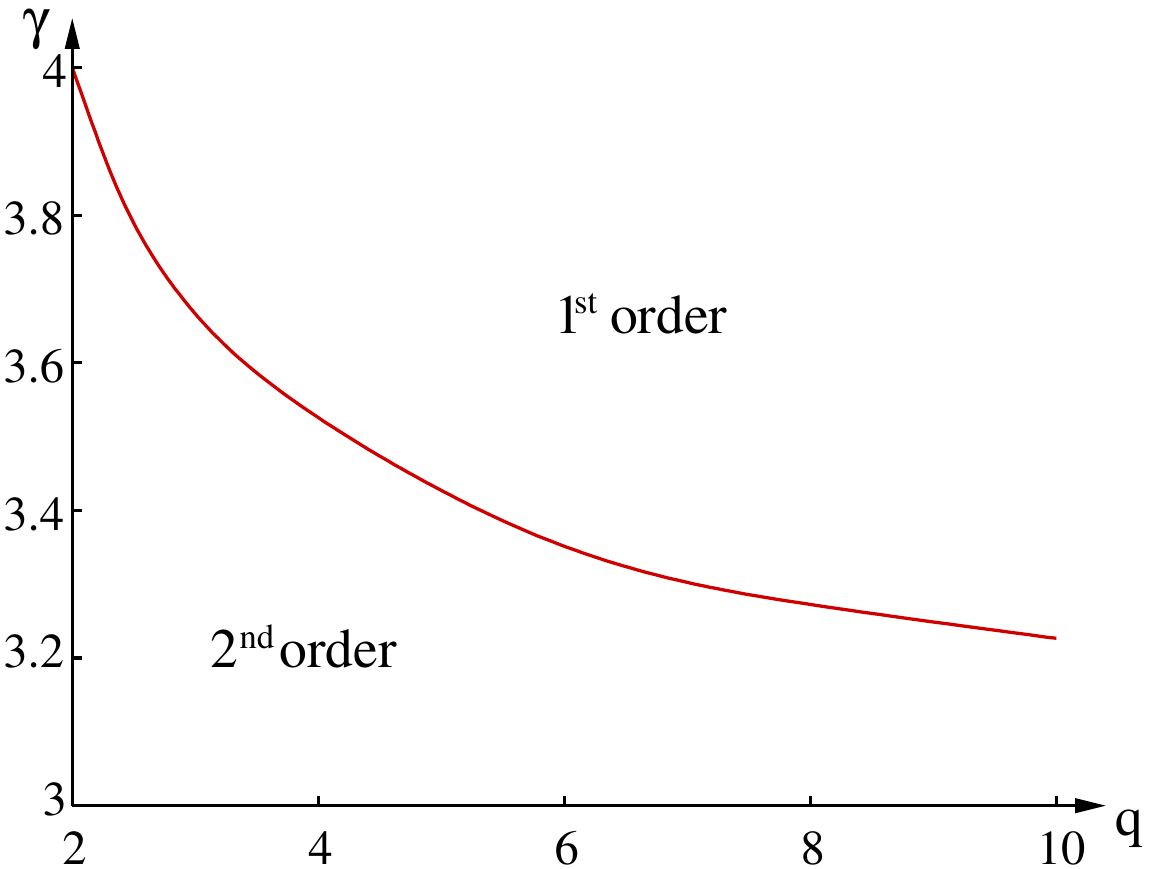}
    \end{minipage}\hspace{.3in}
    \begin{minipage}[b]{0.4\linewidth}
\caption{Regions of first- and second-order phase transitions for the $q$-state Potts model on scale-free network with a degree exponent $\gamma$. In the second-order regime, i.e., below the upper critical value $gamma^{u}$ the singularities are $\gamma$ dependent \cite{igloi-2002-66}.}
\label{SFPottsGamma}
    \end{minipage}
\end{figure}

An equivalent statistical model to simulate optimal cooperation in physics is the ferromagnetic random bond Potts model at $q\rightarrow \infty$ limit. As we mentioned in Section \ref{GeneralPotts} the phase transition of this model in regular lattices with homogeneous couplings $J_{ij}=J$ is of first order. However, in theoretical investigations of cooperative behavior one is usually interested in the influence of disorder on collective phenomena. Putting this model on a complex network, the inhomogeneities of the lattice play the role of some kind of disorder and it is expected that the value of the latent heat is reduced or even the transition is smoothened to a continuous one. This type of scenario is indeed found in a mean field treatment \cite{igloi-2002-66}, in which the phase transition depends on the scale-free degree exponent $\gamma$ as we have discussed in Section \ref{CPIntro}. It is of first-order for $\gamma>\gamma(q)$ and becomes continuous at the unconventional mean field regime for $\gamma_c<\gamma<\gamma(q)$, where $3<\gamma(q)<4$ and $\gamma_c=3$ (see Fig.\ref{SFPottsGamma}). On the other hand in an effective medium Bethe lattice approach, such a study leads to the fact that the phase transition of the $q$-state Potts model is of first order if $q\geq 3$  and the second moment of degrees $\langle k^2 \rangle$ is finite. For scale-free networks it is true only if $\gamma>3$. Consequently $\gamma(q)=\gamma_c=3$, thus the unconventional mean-field regime is absent in this treatment\cite{Dorogovtsev2004}.

Another way to introduce disorder into the system is through the strength of interactions. The interactions were considered homogeneous so far, however, in realistic situations the disorder is inevitable, and has a strong influence on the nature of the phase transition. As we have discussed in Section \ref{RelevanceIrrelevanceCriterion} in regular lattices for a second-order transition Harris-type relevance-irrelevance criterion (see Eq.\ref{HarrisCrit}) can be used to decide about the stability of the pure system's fixed point in the presence of weak disorder. On the contrary for a first-order transition such type of criterion does not exist. In this case rigorous results assert that in two dimensions for any type of continuous disorder the originally first order transition softens into a second order one \cite{PhysRevLett.62.2503}. In three dimensions there are numerical investigations which have shown\cite{PhysRevLett.74.422,PhysRevB.61.3215,PhysRevE.64.036120,Chatelain2005,Mercaldo2005,mercaldo:026126} that this kind of softening takes place only for sufficiently strong disorder.

In this chapter we consider interacting models with random interactions on complex networks and in this way we study the combined effect of network topology and bond disorder. The particular model we consider is the random bond ferromagnetic Potts model for large value of $q$ what we have defined in Section \ref{Potssqinf}. This model besides its relevance in ordering-disordering phenomena and phase transitions has an exact relation with an optimal cooperation problem (see Section \ref{CombinOpt} and \cite{dAuriac2002,dAuriac2004}). This optimization problem can be interpreted in terms of cooperating agents which try to maximize the total sum of benefits received for pair cooperations plus a unit support which is paid for each independent projects. For a given realization of the interactions the optimal state is calculated exactly by a combinatorial optimization algorithm which works in strongly polynomial time. The optimal graph of this problem consists of connected components (representing sets of cooperating agents) and isolated sites. The topology of the optimal set which is temperature (support) dependent contains information about the collective behavior of the agents. In the thermodynamic limit one expects to have a phase transition in the system, which separates the ordered (cooperating) phase with a giant clusters from a disordered (non-cooperating) phase, having only clusters of finite extent.

The structure of the chapter is the following. The model with the applied random edge degradation is presented in Section  \ref{sec_model}. The solution for homogeneous non-random evolving networks can be found in Section \ref{sec_hom}, whereas numerical study of the random model on the Barab\'asi-Albert network is presented in Section \ref{NumMethSFRBPM}. Our conclusions are summarized in Section \ref{SumSFRBPM}.

\section{The Random Bond Potts model}
\label{sec_model}

The random bond Potts model is a special case of the general $q$-state Potts model what we have discussed in details earlier in Section \ref{GeneralPotts} and \ref{RandomBondModels}. In the absence of external field this model is defined by the Hamiltonian:
\begin{equation}
\mathcal{H}=-\sum_{\langle i,j\rangle}J_{ij}\delta(\sigma_i,\sigma_j)
\label{HRBPM}
\end{equation}
where $\sigma_i=0,1...q-1$ are the Potts-spin variables. As we explained in Section \ref{Potssqinf}, in the limit $q\rightarrow \infty$ the partition function can be written in a random cluster representation (see Eq.\ref{PottsZHTL2}) in terms of its subgraphs, which is dominated by the largest term. This so-called optimal set $\phi^*(G)$ is proportional to the free-energy per site of the system and can be recognized as a submodular function. This property gives the possibility to calculate the optimal set in strongly polynomial time using combinatorial optimization. This method and the related iterational algorithm were discussed in details in the previous Section \ref{CombinOpt}.

In our case the couplings, $J_{ij}>0$, are ferromagnetic and they are either equal, $J_{ij}=J$, which defines a homogeneous networks, or they are identically and independently distributed random variables. Here we use a quasi-continuous distribution:
\begin{equation}
P(J_{ij})= \frac{1}{l} \sum_{i=1}^l \delta\left[J\left(1+\Delta\frac{2i-l-1}{2l}\right)-J_{ij}\right] \label{eq:distr}
\end{equation}
which consists of large number of $l$ equally spaced discrete values within the range $J(1 \pm \Delta/2)$. Here $0 \le \Delta \le 2$ measures the strength of disorder. 

Following from earlier calculations on two and three dimensional regular lattices \cite{PhysRevLett.90.190601,PhysRevE.69.056112,Mercaldo2005,mercaldo:026126}, the optimal graph at low temperatures is compact and the largest connected subgraph contains a finite fraction of the sites, $m(T)$, which is identified by the order parameter of the system. In the other limit, for high temperatures, most of the sites in the optimal set are isolated and the connected clusters have a finite extent. Their typical size is used to define the correlation length, $\xi$. Between the two phases there is a sharp phase transition in the thermodynamic limit. Its order depends on the dimension of the lattice and the strength of disorder, $\Delta$.

As already mentioned in the introduction this optimization can be recognized as an optimal cooperation problem in which the agents, which cooperate with each other in some projects, form connected components. Each cooperating pair receives a benefit represented by the weight of the connecting edge (which is proportional to the inverse temperature) and also there is a unit support to each component, i.e. for each projects. Thus by uniting two projects the support will be reduced but at the same time the edge benefits will be enhanced. Finally one is interested in the optimal form of cooperation when the total value of the project grants is maximal.

In the following section we solute exactly this optimization problem for evolving networks with homogeneous interaction, when $\Delta=0$ and each pair of spins receive the same amount of profit. 

\section{Exact solution for homogeneous evolving networks}
\label{sec_hom}

In regular $d$-dimensional lattices the solution of the optimization problem is simple as we presented in Section \ref{Potssqinf}. There are only two distinct optimal sets, which correspond to the $T=0$ and $T \to \infty$ solutions, respectively. In the proof we make use of the fact, that any edge of a regular lattice, $e_1$, can be transformed to any another edge, $e_2$, through operations of the automorphy group of the lattice. Thus if $e_1$ belongs to some optimal set, then $e_2$ belongs to an optimal set, too. Furthermore, due to submodularity the union of optimal sets is also an optimal set, from which follows that at any temperature the optimal set is either $\varnothing$ or $E$. By equating the free energies of the two phases we obtain the position of the transition point as: $T_c^{(0)}=Jd/(1-1/N)$ whereas the latent heat is maximal: $\Delta e/T_c^{(0)}=1-1/N$. (The superscript in $T_c^{(0)}$ is related to the non-random system, i.e. when the strength of disorder in Eq.\ref{eq:distr} is $\Delta=0$.)

In the following we consider the optimization problem in homogeneous evolving networks which are generated by the following rules: we start with a complete graph with $2\mu$ vertices and at each timestep we add a new vertex which is connected to $\mu$ existing vertices. In definition of these networks there is no restriction in which way the $\mu$ existing vertices are selected. These could be chosen randomly, as in the Erd\H os-R\'enyi model (see Section \ref{ERmodel}), or one can follow some defined rule, like the preferential attachment in the Barab\'{a}si-Albert network (Section \ref{BAmodel}). In the following we make some theoretical consideration to show that for such networks the phase-transition point is located at $T_c^{(0)}=J\mu$ and for $T<T_c^{(0)}$ ($T>T_c^{(0)}$) the optimal set is the fully connected diagram (empty diagram), as for the regular lattices. Furthermore, the latent heat is maximal: $\Delta e/T_c^{(0)}=1$.

In the proof we follow the optimal cooperation algorithm (Alg.\ref{algOC}) outlined in Section \ref{CombinOpt}, and in application of the algorithm we add the vertices one by one in the same order as in the construction of the network. We use a mathematical induction in our demonstration as follows:
\begin{enumerate}[I.]
\item First we note, that the statement is true for the initial graph, which is a complete graph. Consequently it is regular. Thus the optimal set following from Eq.\ref{qinfZ} can be of two kind:
\begin{itemize}
 \item If $T\leq T_c$, it is fully connected, and has a free-energy: $-\beta 2\mu f=1+\mu(2\mu-1)\beta J$
\item If $T\geq T_c$ then the optimal set is empty and the free-energy is: $-\beta 2\mu f=2\mu$
\end{itemize}
thus the transition point is at $T_c^{(0)}=J\mu$ which is calculated at the phase coexistence point where $1+\mu(2\mu-1)\beta J=2\mu$.

\item Then we suppose that this property is satisfied after $n$ steps and add a new vertex, $v_0$. Here we utilize an important point of the optimization method, defined in Section \ref{OCMethodFE}. Due to the submodularity of $-\phi(G)$ each connected component in the optimal set are contracted into a new vertex in the network with effective edge weights being the sum of individual weights in the original representation. Then to discuss the behaviour of the optimal set, we investigate the two cases, $T \le T_c^{(0)}$ and $T \ge T_c^{(0)}$ separately again: 
\begin{itemize}
\item If $T \le T_c^{(0)}$, then according to our statement all the edges, which were included in earlier steps, are in the optimal set and all vertices of the original graph are contracted into a single vertex, $s$, which has an effective weight
\begin{equation}
\mu \times J/T > \mu J/T_c^{(0)}=1
\end{equation}
to the new vertex $v_0$. Following the method and using the capacities, defined in Eq.\ref{OSweights1}-\ref{OSweights3}, it is straight forward that in the optimal set $s$ and $v_0$ are connected. Consequently the newly added $\mu$ number of edges proceeding from $v_0$ are in the optimal set also and the graph remains connected. 

\item If $T \ge T_c^{(0)}$, then all vertices of the original graph are disconnected, which means that for any subgraph, $S$, having $n_s \le n$ vertices and $e_s$ edges one has: $n_s \ge e_s JT +1$. This provides $e_s(T)\leq 1$ and becomes equal only at the phase coexistence point $T_c^{(0)}$ where the first edge appears and the graph above becomes fully connected as we mentioned in the previous point. 

Then we add the new vertex $v_0$ with $\mu$ edges. Let us denote by $\mu_s \le \mu$ the number of edges between $v_0$ and the vertices of $S$. One has
\begin{equation}
\mu_s \times J/T \le \mu \times J/T \le \mu J/T_c^{(0)}=1
\end{equation}
so that for the composite $S+v_0$ we have: $n_s+1 \ge e_s JT +1 + \mu_s J/T$, which proves that the vertex $v_0$ will not be connected to any subset $S$ and thus will not be contracted to any vertex.
\end{itemize}
\end{enumerate}

This result, i.e. a maximally first-order transition of the large-$q$ state Potts model holds for a wide class of evolving networks, which satisfies the construction rules presented above. This is true, among others, for randomly selected sites, for the Barab\'asi-Albert evolving network which has a degree exponent $\gamma=3$ and for several generalizations of the Barab\'asi-Albert network\cite{albert-2002-74} including nonlinear preferential attachment, initial attractiveness, etc. In these latter network models the degree exponent can vary in a range of $2<\gamma<\infty$. It is interesting to note that for uncorrelated random networks with a given degree distribution the $q$-state Potts model is in the ordered phase\cite{igloi-2002-66,Dorogovtsev2004} for any $\gamma \le 3$. This is in contrast with evolving networks in which correlations in the network sites result in the existence of a disordered phase for $T>T_c^{(0)}$, at least for large $q$. 

\section{Effective calculation of free energy}
\label{NumMethSFRBPM}

To locate the critical point of the random bond Potts model and calculate the critical properties of the system we need to determine some thermodynamical quantities as a function of the temperature $T$. We can do it iteratively but by exploiting some special features of the free energy we can find a more efficient and precise method that will be the subject of the first part of this section. Following that we then discuss some numerical results of the critical behaviour of the $q\rightarrow \infty$ random bond Potts model applied on scale-free networks.

\vspace{.2in}

The free energy of the Potts model at $q\rightarrow \infty$ limit can be recognized as a submodular function of the form Eq.\ref{qinfZ} which depends on the number of connected components and the sum of edge weights included in the optimal set. As we have shown for homogeneous systems in regular lattices and even in evolving networks, the system has a strictly first order phase transition between the fully connected and the empty graph. The critical point is located at the temperature where the free energy density, which is a linear function of $T$, has a breaking point (see Fig\ref{PottsFE}).

When we define interactions with random capacities, the phase transition between the fully connected and the empty states is modified. A qualitative explanation holds in the fact that the edges with different capacities are excluded continuously from the optimal set as the temperature is varied and not all at the same $T$ as for the homogeneous case. Consequently there is no phase coexistence between the two states and the phase transition softens into a continuous one. At the thermodynamical limit $N\rightarrow \infty$ with continuous edge weight distributions, the free energy changes smoothly. However this is not true at finite size systems with quasi-continuous distributions, where the average free energy functions, calculated over different random realizations has only physical relevance.

In finite systems the free energy is a piecewise constant function of the temperature, with monotonously increasing slope versus the temperature. For a given optimal set $G_1$ it is:
\begin{equation}
-\beta f(G'_1,\beta)=c(G'_1)+\beta\sum_{i,j\in G'_1}J_{ij}
\end{equation}
where the slope of the related straight line is defined as the sum of the weights $\sum_{i,j\in G'}J_{ij}$ in the optimal set, and the dislocation is equal to the number of connected components $c(G')$.  When some edges leave the optimal set, then the free energy has a breaking point. It then fits on a new line
\begin{equation}
-\beta f(G'_2,\beta)=c(G'_2)+\beta\sum_{i,j\in G'_2}J_{ij}
\end{equation}
related to the new optimal set $G'_2$. The breaking point is located at the temperature where $f(G'_1,\beta^*)=f(G'_2,\beta^*)$, which can be calculated as the intersection of the two straight lines:
\begin{equation}
\beta^*=\dfrac{c(G'_1)-c(G'_2)}{\sum_{i,j\in G'_2}J_{ij} - \sum_{i,j\in G'_1}J_{ij}}
\label{CrossPoint}
\end{equation}
The magnitude of the integer members involved in this calculation increases with the system size and $\beta^*$ becomes a ratio of two large numbers. For homogeneous systems when $\beta^*=N/N_e-1/N_e$, the deviation is given by the second term which is inversely proportional to the number of edges $N_e$. For continuous phase transitions where there are more breaking points, it indicates the same trend following from numerical calculations.
\begin{figure}[htb]
\begin{center}
\includegraphics*[ width=14.0cm]{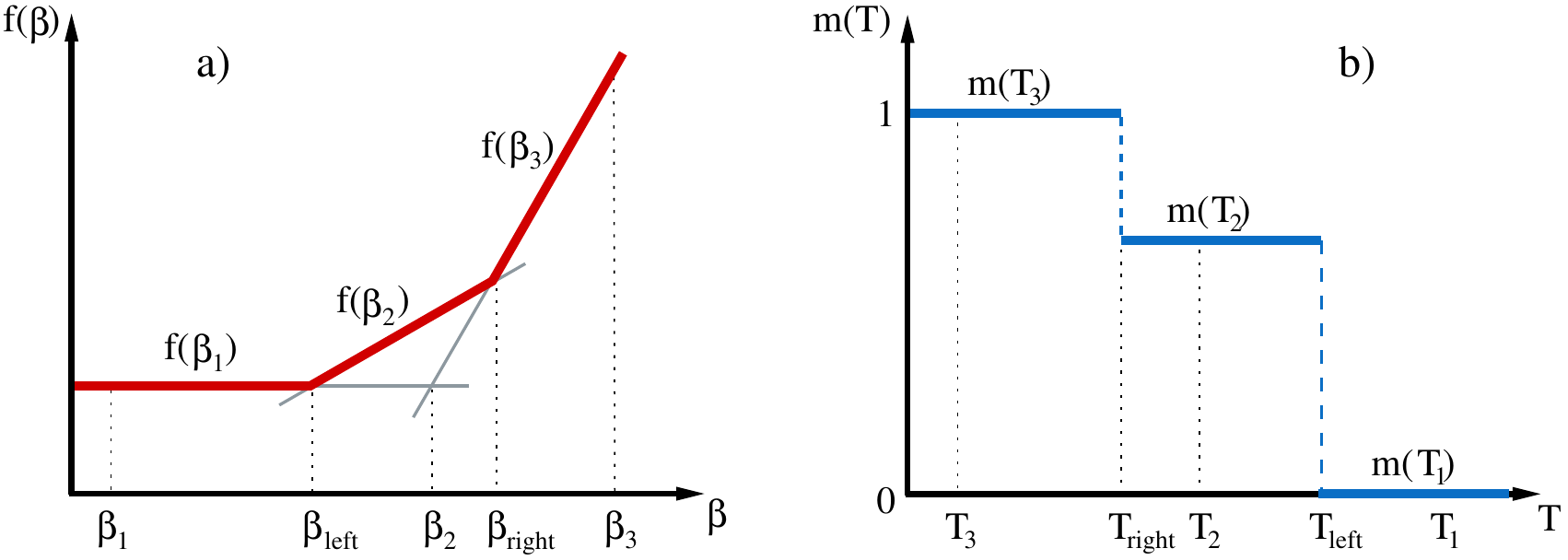}
\caption{Demonstration of free energy calculation of the $q=\infty$ random bond Potts model. (a) For a finite system the free energy is a piecewise linear function of the inverse temperature $\beta$. It is joined by straight lines with breaking points at different $\beta$. (b) The first derivative of the free energy is the system magnetization, which has a steplike behaviour with jumps at the same temperature as the free energy breaking points. The continuous magnetization can be realized over averages of different realizations only.}
{
\label{PottsFECalc}
}
\end{center}
\vspace{-10pt}
\end{figure}

To calculate the free energy for a given realization at a given temperature, calculation have strongly polynomial running time as we discussed in Section \ref{OSImplement}. The simplest method to calculate it for a temperature range, is to scan it with a given discretization frequency. The main problem with this process is that even if we can estimate the execution time, it is not sure that we find all the breaking points and the precision is limited by the resolution of discretization.

Another, more efficient method, can be defined as using the linear behaviour of the free energy, and calculate all its breaking points which fully specify the free energy curve also. This algorithm recursively calculates the intersections of straight lines, which are fitted to the free energy curve at certain temperatures, until it detects all the breaking points. To initialize the algorithm we need to calculate the free energy at three temperatures: at the two initial temperatures $\beta_1$ and $\beta_3$ (see Fig.\ref{PottsFECalc}) and at the temperature $\beta_2$ related to the intersection of the two associated lines $f(\beta_1)$ and $f(\beta_3)$. Then we call the $RecursiveFreeEnergy(f(\beta_1)$,$f(\beta_2)$,$f(\beta_3))$ algorithm defined in Alg.\ref{RecFE}.
\begin{algorithm}
\caption{RecursiveFreeEnergy($f(\beta_1)$,$f(\beta_2)$,$f(\beta_3)$)}
\begin{algorithmic}[1]
\STATE $\beta_{left}\Leftarrow$ Intersection$(f(\beta_1),f(\beta_2))$
\STATE $f(\beta_{left})\Leftarrow$ OptimalCooperation$(\beta_{left})$
\IF{$\big( f(\beta_{left})=f(\beta_1)$ OR $f(\beta_{left})=f(\beta_2) \big)$}
	\STATE $\beta_{left}$ is a breaking point!
\ELSE
	\STATE RecursiveFreeEnergy($f(\beta_1)$,$f(\beta_{left})$,$f(\beta_2)$)
\ENDIF
\STATE
\STATE $\beta_{right}\Leftarrow$ Intersection$(f(\beta_2),f(\beta_3))$
\STATE $f(\beta_{right})\Leftarrow$  OptimalCooperation$(\beta_{right})$
\IF{$\big( f(\beta_{right})=f(\beta_2)$ OR $f(\beta_{right})=f(\beta_3) \big)$}
	\STATE $\beta_{right}$ is a breaking point!
\ELSE
	\STATE RecursiveFreeEnergy($f(\beta_2)$,$f(\beta_{right})$,$f(\beta_3)$)
\ENDIF
\STATE \textbf{return}
\end{algorithmic}
\label{RecFE}
\end{algorithm}

At the first line of the algorithm we calculate the $\beta_{left}$ temperature as the intersection of the first two free energies given in the parameter list, using the Eq.\ref{CrossPoint}. To calculate the free energy $f(\beta_{left})$ we apply the OptimalCooperation() method, introduced in Alg.\ref{algOC}. If the free energy at the crossing temperature $\beta_{left}$ is equal to the free energy at $\beta_1$ or $\beta_2$ then we find an exact breaking point of the original free energy curve. If it is not true, an inner optimal set exists with free energy $f(\beta_1)<f(\beta_{left})<f(\beta_2)$ related to a line with an inner slope. Thus we call again recursively the function $RecursiveFreeEnergy(f(\beta_2)$,$f(\beta_{left}))$,$f(\beta_2)$) with the free energy at $\beta_1,\beta_{left}$ and $\beta_2$. To complete the calculation we need to repeat the same method symmetrically for the right hand side intersection $\beta_{right}$ also.

After the recursion terminated the algorithm gives back all the exact breaking point of the free energy in a given temperature range with an estimated error discussed above. The proof of the method is straightforward and evidently follows from the monotonic behavior of the free energy function.

To calculate the execution time we need to estimate the number of breaking points $n_{br}$. We have found experimentally that in all cases the number of breaking points is polynomial with the size of the lattice.
The number of times we call the function $OptimalCooperation(\beta)$ depends also on $n_{br}$ as $n_{call}=2(n_{br}+1)-1$ thus the number of unuseful calls which are not related to any breaking points is $n_{br}+1$. Consequently the execution time is polynomial in the number of breaking points so finding the exact free energy at all temperature is a polynomial function of the system size.

Since the temperature $\beta^*$ of a breaking point (defined in Eq.\ref{CrossPoint}) is a rational number, it is possible to calculate its value hypothetically with an arbitrary precision if the counter and the denominator are sufficiently large numbers. We realized this case using the GNU Multiple Precision Arithmetic Library (GMP$^\circledR$) in a C++ environment, which is eligible to define arbitrary precise numbers limited only by the available computer memory. This implementation needs high performance, but it is useful for precise - few point - calculations of a system with large size and (or) capacities. However, during average calculations we chose lower precisions where the standard variable types were adaptable. In the following section we are going to review these numerical results, which were calculated using the above defined method on Barab\'asi-Albert scale-free networks.

\section{Numerical study of random Barab\'asi-Albert networks}
\label{sec_SFnumerics}

In this section we study the large-$q$ state Potts model in the Barab\'asi-Albert network. To generate scale-free networks we used Alg.\ref{SFAlg} with preferential attachment. The value of initial connectivity was $\mu=2$ (in Alg.\ref{SFAlg} $\mu=m$), and the network size was varied between $N=2^6$ and $N=2^{12}$. The interactions are independent random variables taken from the quasi-continuous distribution in Eq.\ref{eq:distr} having $l=1024$ discrete peaks and we fix $J=1$. The advantage of using quasi-continuous distributions is that in this way we avoid extra, non-physical singularities, which could appear for discrete (e.q. bimodal) distributions\cite{PhysRevE.69.056112}. For a given size we have generated $100$ independent networks and for each we have $100$ independent realizations of the disordered couplings.

\subsection{Magnetization and structure of the optimal set}

As we have seen the free energy of one realization of random capacities, has breaking points at certain temperatures, consequently its first derivative, the magnetization, is discontinuous at these temperatures (see Fig.\ref{PottsFECalc}.b). To obtain the continuous magnetization curve in the non-homogeneous case, we need to calculate $m(T)$ over averages of random realizations with different breaking points.

\begin{figure}[htb]
\begin{center}
\includegraphics*[width=10.0cm]{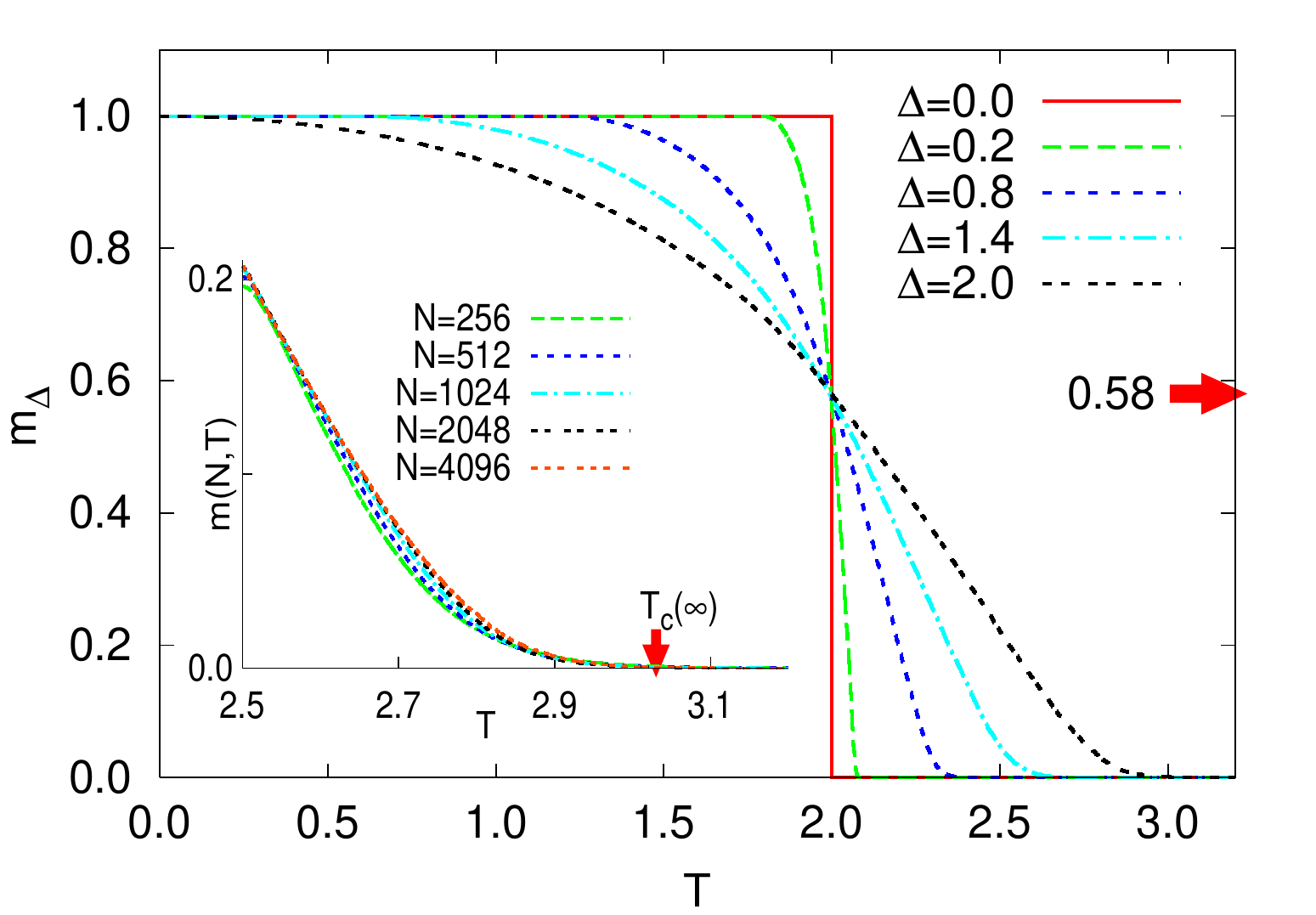}
\caption{Temperature dependence of the average magnetization
in a Barab\'asi-Albert network of $N=1024$ sites for different strength of the disorder, $\Delta$. At
$T=T_c^{(0)}=2$ the magnetization is independent of $\Delta>0$ and its value is indicated by
an arrow. Inset: The average magnetization for uniform disorder, $\Delta=2$, close to the transition point
for different finite sizes. The arrow indicates the critical point of the infinite system.}
\label{SFRBPM_1}
\end{center}
\vspace{-15pt}
\end{figure}

The magnetization which was defined $m(T)=\mathcal{N}/N$, where $\mathcal{N}$ is the size of the largest connected components (see Section \ref{sec_model}), has a modified definition in numerical calculations. It is due to the fact that $\mathcal{N}(T>T_c)=1$ in the disordered phase, consequently the magnetization would be $m(T>T_c)=1/N$ in this phase, instead of $m(T>T_c)=0$. However, if we introduce the magnetization as:
\begin{equation}
m(T)=\dfrac{\mathcal{N}(T)-1}{N-1}
\end{equation}
then $m(T)$ behaves like the order parameter by definition and its deviation is negligible for large size enough.

In Fig.\ref{SFRBPM_1} the temperature dependence of the average magnetization is shown for various strengths of disorder $\Delta$, in a Barab\'{a}si-Albert network of $N=1024$ sites. One can see that the sharp first-order phase transition of the pure system with $\Delta=0$ is rounded and the magnetization has considerable variation within a temperature range of $\sim \Delta$. The phase transition seems to be continuous even for weak disorder. Close to the transition point the magnetization curves for uniform disorder ($\Delta=2$) are presented in the inset of Fig.\ref{SFRBPM_1}, which are calculated for different finite systems.

Some features of the magnetization curves and the properties of the phase transition can be understood by analyzing the structure of the optimal set. For low temperature enough this optimal set is fully connected, i.e. the magnetization is $m=1$, which happens for $T<T_c^{(0)}-\Delta$. Indeed, the first (least connected) sites with $k=\mu=2$ (i.e. those which have only outgoing edges) are removed from the fully connected diagram, if the sum of the connected bonds is $\sum_{i=1}^{\mu} J_i < T$, which happens when the temperature reaches the range $T\geq T_c^{(0)}-\Delta$. From a similar analysis, it follows that the last edges leave the optimal set as the temperature approaches $T\leq T_c^{(0)}+\Delta$ and the optimal set becomes empty for any finite system for $T>T_c^{(0)}+\Delta$, where the magnetization $m=0$.

\begin{figure}[htb]
\begin{center}
\includegraphics*[width=10.0cm]{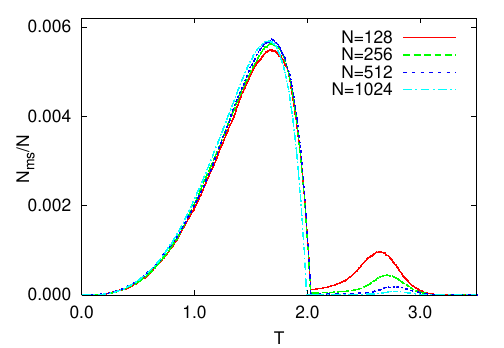}
\caption{Fraction of intermediate size clusters as a function of the temperature. Here $N_{MS}$ denotes the number of middle size clusters which are neither included into the largest component nor isolated one-site clusters.}
\label{SFRBPM_2}
\end{center}
\vspace{-15pt}
\end{figure}

The magnetization can be estimated for $T-(T_c^{(0)}-\Delta) \ll 1$ at the vicinity of the lower temperature bound. Since the distribution of edges is uniform, its probability density function is $p(J_{i})=1/\Delta$, thus the probability that the sum of capacities of arbitrary chosen $n$ edge is smaller than $T$ can be calculated as:
\begin{equation}
P(J_1+J_2+...+J_n<T)=\dfrac{1}{\Delta^n}\int_{T_c-\Delta}^T \int_{T_c-\Delta}^{T-J_1}... \int_{T_c-\Delta}^{T-J_1-...-J_{n-1}}dJ_n ... dJ_1
\label{EdgeLeaveProb}
\end{equation}
For the numerically studied model with $\Delta=2$, as the temperature is increasing, the less connected sites with $\mu=n=2$ number of edges become unconnected first when the capacities of its two edges become $J_1+J_2<T$. Following from Eq.\ref{EdgeLeaveProb} it occurs with the probability $P(J_1+J_2<T)=T^2/8$ whereby the magnetization starts to decrease from $1$ as:
\begin{equation}
m(T) \approx 1-\frac{T^2}{8} 
\label{mEstimat}
\end{equation}
which is indeed a good approximation for $T<1$. 

In the temperature range $T_c^{(0)}-\Delta<T<T_c^{(0)}+\Delta$ typically the sites are either isolated or belong to the largest cluster. There are also some clusters with an intermediate size, which are dominantly two-site clusters for $T<T_c^{(0)}$ and their fraction is less than $1\%$, as shown in Fig.\ref{SFRBPM_2}. The fraction of two-site clusters for $\Delta=2$ and $T<T_c^{(0)}=2$ can be estimated as follows. First, we note that since they are not part of the biggest cluster they can be taken out of a fraction of $p_1 = 1-m(T)$ sites. Before being disconnected a two-site cluster has typically four bonds and three of them connect to the biggest cluster, denoted by $J_1$, $J_2$ and $J_3$. When it becomes disconnected we have $J_1+J_2+J_3<T$, which happens with a probability $p_2=T^3/(6\Delta^3)=T^3/48$ calculated from Eq.\ref{EdgeLeaveProb}. At the same time the coupling within the two-site cluster should be $J_4>T$, which happens with probability $p_3=(T_c^{(0)}-T)/\Delta=(2-T)/2$. Thus the fraction of two-site clusters is approximately:
\begin{equation}
n_2 \approx p_1\times p_2 \times p_3 \approx T^5 (2-T)/768
\end{equation}
where we used the substitution of Eq.\ref{mEstimat}. This form describes well the general behavior of the distribution in Fig.\ref{SFRBPM_2} when $T<T_c^{(0)}$. In the temperature range $T>T_c^{(0)}$ the intermediate clusters have at least three sites and their fraction is negligible, which is seen in Fig.\ref{SFRBPM_2}. Consequently the intermediate size clusters do not influence the properties of the phase transition in the system.

We note an interesting feature of the magnetization curves in  Fig.\ref{SFRBPM_1} that they cross each other at the transition point of the pure system, at $T_c^{(0)}=2$, having a value of $m(T_c^{(0)})=0.58$, for any strength of disorder. This property follows from the fact that for a given realization of the disorder the optimal set at $T=T_c^{(0)}$ only depends on the sign of the sum of fluctuations of given couplings (c.f. some set of sites are connected (disconnected) to the giant cluster only for positive (negative) accumulated fluctuations) and does not depend on the actual value of $\Delta>0$. In other words at this point the cooperation depends on the fact that a collaboration is favorable or not and does not depend on that intensity. However, the slope of the order parameter curves are different at this point since they are not the same functions of the temperature.

\begin{figure}[htb]
\begin{center}
\includegraphics*[width=10.0cm]{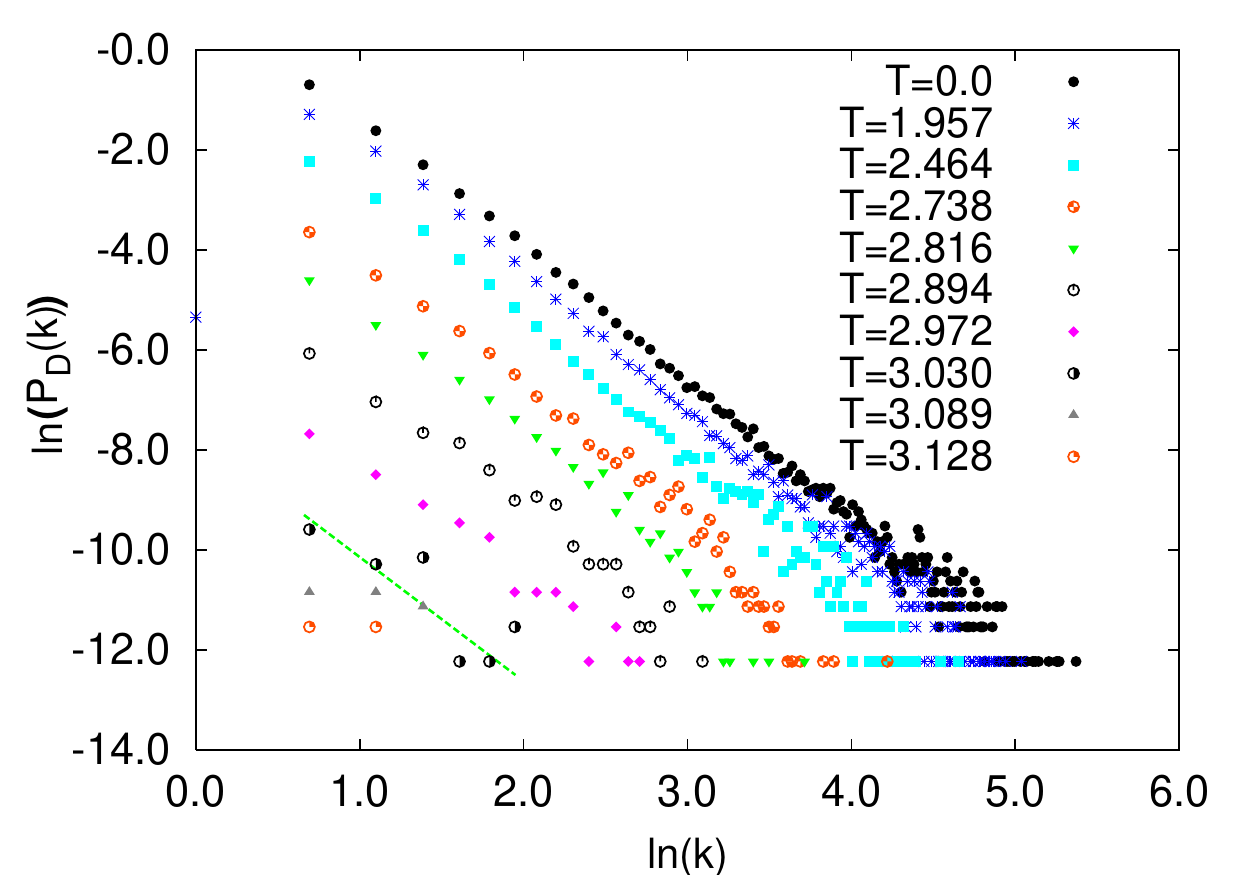}
\caption{Degree distribution of the largest cluster at different temperatures in a finite network with $N=2048$. The dashed straight line indicates the range of the critical temperature.}
\label{SFRBPM_3}
\end{center}
\vspace{-15pt}
\end{figure}

In the ordered phase, $T<T_c$, the largest connected cluster contains a finite fraction $m(T)<1$ of sites. Since the backgrounding graph has a scale-free behaviour, the question arises about the topology of this cluster as a function of temperature. 
We have analyzed the degree distribution of this giant cluster in Fig. \ref{SFRBPM_3}, and found a scale-free behavior for any temperature $T<T_c$ with the same degree exponent, $\gamma=3$, as for the original Barab\'{a}s-Albert network. 

Following from earlier works, it was shown that networks with scale-free degree distribution shows a special behaviour against destruction. Those structure have high tolerance against charges, if the sites are canceled randomly, but break down rapidly if always the highest connected nodes are removed \cite{Albert2000c,albert-2002-74}. During the evolution of the optimal set the highly connected sites, which hold together the cluster and induce scale-free behaviour, are secured by their large number of bonds. As the temperature increases always the weakest points with small degree and(or) capacity are detached, thus the core of the connected component retains its scale-free features until it splits up at the critical point.

We can thus conclude the following picture about the evolution of the optimal set. This is basically one large connected cluster with ${\cal N}$ sites, immersed in the see of isolated vertices. With increasing temperature more and more loosely connected sites are dissolved from the cluster, but for $T<T_c$ we have ${\cal N}/N=m(T)>0$ and the cluster has the same type of scale-free character as the underlaying network. On the contrary, above the phase-transition point, $T_c^{(0)}+\Delta>T>T_c$, the large cluster breaks up for finite extent, ${\cal N} < \infty$, and it loose its scale-free feature.

The order of the transition depends on the way how ${\cal N}$ behaves close to $T_c$. A first-order transition, i.e. phase-coexistence at $T_c$ does not fit to the above scenario, thus it takes place only in homogeneous case. Indeed, as long as ${\cal N} \sim N$ the same type of continuous erosion of the large cluster should occur in the system, i.e. the transition is of second order for any strength $\Delta>0$ of the disorder. Approaching the critical point one expects the following singularities of the magnetization and the largest correlated cluster: 
\begin{equation}
m(T) \sim (T_c-T)^{\beta} \qquad \mbox{and} \qquad \mathcal{N} \sim (T-T_c)^{-\nu'}
\end{equation}
on different sides of the critical point. Finally, at $T=T_c$ the large cluster is supposed to have ${\cal N} \sim N^{1-x}$ sites, with exponent $x=\beta/\nu'$. 

\subsection{Distribution of the finite-size transition temperatures}
\label{sec_distr}

\begin{figure}[htbp]
\hspace*{-.7in}
    \begin{minipage}[b]{0.5\linewidth}
      \includegraphics*[ width=10.0cm]{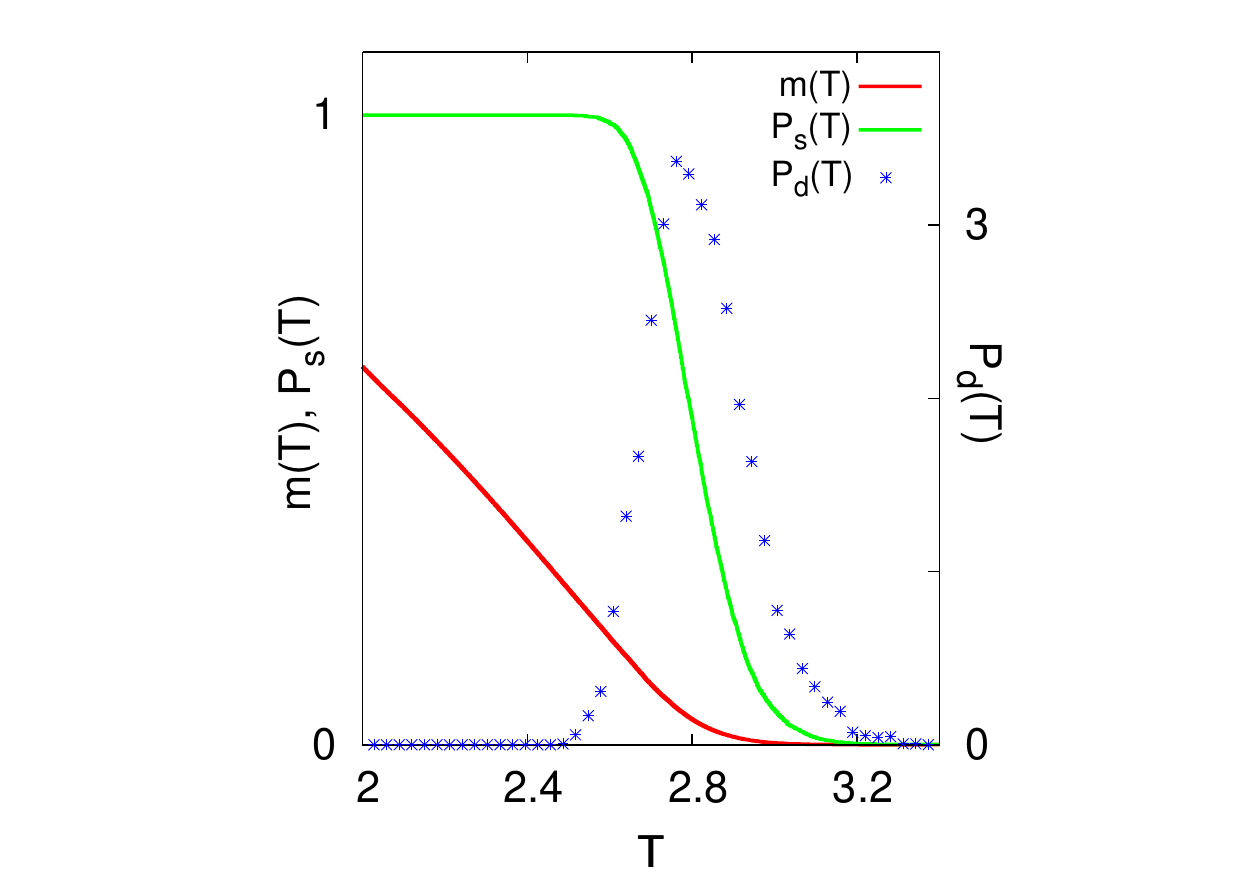}
    \end{minipage}\hspace{.8in}
    \begin{minipage}[b]{0.4\linewidth}
\vspace*{-.5in}
\caption{Demonstration of the magnetization $m(T)$ and the distribution $P_d(T)$ of the finite-size transition temperature at the vicinity of the phase transition point. $P_s(T)$ gives the probability that a sample survives until a given temperature. By definition $P_d(T)$ is the probability mass function of $P_s(T)$ so the peak of $P_d(T)$ distribution is related to point with the highest slope of the $P_s(T)$ curve.  The distributions were generated from average measurements of the size $N=512$ and $\Delta=2$. \vspace{.6in}}
\label{SFRBPM_4_1}
    \end{minipage}
\vspace{-15pt}
\end{figure}

The first step in the study of the critical singularities is to locate the position of the phase-transition point. In this respect it is not convenient to use the magnetization which approaches zero very smoothly (see the inset of Fig.\ref{SFRBPM_4_1}) so that there is a relatively large error by calculating $T_c$ in this way. 

One might have, however, a better estimation by defining for each given sample, say $\alpha$, a finite-size transition temperature $T_c(\alpha,N)$, as has been made for regular lattices\cite{PhysRevE.69.056112,Mercaldo2005,mercaldo:026126}. For regular lattices it is defined as the temperature which separates two qualitatively different regime, in which the largest correlated cluster percolates from that in which it does not. However the cluster percolation is meaningless in case of scale-free networks, and the transition here is continuous between the two states, so we define this finite-size transition point as the temperature when the largest connected component becomes smaller than a given size. For a network we use a condition for this size: ${\cal N}(T) \simeq A N^{1-x}$, in which $x$ is the magnetization critical exponent and $A=O(1)$ is a free parameter, from which the scaling form of the distribution is expected to be independent. 

First to allocate the exponent $x$ we made a self-consistent calculation as follows. We assumed that the critical size of the largest connected component which identifies $T_c(\alpha,N)$ is smaller than any power of the system size so we have started with a logarithmic initial condition, ${\cal N}(T) \simeq A \ln N$, which means formally $x_1=1$ (${\cal N}(T) \simeq A N^0$). Then for a fixed $A$ we have determined the distribution of the finite-size transition temperatures and at their average value we have obtained an estimate for the exponent, $x=x_2$. Following that the whole procedure is repeated with $x_s=x_2$, etc. until a good convergence is obtained. Fortunately the distribution function, $P(T_c,N)$, has only a weak $x$-dependence thus it was enough to make only two iterations whereafter we have obtained $x_2=0.69$. Then in the next step the critical exponents were converged within the error of the calculation and they were found to be independent of the value of $A$, which has been set to be $A=1,2$ and $3$.

\begin{figure}[htb]
\begin{center}
\includegraphics*[width=15.0cm]{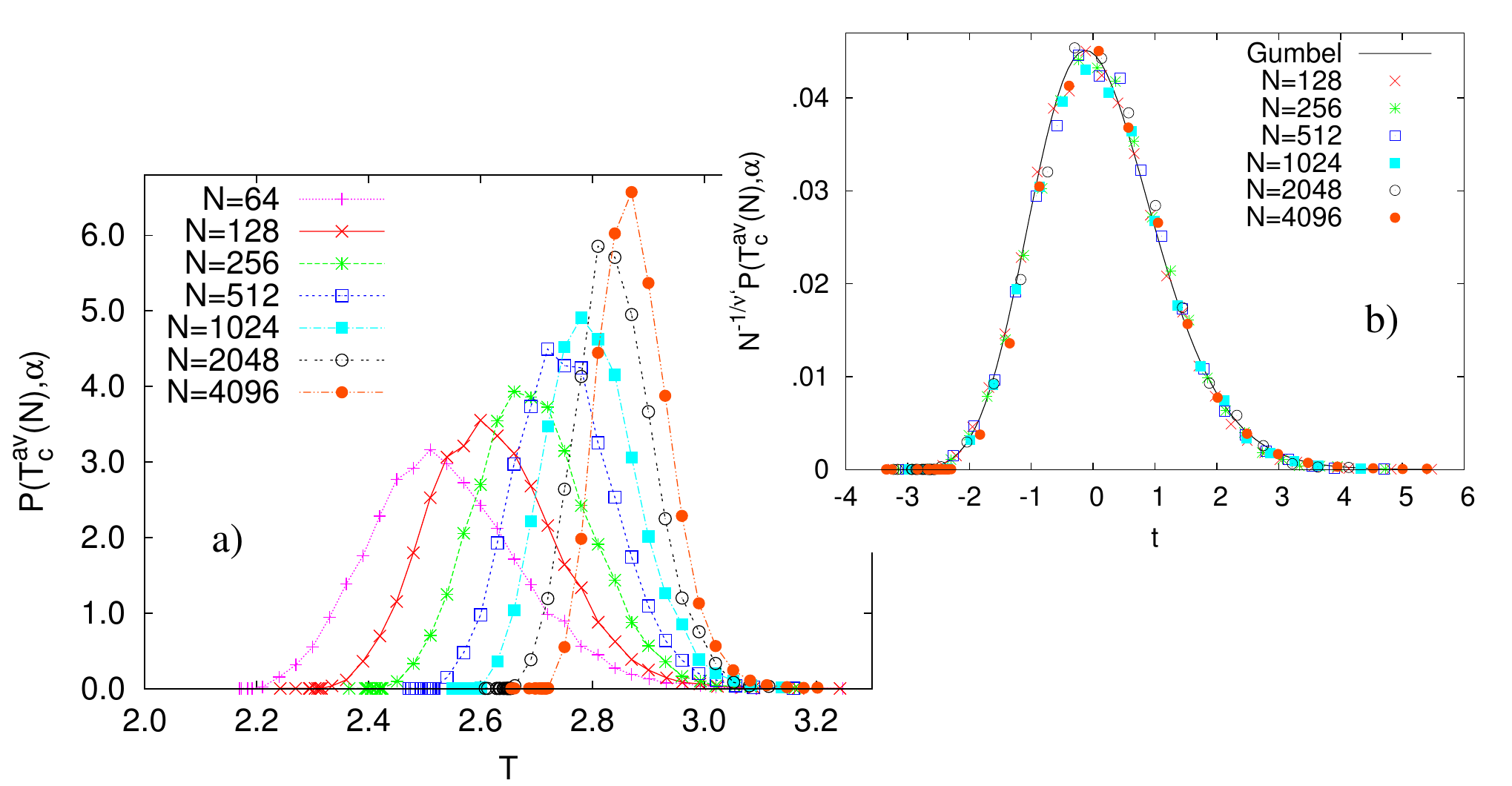}
\caption{(a) Distribution of the finite-size transition temperatures for different sizes of the Barab\'asi-Albert network. (b) Scaling collapse of the data in terms of $t=(T_c(N)-T^{av}_c(N)+AN^{-1/\widetilde{\nu}'})/\Delta T_c(N)$, using the scaling form in Eqs.\ref{T_c_L} and \ref{DT_c_L} with $\nu'=3.8(2)$, and $\tilde{\nu}'=5.6(2)$.}
\label{SFRBPM_4}
\end{center}
\vspace{-15pt}
\end{figure}

The distribution of the finite-size critical temperatures calculated with $x_2=0.69$ and $A=3$ are shown in Fig.\ref{SFRBPM_4} for different sizes of the network. One can observe a shift of the position of the maxima as well as a shrinking of the width of the distribution with increasing size of the network. The shift of the disordered average value, $T^{av}_c(N)$, is asymptotically given by \cite{Monthus2005}:
\begin{equation}
T^{av}_c(N)-T_c(\infty) \sim N^{-1/\widetilde{\nu}'}\;,
\label{T_c_L}
\end{equation}
whereas the width, characterized by the mean standard deviation, $\Delta T_c(N)$, scales with another exponent, $\nu'$, as:
\begin{equation}
\Delta T_c(N) \sim N^{-1/{\nu'}}\;.
\label{DT_c_L}
\end{equation}

\begin{figure}[htb]
\begin{center}
\includegraphics*[width=10.0cm]{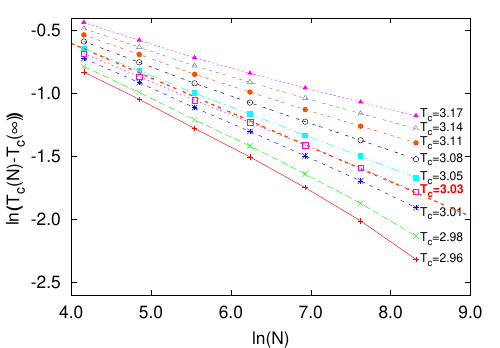}
\caption{Shift of the average finite-size transition temperatures, $T^{av}_c(N)-T_c(\infty)$, vs. $N$ in a log-log scale plotted for different values of $T_c(\infty)$. The lines connecting the points at the same $T_c(\infty)$ are guide for the eye. At the true transition point the asymptotic behavior is linear which is indicated by a dotted straight line.}
\label{SFRBPM_5}
\end{center}
\vspace{-15pt}
\end{figure}

First to determine the $\widetilde{\nu}'$ exponent and the $T_c(\infty)$ transition point we applied two independent methods. Using Eq.\ref{T_c_L} from a three-point fit we have obtained $\widetilde{\nu}'=3.8(2)$ and $T_c(\infty)=3.03(2)$. Another possible way is by plotting the difference $T^{av}_c(N)-T_c(\infty)$ vs. $N$ in a log-log scale for different values of $T_c(\infty)$, see Fig.\ref{SFRBPM_5}. At the true transition point according to Eq.\ref{T_c_L} there is an asymptotic linear dependence, which is indeed the case around $T_c(\infty)=3.03(2)$ and the slope of the line is compatible with $1/\widetilde{\nu}'=.27(1)$. 

For the width exponent, $\nu'$, we obtained from Eq.\ref{DT_c_L} with two-point fit the estimate: $\nu'=5.6(2)$. With these parameters the data in Fig.\ref{SFRBPM_4}.a can be collapsed to a master curve as shown in the Fig.\ref{SFRBPM_4}.b. This master curve looks not symmetric, at least for the finite sizes used in the present calculation. It could be explained by the definition of the finite-size transition temperature distribution, which is actually a distribution of maximum values. Following from this extreme statistic feature it can be well fitted by a modified Gumbel distribution:
\begin{equation}
G_{\omega}(-y)=\dfrac{\omega^{\omega}}{\Gamma(\omega)}\Big(e^{-y-e^{-y}}\Big)^{\omega}
\end{equation}
with a parameter $\omega=4.2$. We note that the same type of fitting curve has already been used in \cite{Monthus2005}. For another values of the initial parameter, $A=1$ and $2$ the estimates of the critical exponents as well as the position of the transition point are found to be stable and stand in the range indicated by the error bars.

The equations in Eqs.\ref{T_c_L} and \ref{DT_c_L} are generalizations of the relations obtained in regular $d$-dimensional lattices~\cite{PhysRevE.52.3469,PhysRevLett.77.3700,PhysRevLett.79.5130,PhysRevLett.81.22,PhysRevE.58.2938,PhysRevLett.81.252} in which $N$ is replaced by $L^d$, $L$ being the linear size of the system and therefore instead of $\nu'$ and $\tilde{\nu}'$ we have $\nu=\nu'/d$ and $\tilde{\nu}=\tilde{\nu}'/d$, respectively. Generally at a random fixed point the two characteristic exponents are equal and satisfy the relation\cite{PhysRevLett.57.2999} $\nu'=\tilde{\nu}' \ge 2$. This has indeed been observed for the $2d$ \cite{PhysRevE.69.056112} and $3d$\cite{Mercaldo2005,mercaldo:026126} random bond Potts models for large $q$ at disorder induced critical points. On the other hand if the transition stays first-order there are two distinct exponents~\cite{PhysRevB.51.6411,Monthus2005a} $\tilde{\nu}'=1$ and $\nu'=2$.

Interestingly our results on the distribution of the finite-size transition temperatures in networks are different of those found in regular lattices. Here the transition is of second order but still there are two distinct critical exponents, which are completely different of that at a disordered first-order transition. For our system $\nu'>\tilde{\nu}'$, which means that disorder fluctuations in the critical point are dominant over deterministic shift of the transition point. Similar trend is observed about the finite-size transition parameters in the random transverse-field Ising model \cite{Igloi2007}, the critical behavior of which is controlled by an infinite randomness fixed point. In this respect the random bond Potts model in scale-free networks can be considered as a new realization of an infinite disorder fixed point.

\subsection{Size of the critical cluster}
\label{cluster}

\begin{figure}[htb]
\begin{center}
\includegraphics*[width=10.0cm]{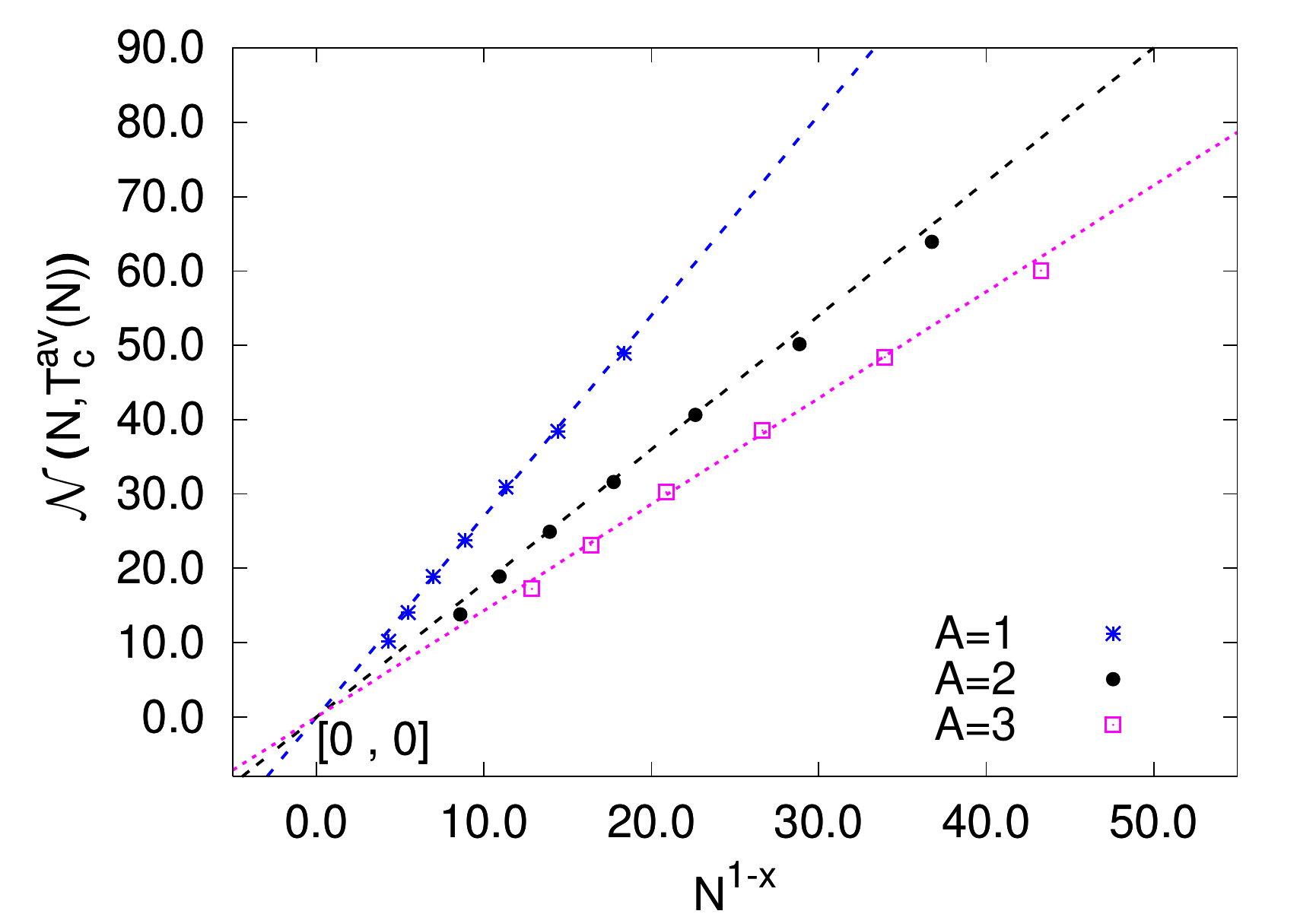}
\caption{Size dependence of the critical cluster at the average finite-size critical temperature as a function of $N^{1-x}$ with $x=.65$. The date points for different initial parameters $A$ are well described by straight lines, which are guides to the eye.}
\label{SFRBPM_6}
\end{center}
\vspace{-15pt}
\end{figure}

Having the distribution of the finite-size transition temperatures we have calculated the size of the largest cluster at $T^{av}_c(N)$, which is expected to scale as ${\cal N}[N,T^{av}_c(N)] \sim N^{1-x}$. Then from two-point fit we have obtained an estimate for the magnetization exponent: $x=.66(1)$. We have also plotted ${\cal N}[N,T^{av}_c(N)]$ vs. $N^{1-x}$ in Fig.\ref{SFRBPM_6} for different initial parameters $A$. Here we have obtained an asymptotic linear dependence with an exponent $x=.65(1)$, which agrees with the previous value within the error of the calculation.

\section{Summary}
\label{SumSFRBPM}

In this chapter we have studied the properties of the Potts model with large value of $q$ on evolving scale-free networks, such as the Barab\'{a}si-Albert network, both with homogeneous and random ferromagnetic couplings. This problem is equivalent to an optimal cooperation problem in where the agents try to optimize the total sum of the benefits coming from pair cooperations (represented by the Potts couplings) and the total sum of the support which is the same for each cooperating projects (given by the temperature of the Potts model). The homogeneous problem is shown exactly to have two distinct states: either all the agents cooperate with each other or there is no cooperation between any agents. There is a strongly first-order phase transition: by increasing the support the agents stop cooperating at a critical value.
 
In the random problem, in which the benefits are random and depend on the pairs of the cooperating agents, the structure of the optimal set depends on the value of the support. Typically the agents are of two kinds: a fraction of $m$ belongs to a large cooperating cluster whereas the others are isolated, representing one man's projects. With increasing support more and more agents split off from the cluster, thus its size, as well as $m$ is decreasing, but the cluster keeps its scale-free topology. For a critical value of the support $m$ goes to zero continuously and the corresponding singularity is characterized by non-trivial critical exponents. This transition, as shown by the numerically calculated critical exponents for the Barab\'asi-Albert network, belongs to a new universality class. One interesting feature of it is that the distribution of the finite-size transition points is characterized by two distinct exponents and the width of the distribution is dominated over the shift of the average transition point, which is characteristic at an infinite disorder fixed point \cite{Igloi2007}.

An interesting question arises concerning to the validity of the Harris criterion on complex networks in case of spin systems which exhibits a second-order phase transition in the absent of disorder. This question have been studied on different complex structures like lattices with aperiodically distributed interactions, hierarchical lattices where an analogous criterion (also called Harris-Luck criterion \cite{Janke2004a}) of relevance of the geometric fluctuations have been found for the critical (second-order) behaviour \cite{PhysRevE.62.7773,Luck1993,Moody1997}. In small-world networks this modified relevance criterion can be seen also which in this case depends on the strongness of the long-range interactions \cite{Ben-Naim2004}. In case of scale-free networks we believe that this criterion is satisfied also however not in its original form since it depends on the dimension which is defined here as the fractal dimension of the network. Probably here the relevance-irrelevance criterion would depend on the degree exponent of the network also, which determine the density long-range interactions in the network.

The above investigated optimal cooperation model is a good approximation of a large set of cooperation process in physics, sociology or economy etc. However, we need to keep in mind that the optimal cooperation algorithm is a Monte Carlo method to calculate an equilibrium state of a system of cooperating agents. At this state the free energy of the total system is minimized and the condition of
detailed balance is satisfied. However in a real social system this state is unreachable, since the system is always out of
equilibrium and the detailed balance condition is not held. Consequently the optimal cooperation method does not describe the real social systems as they are, but it provides important predictions about the system. It gives the best cooperated situation with the ”lowest energy” what the system could ever reach. In real systems the backgrounding network is changing in time plus there are some important social parameters which influence the agents, and then they do not take the optimal decisions always. Consequently a social system is always in an excited ”higher energy” state which can be close to equilibrium, but never the same in deed. In these excited states the sub-modular property of the free energy is expected to hold, but no effective algorithm has found so far to calculate these states. Moreover, even a social system would evolve toward to the equilibrium state, the time scale we can observe it is too short to see this relaxation.

At the same time, the structure of this kind of interacting networks usually is well described by scale-free graphs. Thus the common consideration of these two models gives a possible relevance in real applications even the aim of our study was to understand the problem from a fundamental point of view.

All the results of this chapter were published in a recent paper by M\'arton Karsai, Jean-Christian Angl\`es d'Auriac and Ferenc Igl\'oi in the Physical Review E \textbf{76} 041107 in 2007 \cite{Karsai2007}. This article gave also the ground for the presentation of some parts in the previous chapter. 

%% file: PottsPerco.tex
\chapter{Density of critical clusters in strips of a strongly disordered system}
\section{Introduction}

Every experimental system is geometrically constrained and finite therefore shape a surface, what leading to the fact that we have to discriminate between its bulk and surface properties. This description is justified as long as the correlation length is much smaller than the system size. However, at the critical point where correlation length is divergent and domains of correlated sites appear in all length scales, it is more appropriate to describe the position dependent physical properties of the system by density profiles rather than bulk and/or surface observables.

The correlated domains are most easily visualized for percolation \cite{Stauffer2003}, in which the domains are the connected clusters, as we have seen in Section \ref{percolation}. At the same time in discrete spin models, such as the Ising and the Potts models, domains of correlated spins can be identified in different ways, for example using geometrical clusters\cite{Fisher1967} (also called Ising or Potts clusters) which are domains of parallel spins. In two dimensions, geometrical clusters percolate the sample at the critical temperature and their fractal dimension $d_f$ can be obtained through conformal invariance\cite{PhysRevLett.62.1067,PhysRevLett.63.2536,Vanderzande1992,Janke2004}. This $d_f$ is generally different from the fractal dimension of Fortuin-Kasteleyn representation \cite{Kasteleyn1969} which is another possible method to represent correlated clusters. The domains here are identified by graphs of the high-temperature expansion. Its fractal dimension is directly related to the scaling dimension of the magnetization as $d=d_f+x_b$, where $d$ denotes the spatial dimension of the system (see Section \ref{FiniteSC}). From a geometrical cluster, the Fortuin-Kasteleyn cluster is obtained by removing bonds with a probability, $1-p=e^{-K_c}$, where $K_c$ being the critical value of the coupling (see Section \ref{RandomClust}).

In finite geometries such as strips or squares, one is interested in the spanning probability of critical correlated clusters and universal behaviour of different crossing problems. For two dimensional percolation many exact and numerical results have been obtained in this field \cite{Cardy1992,PhysRevLett.69.2670,Langlands1992,Aizenman1998,PhysRevB.57.R8075,Smirnov2001,Duplantier2003,Kleban2003,Dubedat2006}. However, another interesting problem is the density of clusters in restricted geometries\cite{Simmons2007}, which is defined by the fraction of samples for which a given point belongs to a cluster with some prescribed property, such as touching the edges of infinite and half-infinite strips, squares, etc. This latter problem is analogous to the calculation of order parameter profiles in restricted geometries, which has been intensively studied through conformal invariance and numerical methods\cite{Burkhardt1985,Burkhardt1987,PhysRevB.36.2080,PhysRevLett.65.1443,PhysRevLett.66.895,Burkhardt1991,Burkhardt1994,Turban1997,PhysRevLett.78.2473,PhysRevB.57.7877,PhysRevB.61.14425,Karevski2000,Bilstein2000,PhysRevB.62.6360,Turban2001}.


This density behaviour of critical clusters is expected to hold in any two dimensional systems which represent continuous phase transition where at the critical point the correlation length is divergent and the system becomes scale and conformal invariant. In case of the q-state Potts model, following from exact results \cite{PhysRevLett.62.2503} in two dimension a continuous phase transition takes place if the number of spins states is $q\leq 4$ and the conformal invariance behaviour is expected to hold.


%


In case of disorder the correlated clusters are defined also. In such systems the physical observables (magnetization profile, cluster density, etc.) are averaged  over quenched disorder. In isotropic cases conformal symmetry is expected to hold at the critical point so that average operator profiles and average cluster densities are expected to be invariant under conformal transformations. Among disordered systems an interesting class is represented by such models in which the transition in the pure version is of first-order, but in the disordered version the transition softens into second order (see Section \ref{EffectsOfDisorder}) \cite{Cardy1999}. This type of random fixed point can be found, among others, in the two-dimensional random bond Potts model for $q>4$ \cite{PhysRevLett.79.2998,PhysRevLett.80.1670,PhysRevE.58.R6899,PhysRevE.60.3853,PhysRevB.60.3428,Jacobsen1998}. As we have seen in Section.\ref{RandomClust}, this model can be represented by Fortuin-Kasteleyn random cluster method where the critical properties are influenced by strong-disorder effects. 

When $q<\infty$, in the partition function of the random cluster representation the different clusters are needed to be considered with the related Boltzmann weights. However, in the $q=\infty$ limit for a given realization of bond disorder its high-temperature expansion is dominated by a single graph, the so-called optimal diagram where the weights can be only zero or one thus the calculation of the optimal graph is simpler (see Section \ref{Potssqinf}, \ref{RandomBondModels} and \ref{sec_model}). Clusters in the optimal diagram are isotropic, and the density of clusters is obtained through averaging over disordered realizations.

Very recently the density of critical percolation clusters is studied in different two-dimensional restricted geometries and presumably exact expressions are obtained through conformal invariance. The same results are expected to hold for the densities of Fortuin-Kasteleyn clusters in two-dimensional conformal invariant systems also. In this chapter we study the density of critical clusters of the random bond Potts model in large-$q$ limit represented by Fortuin-Kasteleyn clusters and compare our numerical averages to exact analytical results obtained for percolation through conformal invariance. First we shortly introduce some exact analytical results calculated for percolation in strips \cite{Simmons2007}, then we compare them to numerical results of critical density profiles, calculated in the disordered Potts system. We close this chapter with our conclusions.

\section{Density of critical percolation clusters in strips}
\label{ConformDensityDef}

Critical cluster density profiles can be calculated exactly for two-dimensional percolation systems through conformal invariance. Here we recapitulate some of these results which are valid for an infinite strip of width $L\gg 1$ in the continuum limit. The position of a point is described by the distance $l$ from the strip boundary ($l \gg 1$), and the scaling variable is defined as $y=l/L=\mathcal{O}(1)$ which is between $0 < y \leqslant 1$.

First we consider the density of clusters which span the sides of an infinite strip. We proceed from a complex half plane and determine the critical density profiles using conformal transformations. It has been shown \cite{Belavin1984,PhysRevB.61.14425,Kleban2006,Simmons2007} that in an upper half plane the density $\rho$ of clusters, which are connected to the boundary in an interval $(x_1,x_2)$ is:
\begin{equation}
\rho(z,x_1,x_2)\propto (z-\overline{z})^{-x_b}F\bigg( \dfrac{(x_2-x_1)(\overline{z}-z)}{(\overline{z}-x_1)(x_2-z)} \bigg)
\label{clustDensSI}
\end{equation}
where $x_b$ is the scaling dimension of the bulk order parameter and $z\in \mathbb{C}$ is a point on the complex half plane (see Section \ref{ConfInv}). Here $F(\eta)$ takes one of the two forms of:
\begin{equation}
F_{\pm}(\eta)=\bigg( \dfrac{2-\eta}{2\sqrt{1-\eta}}\pm 1 \bigg)^{2x_s}
\end{equation}
where $x_s$ denotes the scaling dimension of the surface order parameter \cite{Kleban2006}. These results are exact in case of critical percolation where $x_b=5/48$ and $x_s=1/6$, however we hypothetically assume that they are valid for another conform invariant systems with different values of $x_b$ and $x_s$.

If in Eq.\ref{clustDensSI} we parametrize $z$ as $z=e^{i\theta}$ and set $x_1\rightarrow 0$ and $x_2\rightarrow \infty$, then the argument of $F(\eta)$ in Eq.\ref{clustDensSI} can be written as $\eta=1-e^{2i\theta}$. Using this parametrization the critical densities can be written of the form:
\begin{eqnarray}
\rho_+ (r,\theta,x_1\rightarrow 0,x_2\rightarrow \infty) &\propto & (r \mbox{sin}\theta)^{-x_b} [\mbox{cos}(\theta/2)]^{x_s} \label{HInfPlrhoIntervrpl}\\
\rho_- (r,\theta,x_1\rightarrow 0,x_2\rightarrow \infty) &\propto & (r \mbox{sin}\theta)^{-x_b} [\mbox{sin}(\theta/2)]^{x_s}
\label{HInfPlrhoIntervrmin}
\end{eqnarray}
where $\rho_+(r,\theta)$ denotes the critical density attached to the positive real axis and $\rho_-(r,\theta)=\rho_+(r,\pi-\theta)$ is the density of clusters attached to the negative real axis on the semi-infinite complex plane at a distance $r$ from the boundary.

The density of clusters connected to an arbitrary point of the boundary is independent of $x_1$ and $x_2$. Following from the exact result of Fisher and de Gennes \cite{Fisher1978} this is related to the order parameter profile varies as $\langle\phi(z,\overline{z})\rangle \propto (z-\overline{z})^{-x_b}$ which can be also written as:
\begin{equation}
\rho_e(r,\theta) \propto (r \mbox{sin} \theta)^{-x_b}
\label{HInfPlrhoArb}
\end{equation}
The densities, introduced in Eq.\ref{HInfPlrhoIntervrpl},\ref{HInfPlrhoIntervrmin} and \ref{HInfPlrhoArb}, are unnormalized, however close to the boundary they are all dominated by the fixed boundary so they may be normalized relative to one another.

Next we map the half infinite plane onto an infinite strip, following some conformal transformation defined in Eq.\ref{CISemiInf} as:
\begin{equation}
w(z)= x+iy = \dfrac{1}{\pi} \mbox{ln} (z)=\dfrac{1}{\pi}(\mbox{ln}r+i\theta) \qquad
\begin{cases}
r & = e^{\pi x} \\
\theta & = \pi y
\end{cases}
\end{equation}
where $x \in (-\infty ,\infty )$ and $y\in (0,1)$. In the new geometry, the critical densities are transformed also. The density profile of clusters which touch at least one of the boundary (or both) irrespective to the other, say as $y=l/L\rightarrow 0$ (or $y=l/L\rightarrow 1$), can be deduced from the density defined in Eq.\ref{HInfPlrhoIntervrpl} (or Eq.\ref{HInfPlrhoIntervrmin}). They are given by:
\begin{equation}
\rho_0(y)\propto (\mbox{sin} \pi y)^{-x_b} \bigg( \mbox{cos} \dfrac{\pi y}{2} \bigg)^{x_s} \qquad , \qquad \rho_1(y)\propto (\mbox{sin} \pi y)^{-x_b} \bigg( \mbox{sin} \dfrac{\pi y}{2} \bigg)^{x_s} \qquad
\label{Rho01}
\end{equation}
They are related to each other as $\rho_1(y)=\rho_0(1-y)$ and they are analogous to the order parameter profile in the system with fixed-free boundary conditions \cite{PhysRevLett.66.895,Burkhardt1991}.

The density contains clusters which touch the boundary either at $l=1$ or $l=L$ or both, arising from Eq.\ref{HInfPlrhoArb} of the form:
\begin{equation}
\rho_e(y) \propto (\mbox{sin} \pi y)^{-x_b}
\label{Rhoe}
\end{equation}
which corresponds to the order parameter profile with parallel fixed boundary conditions \cite{Burkhardt1985}.

In a simple manner we can find the density of spanning clusters $\rho_b(y)$, which touch both sides of the infinite strip. Adding $\rho_0$ and $\rho_1$ includes all configurations that touch either boundaries, but double counts the clusters that touch both sides of the strip. Subtracting $\rho_e$ leaves only those clusters that touch both boundaries. Thus a simple relation arises like \cite{Simmons2007}:
\begin{equation}
\rho_b(y)=\rho_0(y)+\rho_1(y)-\rho_e(y)
\end{equation}
which leads to the expression for $\rho_b(y)$ given by:
\begin{equation}
\rho_b(y)\propto (\mbox{sin}\pi y)^{-x_b}\bigg[\bigg( \mbox{cos}\dfrac{\pi y}{2}\bigg)^{x_s}+\bigg( \mbox{sin}\dfrac{\pi y}{2}\bigg)^{x_s}-1\bigg]
\label{Rhob}
\end{equation}

The results in Eqs.\ref{Rho01}, \ref{Rhoe} and \ref{Rhob} are derived for percolation, however it is expected that they also hold for the densities of Fortuin-Kasteleyn clusters in two-dimensional conformal invariant systems. During this work we refer to the analogous expressions for order parameter profiles, assume that Eq.\ref{Rho01} and Eq.\ref{Rhoe} are valid for conformal invariant models since they follow from conformal consideration. On the other hand, the third density $\rho_b(y)$ in Eq.\ref{Rhob} is expected to be valid also, since it can be expressed using the two previous ones.

\section{Clusters in the optimal set of the random bond Potts model}

The investigated strongly disordered system, which can be represented by Fortuin-Kasteleyn cluster method and presents conformal invariant behaviour at its criticality is the random bond Potts model in large $q$-limit, as we mentioned in the introduction. The Hamiltonian of the model is defined in Eq.\ref{HRBPM} and according to our earlier recognitions, the partition function is dominated by an optimal set, which analogous to the set of connected bonds in percolation problems. The phase transition of the model is of first order at the pure case, when each bond have the same weight $J_{ij}=J$ and it softens to a second oder transition for an arbitrary small disorder. Here we are going to study the disordered case of this model and investigate the density of clusters in the optimal set in strip geometry.

In our system the $d_f$ fractal dimension of the infinite cluster at the critical point and the $x_b$ scaling dimension of the bulk magnetization are related to the golden-mean ratio, $\varPhi=(1+\sqrt{5}/2)$ as \cite{PhysRevE.69.056112}:
\begin{equation}
d_f=1+\dfrac{\varPhi}{2} , \qquad x_b=1-\dfrac{\varPhi}{2}=\dfrac{3-\sqrt{5}}{4}
\label{x_bulk}
\end{equation}
where we used Eq.\ref{fractdimxb}. The analogous surface quantities are conjectured to be:
\begin{equation}
d_f^s=\frac{1}{2} , \qquad x_s=\dfrac{1}{2}
\label{x_surf}
\end{equation}

During our study we use a discrete bimodal form of disorder, when the reduced couplings $K_{ij}=\beta J_{ij}$ (where $0\leqslant K_{ij} \leqslant 1$) can take values following a distribution of:
\begin{equation}
P_b(K_{ij})=p\delta(K-\Delta-K_{ij})+(1-p)\delta(K+\Delta-K_{ij})
\end{equation}
where $K$ denotes the average reduced couplings and we choose $p=1-p=1/2$, which defines the distribution symmetrical. Generally we study the critical point of the system which is located at $K=K_c=1/2$ following from self-duality (see Section \ref{RandomBondModels} and \cite{PhysRevB.23.3421}) and it is independent of the strength of disorder: $0\leqslant \Delta \leqslant 1/2$. Note that the pure system is obtained for $\Delta=0$, whereas for $\Delta=1/2$, when just strong bonds are presented in the system, we get back the traditional percolation problem.

\begin{figure}[htbp]
\begin{center}
      \includegraphics*[ width=9.0cm]{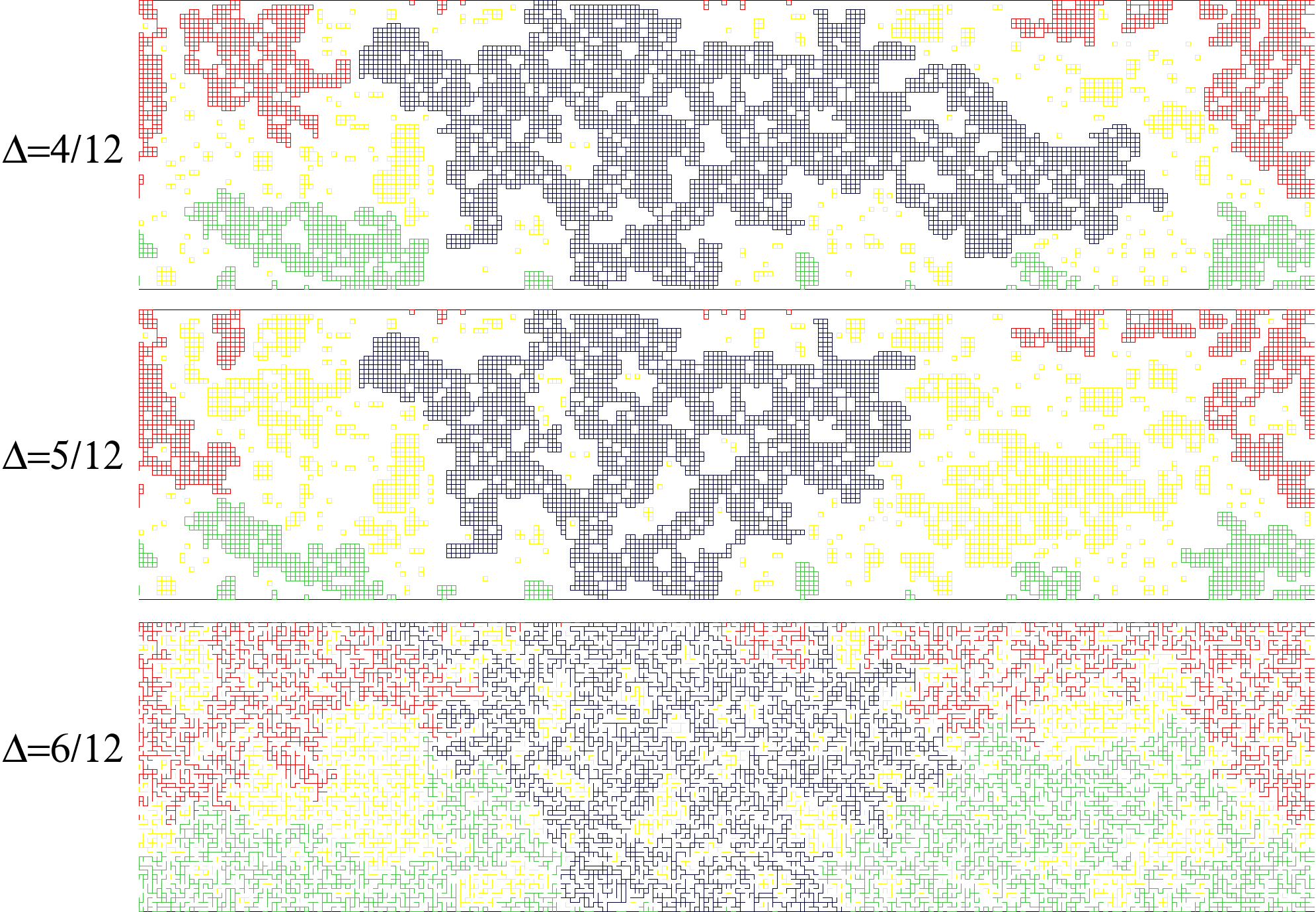}
\caption{Optimal diagrams of the random bond Potts model for three realizations of the disorder of different strength. Lower panel, $\Delta=1/2$, standard bond percolation; middle panel, $\Delta=5/12$; upper panel, $\Delta=4/12$. We have used different colors to visualize the different type of clusters in the optimal diagram: spanning clusters [black], clusters which touch only the upper (lower) boundary of the strip [red (green)] and clusters which have no common points with the boundary [yellow]. Note that the breaking-up length is increasing with decreasing $\Delta$.}
\label{PottsPercoFig1}
\vspace{-10pt}
\end{center}
\end{figure}

The evaluation of the optimal diagram with decreasing values of $\Delta$ is shown in Fig.\ref{PottsPercoFig1}. Here one can see that with decreasing $\Delta$, the clusters become more compact. To characterize this effect we define a length-scale, the so-called breaking-up length, $l_b$: in a finite system of linear size $L<l_b$ the optimal diagram is typically homogeneous (either empty of fully connected) and for $L>l_b$ it contains both empty and connected parts. If we want to determine $l_b$, we fix the size of the system, $L$, and for a given distribution of the bonds decrease $\Delta$ until $\Delta_b$, where the optimal set becomes fully connected. Repeating this calculation for several realizations of the disorder we obtain an average value $\overline{\Delta}_b$, to which the breaking-up length is just $L=l_b$. $l_b$ is a rapidly increasing function of $1/\Delta$, for small $\Delta$ it behaves as\cite{PhysRevE.69.056112}:
\begin{equation}
l_b \approx l_0 \exp\left[ A \left( \frac{K}{\Delta} \right)^2 \right]\;.
\label{eq:l_b}
\end{equation}
In a numerical calculation on a finite sample one should have the relation, $L \gg l_b$, thus $\Delta$ should be not too small. On the other hand one should also be sufficiently far from the percolation limit, $\Delta=1/2$, in order to get rid of cross-over effects. This means that the optimal choice of $\Delta$ is a result of a compromise, which in our case seems to be around $\Delta=5/12$, when the typical breaking-up length is about $l_b \sim 14$. Most of our studies are made for this value, but in order to check universality, i.e. disorder independence of the results, we have made also a few calculations for $\Delta=21/48$ also.

To calculate the optimal diagram of a given realization of disorder we use the combinatorial optimization algorithm that we defined in Section \ref{CombinOpt} and which works in strongly polynomial time. Application of this method
make possible to obtain the exact optimal diagram for comparatively large finite systems. In order to have an effective strip geometry we have considered lattices of rectangle shape with an aspect ratio of four. The strips have open boundaries along the long direction and periodic boundary condition was used in the other direction. We mention that the same geometry has been used before for percolation, too \cite{Simmons2007}. The width of the lattices we considered are from $L=32$ up to $256$ (i.e. the largest systems contained 262144 sites) and for each sizes we have considered $1000$ samples, except for $L=64$ and $128$, when we had $1509$ and $2128$ samples, respectively. These calculations were performed on a cluster of 16 quadricore processors during more than a month.

\section{Densities of critical clusters}

In the following section we are going to present and analyze our average numerical results of density of critical clusters of the random bond Potts model and compare those to the exact analytical results obtained by conformal mapping for critical percolation \cite{Kleban2006,Simmons2007}.

\subsection{The $\rho_b(y)$ cluster density}

\begin{figure}[htbp]
\begin{center}
      \includegraphics*[ width=10.0cm]{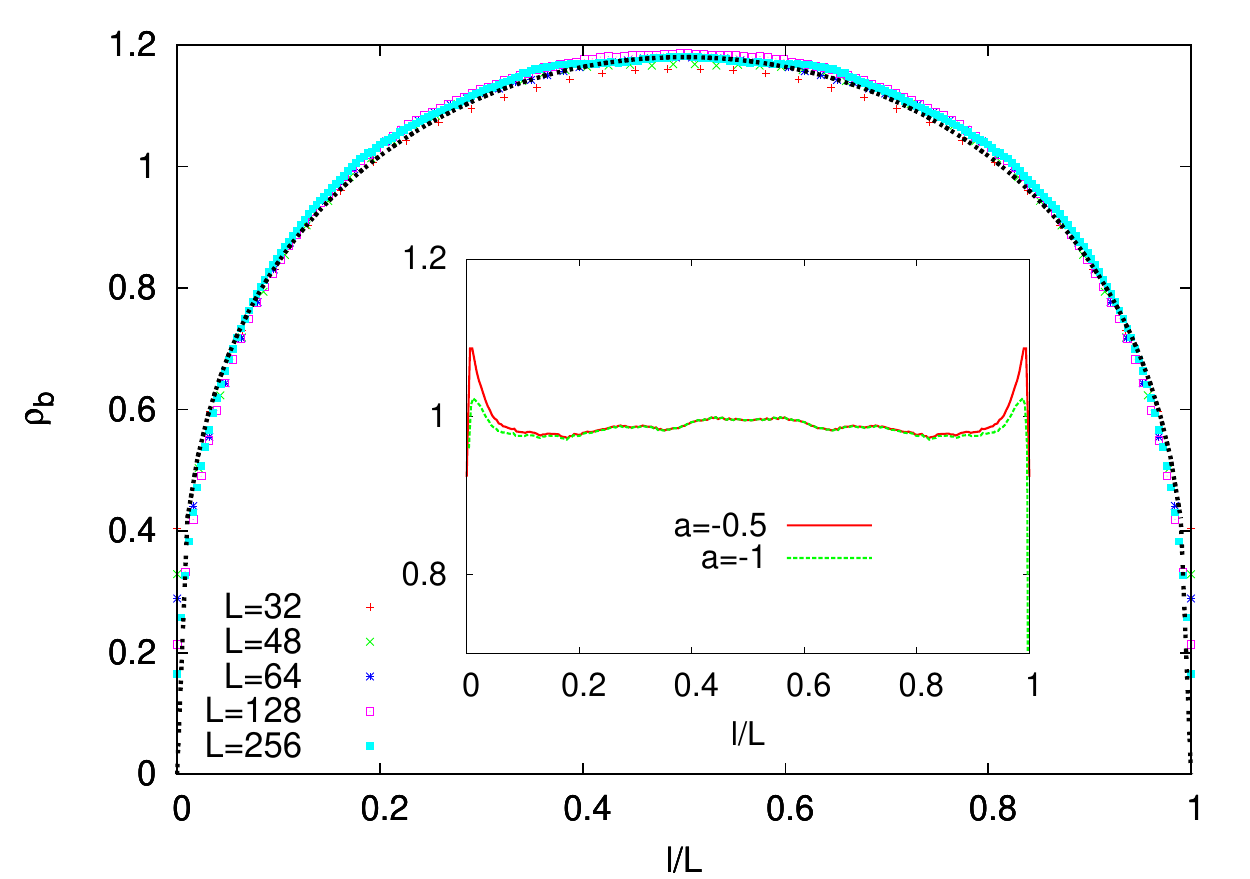}
\caption{Normalized density profiles $\rho_b(l/L)$, of the random bond Potts model for different widths. The standard deviation of the calculated points is smaller than the size of the symbols. The dashed line indicates the conformal result in Eq.\ref{Rhob} with the conjectured exponents in Eq.\ref{x_bulk} and Eq.\ref{x_surf} using the boundary parameter $a=0.$ defined in Eq.\ref{y_a}. In the inset the ratio of simulation to theoretical results are presented for $L=256$ with two different boundary parameters: $a=-0.5$ and $a=-1.0$, see Eq.\ref{y_a}}
\label{PottsPercoFig2}
\vspace{-10pt}
\end{center}
\end{figure}

We start to study the density of crossing clusters, $\rho_b(l/L)$, using the scaling form of which is conjectured in Eq.\ref{Rhob} for conformal invariant systems. For the random bond Potts model the numerically calculated normalized densities $\rho_b(l/L)$, for different widths are shown in Fig.\ref{PottsPercoFig2}. The data for different widths fit to the same curve and the finite breaking-up length $l_b$, seems to have only a small effect.

\begin{figure}[htbp]
\begin{center}
      \includegraphics*[ width=10.0cm]{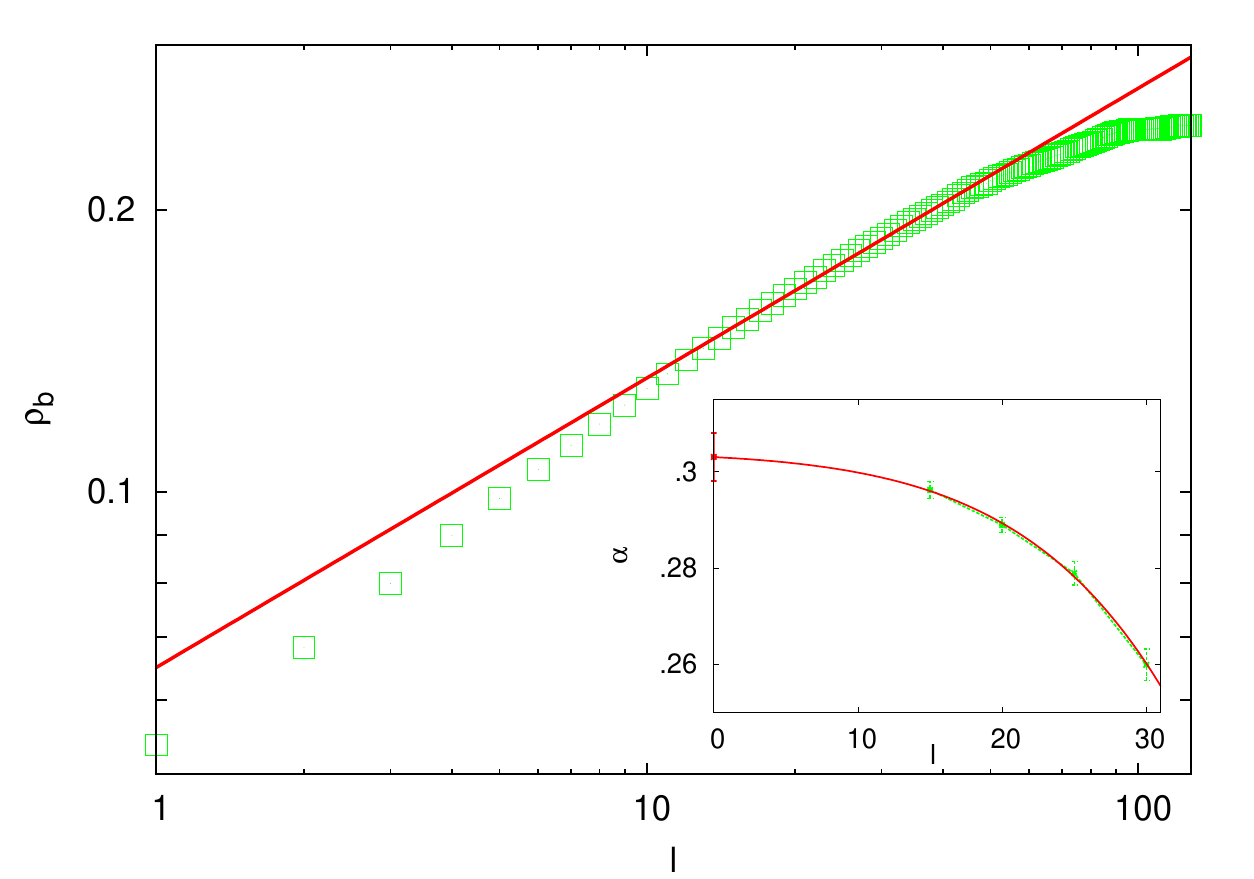}
\caption{Density profile, $\rho_b(l)$, of the random bond Potts model for $L=256$ close to the surface in log-log plot. The straight line has a slope $x_s-x_b=0.309$. Inset: estimates of the slope using different windows of the fit, see the text. Here the full line indicates a parabolic fit.}
\label{PottsPercoFig3}
\vspace{-10pt}
\end{center}
\end{figure}

In the surface region $l \ll L$, but $l > l_b$, one expects from scaling theory:
\begin{equation}
\rho_b(l) \sim l^{x_s-x_b} 
\end{equation}
which is in accordance with the limiting behavior of the conformal prediction in Eq.\ref{Rhob}. In Fig.\ref{PottsPercoFig3} we have presented $\rho_b(l)$ in a log-log plot in the surface region for the largest finite system. Indeed, for $l>l_b$ the points are well on a straight line, with a slope of which is compatible with the conjectured value: $x_s-x_b=0.309$. We have also estimated the asymptotic slope of the curve by drawing a straight line through the points in a window $[l_b+l/2,l_b+3l/2]$ by least square fit. Fixing $l_b=15$ the estimates with varying $l$ seem to have a $\sim l^2$ correction (see the inset of Fig.\ref{PottsPercoFig2}) and the extrapolated slope is $x_s-x_b=0.303(8)$ in agreement with the conjectured values in Eq.\ref{x_bulk} and Eq.\ref{x_surf}.

We have also checked the conjectured form of the profile in Eq.\ref{Rhob} using the scaling exponents in Eq.\ref{x_bulk} and Eq.\ref{x_surf}, which indeed fits very well the scaling curve for the random bond Potts model for the whole profile. We calculated the ratio of the simulation to the theoretical results. However, when we compare the numerical results on a finite lattice to the results of continuum theory, the question arises due to finite-size and edge effects, of what value of the continuous variable $y$ should correspond to the lattice variable $l$, specially close to the strip boundaries \cite{Simmons2007}. The density at the vicinity of the boundaries at $l=1$ and $l=L$ goes to zero, which suggests a simple assignment to the continuous variable like $y=l/L$ as we applied generally. However we can get a better fit to the data near to the boundaries by assuming that the effective position of the boundary is at a distance $a$ (in units of the lattice spacing) beyond the lattice boundaries. This is accomplished by defining a continuum variable $y$ as:
\begin{equation}
y=(l+a)/(L+2a)\;,
\label{y_a}
\end{equation}
In such a way $l=-a$ corresponds to $y=0$ and $l=L+a$ corresponds to $y=1$. By varying $a$ from $0$ to $-1$ one obtains a better fit in the boundary region, but at the same time the bulk part of the profile remains practically unchanged. As seen in the inset of Fig.\ref{PottsPercoFig3} in the bulk part of the profile the non-systematic fluctuation of the ratio around unity is typically 1-2\%.

\subsection{The $\rho_0(y)$ cluster density}

\begin{figure}[htbp]
\begin{center}
      \includegraphics*[ width=9.5cm]{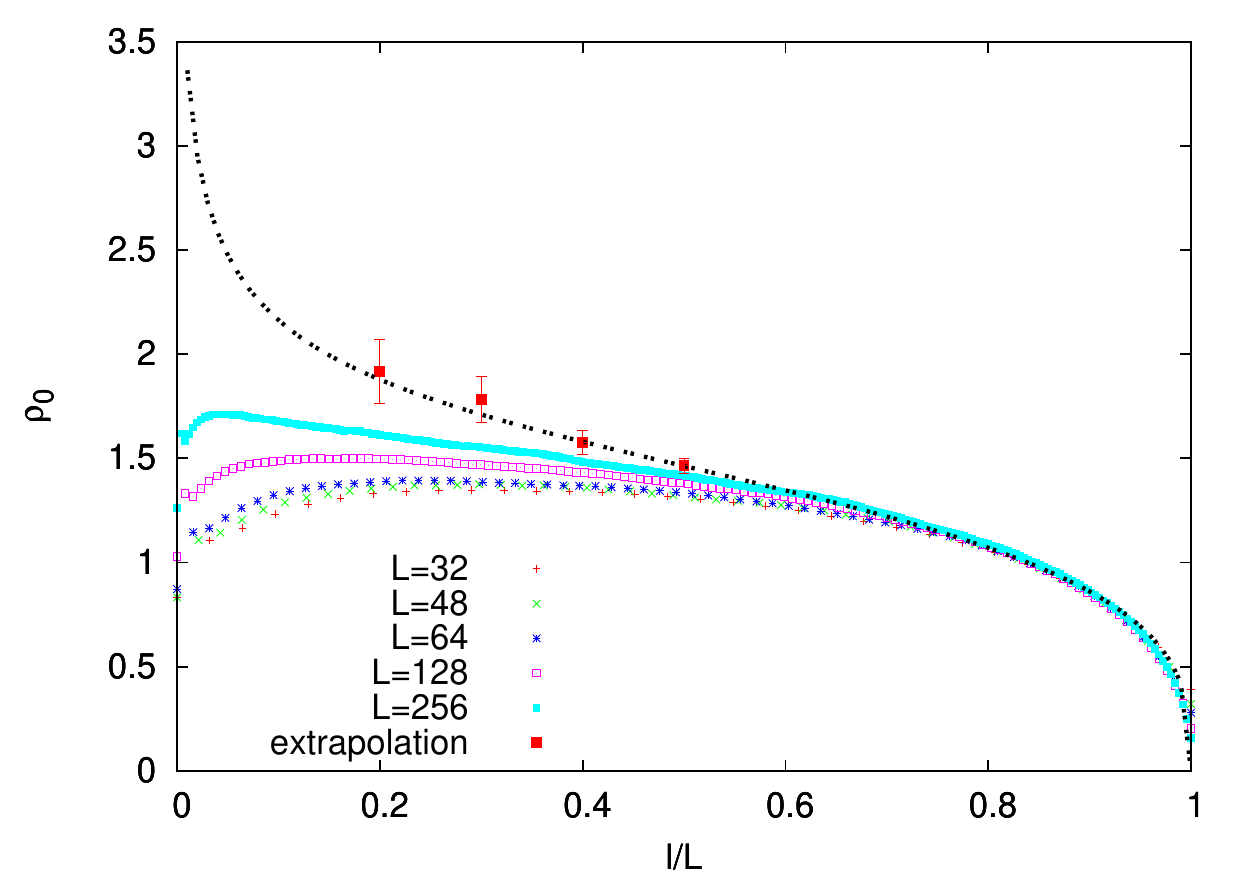}
\caption{Density profiles, $\rho_0(l/L)$, of the random bond Potts model for different widths which approach the same scaling curve at the free boundary $y=1$. The standard deviation of the calculated points is smaller than the size of the symbols. The extrapolated values for $l/L \ge 0.5$ are denoted by red squares. The dashed line indicates the conformal result in Eq.\ref{Rho01} with the conjectured exponents in Eqs.\ref{x_bulk} and \ref{x_surf}.}
\label{PottsPercoFig4}
\vspace{-10pt}
\end{center}
\end{figure}

At second we consider the density $\rho_0(l/L)$ of those clusters which are touching one boundary of the strip including the spanning clusters, compared to the scaling form of which is conjectured in Eq.\ref{Rho01}. This density is analogous to the order parameter profile in the system with fixed-free boundary conditions \cite{PhysRevLett.66.895,Burkhardt1991}. The numerically calculated densities are shown in Fig.\ref{PottsPercoFig4} for different $L$ widths. 

The profiles at the fixed boundary $y=0$, are perturbed by surface effects, which are due to the presence of the finite breaking-up length. Indeed in terms of the scaled variable $y=l/L$, the size of the perturbed surface region, $\tilde{y}_L$ is a decreasing function of $L$. On the other hand at the contrary boundary $y=1$, where the profiles are not influenced by the fixed surface at $y=0$ and the clusters touching both boundaries dominate, the densities approach the same scaling curve, which in the vicinity of the boundary behaves as $\rho_0(y) \sim (1-y)^{x_s-x_b}$. Comparing the scaling curve with the conformal prediction in Eq.\ref{Rho01} we obtain an overall good agreement for  $1 \ge y>0.5$. In the region $y \le 0.5$, where the finite-size profiles deviate more strongly from each other we used an extrapolation procedure. At a fixed $y$ we have plotted $\rho_0(y)$ as a function of $1/L$ and from this we have estimated the value of the scaled curve as $L \to \infty$. With this method we have obtained estimates in the region $0.2 \ge y$, which are in agreement with the conformal result as seen in Fig.\ref{PottsPercoFig4}.

Turning back to the finite-size dependence of the densities at the fixed surface, $y=0$, we note that in the continuum limit, in case of $l_b \ll l \ll L$, the scaling form of the density is described by the result of Fisher and de Gennes \cite{Fisher1978}: $\rho_0(l) \sim l^{-x_b}$ as we discussed in Section \ref{ConformDensityDef}. However by approaching the breaking-up length, $l_b$, the increase of the profile is stopped and for $l<l_b~~\rho_0(l)$ starts to decrease. This is due to the structure of the connected clusters close to the surface. As seen in Fig.\ref{PottsPercoFig1} the number of touching sites in a cluster is comparatively smaller for the random bond Potts model with $\Delta<1/2$ (upper and middle panel of Fig.\ref{PottsPercoFig1}), than for percolation with $\Delta=1/2$ (lower panel of Fig.\ref{PottsPercoFig1}). Also for finite widths the small and medium size touching clusters are rarely represented for the random bond Potts model. By approaching with $l$ the other, free side of the strip the crossing clusters start to bring the dominant contribution to the density, $\rho_0(l/L)$, which is then well described by the conformal formula.

\subsection{The $\rho_e(y)$ cluster density}

Finally we consider the density of points in such clusters which are touching either the boundary at $l=1$ or at $l=L$ or both, which is denoted by $\rho_e(l/L)$ and the conjectured conformal formula is given in Eq.\ref{Rhoe}. This density is analogous to the order parameter profile with parallel fixed spin boundary conditions\cite{Burkhardt1985}.

\begin{figure}[htbp]
\begin{center}
      \includegraphics*[ width=10.0cm]{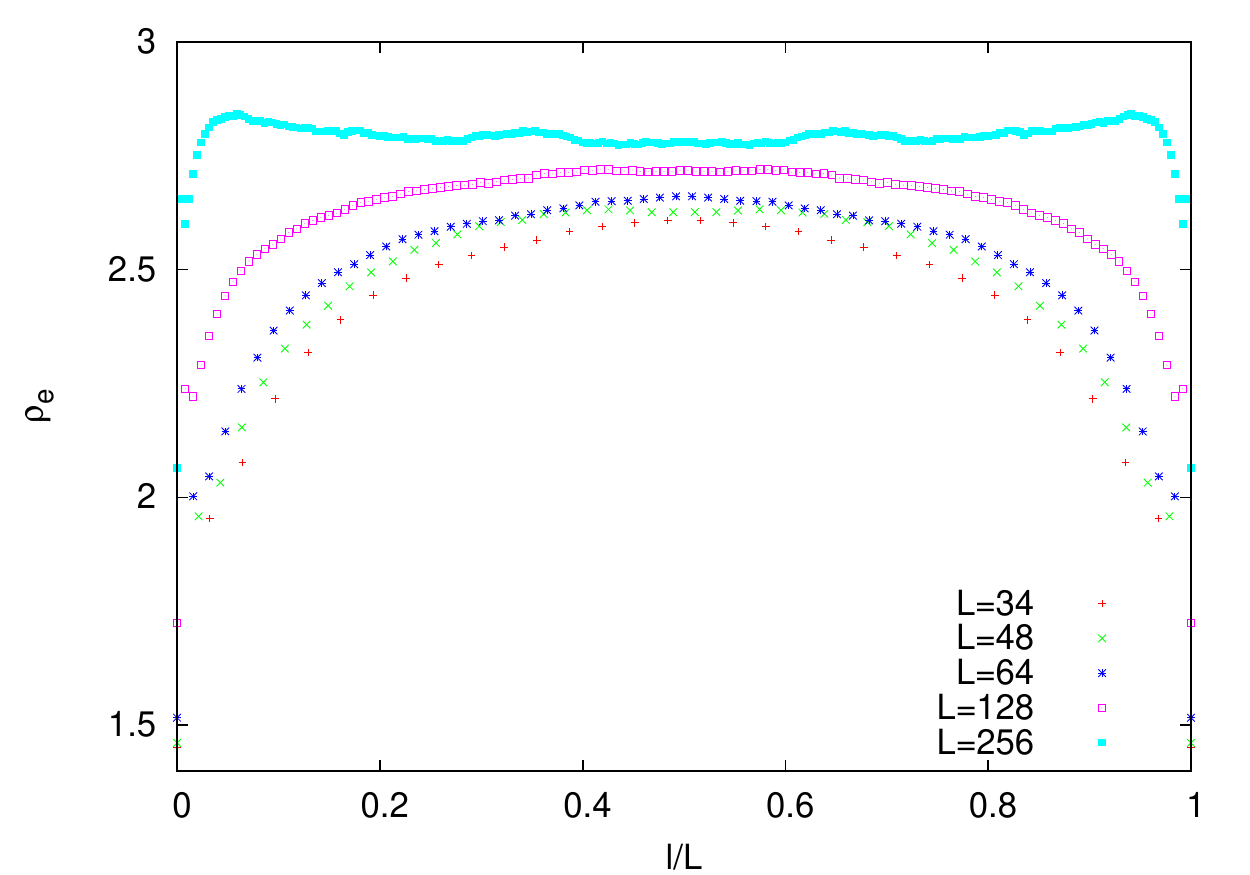}
\caption{Density profiles $\rho_e(l/L)$ for different widths $L$. These densities are strongly perturbed by the finite breaking-up length $l_b$ at both boundaries and fit with the expected conformal predictions only at the thermodynamical limit.}
\label{PottsPercoFig5}
\vspace{-10pt}
\end{center}
\end{figure}

\begin{figure}[htbp]
\begin{center}
      \includegraphics*[ width=10.0cm]{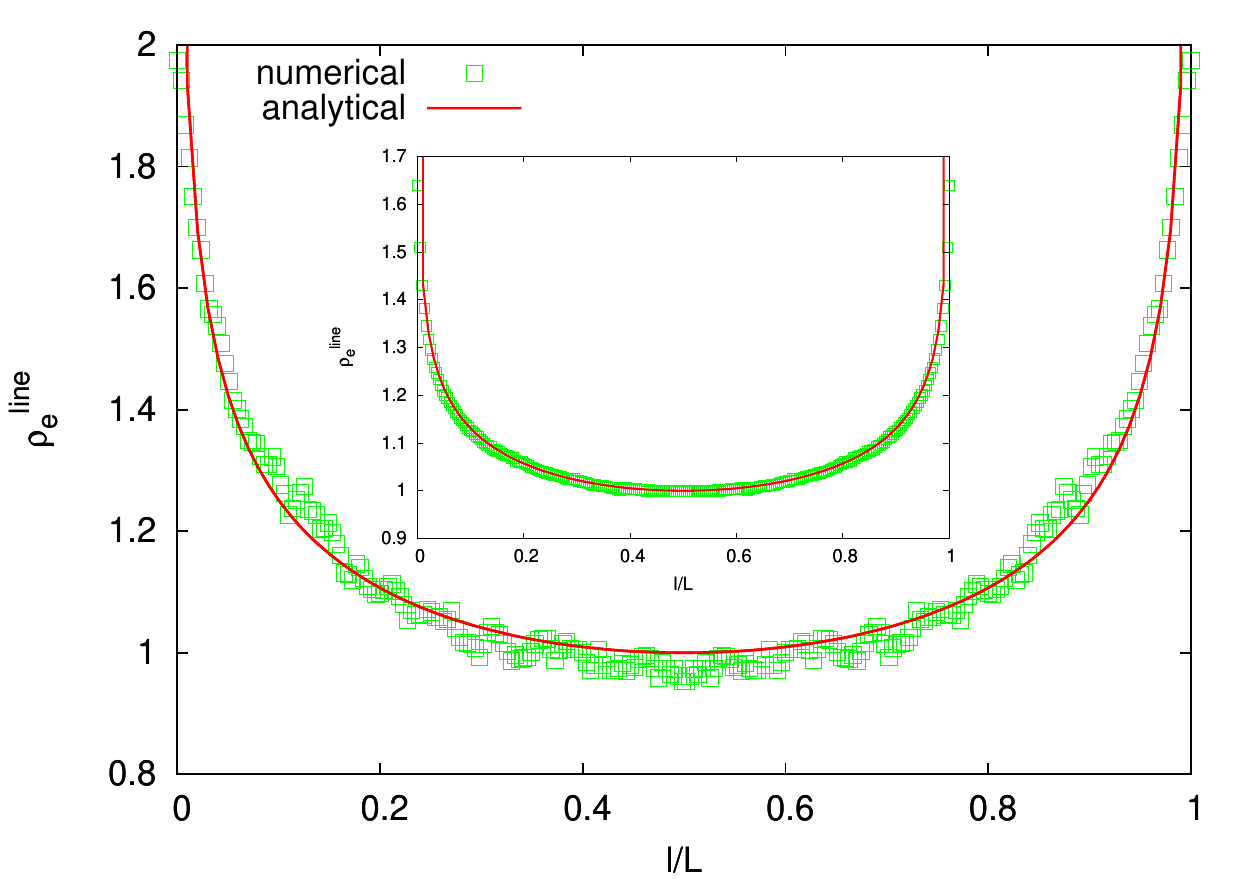}
\caption{Density profile along a vertical line with two touching boundary points, $\rho_e^{\rm line}(l/L)$, for the random bond Potts model. The solid curve indicates the conformal result in Eq.\ref{Rhoe} with the conjectured exponents in Eq.\ref{x_bulk} and Eq.\ref{x_surf}. In the inset the same quantity is shown for percolation. Here in the analytical formula in Eq.\ref{Rhoe} we use $x_b=5/48$ and $x_s=1/3$, the scaling dimension values of the bulk and surface magnetization for percolation. In both figures the boundary parameter in Eq.\ref{y_a} is $a=0$.}
\label{PottsPercoFig6}
\vspace{-10pt}
\end{center}
\end{figure}

For the random bond Potts model this density is strongly perturbed by the finite breaking-up length at both boundaries as can be seen in the Fig.\ref{PottsPercoFig5}. In this case we did not try to perform an extrapolation and conclude that even larger finite systems would be necessary to test the conformal predictions in a direct calculation. In order to try to test the result in Eq.\ref{Rhoe} we studied another density which is defined on crossing clusters, so that one expects it to be represented correctly in smaller systems, too. Here we define a density, $\rho^{\rm line}_e(l/L)$, in crossing clusters and consider points only in such vertical lines, where at both ends of the given line the cluster touches the boundaries. Since $\rho^{\rm line}_e(l/L)$ is related to the operator profile with fixed-fixed boundary conditions we expect that it has the same scaling form as the previously defined density, $\rho_e(l/L)$. In Fig.\ref{PottsPercoFig6} we show the calculated densities for the random bond Potts model, which are compared with the analytical prediction in Eq.\ref{Rhoe}. A similar analysis for percolation is shown in the inset of Fig.\ref{Rhoe}. In both cases we found that the numerical and analytical results for this type of profile are in satisfactory agreement, although the statistics of the numerical data is somewhat low, since just a fraction of $\sim L^{-2x_s} \sim L^{-1}$ lines can be used in this analysis. The non-systematic fluctuation of the numerical data is less than 1\% for percolation and about 3\% for the random bond Potts model.

\section{Summary}

In this chapter we have studied the density of critical clusters in a model with critical properties of which are dominated by disorder effects. Our study is motivated by a recent investigation about ordinary percolation in \cite{Simmons2007} in which the densities are calculated in the continuum approximation through conformal invariance. Here we have suggested the generalization of these analytical results for another conformal invariant system in Eqs.\ref{Rhob}, \ref{Rho01} and \ref{Rhoe}. To test these predictions we have studied numerically the density of Fortuin-Kasteleyn clusters in infinite strips of the two-dimensional random bond Potts model in the large-$q$ limit. This model is expected to be conformal invariant, which means that average quantities which are related to Fortuin-Kasteleyn clusters (such as correlation function and magnetization densities) are invariant under conformal transformations.

In the actual work we have calculated the density of points of different type of clusters (crossing clusters, clusters which touch one boundary of the strip, etc.) in analogy with a related study of percolation in \cite{Simmons2007}. We can thus conclude that all these critical densities we considered here for the random bond Potts model are found to be in agreement with the conformal predictions in Eqs.\ref{Rhob}, \ref{Rho01} and \ref{Rhoe} in which we have used the scaling dimensions of the random bond Potts model in Eq.\ref{x_bulk} and Eq.\ref{x_bulk}. From an analysis of the profile, $\rho_b(l)$, close to the boundary we have obtained new accurate estimate of the critical exponent, $x_s-x_b$, giving further support of the conjecture in Eq.\ref{x_bulk} and Eq.\ref{x_surf}.  As far as the full profiles are considered they are well described by the conformal continuum predictions at least for lengths which are larger than the breaking-up length, $l_b$. Consequently our study has given support for the possible validity of the conjectured results.

Our investigations can be extended into different directions. For two-dimensional classical systems one can study the density of Fortuin-Kasteleyn clusters in the $q$-state Potts model, both without disorder (for $q \le 4$) and in the presence of disorder (for general value of $q$) and one can consider another type of geometries (semi-infinite strip, square, etc) as well. 

The above presented results were published in a recent paper wherefrom some parts of the corresponding chapter were taken. This paper was written by M\'arton Karsai, Istv\'an Kov\'acs, Jean-Christian Angl\`es d'Auriac and Ferenc Igl\'oi, and it was published in the Physical Review E, \textbf{78}, 061109 in 2008 \cite{Karsai2008}. This article contains results of the quantum random transverse-field Ising chain which are not included into the dissertation.

%% file: TAFIM.tex
\chapter{Non-equilibrium dynamics of triangular antiferromagnetic Ising model at $T=0$}

\section{Introduction}

The non-equilibrium relaxation of critical systems is an interesting subject that was studied intensively experimentally and theoretically in the last few decades \cite{Calabrese2005,Melin2006,PhysRevLett.62.1512,PhysRevLett.94.245701,Melin2005}. Moreover, dynamics out of equilibrium present some special features if the system is constrained by different rules e.g. arising from the non-homogeneous interactions or induced by geometrical frustrations. In geometrically frustrated systems structural properties forbid to minimize spontaneously the energy which leads to highly degenerated ground state where the system has non-zero residual entropy even at $T=0$.

One of the simplest model which is capable to study such behaviour is the antiferromagnetic Ising model on triangular lattice \cite{PhysRev.79.357}, where the frustration arises from triangular structure and the antiferromagnetic couplings. This system was one of the first exactly solved spin model in the 50s \cite{Houtappel1950,PhysRev.79.357,Stephenson1970}. The equilibrium critical behaviour of this model was carefully studied at $T=0$ where the ground state dynamics is governed by fluctuation of spins where flipping does not cost any energy. At zero temperature in absence of external field the two-dimensional system is in a critical phase since the correlation function has an algebraic decay and by increasing the external magnetic field at $h=h_c$ it passes through a Kosterlitz-Thouless phase transition \cite{Blote1982,Blote1993,PhysRevB.43.8751}. A similar type of critical behaviour was earlier found for the XY model in two dimension \cite{Kosterlitz1973}.

The non-equilibrium dynamical behaviour is influenced by motions of topological defect (plaquettes of three identically orientated spins). However, controversial results have been found about the characteristic rules of these motions that suggested different dynamical descriptions. In one way the defects were found not to be independent and joined by a Coulomb-force like interaction. Their motions are slowed down by a logarithmic factor which is then observable in dynamical quantities. Such a phenomena has been found in pure systems with topological defects like the XY model. This logarithmic corrected dynamical behaviour was observed by Moore et.al. \cite{Moore1999} who deduced this implications from the time dependency of average defect densities. They suggested that the dynamical exponent $z$ was equal to two with an additional size-dependent logarithmic term. However, this behaviour is not supported by the numerical results of Kim et.al. \cite{PhysRevE.68.066127,Kim2007} who found the defects are moving like a random walker in limited directions. They examined the two-points and two-times correlation functions and observed a subdiffusive behaviour where the spreading of correlation was governed by and effective dynamical exponent $z\simeq 2.33$.

Our aim is to find independent evidences to decide which paradigm is relevant to describe the non-equilibrium critical dynamical behaviour of the triangular Ising antiferromagnet at $T=0$ in absence of external field or when the $h=h_c$ critical field is applied. In earlier works the dynamical exponent was estimated through exponent combinations or its value was strongly influenced by the choice of some parameters. Here we introduce a quantity which depends only on the dynamical exponent $z$ which then enables us to measure and analyze its asymptotic behaviour directly. This quantity is defined as the relaxation time when the system reaches its energy minima, after it was initiated from a random $T\rightarrow\infty$ state and quenched to $T=0$. This relaxation time characterize the length of the non-equilibrium aging regime and possible to measure sample-by-sample. 

After a brief theoretical introduction, we are going to study the distribution of this relaxation time which directly relates to the aging dynamics of the system thus we can use its true dynamical features to conclude. We also measured the two-times autocorrelation function in equilibrium and in the aging regime in the case of different waiting times. We examine our numerical results using a similar method as in \cite{Walter2008} which will be the subject of the second and the third sections. We close this chapter with a short summary of our conclusions. The considerations presented in this chapter are new scientific results which have not been published, however the publication is in progress currently.

\section{The model}
\label{GeomFrustr}

The Hamiltonian of the triangular Ising antiferromagnet can be deduced from Eq.\ref{IsingHGen} with $J<0$ antiferromagnetic exchange energy. Since the underlying lattice is triangular, on a single plaquette (three neighbouring spins) two spins can align antiparallel satisfying the antiferromagnetic rule but the third spin is frustrated since it cannot simultaneously minimize the energy of interactions with both of the other spins. So it chooses one from the two possible equal energy state and aligns antiparallel with one spin but parallel with the other where the coupling then stays unsatisfied (ferromagnetic). The lowest-energy configurations of the system is that in which every elementary triangle are maximally satisfied.

\begin{figure}[htb]
\begin{center}
\includegraphics*[ width=9.0cm]{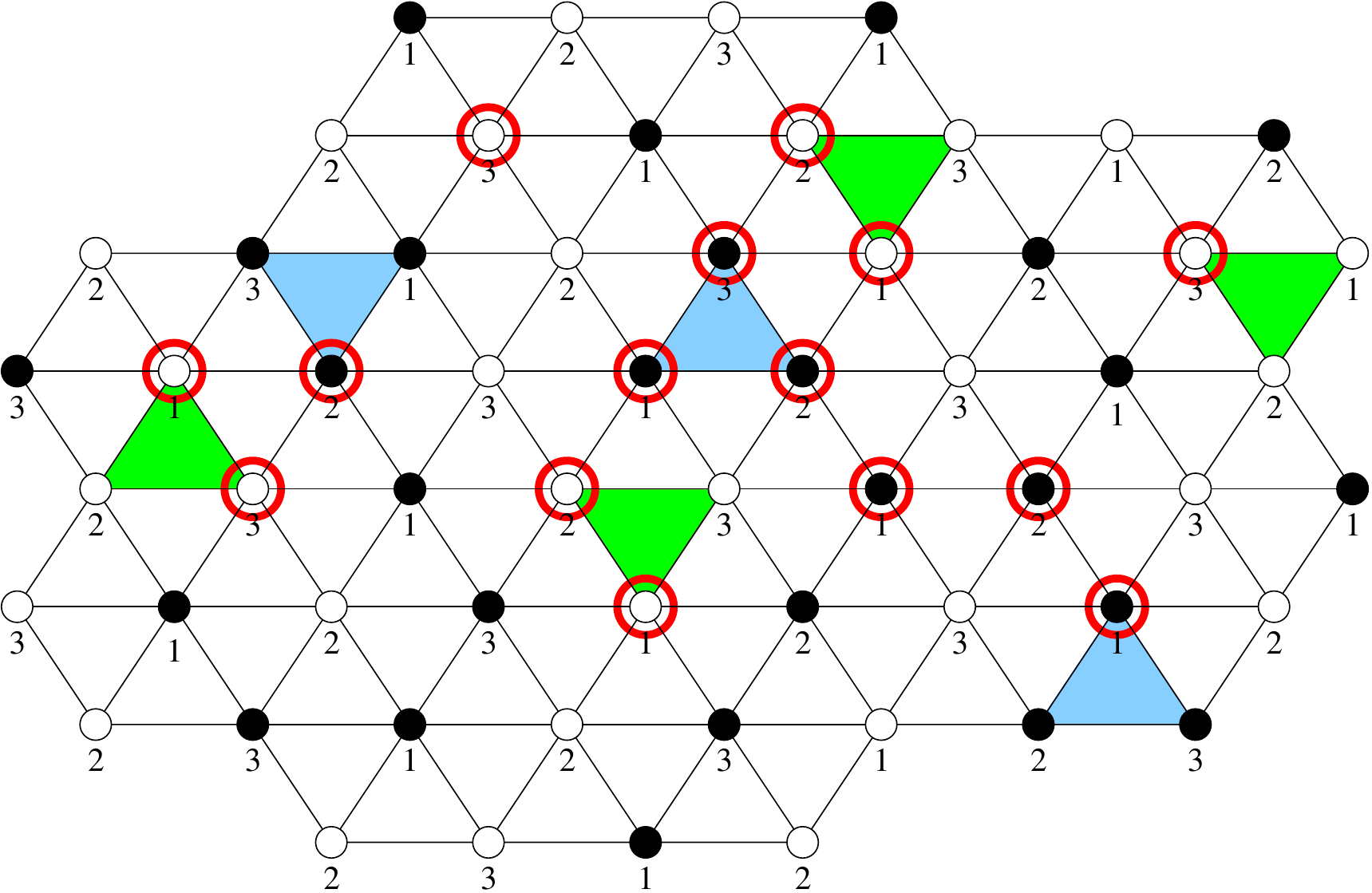}
\caption{Demonstration of the triangular Ising antiferromagnet. The spins in different states are located at the lattice sites and signed by filled or empty circles. The triangular lattice can be decomposed to three sublattices of non-neighbouring vertices, denoted here by numbers. The defects of different sign (green or blue triangles) are induced on a plaquette by three neighbouring spins at a common state. The loose spins which flip without changing the energy are signed by a red ring around.}
{
\label{TAFIM}
}
\end{center}
\vspace{-10pt}
\end{figure}

The triangular antiferromagnetic Ising model in two dimensions was one of the first solved Ising system investigated by Houtappel in 1950 who gave the partition function of the model \cite{Houtappel1950}. At the same year, Wannier \cite{PhysRev.79.357} calculated exactly the density of residual entropy $S(T)$ at $T=0$ of the form:
\begin{equation}
S(0)=\dfrac{2}{\pi}\int_{0}^{\pi/3}\mbox{ln}( 2 \mbox{ cos}( x))dx\simeq 0.3383
\label{SWannier}
\end{equation}
This non-vanishing entropy suggests that the system in the thermodynamical limit and at $T=0$ has infinite number of ground states and the energy minima state is still exponentially degenerated in case of finite size. To calculate thermodynamical averages, we need to pass through this equilibrated configuration space of possible ground states. If we derive the spin-spin correlation function averaging over this configuration set, it shows an algebraic decay with the distance, rather than an exponential behaviour, which denotes that the ground state at $T=0$ is critical. Following from calculations of Stephenson \cite{Stephenson1970} the spin-pair correlation is described by the form along the three major axes as:
\begin{equation}
\Gamma_{eq}(r)\sim r^{-\eta}\mbox{ cos}\bigg(\dfrac{2\pi r}{3} \bigg)
\end{equation}
where $\eta=1/2$. However the lattice can be divided into three equivalent sublattices of non-neighbouring spins as we demonstrated on Fig.\ref{TAFIM}. Within a sublattice the equilibrium correlation function has a clear singular behaviour as $\Gamma_{eq}(r)\sim r^{-\eta}$ with the same exponent and in agreement with Eq.\ref{CF} since the $d$ spatial dimension is two.

Let us remark that it is possible to choose an appropriate boundary shape where the system can have only one ground state and it becomes not degenerated \cite{Millane2003}. This special behavior arises for parallelogram shape geometries with free boundary condition, however a lattices with periodical boundary conditions, what we use during our studies, do not induce this special feature.

If we include a sufficiently small external field $H \ll J$ then the configuration space keeps equilibrated. However, by increasing $H$ at the range $H\sim J$ the external field becomes dominant and then some ordered states turn to be more favorable and long-range order evolves in the system.

In zero temperature limit in order to define dynamics, we can let the external field to scale with the temperature as $h=H/T$ thus when we consider the $T\rightarrow 0$ and $H\rightarrow 0$ limit $h$ remains finite, while the nearest-neighbour couplings does not, $J/T\rightarrow -\infty$ \cite{PhysRevB.43.8751}. It was originally suspected that the reduced field $h$ is relevant so the system would immediately enter the ordered phase when $h\neq 0$ however it was not supported by some later analysis \cite{Nienhuis1984}. It turned out the the antiferromagnetic triangular Ising model is exactly mapped to the triangular solid-on-solid model describing the equilibrium shape of a $(1,1,1)$ surface of a cubic crystal \cite{Blote1982,PhysRevB.43.8751}. Due to this mapping one can express the so-called height-height correlation function of the solid-on-solid model in terms of Ising correlation. Following from a non-exact renormalization mapping of Nienhuis \cite{Nienhuis1987} this correlation has the same asymptotic behaviour as the discrete Gaussian model and the lattice Coulomb gas model. Using these mappings the critical behaviour of the model in equilibrium was found to be formally equivalent to the one of XY model  which maps exactly to the Gaussian model \cite{Nienhuis1984} and has a \textit{Kosterlitz-Thouless phase transition} \cite{Kosterlitz1973}. Consequently the same type of critical behaviour would be expected for the triangular Ising antiferromagnet also.

\begin{figure}[htb]
  \begin{center}
    \begin{minipage}[b]{0.58\linewidth}
\includegraphics*[ width=8cm]{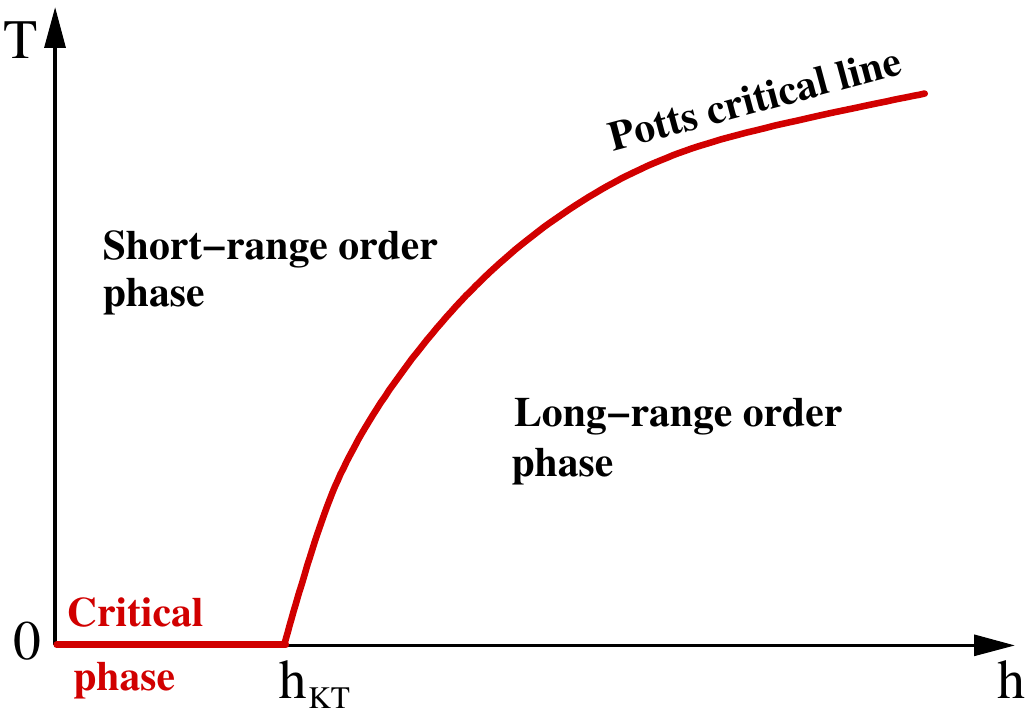}
    \end{minipage}\hspace{.2in}
    \begin{minipage}[b]{0.30\linewidth}
      \caption{Phase diagram of the two dimensional triangular Ising antiferromagnet. The system at $T=0$ for weak external field is in a critical phase and pass through a Kosterlitz-Thouless phase transition at $h=h_{KT}$. If $T>0$, the short-range order and the long-range order phase are separated by a three-state Potts critical line which saturates at $T=0$ to the $h_{KT}$ phase transition point. \qquad\qquad\qquad\qquad\qquad\qquad\qquad}
\label{tafimPhase}
    \end{minipage}
  \end{center}
\end{figure}

The phase diagram of the model is shown in Fig.\ref{tafimPhase}. At zero temperature and zero field, no long-range order arises in the system at least in the isotropic case \cite{PhysRevE.69.036127}, but we are at the critical phase since the correlation function has an algebraic decay (see above). Moving on the $T=0$ line, if we introduce a weak external field $h$, at a finite $h_{KT}$ value the system undergoes the expected Kosterlitz-Thouless phase transition of infinite order \cite{Kosterlitz1973} and arrives to the long-range order phase, where the residual entropy vanishes out. Here the so-called loose spins (that can flip without changing the energy) are condensed on one of the three sublattices, and directed by the external field. The resulting ordered state has spin of one sign on two of three sublattice and spins of the opposite sign on the third sublattice \cite{PhysRevB.43.8751}. Consequently this state is threefold degenerated only. For $T>0$ at zero field the system does not have any phase transition and it is in its short-range order phase where the correlation function shows exponential decay. For temperatures larger than zero the long and short-range order phases are separated by a critical three state Potts line which arrives to the $h_{KT}$ point at $T=0$ \cite{PhysRevE.69.036127}. The critical point is located at $h_{KT}=0.266(10)$ which comes from means of transfer matrix calculations and phenomenological renormalization \cite{Blote1993}. Following from these calculations, at the critical phase T = 0 along the critical line $0 \leq h \leq h_c$ some exponents were found to be constant, whereas others are $h$-dependent.

A general method to classify the ground states of the triangular Ising antiferromagnet in equilibrium at the criticality is to map each ground state to its corresponding dimer configuration \cite{Dhar2000}. Such a configuration is composed by strings of a dimer which can be established as a projection of the dual lattice of the given and a standard ground state configuration. These strings cannot intersect each other and their density can rank the different ground states into different density sectors. The string density is conserved during the time evolution of a ground state, consequently a ground state belonging to a certain sector cannot evolve into another ground state of a different density sector. In this sense, the dynamics within the ground state manifold is nonergodic, but for each sector with nonzero string density it is ergodic within the sector itself. The maximum of the residual entropy density in Eq.\ref{SWannier} belongs to the density value $2/3$ that defines the so-called dominant sector which contains the most ground states \cite{Dhar2000,Kim2007}. This sector plays a special role in case of non-equilibrium dynamics what we discuss in the next section.

\section{Dynamical behaviour}

In Section \ref{CritDyn} we already mentioned that a system that we initiate from a disordered state at time $t=0$, passes through a non-equilibrium relaxation toward its equilibrium phase. In an intermediate state during this relaxation the non-equilibrium two-point correlation function (defined as Eq.\ref{CFDef}) has the form:
\begin{equation}
\Gamma(r,t)\sim r^{-(d-2-\eta)}f\bigg(\dfrac{r}{\xi(t)}\bigg)
\label{noneqCF}
\end{equation}
where $d$ is the spatial dimension and $\eta$ is the equilibrium correlation exponent. This scaling form holds for arbitrary ratio of $r/\xi(t)$ if $r,\xi(t)\gg a$ where $a$ is the lattice constant. Moreover the first term of Eq.\ref{noneqCF} gives back the equilibrium correlation function as it is expected in the time limit $t\rightarrow \infty$ when $f(0)=1$.

The length $\xi(t)$ in Eq.\ref{noneqCF} is not the equilibrium correlation length which evolves during coarse-graining processes since here it is time dependent. During the non-equilibrium dynamics some correlation arises between the originally uncorrelated spins which induces dynamically increasing correlated domains in the system. We introduce here $\xi(t)$ as a time dependent length scale to characterize the size of these domains during the relaxation process. This length scales as $\xi\sim t^{1/z}$ for large $t$ with the same usual exponent $z$ which characterizes temporal correlation in equilibrium (see Section \ref{CritDyn}). Consequently this exponent $z$ keeps independent of the non-equilibrium initial state which can affect only the function $f(r/\xi(t))$.

\subsection{Non-equilibrium relaxation}

Following from earlier results found in the literature, the non-equilibrium dynamical behaviour of frustrated systems and related models has many interesting features. The fact that the dynamics depend on the initial state wherefrom the system was started is a very important character. This initial state dependence was observed for the XY model \cite{Bray1994,Bray1994a,Bray2000} which has the same critical behaviour than the triangular Ising antiferromagnet as we have seen earlier in Section \ref{IsingModel}. The dynamical behaviour is different if we start from a defect-free state where the growing length scale satisfies Eq.\ref{dynCorrL} with $z=2$. On the other hand if the system is started from a random initial state and quenched to any temperature between $0<T\leq T_{KT}$ (where $T_{KT}$ being its Kosterlitz-Thouless critical temperature) the growing length scale exhibits a diffusive growth during the non-equilibrium relaxation which is slowed down with a logarithmic factor as
\begin{equation}
\xi(t)\sim \bigg( \dfrac{t}{\mbox{ln}(t)}\bigg)^{1/z}
\label{logCorrection}
\end{equation}
with the same $z$ exponent value \cite{Bray1994,Bray1994a,Bray2000,He2009}.

An initial state dependence has been found for the triangular Ising antiferromagnet also. If the initial state is a ground state which belongs to a certain sector (see above) then the system is driven by only loose spin fluctuations and cannot pass to another sector. These ground state dynamics induce a growing correlation length scale which is expected to satisfy Eq.\ref{dynCorrL} with the equilibrium dynamical exponent $z$. However if we start the system from a random initial state it spontaneously evolves into the dominant sector with string density of $2/3$ \cite{Kim2007}. In this case at the disordered initial state many defects take place in the system (see Fig.\ref{TAFIM}). If we quench the magnet to $T_c=0$, first it passes through a non-equilibrium aging regime where the defects are moving by loose spin fluctuations and vanish out of the system by pair annihilation. When the $\rho(t)$ defect density relaxes to zero, the system there reaches its energy minima, which after the ground state dynamics govern the system toward the equilibrium state.

There are different points of view in the literature about the zero temperature dynamics in the aging regime of the system. First it was studied by Moore et.al. \cite{Moore1999}  in absence of external field, who recognized an entropy-driven Coulomb force acting between the topological defects of opposite sign. They found that the defect density is evolving as $\rho(t)\sim \mbox{ln}(t)/t$ with logarithmic correction, expected from the dynamical behaviour of the XY model.

However this result is not supported by some numerical simulations of Kim et.al. \cite{PhysRevE.68.066127,Kim2007} who questioned the propositions of Moore. According to them, other degrees of freedom are important during the out of equilibrium dynamical evolution since they found a random walk like defect motion moving in certain directions governed only by loose spin fluctuations. They did not see logarithmic correction of the defect densities but they found a subdiffusive and a diffusive growth of the correlation length of the form of Eq.\ref{dynCorrL} with different exponent in case of the two initial states. They located an exponent $z_{\mathit{eff}}\simeq 2.33$ during the aging regime when they started the system from a disordered initial state, and $z=2$ for diffusive growth, after the system had relaxed to its energy minima state and entered the dominant sector of the ground state manifold.

Another model was investigated by Walter and Chatelain \cite{Walter2008}, who examined the fully frustrated Ising model. This system belongs to the same static universality class as the triangular Ising antiferromagnet. Consequently the same asymptotic behaviour was expected for the two models. They observed the same subdiffusive behaviour with effective $z_{\mathit{eff}}$ exponent as Kim et.al at the pre-asymptotic range before the system reaches the aging regime. However at the aging regime in agreement with the XY model they found the diffusive description to describe better the asymptotic behaviour with logarithmic correction and $z=2$.

According to the previous studies the out-of-equilibrium critical behaviour of the triangular Ising antiferromagnet is not fully settled. In the present study we investigate the non-equilibrium dynamical behaviour of this model at $T_c=0$ in the aging regime. However our study is irrespective to the earlier methods and our goal is to find independent evidences of one of the above presented opposite descriptions. All the previous results deduced the dynamical exponent $z$ through some exponent combinations, or the results were dependent on the choice of some parameter like the waiting time of the autocorrelation. Here we allocate the dynamical exponent in a direct way by measuring a quantity which depends only on the exponent $z$. Another feature of our study is that for the triangular Ising antiferromagnet we consider not only the zero field model, but the Kosterlitz-Thouless transition at $h = h_{KT}$ as well. We compare the numerical results in the two cases in order to decide if the two transitions are in the same dynamical universality class. We executed long-time numerical calculations to examine the non-equilibrium relaxation and the autocorrelation function scaling in equilibrium and in the aging regime also.

\section{Measurements of the dynamical exponent}

\subsection{Microscopic dynamical rules}
\label{MicdDynRules}

In the performed Monte Carlo simulations we used random update rules. The dynamics of the Ising antiferromagnet are governed by a modified Glauber dynamics where the new state of a chosen spin $i$ is irrespective to its current value and takes its new state according to the probabilities given by:
\begin{equation}
P(\uparrow)=\dfrac{e^{\beta J\sum_i \sigma_i +h}}{e^{\beta J\sum_i \sigma_i +h}+e^{-\beta J\sum_i \sigma_i -h}} \qquad \mbox{and} \qquad P(\downarrow)=\dfrac{e^{-\beta J\sum_i \sigma_i -h}}{e^{\beta J\sum_i \sigma_i +h}+e^{-\beta J\sum_i \sigma_i -h}}
\label{TAFIMFlip}
\end{equation}
where we choose $J<0$ since the system is antiferromagnetic and the summation runs over the $\sigma_i$ neighbours of the current spin $i$. In the given limit when $T\rightarrow 0$ and $\beta J\rightarrow 0$, the time evolution is defined by the Alg.\ref{ZeroTempDyn}. Using this zero temperature dynamics, only spin-flips which do not increase the system energy are accepted, so $\Delta E\leqslant 0$. Let us denote that at zero field in case of loose spins with $\sum_i\sigma_i=0$, the flip probabilities become $P(\uparrow)=P(\downarrow)=1/2$.

\begin{algorithm}
\caption{ZeroTemperatureDynamics($Lattice$,$N$,$N_{MCS}$,$h$)}
\begin{algorithmic}[1]
\FOR{1 to $N_{MCS}$}
\FOR{1 to  $N$}
\STATE Choose a spin $i$ randomly
\IF {$\sum_i\sigma_i > 0$}
\STATE align the spin down
\ELSIF {$\sum_i\sigma_i < 0$}
\STATE align the spin up
\ELSIF {$\sum_i\sigma_i = 0$}
\STATE align the spin using the probabilities $P(\uparrow,\downarrow)=\frac{e^{\pm h}}{e^h+e^{-h}}$
\ENDIF
\ENDFOR
\ENDFOR
\end{algorithmic}
\label{ZeroTempDyn}
\end{algorithm}

\subsection{Sample dependent relaxation time}
\label{RelaxTimeSection}

To determine the dynamical exponent $z$ in the aging regime we need to define a quantity which clearly depends only on this exponent. An appropriate quantity is the non-equilibrium relaxation time $t_r$ which characterizes the size of the aging regime and that we define like the time when the system reaches its energy minima after it was started from a disordered initial state.  When $t<t_r$ we are in the aging regime and the system is governed by non-equilibrium relaxations, however when $t\geqslant t_r$ the defect density becomes $\rho(t)=0$ and the system evolves into a ground state. This relaxation time depends on the system size since it measures the time when $\xi(t)$, the time dependent characteristic length scale becomes $\xi(t_r)=L$. If we calculate the distribution of $t_r$ over different disordered samples with size $L$, it is expected to hold an exponential form as:
\begin{equation}
P_L(t_r)\propto e^{-t_r/\tau(L)}
\end{equation}
where $\tau(L)$ is defined as the size dependent characteristic time. As we have seen in Section \ref{CritDyn} and following from  numerical observations this characteristic time goes as:
\begin{equation}
\tau_1(L)\sim L^z
\label{tauLin}
\end{equation}
consequently an effective exponent can be recognized as $\tau(2L)/\tau(L)=2^{z_{\mathit{eff}}}$. 

On the other hand theoretical consideration (derived from \ref{logCorrection}) suggest an asymptotic behaviour as
\begin{equation}
\tau_2(L) \sim L^z \mbox{ln}\bigg(\dfrac{L}{L_0}\bigg)
\label{tauLog}
\end{equation}
which leads to the form $\tau(2L)/\tau(L)=2^z\mbox{ln}(2L/L_0)/\mbox{ln}(L/L_0)$ where $L_0$ is an $\mathcal{O}(1)$ constant,  . Confront the two case the effective dynamical exponent can be deduced of the form:
\begin{equation}
z_{\mathit{eff}}(L)=z+\dfrac{1}{\mbox{ln}(L/L_0)}
\label{zeffDef}
\end{equation}

To numerically verify these results we made high performance computing to calculate the distribution of the relaxation time $t_r$ for different system sizes $L$. We started each sample from a random initial state and measure the time when the system energy reached its possible minima value $E_{min}=-L^2$.

\begin{figure}[htb]
\begin{center}
      \includegraphics*[width=12.0cm]{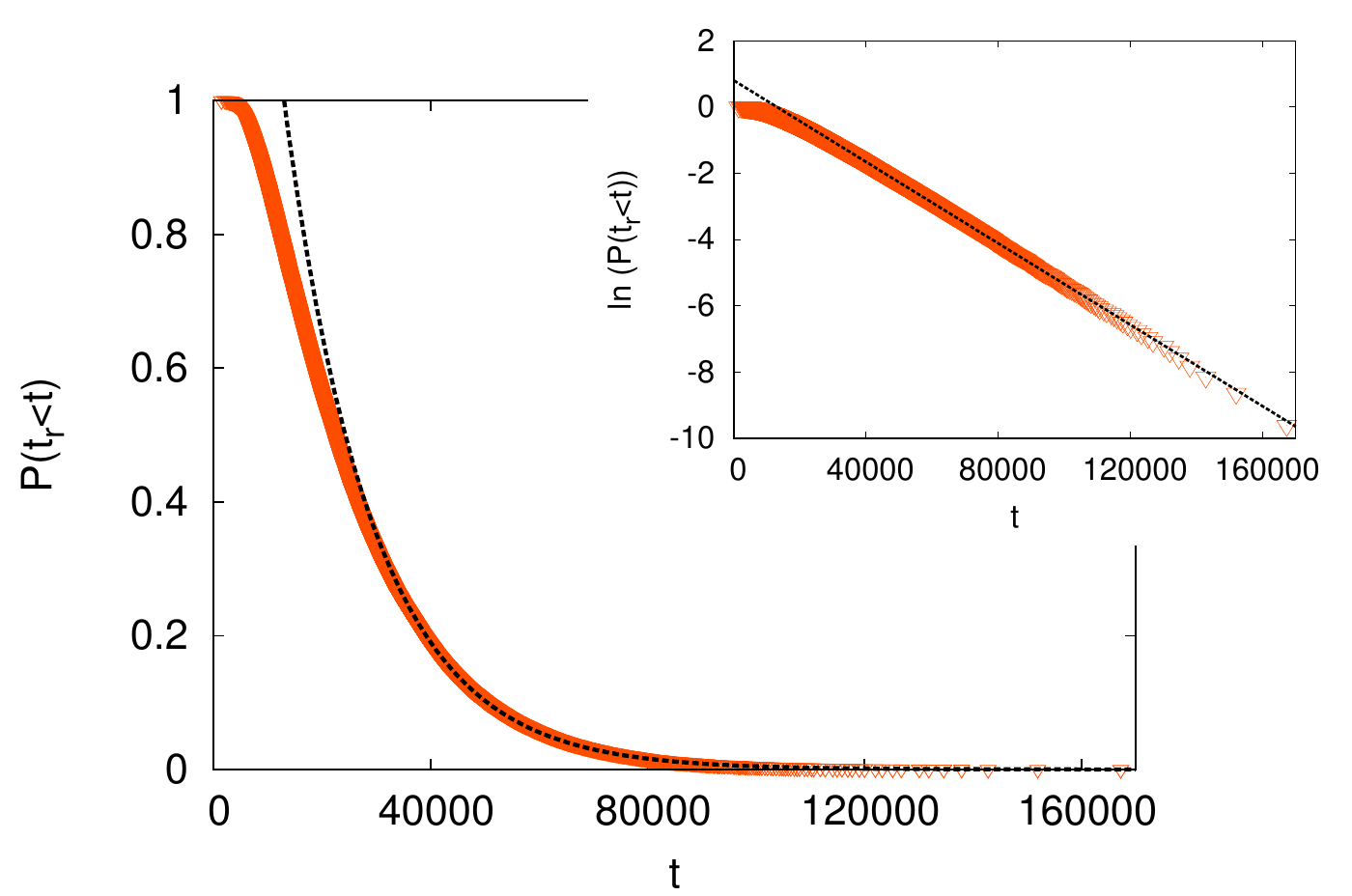}
      \caption{The $P(t_r<t)$ distribution of the relaxation time $t_r$ of a system with linear size $L=192$. The distribution contains data of $90000$ independent realizations. The tail of the distribution ($t>40000$) has an excellent fit with an exponential curve (black line) which is presented in the inset on a lin-log scale also.}
	\label{ERelax}
  \end{center}
\vspace{-10pt}
\end{figure}

We investigated the model on lattices with periodical boundary conditions with linear sizes between 33 and 768. The averages were computed over large number of random samples between 2000 to 200000 depends on the current system size.

In order to estimate the effective $z$ exponents for different sizes, first we calculated the characteristic time $\tau(L)$ as the average relaxation time $\tau(L)=\langle t_r(L)\rangle$ over all samples of a given size. In the following we are going to use this definition but let us denote that $\tau(L)$ can be derived from the fitting of the tail of the $P(t_r<t)$ distribution, which gives an analogous result. To evaluate the effective values of the dynamical exponent we used the expression 
\begin{equation}
z_{\mathit{eff}}(L)=\dfrac{\mbox{ln}\langle t_r(L)\rangle-\mbox{ln}\langle t_r(L')\rangle}{\mbox{ln}(L)-\mbox{ln}(L')}
\label{zeff}
\end{equation}
which comes straightforward form the previous methods and we found the values presented in Table \ref{table1}. 
\begin{center}
\begin{table}
\begin{center}
\begin{tabular}{|c|c|c|r|c|c|r|}  \hline
&\multicolumn{3}{|c|}{ATIM($H=0$)}&\multicolumn{3}{|c|}{ATIM($H=H_{KT}$)}\\ \hline
L&$log_2\langle t_r \rangle$ & $z_{eff}$ & samples & $log_2\langle t_r \rangle$ & $z_{eff}$ &samples\\ \hline
33  & 8.80  & 2.38 & 200000 & & & \\
48  & 10.09 & 2.33 & 200000 & 10.17 & 2.35 & 200000 \\
96  & 12.42 & 2.29 & 200000 & 12.52 & 2.29 & 200000 \\
192 & 14.71 & 2.24 & 89848  & 14.81 & 2.25 & 103799 \\
384 & 16.95 & 2.15 & 35271  & 17.06 & 2.13 & 38414 \\
768 & 19.10 &      & 1667   & 19.19 &      & 1701 \\ \hline
\end{tabular}
\end{center}
\caption{Average of the sample dependent relaxation time $\langle t_r \rangle$ for different finite systems of the triangular antiferromagnetic Ising model at zero field and at the field induced Kosterlitz-Thouless transition point, $H=H_{KT}$. The effective dynamical exponent calculated from two consecutive sizes, see Eq.(\ref{zeff}) is also indicated together with the number of samples used in the simulation.}
\label{table1}
\end{table}
\end{center}

To examine the asymptotic behaviour of the characteristic time and find the real dynamical exponent we fitted function of two forms defined in \ref{tauLin} and \ref{tauLog} on the calculated numerical values. We tried to find the best fit of this two functions, $\tau_i(L)$, to the numerical averages $\langle \tau(L)\rangle$ by minimizing their deviation from the numerical values of different sizes as:
\begin{equation}
\mbox{min} \Bigg\{ \sum_{L=33,48,96,192,384,768}\bigg| \dfrac{\tau_i(L)-\langle \tau(L)\rangle}{\tau_i(L)+\langle \tau(L)\rangle} \bigg| \Bigg\}
\end{equation}
with given parameters $z$, $L_0$ and a constant factor $A$. In zero field the best parameter set were found as:
\begin{eqnarray}
\tau_1(L) &=& 0.18 L^{2.26}
\label{tau2FitLin}\\
\tau_2(L) &=& 0.183 L^{2} \mbox{ln}\bigg(\dfrac{L}{3.6}\bigg) 
\label{tau2Fit}
\end{eqnarray}

\begin{figure}[htb]
\begin{center}
\includegraphics*[width=10.0cm]{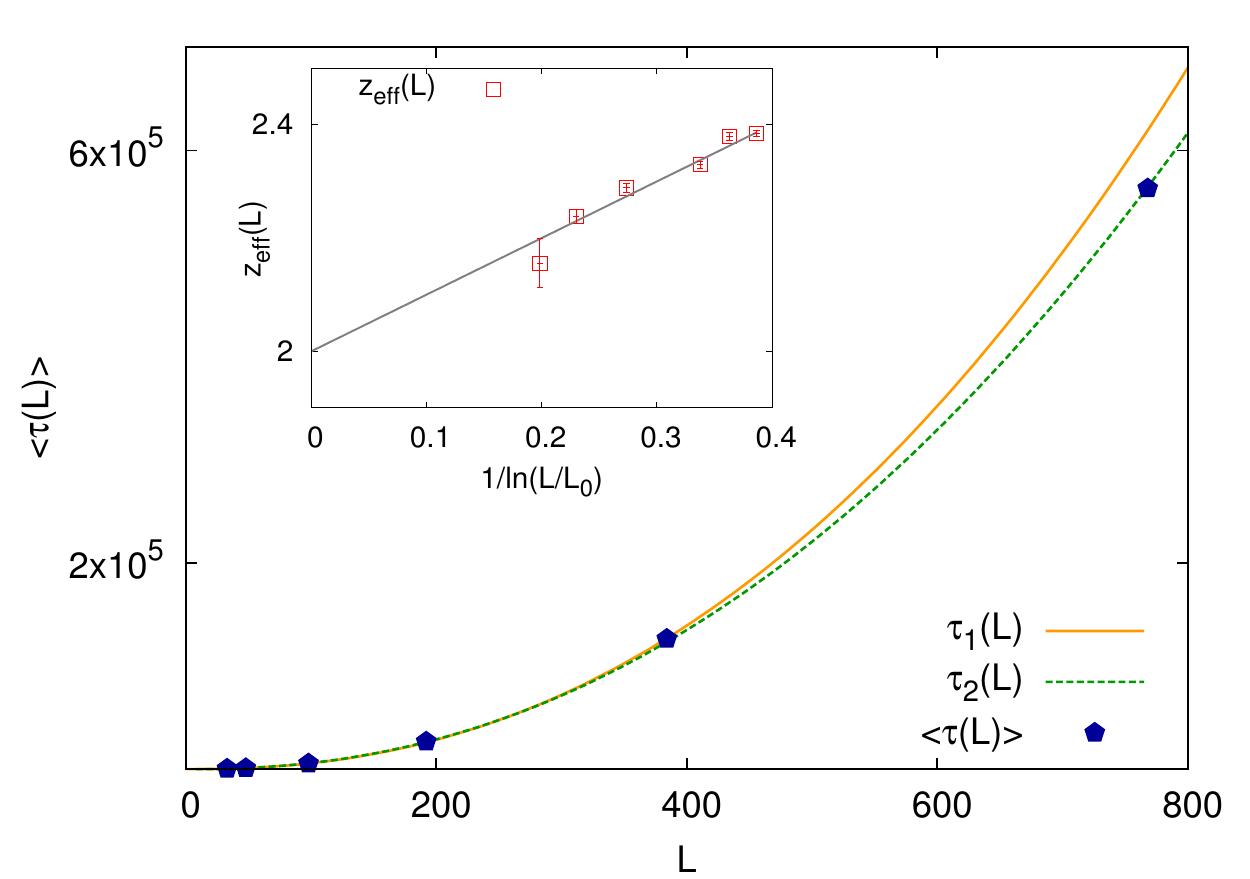}
\caption{Fitting of the size dependent average characteristic time $\langle\tau(L)\rangle$ with the curves defined in Eq.\ref{tauLin} and Eq.\ref{tauLog}. The best parameters are in Eq.\ref{tau2FitLin} and Eq.\ref{tau2Fit} respectively. This calculation gives an evidence that the triangular Ising antiferromagnet has a diffusive dynamical behavior with logarithmic correction. The estimated error are smaller than character size. In the inset extrapolation of the effective dynamical exponents in Table \ref{table1} are given as a function of $1/ \mbox{ln}(L/L_0 )$ with $L_0 = 2.5$.}
\label{TauFit}
  \end{center}
\vspace{-10pt}
\end{figure}

The Fig.\ref{TauFit} presents the two analytical curve fitting on numerical values. It is well seen that the function with logarithmic correction and exponent $z=2$ has much better fit. We denote that the value $z=2$ in Eq.\ref{tau2Fit} was not imposed but found directly, performing an accurate scan on the parameter set $z,L_0$ and $A$. It is possible since the number of fitted points is not too large. The function to minimize shows is slowly varying in the range $\left[1.96,2.04\right]$ with a minimum located precisely at $z=2$, outside this range, it is rapidly varying. It is also shown that measurements with large size are needed to discriminate the two asymptotic behavior. We tried to extrapolate the dynamical exponent $z$, via the calculated effective values using the scaling $z_{\mathit{eff}}(L)\sim 1/\mbox{ln}(L/L_0)$ and we also found the asymptotic value $z=2$. Repeating this analysis in case of $h=h_{KT}$ external field the best fits are $\tau_1(L) = 0.18 L^{2.28}$ and $\tau_2(L) = 0.20 L^{2} \mbox{ln}(L/3.9)$, and again the logarithmically corrected scaling variable provides a better fit to the numerical values. In both cases the relative error of the fits were 5 to 10-times smaller for the second scenario.

Following from these considerations we can proposed that the dynamics in the aging regime has a diffusive behaviour with exponent $z=2$ and logarithmic correction rather than some subdiffusive growth characterizes the system even at zero field and at the critical external field $h=h_{KT}$.

\section{Scaling of the autocorrelation}

As we have seen in Section \ref{CritDyn} during the non-equilibrium relaxations, the dynamics can be decomposed to two parts (see Eq.\ref{AutCorrDecomp}) where the two-time autocorrelation $A(s,t)$ after a first short-time period becomes a homogeneous function of $t/s$ of the form of Eq.\ref{AutCorrAgscaling}. Following this aging regime when the system reaches its equilibrium, the autocorrelation is expected to behave as Eq.\ref{AutCorrEq}. In this section we are going to study the autocorrelation function of the triangular Ising antiferromagnet in equilibrium and in the aging regime using the method as it was used for the fully frustrated Ising model by Walter and Chatelain \cite{Walter2008}. We are going to discuss the case of zero external field and when a critical field $h=h_c$ is applied on the system. First we examine the equilibrium case where we deduce the dynamical exponent through exponent combination then we study the scaling behaviour in the non-equilibrium regime with linear and logarithmic corrected scaling variables. 

\subsection{Equilibrium scaling}

The scaling behaviour of the two-time autocorrelation function $A(s,t)$ (defined in Eq.\ref{autocorrDef}) has been shortly discussed in Section.\ref{CritDyn}. As we saw there, the autocorrelation is invariant under time-translation since it depends only on the time difference $(t-s)$ and it is independent on the choice of the waiting time $s$. Moreover due to the divergence of the relaxation time at the critical point an algebraic decay of $A_{eq}(t,s)$ is expected of the form $A_{eq}(t,s)\sim(t-s)^{-a_c}$ with exponent $a_c=2\beta/\nu z$ where $2\beta/\nu=1/2$ in case of zero field \cite{Walter2008,Calabrese2005,Kim2007} and it is expected to be $2\beta/\nu=4/9$ \cite{Blote1993} in critical external field. This scaling behaviour enables us to estimate the dynamical exponent $z$ in an indirect way by calculating the average autocorrelation function in equilibrium.

\begin{figure}[htb]
\begin{center}
\includegraphics*[width=7.0cm]{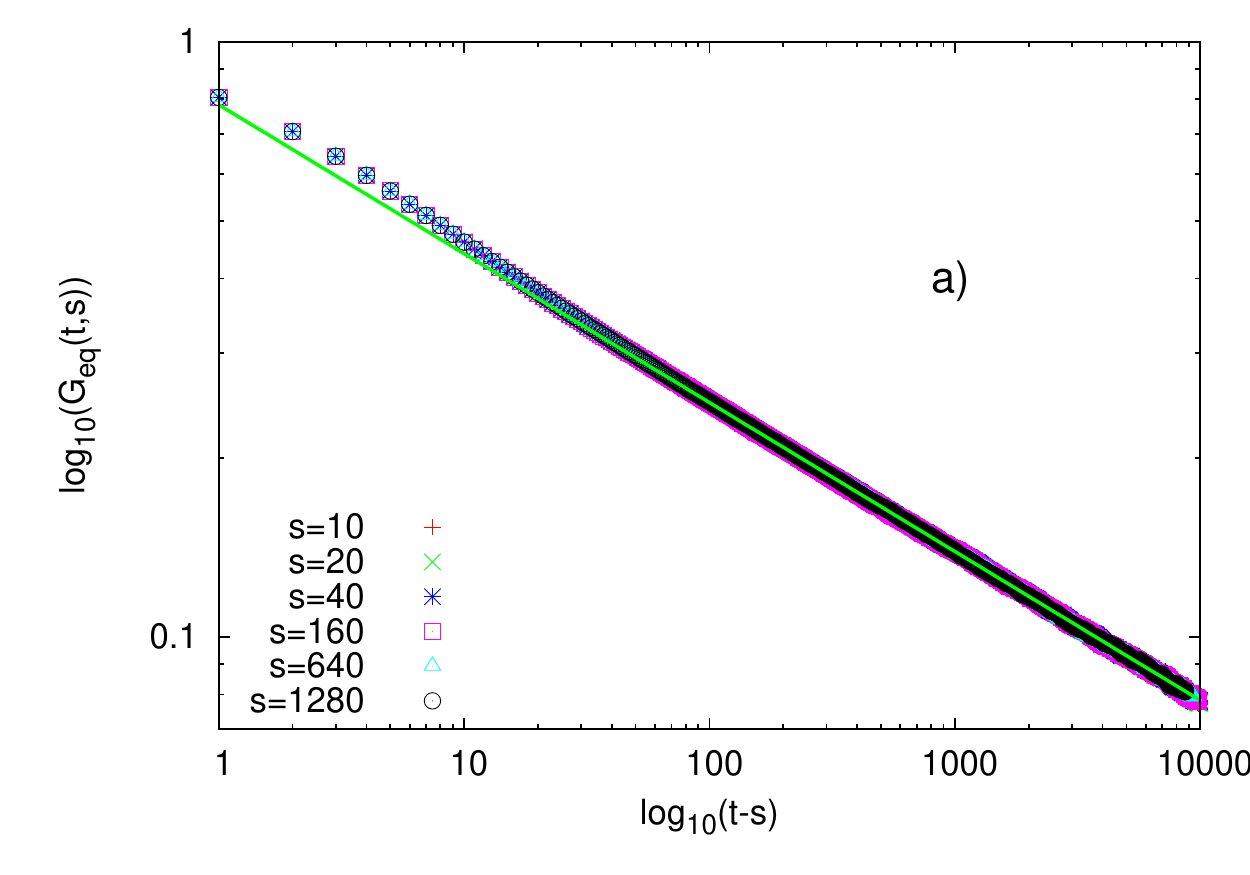}\hspace{.1in} \includegraphics*[width=7cm]{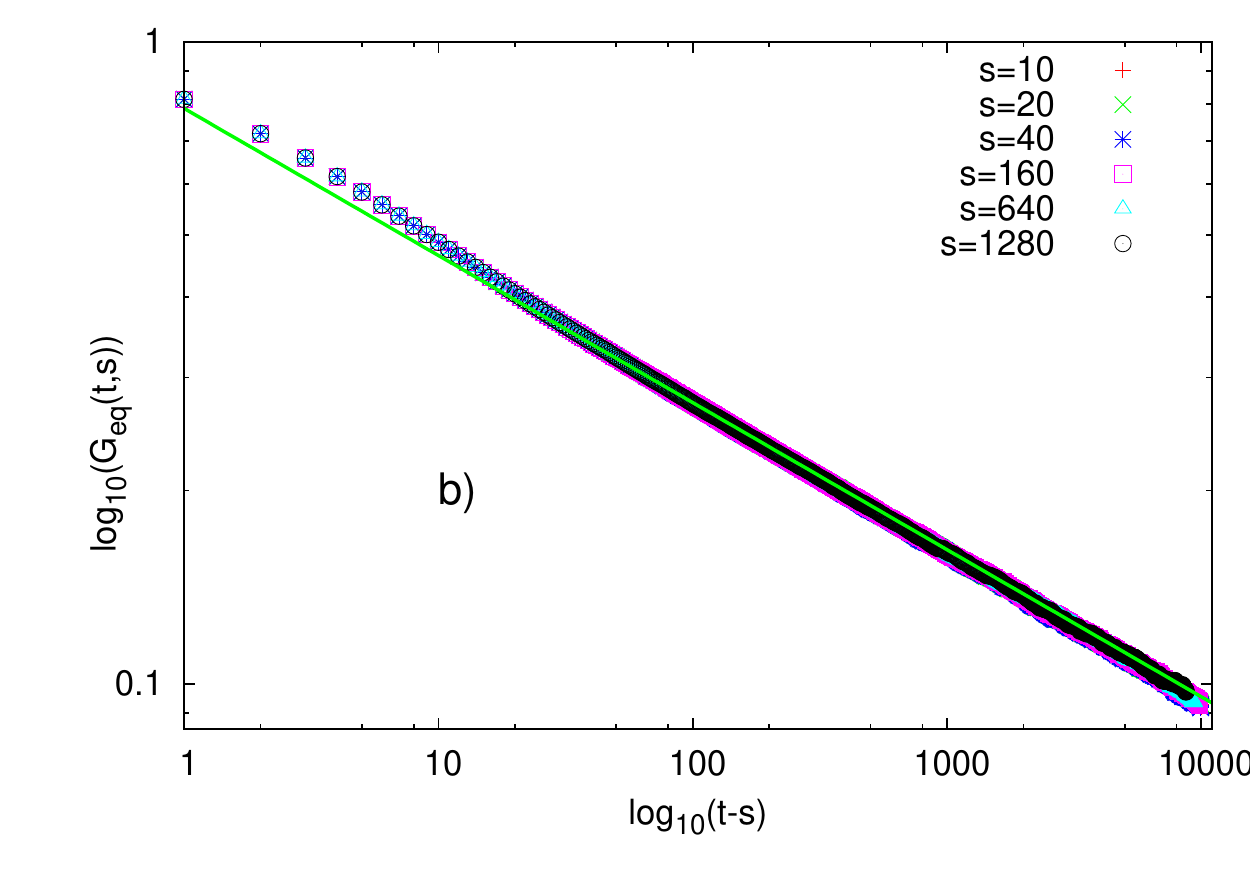}
\caption{Autocorrelation in equilibrium with different waiting times $s$. (a): The data were calculated on lattice with linear size L=192 and periodic boundary conditions, average over 1300 equilibrated samples. The function has a clear algebraic decay for sufficiently large waiting time $(t-s)>40$. The asymptotic slope of the curve is given by $a(h=0)=0.2501(10)$. (b): the same analysis at $h=h_{KT}$. The asymptotic slope of the curve is given by $a(h=h_{KT})=0.229(5)$.}
\label{AEq}
  \end{center}
\vspace{-10pt}
\end{figure}

To examine the equilibrium dynamical behaviour first we needed to evolve the system into a homogeneous ground state where only the loose spins fluctuations influence the dynamics. To reach this kind of equilibrated states for each sample, we started the system from different random initial states, quench it to $T=0$ and evaluate it following by the dynamical rules defined in Section \ref{MicdDynRules}. After the system reached its energy minima state at time $t_r$ we let it evolve until $t=1.5*t_r$. This time was then defined as the zero time point $s=t=0$ for the autocorrelation. 

The average autocorrelation function with different waiting times is shown in Fig.\ref{AEq}. The calculations were performed on a lattice with linear size $L=192$ and periodic boundary conditions, averaging over 1300 (1200) samples in zero (critical) field. One can clearly seen that $A_{eq}(t,s)$ decays as a power-law for sufficiently large separation time. We used the least squares method to fit the sample average of $A_{eq}(t,s)$ for  $(t-s)>40$, where we did not take in account the statistical error of the average calculation. In the absent of ordering field the slope of the fitted straight line is $-0.250122 \pm 4.917e-05$ with an offset $-0.105929\pm 0.0001742$ which suggest a very good systematic error. It leads to estimate $a_c\simeq 0.25$ and the dynamical exponent $z\simeq 2$ which is in complete agreement with our previous results in Section \ref{RelaxTimeSection}. In case of critical field the slope is $-0.229051 \pm 0.0001075$ with an offest $-0.103873 \pm 0.0003036$ (see Fig.\ref{AEq}) which is also close to the conjectured exact value: $a_c=2/9$ and suggests $z\simeq2$ for the equilibrium dynamical exponent as it is expected. The time-translation invariance is well presented too since all the calculated averages collapse together independently from the waiting time $s$.

\subsection{Non-equilibrium scaling}

The autocorrelation function of homogeneous ferromagnets in the aging regime after a short time relaxation ($t,s\gg 1$) becomes a homogeneous function of $t/s$ and is expected to scale as Eq.\ref{AutCorrAgscaling} \cite{Walter2008}. This asymptotic behaviour was found for a large class of different models \cite{Godreche2000,Godreche2000a,Calabrese2005}. However for the XY model the relaxation in the aging regime is slowed down by a logarithmic factor and the correlation length grows as Eq.\ref{logCorrection} so the autocorrelation function has another scaling variable and scales as:
\begin{equation}
A_{ag}(s,t)\sim s^{-2\beta/\nu z}\widetilde{A}\bigg(\dfrac{t\hspace{.05in} \mbox{ln}(s)}{s\hspace{.05in}\mbox{ln}(t)}\bigg)
\label{AutoCorrLogScale}
\end{equation}
which scaling form have been observed for the fully frustrated Ising model also \cite{Walter2008}.

\begin{figure}[htb]
\begin{center}
\includegraphics*[width=7.0cm]{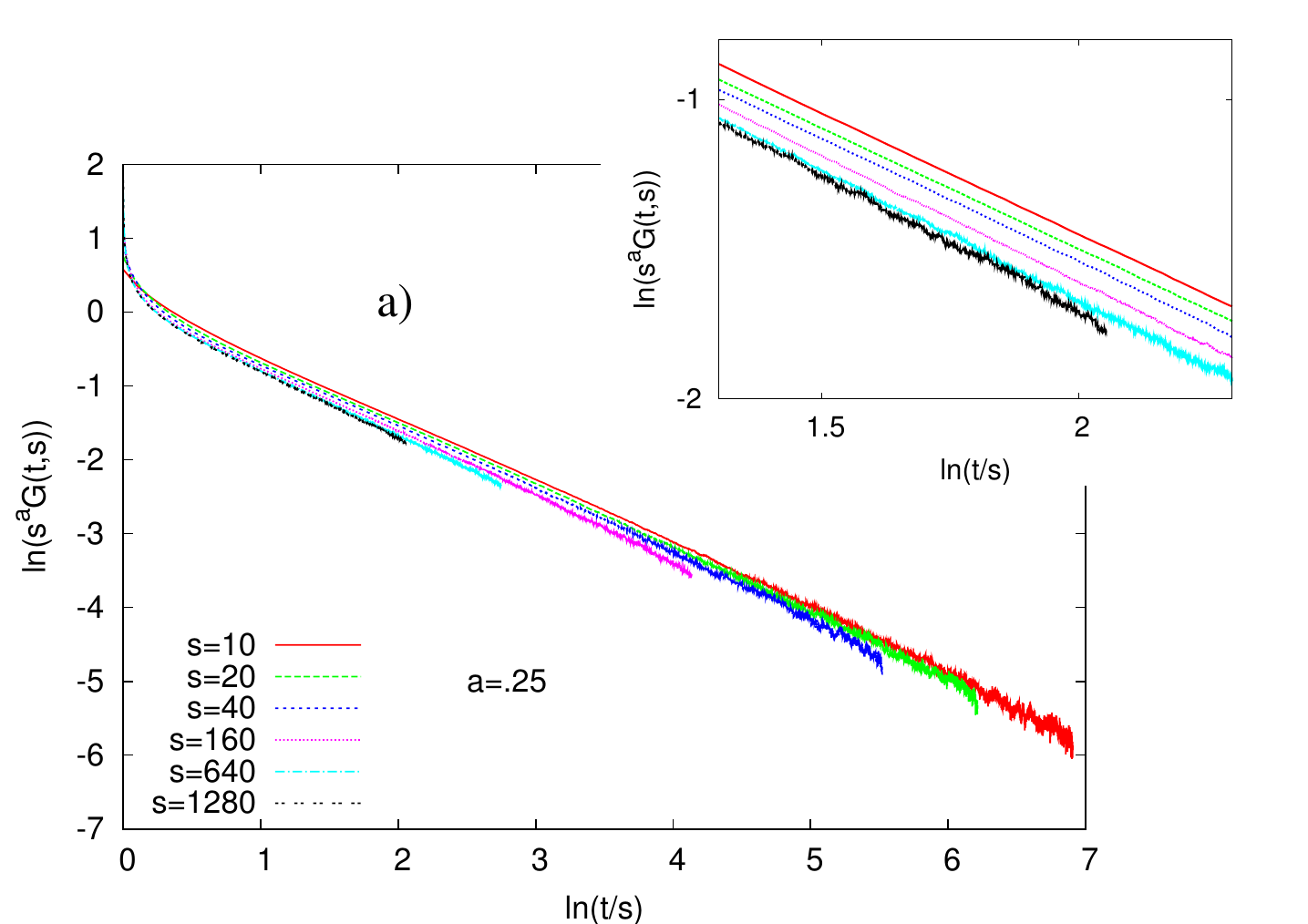}\includegraphics*[width=7.0cm]{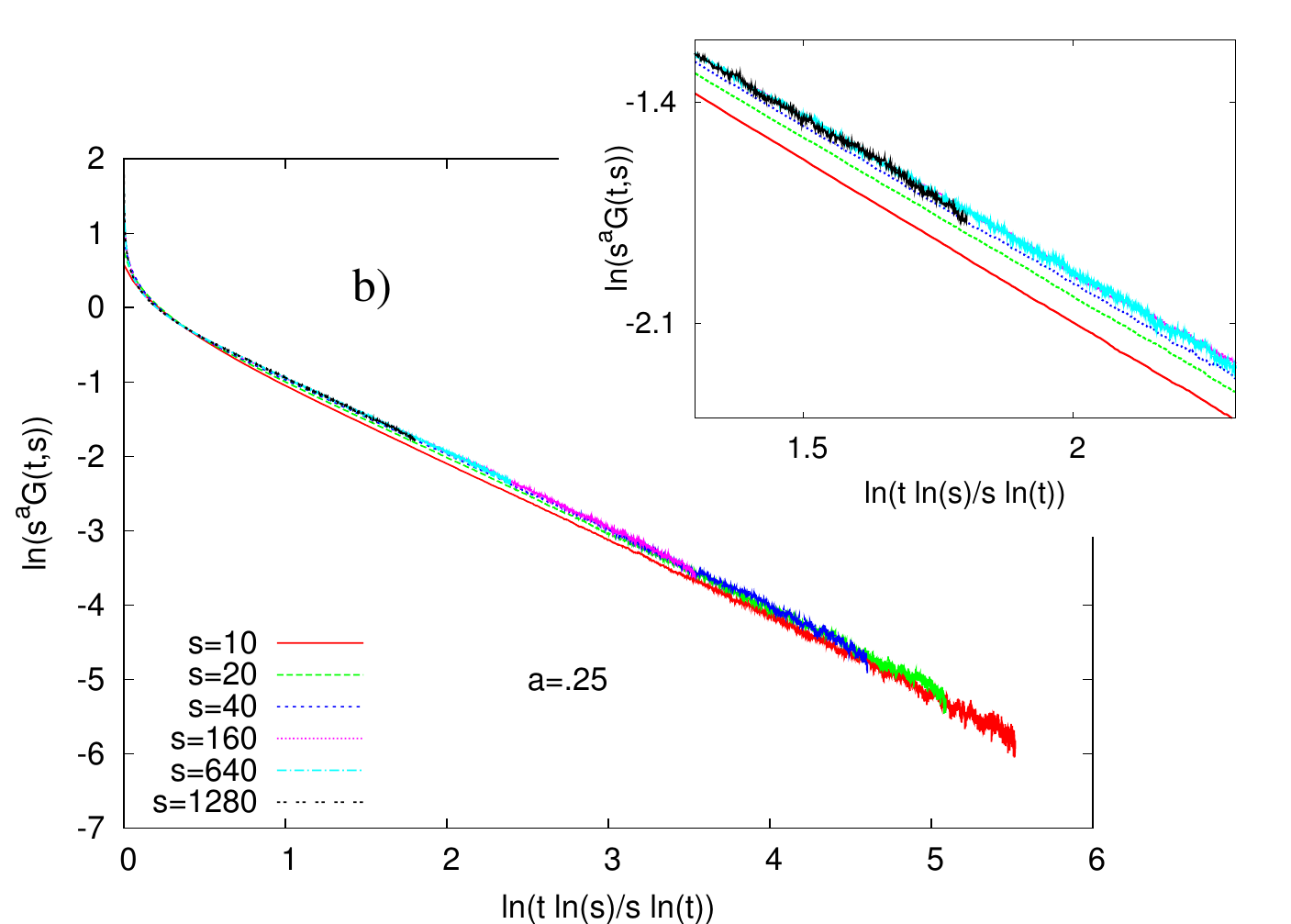}
\caption{(a): $s^{a_c}A_{ag}(t,s)\sim t/s$ and (b): $s^{a_c}A_{ag}(t,s)\sim t\mbox{ln}(s)/s\mbox{ln}(t)$ scaling of the autocorrelation function in the aging regime in zero external field. The system size is $L=192$, the lattice has periodic boundary conditions and the averages were calculated over 3000 realizations. We applied the exponent $a_c=0.25012$ that we considered for equilibrium.}
\label{ANeqh0}
  \end{center}
\vspace{-10pt}
\end{figure}

Here we are going to study the scaling of the autocorrelation function in the aging regime of the triangular Ising antiferromagnet to determine which scaling scenario, the one in Eq.\ref{AutCorrAgscaling} or the other Eq.\ref{AutoCorrLogScale}, is more compatible with the numerical data. In order to do so we search for the better collapse of the scaling function $s^{a_c}A_{ag}(t,s)$ using the scaling variable $t/s$ or the one of $t\mbox{ln}(s)/s\mbox{ln}(t)$ with logarithmic correction. In our numerical simulations we chose the same lattice conditions as for the equilibrium case and calculated averages over 3000 realizations. The numerical results of different waiting times are presented on Fig.\ref{ANeqh0}.a and Fig.\ref{ANeqh0}.b  with respect to the two types of scaling variables. Here we used the exponent $a_c=2\beta/\nu z=0.25012$ in case of zero field which was determined in the previous section for equilibrium.

On Fig.\ref{ANeqh0}.a when we choose the scaling variable for $t/s$, the autocorrelation functions are not collapsed and they scale over by the waiting time $s$ except the curves of $s=640$ and $s=1280$ which are closer to each other. This could be a sign of an asymptotic behavior, however we need to calculate with much larger waiting time and system size to verify this trend.

On the other hand when the scaling variables are chosen as $t\mbox{ln}(s)/s\mbox{ln}(t)$, we find a very good collapse if the waiting time was $s>40$ (see Fig.\ref{ANeqh0}.b). Here the difference between the scaling curves is in the range of the magnitude of the statistical fluctuations. Since the scaling becomes better as we increase the waiting time and the curves present an excellent collapse for sufficiently large $s$, we can take that proposition to be the true asymptotic behaviour. Consequently our data are in favour of logarithmic corrections in the scaling function of the autocorrelation function at H = 0. We have then estimated the autocorrelation exponent, $\lambda/z$, from the slope of the linear part of Fig.\ref{ANeqh0}.b for the three largest waiting times and we obtained: $\lambda/z = 1.003(10)$. That value corresponds to $\lambda=d$, as obtained for the fully frustrated Ising model  \cite{Walter2008}.

The non-equilibrium autocorrelation function at the $h=h_c$ Kosterlitz-Thouless phase transition point presents the same type of scaling as it was seen in zero external field. Here we used the exponent $a_c=0.229051$ which was determined for in the previous section for equilibrium at $h=h_c$. The scaling collapse for different waiting time is much better with logarithmically corrected scaling variable. We have also measured the autocorrelation exponent with the result: $\lambda/z = 0.93(2)$, what is somewhat
smaller than those found in the other case.

As a conclusion we can point out that the the similar rules govern the dynamics of the triangular Ising antiferromagnet at its criticallity as the XY model and supports the assumption of dynamics with logarithmic correction. Let us denote that the same asymptotic behavior was found for the fully frustrated Ising model \cite{Walter2008} also, which suggests that beyond the two models belong to the same static universality class, their dynamical behavior is similar too.

\begin{figure}[htb]
\begin{center}
\includegraphics*[width=7.0cm]{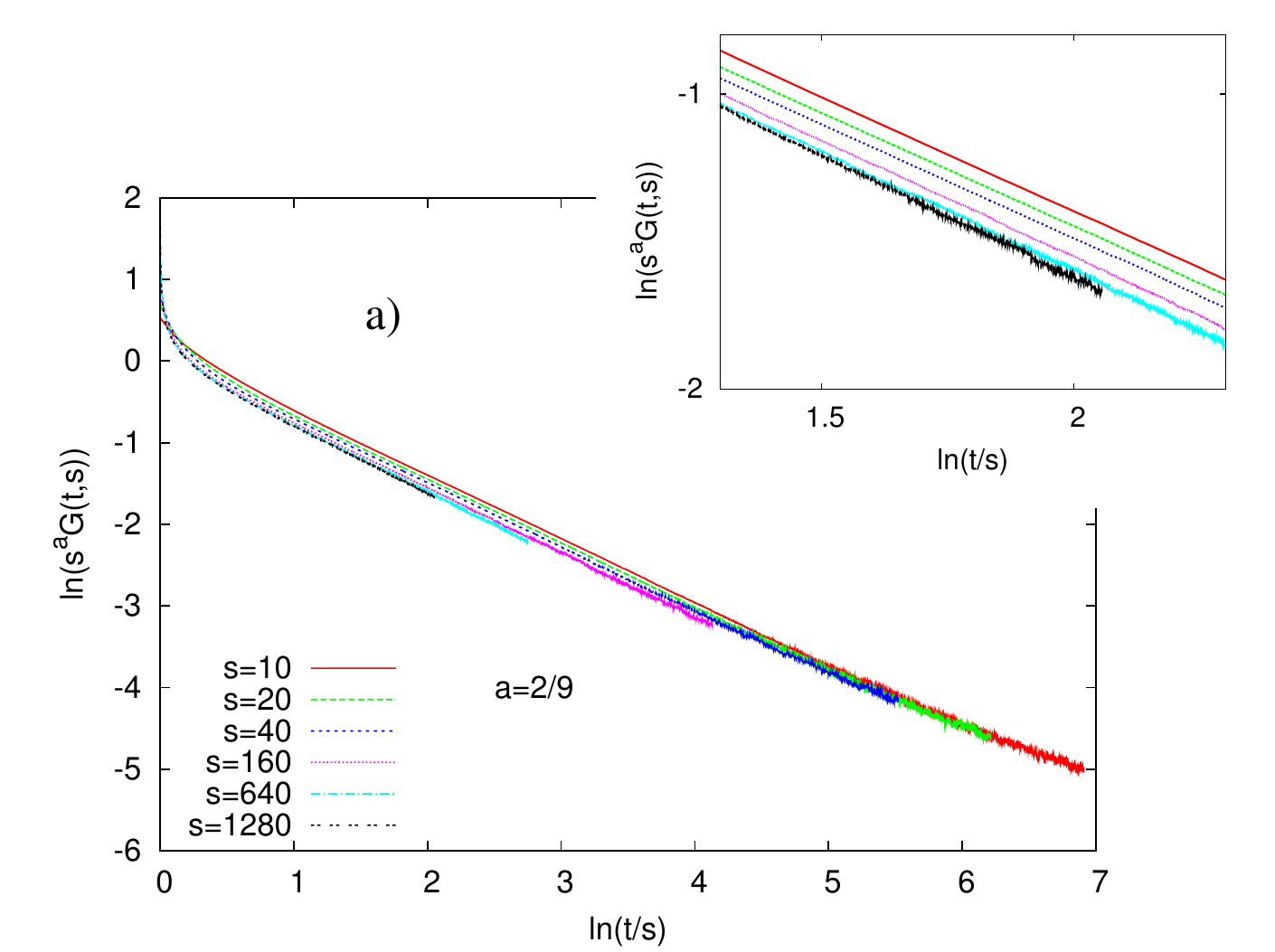}\includegraphics*[width=7.0cm]{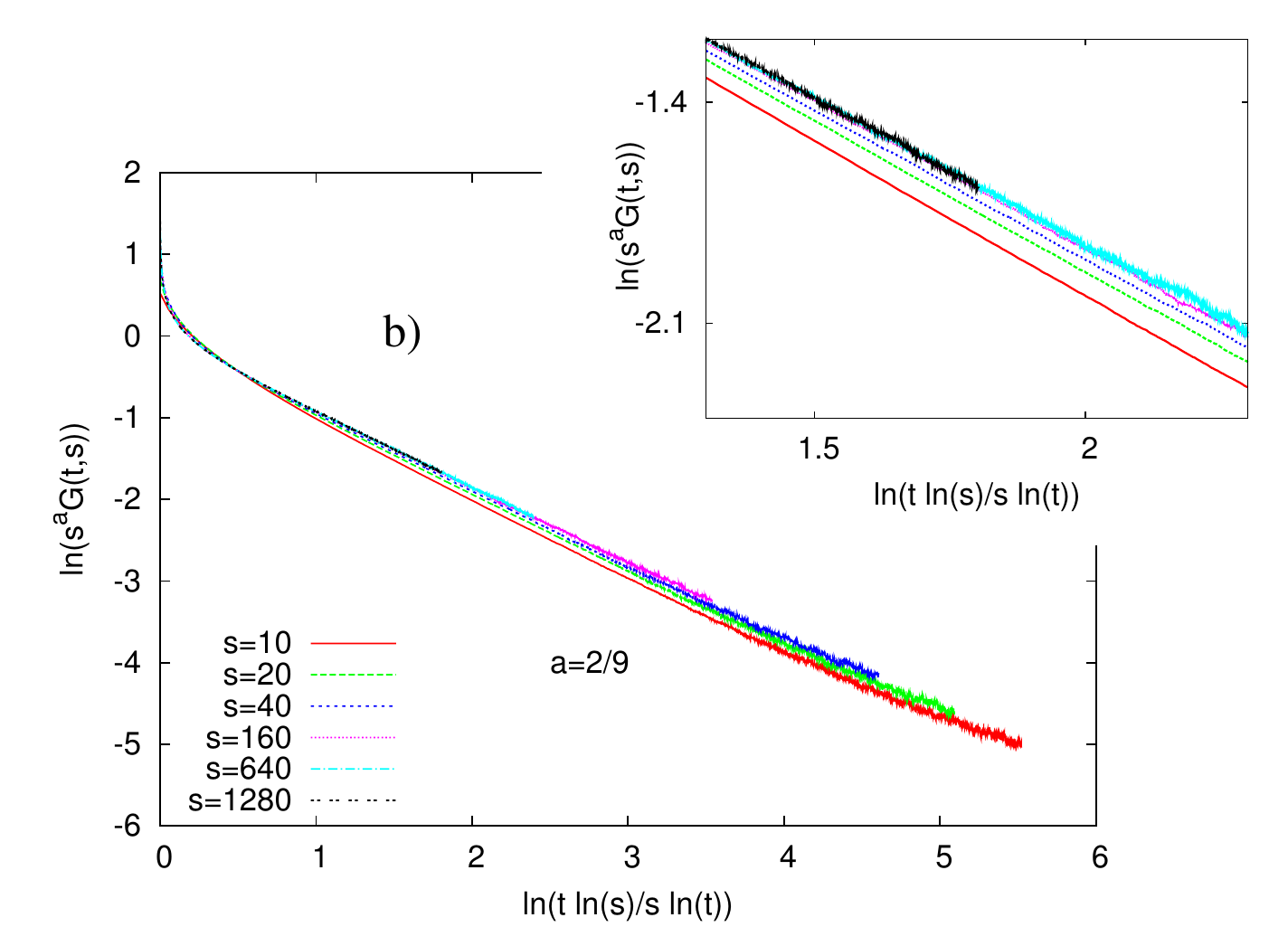}
\caption{(a): $s^{a_c}A_{ag}(t,s)\sim t/s$ and (b): $s^{a_c}A_{ag}(t,s)\sim t\mbox{ln}(s)/s\mbox{ln}(t)$ scaling of the autocorrelation function in the aging regime in $h=h_c$ external field. The system size is $L=192$, the lattice has periodic boundary conditions and the averages were calculated over 5000 realizations. We applied the exponent $a_c=0.229051$ that we considered for equilibrium.}
\label{ANeqhc}
  \end{center}
\vspace{-10pt}
\end{figure}

\section{Summary}

In this chapter we studied the dynamical behaviour of the antiferromagnetic Ising model on triangular lattice at zero and field and at the $h=h_c$ Kosterlitz-Thouless phase transition point. We used zero temperature Glauber dynamics to perform long-time running computer simulations in order to calculate numerical averages.

Our aim was to choose the relevant description between the one of Moore et.al. \cite{Moore1999} who saw a non-equilibrium relaxation slowered by a logarithmic factor or the one of Kim et.al \cite{PhysRevE.68.066127,Kim2007} who believed that the relaxation dynamics were governed by subdiffusive scalings with an effective exponents $z=2.33$.

In order to do so, we defined a new measurable quantity, the $t_r$ relaxation time, which scales clearly with the dynamical exponent $z$ and characterizes the length of the aging regime. From its distribution we could conclude that the non-equilibrium dynamics of the triangular Ising antiferromagnet in the aging regime presents an analogous behaviour as the XY model.

Using this result we can estimate the asymptotic behaviour of the density of defects, $\rho(t)$, in the systems. To be concrete we consider the model at $h=0$ for which a defect is represented by an elementary triangle having all the three spins in the same state. In a finite sample of linear size, $L$, and sample dependent first passage time, $t_r$, at $t=t_r$ there is one (or $O(1)$) defect, 
consequently the density of defects at $t=\langle t_r(L) \rangle$ is $\rho(t) \sim L^{-2}$. Using the relation in Eq.\ref{logCorrection} and replacing $\xi$ by $L$ we arrive to the result: $\rho(t) \sim \log(t/t_0)/t$, in agreement with the conjecture in \cite{Moore1999}.

The analogy with the nonequilibrium scaling of the $XY$-model is found to be valid for the autocorrelation function, too. Both at zero field and at $h=h_c$ optimal scaling collapse of the autocorrelation function is found if the scaling variable is taken in the same form, as in the $XY$-model. At $h=0$ as for the FFIM \cite{Walter2008}, the autocorrelation exponent is found to be $\lambda=d$, like in mean-field theory, however with logarithmic corrections. On the other hand the field-induced Kosterlitz-Thouless transition point the autocorrelation exponent is found somewhat smaller, than $d$. Consequently in the ATIM with varying field the dynamical exponent is presumably constant, $z=2$, the scaling functions contain same type of logarithmic corrections, but the autocorrelation function weakly depends on the value of $h$.

The publication of the new scientific results presented in this chapter is in progress \cite{Karsai2009}.

%% file: Discussion.tex
\chapter{Discussion}

In my PhD thesis I studied the cooperative behaviour of complex interacting many body systems using the methods of statistical and numerical physics. I completed studies in four different subjects and I summarized my new scientific results in the current dissertation as follows. 

To outline the relevant scientific background of my work I reviewed the related models and theoretical considerations of the subject of phase transitions, universality and critical phenomena in case of homogeneous and disordered models. I summarized the basic features of network science as I defined the important properties of regular graphs and reviewed the main models of random networks. I discussed the basic concepts of Monte Carlo techniques and introduced the applied numerical methods. After this introductory section I discussed my new scientific results in the case of the following investigated problems.

\vspace{.1in}

First I studied non-equilibrium phase transitions in weighted scale-free networks where I introduced edge weights and rescaled each of them by a power of the connectivities, thus a phase transition could be realized even in realistic networks having a degree exponent $\gamma \leq 3$. The investigated non-equilibrium system was the contact process which is a reaction-diffusion model belonging to the universality class of directed percolation. 

\begin{description}

\item[$\mathcal{I}/ a:$] First I gave the dynamical mean-field solution of the model in scale-free networks and I located the previously known three critical regimes. The first regime is the one where the mean-field behaviour is conventional. Then this solution becomes unconventional since critical properties are $\gamma$ dependent. In the third regime, the system is always in an active phase.

\item[$\mathcal{I}/ b:$] I also made some theoretical considerations to generalize recent field-theoretical results about finite-size scaling which are expected to be valid above the upper critical dimension i.e. in the conventional mean-field regime. I introduced in a simple way the volume of the network into the scaling functions that I derivate for two cases, where the infection was initiated from a typical connected site or from the most connected vertex of the network.

\item[$\mathcal{I}/ c:$] I executed high performance numerical simulations in order to simulate the contact process in the conventional mean-field regime. I located the phase transition point and I calculated the finite-size scaling exponents for typical and maximally connected site that I found to be in good agreement with the field-theoretical predictions. I also determined the correlated volume exponent which was found equal in the two cases and compatible with the theoretical expectations. Finally I analyzed the dynamical scaling of the system at the critical point for the two above mentioned cases. Through extrapolations I considered the related exponents to be different in the case of typical and maximally connected sites but compatible with the mean-field and finite-size scaling predictions.
\end{description}

The second problem I investigated was the ferromagnetic random bond Potts model with large values of $q$ on evolving scale-free networks. This problem is equivalent to an optimal cooperation problem, where the agents try to find an optimal situation where the benefits of pair cooperation (here the Potts couplings) and total sum of the support, which is the same for all projects (introduced here as the temperature), is maximized. I examined this model using a combinatorial optimization algorithm on scale-free Barab\'asi-Albert networks with homogeneous couplings and when the edge weights were independent random values following a quasi-continuous distribution with different strength of disorder.

\begin{description}
\item[$\mathcal{II}/a:$] As a first point I gave the exact solution of the system for a wide class of evolving networks with homogeneous couplings. The phase transition was found to be strictly first-order and the critical point was determined through simple theoretical considerations. 

\item[$\mathcal{II}/b:$] By numerical calculations I examined the magnetization curve for different strength of disorder and found the theoretically predicted first-order phase transition in the pure case. However, the phase transition softened to a continuous one for any strength of disorder larger than zero. I examined the structure of the optimal set and I found out its structural behaviour was altered by the temperature.

\item[$\mathcal{II}/c:$] I studied also the critical properties of the system and I calculated the distribution of the finite-size critical temperatures for different sizes in case of maximal disorder. I deduced by iterative calculations the critical magnetization exponent and I located the critical temperature using two independent methods. The scaling of the finite-size transition points distribution was characterized by two distinct exponents, that I located and used to consider a scaling collapse of the distribution curves.

\item[$\mathcal{II}/d:$] I also deduced the critical magnetization exponent by two-point fits using the average size of the critical cluster. I obtained compatible values with previous results within the error of the calculations.
 \end{description}

The third examined problem was related to the large-$q$ sate random bond Potts model also. Here I examined the critical density of clusters which touched a certain border of a perpendicular strip like geometry. Following from conformal prediction I expected the same density behavior as it was exactly derived for critical percolation in infinite strips. I calculated averages by the above mentioned effective combinatorial optimization algorithm and I compared the numerical means to the expected theoretical curves.

\begin{description}
\item[$\mathcal{III}/a:$] During my study I used a bimodal form of disorder for the random couplings, which intensity influenced the breaking-up length of the critical clusters. I allocate an appropriate value of the strength of the disorder which set the breaking up length large enough for relevant measurements but small enough to keep away from the percolation limit.

\item[$\mathcal{III}/b:$] First, I examined the critical densities of spanning clusters which touch both boundaries along the strip geometry. I found a good agreement between the predicted conformal values and the calculated averages of different linear size. I checked the validity of a combination of the bulk and surface magnetization exponents coming from scaling predictions through the study of the density behaviour close to the boundaries and I found reasonable accordance. I also applied a correction close to the boundaries in order to obtain a better fit between the calculated and predicted curves.

\item[$\mathcal{III}/c:$] Second, I considered the density of the clusters which are touching one boundary of the strip. This density is analogous to the order parameter profile in the system with fixed-free boundary conditions. This profile close to the fixed boundary was strongly perturbed by surface effects, which are due to the presence of the finite breaking up length. However, at the free boundaries the density curves approached a scaling curve which sat on the predicted conformal function. Close to the fixed boundary I estimated the asymptotic behaviour of the scaling curve through extrapolation and I obtained values in agreement with the conformal results.

\item[$\mathcal{III}/d:$] Finally I considered the density of points in clusters that are touching one of the boundaries or both. This density is analogous to the order parameter profile with parallel fixed spin boundary conditions. This density profile was found to be strongly perturbed by the breaking-up length at both boundaries thus I studied another density which was defined on crossing clusters only. However, it was supposed to be related to the same operator profile and expected to present the same scaling form. I performed the same calculation for percolation also and I found that the numerical and analytical results for this type of density profile were in good agreement in both cases.
\end{description}

The last investigated problem was the antiferromagnetic Ising model on two-dimensional triangular lattice at zero temperature in the absence of external field. This model was intensively studied during the last few decades, since it shows exotic features in equilibrium due to its geometrical frustration. However contradictory explanations were published in the literature about its non-equilibrium dynamical behaviour as it was characterized by a diffusive growth with logarithmic correction or by a subdiffusive dynamics with effective exponents. My aim was to find independent evidences for one of the explanation and examine the dynamical behavior in the aging regime.

\begin{description}
\item[$\mathcal{IV}/a:$] To study the non-equilibrium behaviour of the system I introduced a new quantity, the non-equilibrium relaxation time $t_r$, which depends only on the dynamics and that is capable to determine directly the true value of the dynamical exponent $z$. The analysis of the distribution provided a strong proof for the dynamics in the aging regime to be governed by an exponent $z=2$ with logarithmic correction as it was found in related problems like the two-dimensional XY model and the fully frustrated Ising model.

\item[$\mathcal{IV}/b:$] Following some recent published methods I also studied numerically the two-time autocorrelation function in the equilibrium regime with different waiting times, after the system was relaxed into an equilibrated state. Here I found the expected time-translational invariant behaviour of the autocorrelation function and I located the same dynamical exponent value as above through some exponent combination.

\item[$\mathcal{IV}/c:$] Finally I considered the scaling of the autocorrelation function of different waiting times in the aging regime where I used two kind of scaling variables. The one without logarithmic correction did not show a good scaling collapse which suggested that this variable choice was inconvenient. However the scaling with a logarithmically corrected variable presented an asymptotically good scaling collapse which proved the validity of the logarithmic corrections during the non-equilibrium dynamics in agreement with the previous considerations.
\end{description}

As a summary of my considerations that I obtained during my PhD work and reviewed above in the current thesis, I can declare that I completed studies of recent physical problems in order to describe cooperative behaviour in complex systems. My motivation was to give a relevant description of the related models and complete our knowledge about this segment of science. The validity of my results will be judged in the future as it may become an adaptable description of Nature.

%% file: Appendix.tex
\appendix

\chapter[Appendix A]{Equivalence between the Potts model and the random-cluster model through conditional measures}
\label{RCCondMeas}

Here we are going to discuss an alternative method to see the equivalence between the $q$-state Potts model and the random cluster model. In order to do so we use the notation introduced in Section.\ref{RandomClust}.

When $q\in \mathbb{Z}$ is an integer, the \textit{random-cluster measure} (RC measure), defined in Section \ref{RandomClust}, corresponds to a \textit{Potts measure} $\pi_{\beta,J,q}(\sigma)$ in the absence of external field $h_i=0$ on the sample space $\Sigma$. Let $q\geq 2$ and take the sample space $\Sigma=\left\lbrace 1,2,...,q \right\rbrace^{V} $ thus assign any of $q$ state to each vertex in $V$. Now the relevant probability measure is given by:
\begin{equation}
 \pi_{\beta,J,q}(\sigma)=\dfrac{1}{\mathcal{Z}}e^{-\beta \mathcal{H}(\sigma)} \qquad \mbox{where} \qquad \sigma \in \Sigma
\end{equation}
with $\mathcal{H}$ and $\mathcal{Z}$ were defined as in Eq.\ref{PottsH} and Eq.\ref{PottsZ}. 
\begin{figure}[htb]
\begin{center}
\includegraphics*[ width=10.0cm]{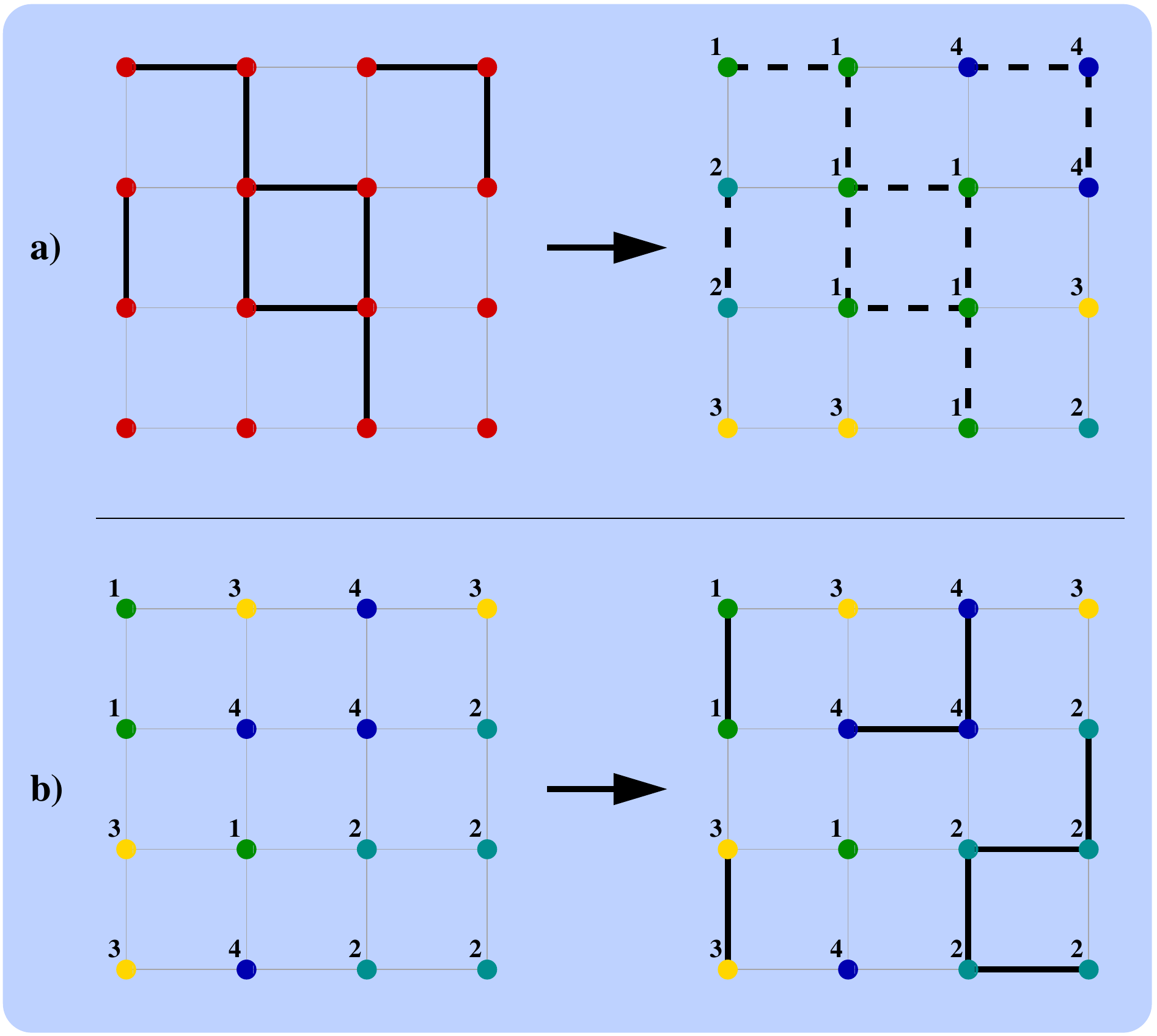}
\caption{Illustration of the conditional measure of $\mu(\sigma,\omega)$ with $q=4$. On the figure a. when $\mu(\cdot | \omega)$ we choose uniformly a spin value from $\left\lbrace 1,2,3,4\right\rbrace$ independently for each cluster of $\omega$. On the contrary case, when $\mu(\cdot | \sigma)$, we begin with a configuration $\sigma$ and attach two neighbouring spins $i$ and $j$ with probability $p$ if $\sigma_i=\sigma_j$ otherwise let them disjoint.}
{
\label{RCmodel}
}
\end{center}
\end{figure}
Hereafter if we construct the two systems on the same probability space we can observe their relationship through their realizations using essentially combinatorial methods \cite{Grimmett2006,Edwards1988}. We consider a product sample space $\Sigma \times \Omega$, where $\Sigma=\left\lbrace 1,2,...,q \right\rbrace^{V} $, $\Omega=\left\lbrace 0,1 \right\rbrace^E$ and let $q\in \left\lbrace 2,3,...\right\rbrace $, $0 \leq p \leq 1$ and measure on a finite graph $G=(V,E)$ as before. Now we define a probability mass function $\mu(\sigma,\omega)$ of the coupled measurement on $\Sigma \times \Omega$
\begin{equation}
 \mu(\sigma,\omega) \propto \prod_{e \in E}\left\lbrace (1-p)\delta(\omega(e),0)+p\delta(\omega(e),1)\delta_e(\sigma_i,\sigma_j) \right\rbrace , \qquad (\sigma,\omega)\in \Sigma \times \Omega
\label{RCmassF}
\end{equation}
where $\delta_e(\sigma_i,\sigma_j)=\delta(\sigma_i,\sigma_j)$ for the edge $e=\langle i,j \rangle \in E$. 

In a conditional measure $\mu(\cdot | \omega)$ if firstly $\omega$ is given, a selection on $\Sigma$ is obtained by choosing uniformly distributed random spins for the $c(\omega)$ number of connecting component (see Fig.\ref{RCmodel}.a). These spins are constant on a given cluster and are independent between clusters. On the contrary if $\sigma$ is given firstly, the conditional measure $\mu(\cdot | \sigma)$ on $\Omega$ is obtained by choosing edges $\omega(e)=1$ with probability $p$ if $\delta_e(\sigma_i,\sigma_j)=1$ (independently between edges), otherwise set $\omega(e)=0$ (see Fig.\ref{RCmodel}.b).

Consequently the probability mass function $\mu$ is a coupling between the RC measure $\phi_{p,q}(\omega)$ on $\Omega$ and the Potts measure $\pi_{\beta,J,q}(\sigma)$ on $\Sigma$, and the parameters are related by $p=1-e^{-\beta}$. At the thermodynamical limit, the phase transition of a Potts model corresponds to a creation of an infinite size cluster in RC model. This special coupling may be used to understand the correlation of the Potts model in a particular simple way at its criticality. 

\chapter[Appendix B]{The continuum and the master-equation approach of Barab\'asi-Albert model}
\label{BASolutions}

Here we are going to discuss shortly the relevant solutions of the Barab\'asi-Albert model. The two presented methods were published by Barab\'{a}si and Albert \cite{Barabasi1999,barabasi-1999} and by Dorogovtsev \textit{et.al.} \cite{Dorogovtsev2000b,379835} and Krapivsky \textit{et.al.} \cite{PhysRevLett.85.4629}. Here we are going to use the notations of Section \ref{BAmodelSec}.

\section{Continuum approach}
Barab\'{a}si and Albert calculated the time evolution of the degree $k_i$ of a given node $i$ in the network following the next considerations \cite{Barabasi1999,barabasi-1999,albert-2002-74}.

At each time step when we add a new point to the network with $m$ edges, $k_i$ is increasing with the probability $\Pi(k_i)$. Suppose that $k_i$ is a real continuous variable and evolving by the dynamical equation:
\begin{equation}
 \dfrac{\partial k_i}{\partial t}=m \Pi(k_i)=m \dfrac{k_i}{\sum_{j=1}^{N-1} k_j}
\end{equation}
where the sum goes through all nodes except the new added vertex with $m$ edges. So $\sum_{j=1}^{N-1} k_j=2mt-m$ and the dynamical equation changes as:
\begin{equation}
\dfrac{\partial k_i}{\partial t}=m \dfrac{k_i}{(2t-1)m}=\dfrac{k_i}{2t-1}\simeq \dfrac{k_i}{2t} \quad \mbox{if} \quad t \gg 1
\end{equation}
If we keep the condition that each node $i$ has a degree $k_i(t_i)=m$ at the time $t_i$ when it is added to the network, the solution of this equation arises as:
\begin{equation}
k_i(t)=m \left( \dfrac{t}{t_i} \right)^{\beta} \qquad \mbox{where} \qquad \beta= \dfrac{1}{2}
\end{equation}
Consequently the degree of every nodes evolves following a power-law. The probability that $P(k_i(t)<k)$ is:
\begin{equation}
P(k_i(t)<k)=P(m \left( \dfrac{t}{t_i} \right)^{\beta} < k)=P(m^{1/\beta}\left(\dfrac{t}{t_i}\right) < k^{1/\beta})=P(t_i>\dfrac{m^{1/\beta}t}{k^{1/\beta}})
\label{eq_k_i_Prob_1}
\end{equation}
Suppose that the nodes were added to the network with equal time difference, and $t_i$ has a constant probability density: $P(t_i)=1/m_0+t$ where $m_0$ is the initial network size. By substituting it to Eq.\ref{eq_k_i_Prob_1} we get:
\begin{equation}
P(t_i>\dfrac{m^{1/\beta}t}{k^{1/\beta}})=1-P(t_i \leq \dfrac{m^{1/\beta}t}{k^{1/\beta}})=1-\dfrac{m^{1/\beta}t}{k^{1/\beta}(m_0+t)}
\end{equation}
Finally after a derivation we can write the degree distribution in the following way:
\begin{equation}
P(k)=\dfrac{\partial P(k_i(t) < k)}{\partial k}= \dfrac{m^{1/\beta}t}{\beta (m_0+t)}\dfrac{1}{k^{(1/\beta+1)}} \qquad \mbox{which goes as } \qquad P(k) \sim \dfrac{m^{1/\beta}}{\beta k^{(1/\beta+1)}}
\end{equation}
if $t\rightarrow \infty$. Since $\beta=\frac{1}{2}$, the degree distribution can be written as $P(k)\sim 2m^2k^{-\gamma}$, where $\gamma=\frac{1}{\beta}+1=3$ is independent of $m$. It is demonstrated in the inset of Figure \ref{BA_DegrDist}, where the rescaled degree distribution with different $m$ fits well on a power-law function with the expected exponent.

\section{Master equation approach}

The method of Dorogovtsev, Mendes and Samukhin \cite{Dorogovtsev2000b,379835} studies the time evolution of $p_k$ fraction of nodes with degree $k$. The probability that a new edge linking to a node which has $k$ neighbors is $\frac{k p_k}{\sum_k k p_k}=\frac{kp_k}{2m}$. Here the denominator is the mean degree of the network, equal to $\sum_k k p_k = 2m$ since every new node introduces $m$ undirected edges and increases with $2m$ the total degrees of nodes \cite{newman-2003-45}. Now the mean number of nodes with degree $k$ which receive an edge when a new nodes is added is $m\frac{kp_k}{2m}=\frac{1}{2}kp_k$ independent of $m$.
\\
\indent
Following the master equation treatment, the $Np_k$ number of nodes with degree $k$ increases with the number of vertices which had $k-1$ neighbors before, but received a new edge in the given step $N\rightarrow N+1$. However $Np_k$ decreases with the number of such nodes which had $k$ degree, but also attached to an edge and has $k+1$ neighbours after. So the number of nodes with degree $k$ changes when we add a new node with $m$ edges and the size increases from $N$ to $N+1$:
\begin{equation}
(N+1)p_{k,N+1}-Np_{k,N}=\frac{1}{2}(k-1)p_{k-1,N}-\frac{1}{2}kp_{k,N} \qquad \mbox{when} \qquad k>m
\end{equation}
where $p_{k,N}$ denotes the probability $p_k$, when the system size is $N$. When $k=m$, the number of nodes which increase $Np_m$ is equal to $1$, the new added node only, since there is no vertex with degree $k<m$, so:
\begin{equation}
(N+1)p_{m,N+1}-Np_{m,N}=1-\frac{1}{2}mp_{m,N} \qquad \mbox{when} \qquad k=m
\end{equation}
Finally give the stationary solution when $p_{k,N}=p_{k,N+1}=p_{k}$:
\[p(k) = \left\{ 
\begin{array}{l l}
  \frac{1}{2}(k-1)p_{k-1}-\frac{1}{2}kp_k & \quad \mbox{if $k>m$}\\
  1-\frac{1}{2}mp_m & \quad \mbox{if $k=m$}\\
\end{array} \right. \]
Rearranging for $p_k$ we get a self consistent solution \cite{Dorogovtsev2000b,PhysRevLett.85.4629} for $p_k$:
\begin{equation}
p(k)=\dfrac{2m(m+1)}{k(k+1)(k+2)}
\end{equation}
which follows a power law $p(k)=P(k)=k^{-\gamma}$ in the limit $k\gg m$ with a fixed exponent $\gamma =3$.
\\
\indent
The \textit{rate equation approach} of Krapivsky \textit{et.al.} \cite{PhysRevLett.85.4629} is completely equivalent with the above discussed solution but applies the dynamical master equation treatment on the average number of nodes with degree $k$ at time $t$.

%% file: Aknowledgement.tex
\chapter{Acknowledgements}

First of all, I would like to express my deepest gratitude to my two supervisors whose understanding, motivation and personal guidance was indispensable during my PhD studies. I am deeply indebted to Prof. Ferenc Igl\'oi for his continuous encouragement, stimulating suggestions and healthy criticism which helped me greatly during my research and writing of this thesis. I am especially beholden to Dr. Jean-Christian Angl\`es d'Auriac, who supported my studies in France from the very first moment. His inspiration and patience was assiduous during our common work and his constant professional approach and accuracy was particularly beneficial to acquire the contrivance of numerical methods.

I wish to express my warm and sincere thanks to the head of the Department of Theoretical Physics at the University of Szeged, Dr. Mih\'aly Benedict, and also his predecessor Dr. Iv\'an Gy\'emant. I am grateful for their personal support and for their reassurance regarding technical conditions during my PhD work. I would like to take this opportunity to thank all the other members of the Theoretical Physics Department for their help and encouragement throughout the years.

I would like to gratefully thank David Baines (Yannick) for the friendly conversations and his enormous assistance as he spared no effort to correct my thesis. I owe a debt of gratitude to the members of the Departement Mati\`ere Condens\'ee et Basses Temp\'eratures at the CNRS-Institut N\'eel (formally called Centre de Recherches sur les Tr\`es Basses Temp\'eratures) and especially to all the PhD students which I shared good times and had fun with.

I would also thank Prof. Heiko Rieger for the opportunity to work at the University of Saarland which period indelibly changed my life and gave me new motivations.

I am also thankful for Prof. Bertrand Berche and Dr. G\'eza \'{O}dor, the referees of my thesis. Their comment and reflections were indispensable to complete my dissertation.

I am beholden to the Minst\`ere Fran\c{c}ais des Affaires \'Etrang\`eres, to the CROUS-International de Grenoble and to the Institut Fran\c{c}ais de Budapest for my research grant and specially to Mme. Marie-Ange Pena, Mme. Noelle Maitre and M. Bob Kaba.

My deepest gratitude goes to my family for their unflagging love and support throughout my life; this dissertation would have been simply impossible without them. I am indebted to my parents who spared no effort to provide the best possible environment for me to grow up in and their constant support during my extended studies. They have always allowed me to have the freedom of choice, and for that I am truly grateful. I am also thankful to my sister Eszter, to my brother M\'aty\'as and to all my friends, as they have always been willing to share their thoughts with me and provide alternative viewpoints which have helped me think in a more holistic fashion.
 
Finally, I am glad to be able to thank Sophie for her love and patience. She has never let me forget the true values of life.

%% file: Declaration.tex
\newpage
\emptypage
\vspace{50mm}
\chapter{Declaration}

The work in this thesis is based on research carried out at the Department of Theoretical Physics at the University of Szeged (Szeged, Hungary), at the CNRS-Institut N\'eel (Grenoble, France) and at the Universit\'e Joseph Fourier. No part of this thesis has been submitted elsewhere for any other degree or qualification and it all
my own work unless referenced to the contrary in the text. 
\vspace{2in}

\noindent \textbf{Copyright \copyright\; 2009 by M\'arton Karsai}.\\
``The copyright of this thesis rests with the author.  No quotations from it should be published without the author's prior written consent and information derived from it should be acknowledged''.

%% file: SummaryEng.tex
\chapter{Summary}

In this PhD thesis I investigated cooperative behaviour in various type of complex many body systems which were applied on scale-free graphs and regular structures. I interested in their static and dynamic critical behaviour in the vicinity of their phase transition point. I completed high performance computations using recent numerical methods in order to characterize their critical properties through numerical averages. Here, following a short introduction to the subject, I shortly review my main results what I developed in four different subjects.

\vspace{.1in}

Cooperative phenomena is an ordinary pattern in various part of science where interacting entities attempt to satisfy some optimal conditions. Such kind of behaviour is observable in sociology, economy, biology or physics etc., where the correlative agents are defined differently but their behaviour presents universal attributes. In this sense the complexity of a system can be defined by its interconnected features arising from properties of its individual parts. Such parts could be either which govern the interactions or the dynamical rules to control the time evolution or the backgrounding structure which also can influence the system behaviour. This kind of combined effects leads to a complex system with non-trivial cooperative behaviour and brings on many interesting questions which give the motivation for advanced studies. 

Correlated systems can be examined efficiently within the frame of statistical physics. The subject of this area is to study systems which depend on random variable, and describe their behaviour obtained from large number of observation using physical terminology. An interesting subject of this segment is the one of phase transitions which has been intensively studied since the beginning of the 20th century. Phase transitions occur in many part of Nature where counteractive processes compete to determine the state of the system controlled by an external parameter. The best-known examples in physics are the liquid-vapor or the ferromagnetic-paramagnetic phase transitions but similar behaviour can be observed in model systems like in different spin models, in epidemic spreading problems or in critical percolation. 

The first relevant approach was the \textit{mean field theory} which gave a phenomenological description of phase transitions and critical phenomena and is capable to describe a wide range of model systems. It was first defined by Pierre Weiss in 1907 for ferromagnetic systems where he assumed that the spins interact with another through a molecular field, proportional to the average magnetization \cite{Weiss1907}. This \textit{molecular field approximation} method neglects the interactions between particles and replaces them with an effective average field which enables to simplify the solution of the problem. At the same time the disadvantage of the method originates from the average molecular field also, since it neglects any kind of fluctuations in the system. Therefore the mean field approximation is valid only in higher dimensional systems or in case of models where the fluctuations are not important.

One of the simplest and most elegant mean field speculations concerning the possible general form of a thermodynamical potential near to the critical point was introduced later by Lev. D. Landau \cite{Landau1936,Landau1980} in 1936. The \textit{Landau theory} allows a phenomenological reproduction of continuous phase transitions based on the symmetry of the order parameter. He assumed that the free energy can be expanded as a power series of the order parameter where the only terms which contribute are the ones compatible with the symmetry of the system.

The observation that the correlation length diverges at the phase transition point and the fluctuations in the system are self-similar in every length scale, led to the recognition of scale invariant behaviour of critical systems. The first comprehensive mathematical approach of such phenomena was given by Leo Kadanoff in the 1960s \cite{RevModPhys.39.395,Kadanoff1966}. A few years later in the beginning of the 1970s based on his results, Kenneth G. Wilson gave the relevant discussion of the subject in his celebrated papers \cite{Wilson1971,Wilson1971b}. He introduced the \textit{renormalization group theory} which serves predictions about critical behaviour in agreement with experimental results and give a possibility to categorize critical systems into universality classes due to them singular behaviour. His investigations were awarded with a Nobel prize in 1982.

In the beginning of the 1970s another theory emerged which completed the knowledge about the universal behaviour around the criticality. The \textit{conformal field theory} was firstly investigated by Polyakov \cite{Polyakov1972}, who capitalized that the group of conformal mappings is equivalent to the group of complex analytical functions in two dimensions \cite{Belavin1984,Henkel1999}. Exploiting this property and the scale invariant behaviour, the partition function of a critical system is derivable which leads to exact calculations of critical exponents. In the case of conform transformations the order parameter correlation functions can be calculated also which then enables to deduce the order parameter profiles along the system boundaries.
\vspace{.1in}

As a first approximation generally the investigated system is assumed to be homogeneous, which condition simplifies its study and the physical description. However, in Nature a substance is often characterized by certain degree of inhomogeneity which could perturb its critical behaviour \cite{Ziman1979}. Frequently stated examples of this abnormality, are the lattice defects and impurities in crystals. In theoretical description this kind of feature is introduced by the concept of disorder which is defined as random distributed values of certain properties of the investigated \textit{disordered model}. These random properties can be the strength of interaction, a random external field or can arises by random dilution. These properties are able to modify the critical features of the system as changing the order of phase transition or shift the critical point and exponents, thus transforming the model into a new universality class.

However, inhomogeneity can arise from the geometrical structure of the backgrounding framework of materials also. Beyond solid-state physics where crystals have a regular geometry, many self-organizing media in the Nature can form random structures. Two Hungarian mathematicians, P\'al Erd\H{o}s and Alfr\'ed R\'enyi defined the first related \textit{random graph model} in their pioneering papers \cite{Erdos1959,Erdos1960,Erdos1961} in the beginning of the 1960s which was the origin of the new science of networks. However, as the informations of real world networks became available by the emergence of large databases, a deeper view into the underlying organizing principles suggested a more complex structure. In the beginning of the 1990s Barab\'asi, Albert and Jeong after they examined the structure of the World Wide Web found an algebraic decay of the network degree distribution. This observation made them realize the real rules which govern the evolution of such kind of \textit{complex networks}. They defined a dynamical growing network model, \textit{the Barab\'asi-Albert model} \cite{Barabasi1999}, where the sites are not connected homogeneously but follow a so-called preferential attachment rule leading to a \textit{scale-free graph} with a power-law degree distribution. This model gave a very good approximation of real complex networks and became popular after many complex systems in science and technology were found to present the same structure \cite{albert-2002-74}. This interdisciplinary behaviour which suggests universal rules behind self-organizing networks keeps this discipline to the frontline of science up to this day.
\vspace{.1in}

Analyzing a complex many body system was difficult earlier since generally these systems have large degrees of freedom and those could be in many possible states. However, by the improvement of computational engines new possibilities appeared and high precision numerical methods were developed to examine the relevant models. It was the base of the new discipline of numerical physics which then became the third pillar of the science beyond the experiments and theory. By using numerical calculations, the statistical description of many body systems became available which then led to a renaissance of the statistical physics. This relation induced the evaluation of \textit{Monte Carlo methods} which provided special computational techniques for statistical physics simulations.

One of the first and most frequently used method which is capable to simulate interacting spin systems is the \textit{Metropolis algorithm}, introduced in 1953 \cite{metropolis:1087}. This so-called single spin flip algorithm evaluate the system through a Markov chain where the system energy change depends on local configurations. However, some other algorithms of \textit{Swendsen-Wang} \cite{PhysRevLett.58.86} and \textit{Wolff} \cite{PhysRevLett.62.361} provided more efficient methods where instead of one-by-one spin flips, complete domains are turned in one step and evaluate the system faster toward its equilibrium.

Beyond that, many other kind of statistical methods were developed in order to calculate some thermodynamical function using mathematical considerations. A recent algorithm which capitalize \textit{combinatorial optimization} is capable to calculate exactly the free energy for models where the free energy function can be recognized as a \textit{submodular function} \cite{dAuriac2002,dAuriac2004}. This iterative method which provided results in strongly polynomial time was applied frequently in course of my work.

\vspace{.1in}

My motivation during my PhD studies was to examine cooperative behaviour in complex systems using the previously mentioned methods of statistical and computational physics. The aim of my work was to study the critical behaviour of interacting many-body systems during their phase transitions and describe their universal features analytically and by means of numerical calculations. In order to do so I completed studies in four different subjects which are presented in the dissertation as follows: 

After a short introduction I summarized the capital points of the related theoretical results. I shortly discussed the subjects of phase transitions and critical phenomena and briefly wrote about the theory of universality classes and critical exponents. Then I introduced the important statistical models which were examined later in the thesis and I gave a short description of disordered models. In the next chapter first, I pointed out the definitions of graph theory that I needed to introduce the applied geometrical structures and I reviewed the main properties of regular lattices and defined their general used boundary conditions. I closed this chapter with a short introduction to complex networks. The next chapter contains the applied numerical methods that I used in the course of numerical studies. I write a few words about Monte Carlo methods and introduce a combinatorial optimization algorithm and its mathematical background. As a last point I describe my own techniques to generate scale-free networks.
\vspace{.1in}

Following this theoretical introduction the obtained scientific results were presented in the following way:

\vspace{.1in}
My first investigated subject was a study of non-equilibrium phase transitions in weighted scale-free networks where I introduced edge weights and rescaled each of them by a power of the connectivities, thus a phase transition could be realized even in realistic networks having a degree exponent $\gamma \leq 3$ \cite{Karsai2006}. The investigated non-equilibrium system was the contact process which is a reaction-diffusion model belonging to the universality class of directed percolation. This epidemic spreading model presents a phase transition between an infected and a recovered state ordered by the ration of the recovering and infecting probability.

\begin{description}

\item[$\mathcal{I}/ a:$] First I gave the dynamical mean-field solution of the model in scale-free networks and I located the previously known three critical regimes. The first regime is the one where the mean-field behaviour is conventional. Then this solution becomes unconventional since critical properties are $\gamma$ dependent. In the third regime, the system is always in an active phase.

\item[$\mathcal{I}/ b:$] I also made some theoretical considerations to generalize recent field-theoretical results about finite-size scaling which are expected to be valid above the upper critical dimension i.e. in the conventional mean-field regime. I introduced in a simple way the volume of the network into the scaling functions that I derivate for two cases, where the infection was initiated from a typical connected site or from the most connected vertex of the network.

\item[$\mathcal{I}/ c:$] I executed high performance numerical simulations in order to simulate the contact process in the conventional mean-field regime. I located the phase transition point and I calculated the finite-size scaling exponents for typical and maximally connected site that I found to be in good agreement with the field-theoretical predictions. I also determined the correlated volume exponent which was found equal in the two cases and compatible with the theoretical expectations. Finally I analyzed the dynamical scaling of the system at the critical point for the two above mentioned cases. Through extrapolations I considered the related exponents to be different in the case of typical and maximally connected sites but compatible with the mean-field and finite-size scaling predictions.
\end{description}

The second problem I investigated was the ferromagnetic random bond Potts model with large values of $q$ on evolving scale-free networks \cite{Karsai2007}. This problem is equivalent to an optimal cooperation problem, where the agents try to find an optimal situation where the benefits of pair cooperation (here the Potts couplings) and total sum of the support, which is the same for all projects (introduced here as the temperature), is maximized. A phase transition occurs in the system between a state when each agents are correlated and a high temperature disordered state. I examined this model using a combinatorial optimization algorithm on scale-free Barab\'asi-Albert networks with homogeneous couplings and when the edge weights were independent random values following a quasi-continuous distribution with different strength of disorder.

\begin{description}
\item[$\mathcal{II}/a:$] As a first point I gave the exact solution of the system for a wide class of evolving networks with homogeneous couplings. The phase transition was found to be strictly first-order and the critical point was determined through simple theoretical considerations. 

\item[$\mathcal{II}/b:$] By numerical calculations I examined the magnetization curve for different strength of disorder and found the theoretically predicted first-order phase transition in the pure case. However, the phase transition softened to a continuous one for any strength of disorder larger than zero. I examined the structure of the optimal set and I found out its structural behaviour was altered by the temperature.

\item[$\mathcal{II}/c:$] I studied also the critical properties of the system and I calculated the distribution of the finite-size critical temperatures for different sizes in case of maximal disorder. I deduced by iterative calculations the critical magnetization exponent and I located the critical temperature using two independent methods. The scaling of the finite-size transition points distribution was characterized by two distinct exponents, that I located and used to consider a scaling collapse of the distribution curves.

\item[$\mathcal{II}/d:$] I also deduced the critical magnetization exponent by two-point fits using the average size of the critical cluster. I obtained compatible values with previous results within the error of the calculations.
 \end{description}

The third examined problem was related to the large-$q$ sate random bond Potts model also. Here I examined the critical density of clusters which touched a certain border of a perpendicular strip like geometry \cite{Karsai2008}. Following from conformal prediction I expected the same density behavior as it was exactly derived for critical percolation in infinite strips \cite{Simmons2007}. I calculated averages by the above mentioned effective combinatorial optimization algorithm and I compared the numerical means to the expected theoretical curves.

\begin{description}
\item[$\mathcal{III}/a:$] During my study I used a bimodal form of disorder for the random couplings, which intensity influenced the breaking-up length of the critical clusters. I allocate an appropriate value of the strength of the disorder which set the breaking up length large enough for relevant measurements but small enough to keep away from the percolation limit.

\item[$\mathcal{III}/b:$] First, I examined the critical densities of spanning clusters which touch both boundaries along the strip geometry. I found a good agreement between the predicted conformal values and the calculated averages of different linear size. I checked the validity of a combination of the bulk and surface magnetization exponents coming from scaling predictions through the study of the density behaviour close to the boundaries and I found reasonable accordance. I also applied a correction close to the boundaries in order to obtain a better fit between the calculated and predicted curves.

\item[$\mathcal{III}/c:$] Second, I considered the density of the clusters which are touching one boundary of the strip. This density is analogous to the order parameter profile in the system with fixed-free boundary conditions. This profile close to the fixed boundary was strongly perturbed by surface effects, which are due to the presence of the finite breaking up length. However, at the free boundaries the density curves approached a scaling curve which sat on the predicted conformal function. Close to the fixed boundary I estimated the asymptotic behaviour of the scaling curve through extrapolation and I obtained values in agreement with the conformal results.

\item[$\mathcal{III}/d:$] Finally I considered the density of points in clusters that are touching one of the boundaries or both. This density is analogous to the order parameter profile with parallel fixed spin boundary conditions. This density profile was found to be strongly perturbed by the breaking-up length at both boundaries thus I studied another density which was defined on crossing clusters only. However, it was supposed to be related to the same operator profile and expected to present the same scaling form. I performed the same calculation for percolation also and I found that the numerical and analytical results for this type of density profile were in good agreement in both cases.
\end{description}

The last investigated problem was the antiferromagnetic Ising model on two-dimensional triangular lattice at zero temperature in the absence of external field \cite{Karsai2009}. This model was intensively studied during the last few decades, since it shows exotic features in equilibrium due to its geometrical frustration. However contradictory explanations were published in the literature about its non-equilibrium dynamical behaviour as it was characterized by a diffusive growth with logarithmic correction \cite{Moore1999} or by a sub-diffusive dynamics with effective exponents \cite{PhysRevE.68.066127,Kim2007}. My aim was to find independent evidences for one of the explanation and examine the dynamical behavior in the aging regime.

\begin{description}
\item[$\mathcal{IV}/a:$] To study the non-equilibrium behaviour of the system I introduced a new quantity, the non-equilibrium relaxation time $t_r$, which depends only on the dynamics and that is capable to determine directly the true value of the dynamical exponent $z$. The analysis of the distribution provided a strong proof for the dynamics in the aging regime to be governed by an exponent $z=2$ with logarithmic correction as it was found in related problems like the two-dimensional XY model and the fully frustrated Ising model.

\item[$\mathcal{IV}/b:$] Following some recent published methods \cite{Walter2008} I also studied numerically the two-time autocorrelation function in the equilibrium regime with different waiting times, after the system was relaxed into an equilibrated state. Here I found the expected time-translational invariant behaviour of the autocorrelation function and I located the same dynamical exponent value as above through some exponent combination.

\item[$\mathcal{IV}/c:$] Finally I considered the scaling of the autocorrelation function of different waiting times in the aging regime where I used two kind of scaling variables. The one without logarithmic correction did not show a good scaling collapse which suggested that this variable choice was inconvenient. However the scaling with a logarithmically corrected variable presented an asymptotically good scaling collapse which proved the validity of the logarithmic corrections during the non-equilibrium dynamics in agreement with the previous considerations.
\end{description}

%% file: SummaryHun.tex
\selectlanguage{magyar}
\chapter*{Összefoglalás}

\addcontentsline{toc}{chapter}{Összefoglalás}
Disszertációmban különböző komplex sok-test rendszerekben kialakuló kooperatív viselkedést vizsgáltam skálamentes hálózatokon és reguláris rácsokon. Elsősorban a tanulmányozott modellek fázisátalakulása során mutatott statikus-, és dinamikus kritikus viselkedésére voltam kíváncsi a fázisátalakulási pont környékén. Az analitikus számolásokat nagy pontosságú numerikus szimulációkból származott eredményekkel vetettem össze, és a főbb következtetéseket referált tudományos folyóiratokban megjelent publikációkban tettem közzé. Munkám eredményeit a disszertációmban fejtettem ki, amit a következőkben egy rövid tárgyi bevezető után té\-zis\-sze\-rű\-en foglalok össze.
\vspace{.1in}

Kooperatív jelenségek a tudomány számos területén megjelennek - elsősorban olyan köl\-csön\-ha\-tó rendszerekben, ahol az egyes egyedek valamilyen optimális állapot elérésére törekednek, és ennek érdekében hajlamosak az együttműködésre. Ilyen típusú viselkedés megfigyelhető a gazdaságban, a szociológiában, a biológiában vagy fizikai rendszerekben is, ahol ugyan a kölcsönható entitásokat eltérő módon definiálják, de hasonló korrelatív viselkedésük a háttérben univerzális törvényeket sejtet. Ezen rendszerek esetén a komplexitás az egyszerre jelenlévő tényezők együttes hatásaként jelenik meg. Ilyen hatások lehetnek az egyedek kölcsönhatását, vagy dinamikai viselkedését leíró törvényszerűségek, vagy egy külső tényező befolyása, esetleg a rendszer speciális geometriai struktúrájából eredő kényszerek. Az ilyen tényezők együttes jelenléte vezet a rendszer komplexitásához, és vet fel olyan érdekes kérdéseket, amelyek mélyebb vizsgálódásra ösz\-tö\-nöz\-nek.

A kooperáló rendszerek hatékonyan vizsgálhatóak a statisztikus fizika keretein belül. Ezen tudományág tárgya olyan jelenségek és modellek vizsgálata, ahol a rendszer állapota véletlen valószínűségi változók függvénye, és célja nagyszámú megfigyelésből származó statisztikai e\-red\-mé\-nyek alapján ezen rendszerek megfelelő fizikai leírása. E tárgykörön belül kiemelkedő figyelmet élvez a fázisátalakulások vizsgálata, amely a XX. század elejétől fogva vált intenzíven kutatott területté. Ilyen jelenségek számos helyen előfordulnak a természetben, ahol egymással versengő folyamatok igyekeznek felülírni a rendszer állapotát egy külső paraméter függvényében. A legjobban ismert példák a folyadék-gőz, vagy a ferromágneses-paramágneses átalakulások, de hasonló viselkedések figyelhetők meg modell rendszerekben is, mint például különböző spin-modellekben, vagy fertőzés terjedési problémákban valamint kritikus perkoláció esetén. 

A téma egyik első releváns tárgyalása az \textit{átlagtér elmélet} volt, amely egy széles körben alkalmazható fenomenológikus közelítéssel szolgált a fázisátalakulások és kritikus jelenségek megértéséhez. Az elméletet Pierre Weiss definiálta először 1907-ben ferromágneses rendszerekre, ahol alapötletként feltételezte, hogy a spinek kölcsönhatása egy átlagtérrel helyettesíthető, ami minden egyes spinre ugyanúgy hat \cite{Weiss1907}. Ez az ún. \textit{molekuláris térközelítés} teljes mértékben elhanyagolja a kölcsönhatási tagban fellépő fluktuációkat, így kellőképp leegyszerűsítve a probléma megoldását. Azonban a modell hátránya szintén a fluktuációk elhanyagolásából származik, miután így alkalmatlan olyan rendszerek kezelésére, ahol ezek hatása fontos szerepet játszik. Így ez a modell csak magasabb dimenziójú rendszerekre alkalmazható, vagy ott, ahol a flukuációk nem befolyásolják drasztikusan a rendszer viselkedését.

Az egyik legegyszerűbb és legelegánsabb átlagtér tárgyalás 1936-ból Lev D. Landau-tól szár\-ma\-zik \cite{Landau1936,Landau1980}. Ő a termodinamikai potenciál általános alakjából kiindulva és a rendparaméter szimmetria tulajdonságait kihasználva sikerrel értelmezte a folytonos fázisátalakulások során fellépő jelenségeket. Feltételezte, hogy a szabad energia analitikus, és a rendparaméter szerint sorba fejthető, ahol csak a rendszer szimmetriáját nem sértő tagok jelennek meg.
\vspace{.1in}

Az a megfigyelés, - hogy a kritikus rendszer korrelációs hossza a fázisátalakulási pont kör\-nyé\-kén divergál, és a rendszerben megjelenő fluktuációk önhasonlóak a hosszúság skálától függetlenül - ez vezetett a kritikus rendszerek skálainvariáns viselkedésének felismeréséhez. A jelenség első átfogó matematikai tárgyalását Leo Kadanoff adta \cite{RevModPhys.39.395,Kadanoff1966}, akinek az eredményeit felhasználva az 1970-es évek elején Kenneth G. Wilson dolgozta ki a \textit{renormalizációs csoportelméletet} \cite{Wilson1971,Wilson1971b}. Ez a hipotézis a kísérleti eredményekkel egybecsengő elméleti jóslásokkal szolgált, melynek segítségével a kritikus rendszerek az őket jellemző szingularitások alapján univerzalitási osztályokba sorolhatóak. Ezekért az eredményekért Wilson 1982-ben megkapta a Nobel-díjat. 

A hetvenes években egy másik teória is napvilágot látott - kiegészítve a már korábban kidolgozott elméleti módszereket. A \textit{konform térelméletet} Polyakov definiálta először \cite{Polyakov1972} kihasználva, hogy a konform-leképezések csoportja ekvivalens a kétdimenziós komplex analitikus függvények csoportjával \cite{Belavin1984,Henkel1999}. Ezt és a skálainvariáns viselkedést felhasználva egy kritikus rendszer állapotösszege felírható, ami aztán alkalmas a kritikus tulajdonságok pontos meghatározására. A konform-leképezések során a rendparaméter korrelációs függvénye szintén kiszámolható, amiből következtethetünk a renparaméter profilok alakjára a rendszer határai mentén. 
\vspace{.1in}

Első közelítésben a vizsgált rendszert homogénnek szokás tekinteni, ez nagyban megkönnyíti annak vizsgálatát, és fizikai leírását. Azonban a természetben az anyagokra jellemző nagyfokú inhomogenitás is fontos szerepet játszhat, amikor azok fázisátalakuláson mennek keresztül \cite{Ziman1979}. Erre a legjobb példa a kristályokban előforduló szennyeződések és rácshibák, melyek jelentékenyen képesek befolyásolni az anyag kritikus viselkedését. Az elméleti leírás során az ilyen inhomogenitásokat valamely paraméteren definiált véletlenszerűséggel szokás bevezetni a vizsgált \textit{rendezetlen modellbe}. Ilyen random paraméter lehet a kölcsönhatás erőssége, vagy megjelenhet egy a rendszerre ható külső térben, vagy pl. definiálható a kölcsönható részecskék véletlenszerű hígításával. A rendezetlenség ilyen fajta bevezetése perturbálhatja a kritikus viselkedést, megváltoztatva a fázisátalakulás rendjét, valamint felülírhatja annak kritikus jellemzőit, így transzformálva a rendszert egy másik univerzalitási osztályba.

Ezen kívül inhomogenitás a rendszer geometriai tulajdonságaiból is eredhet. A szilárdtest fizikából ismert kristályos szerkezeteken túl sok önszervező hálózat a természetben véletlenszerű struktúrákba fejlődik. Az ilyen típusú struktúrákra az első releváns modellt két magyar matematikus, Erdős Pál és Rényi Alfréd definiálta, akik az ún. \textit{véletlen gráf modellben} figyelembe vették az elrendezés lehetséges rendezetlenségét is \cite{Erdos1959,Erdos1960,Erdos1961}. Az ő munkájuk során fejlődött ki a máig széles körben kutatott hálózatok tudománya. Azonban, a technika előrehaladásával a valós hálózatokról egyre több információ vált elérhetővé, így lehetőség nyílott a háttérben meghúzódó valódi rendezőelvek megértésére, ami alapján egy összetettebb viselkedés körvonalazódott ki. A kilencvenes évek elején Barabási-Albert László, Albert Réka és Hawoong Jeong a világháló struktúrájának vizsgálata során figyeltek fel arra, hogy a hálózat fokeloszlása hatványfüggvény szerint cseng le, ami alapján következtettek a hálózat fejlődéséért felelős törvényszerűségekre \cite{Barabasi1999}. Az általuk definiált \textit{Barabási-Albert hálózati modellben}, figyelembe vették, hogy a hálózat időben dinamikusan növekszik, valamint a fejlődés során egy ún. preferált kapcsolódási szabály felelős a rendszer inhomogenitásáért. Ez a két tulajdonság együttes megjelenése vezet a kialakuló hálózat skálamentes viselkedéséhez. Az általuk felállított modell széles körben népszerűvé vált, miután egyre több valós \textit{komplex hálózatban} találtak hasonló struktúrát \cite{albert-2002-74}. Ez a skálamentes tulajdonság az önszervező hálózatok fejlődésénél univerzális törvényszerűségeket mutat a háttérben, mely intenzíven kutatott terület mind mai napig.
\vspace{.1in}

A komplex sok-test rendszerek vizsgálata korábban nehézkes volt, mert az ilyen típusú rendszerek nagy szabadsági fokkal rendelkeznek, így rengetek különböző állapotuk lehetséges. Azonban a számítógépek fejlődésével lehetőség nyílott nagy forrásigényű eljárásokat felhasználva a releváns modellek hatékony vizsgálatára. Ez szolgáltatta az alapot egy új szemléletmód, a numerikus fizika kialakulásához, mely az elmúlt évtizedekben az elméleti és kísérleti tudományon kívül a fizikai harmadik alappillérévé fejlődött. A numerikus módszerek segítségével a kölcsönható sok-test rendszerek statisztikai vizsgálata is lehetővé vált, ez pedig a huszadik század második felétől kezdve a statisztikus fizika reneszánszához vezetett. A statisztikus modellek vizsgálatára kifejlesztett numerikus módszereket közösen \textit{Monte Carlo} metódusoknak nevezzük, melyek később a fizikán kívül számos más tudományágban is alkalmazásra találtak.

Az egyik legelső, és legtöbbet használt módszer az 1953-ban megjelent ún.\textit{ Metropolis algoritmus} \cite{metropolis:1087}, amely kölcsönható spin-rendszerek vizsgálatára alkalmas. Ez a spin-forgató algoritmus a rendszert, mint egy Markov folyamatot kezeli, ahol az energia fejlődése csak lokális konfigurációk függvénye. Ezen túl más módszerek is léteznek ilyen típusú rendszerek vizsgálatára, mint pl. a \textit{Wolff}, vagy a \textit{Swendsen-Wang} algoritmusok \cite{PhysRevLett.58.86,PhysRevLett.62.361}, amelyek nem egyesével forgatják át a spineket, hanem egyszerre teljes domének irányát változtatják meg, így kényszerítve a rendszert egy gyorsabb időfejlődésre.

Azonban ezeken kívül számos más olyan statisztikai módszerek léteznek, melyek szigorú matematikai eljárásokat követve alkalmasak különböző fizikai tulajdonságok, vagy termodinamikai függvények pontos meghatározására. Ilyen, pl. az az algoritmus, amely a \textit{kombinatorikus optimalizáció} módszerét felhasználva alkalmas a szabadenergia egzakt meghatározására olyan spin modellekben, ahol a szabadenergia, mint \textit{szubmoduláris függvény} írható fel \cite{dAuriac2002,dAuriac2004}. Egy ilyen iteratív, erősen polinomiális futásidejű algoritmust alkalmaztunk számos esetben a munkánk során.
\vspace{.1in}

Doktori tanulmányaim fő motivációja - a fent említett elméleti és numerikus módszerek felhasználásával - a kooperatív viselkedés vizsgálata volt komplex rendszerekben. Munkám célja a kölcsönható sok-részecske rendszerekben megjelenő fázisátalakulások és kritikus jelenségek tanulmányozása, és azok univerzális tulajdonságainak leírása volt. Ennek céljából négy különböző témában végeztem kutatásokat, ezek analitikus és numerikus eredményeit disszertációban foglaltam össze a következő struktúrát követve:	

Egy rövid bevezető után ismertetem a témához kapcsolódó fontosabb elméleti eredményeket. Ebben röviden tárgyalom a fázisátalakulások és kritikus jelenségek témakörét, összefoglalóan írok az univerzalitási osztályok, és kritikus exponensek elméletéről, majd bevezetem azokat a modelleket, melyek később a tézisben a vizsgálatok tárgyát fogják képezni. A fejezet végén röviden bevezetem a rendezetlen rendszerek eseteit, és a hozzá kapcsolódó fontosabb elméleti modelleket. A következő fejezetben először azokat az alapvető gráfelméleti definíciókat tekintem át, amelyek szükségesek a fejezet egyszerűbb tárgyalásához, majd ezt követően tárgyalom a használt reguláris geometriai struktúrákat, és azok tulajdonságait. Ezt a fejezetet a komplex véletlen hálózatokba való rövid bevezetőzővel zárom, ahol konkrétan az Erdős-Rényi, Watts-Strogatz és Barabási-Albert hálózati modelleket tárgyalom részletesebben. Az utolsó bevezető jellegű fejezetben a használt numerikus módszereket ismertetem. Itt a Monte Carlo metódusok alapötleteit fejtem ki, majd az alkalmazott spin algoritmusokkal és kombinatorikus optimalizáció módszerével és annak matematikai hátterével foglalkozom. A fejezetben végül a Barabási-Albert hálózat generálására használt saját metódus kerül definiálásra.
\vspace{.1in}

Ezt az általános bevezetőt követően a munkám során kapott tudományos eredmények kerülnek ismertetésre, melyeket itt tézispont szerűen össze is foglalok.
\vspace{.1in}

Elsőként a nemegyensúlyi fázisátalakulásokat vizsgáltam skálamentes hálózatokban, ahol minden élhez a végpontok fokszámának hatványával arányos élsúlyt rendeltem \cite{Karsai2006}. Ez az újraskálázás úgy módosítja a rendszert, hogy a valós esetekhez hasonlóan fázisátalakulás alakulhat ki benne, még $\gamma\leqslant 3$ értékű fokexponens esetén is. A vizsgált nemegyensúlyi folyamat egy reakció-diffúzió modellek közé tartozó ún. kontaktfolyamat volt, ami az irányított perkoláció univerzalitási osztályba sorolható. Ez a fertőzés terjedési modell fázisátalakulást mutat egy fertőzött-aktív és egy tiszta-inaktív fázis között a fertőzési és meggyógyulási valószínűségek hányadosának függvényében. A kontaktfolyamat élsúlyozott skálamentes hálózatokon való vizsgálatai során következő eredményekre jutottam:

\begin{description}

\item[$\mathcal{I}/ a:$] Először a rendszer dinamikai átlagtér megoldását adtam meg, melynek segítségével sikerült azonosítani a korábbi számolásokból várt három fokexponenstől függő kritikus tartományt. Az elsőben a megadott átlagtér megoldás maradéktalanul teljesül. A második az ún. nem-konvencionális átlagtér tartomány, ahol a fázisátalakulás kritikus paraméterei $\gamma$ függővé válnak. Végül a harmadik rezsimben nem történik fázisátalakulás a rendszerben, mivel az mindig az aktív állapotában van.

\item[$\mathcal{I}/ b:$] Ezek után a konvencionális tartományban, tér-elméleti megfontolások alapján vizsgáltam a rendszer véges-méret skálázását. Sikerült egyszerűen bevezetni a rendszer térfogatát a skálafüggvényekbe, ezeket két alapesetre írtam fel: egyrészt, amikor a fertőzés egy átlagos fokszámú pontból indul ki, másrészt, amikor a legjobban összekötött pontból terjed szét.

\item[$\mathcal{I}/ c:$] A kontakt-folyamat vizsgálatára a konvencionális átlagtér tartományban hosszú futásidejű numerikus szimulációkat végeztem. Sikerült lokalizálni a rendszer fázisátalakulási pontját, és megállapítani a véges-méret kritikus exponenseket mindkét fertőzési esetre. Ezek az átlagtér értékekkel jó egyezést mutattak. Szintén kiszámoltam a korrelált térfogati exponenst, ami azonos értéket mutatott mindkét esetben, és jó illeszkedett az elméletből várt eredményekhez. Végül rendszer dinamikus skálázását vizsgáltam a fenti két esetben, ahol egy extrapolációs eljárással sikerült a kapcsolódó dinamikai exponenseket megkapni, amelyek ugyan nem voltak azonosak a maximális és a tipikus fokszámú esetre, de összeegyeztethetőek voltak az átlagtér és véges-méret skálázásból származó elméleti értékekkel.
\end{description}

Az általam vizsgált második probléma a ferromágneses nagy-$q$ állapotú Potts-modell volt, élsúlyozott skálamentes hálózatokon \cite{Karsai2007}. Ez a modell megfeleltethető egy olyan optimális kooperáció problémának, ahol az egyes egyedek a párkölccsönhatásból (itt a Potts kölcsönhatás) és egy külső támogatásból (esetünkben a hőmérséklet) származó bevételeket alapján próbálják megtalálni azt a számukra optimális konfigurációt, amikor a nyereségük maximális. A rendszerben fázisátalakulás jelenik meg egy korrelált és egy rendezetlen magas hőmérsékleti fázis között. Ezt a problémát skálamentes gráfokon egy polinomiális időben megoldható kombinatorikus optimalizációs eljárással vizsgáltam homogén hálózatokon, valamint olyan esetben, ahol az élsúlyok egy kvázifolytonos eloszlást követtek különböző rendezetlenségi erősség mellett.

\begin{description}
\item[$\mathcal{II}/a:$] Elsőként felírtam a probléma egzakt megoldását homogén esetben - nagyszámú dinamikusan fejlődő hálózatra általánosítva. Ekkor a rendszerben végbemenő fázisátalakulást szigorúan elsőrendűnek találtam, ahol egyszerű elméleti meggondolások alapján a kritikus pontot is sikerült határoznom.

\item[$\mathcal{II}/b:$] Számítógépes szimulációk segítségével is sikerült homogén élsúlyozás esetén a fent em\-lí\-tett fázis egybeesést reprodukálni. Azonban a rendszer mágnesezettségét numerikusan vizs\-gál\-va kitűnt, hogy bármilyen erősségű nem-zéró rendezetlenséget bevezetve a rendszerben lévő fá\-zis\-át\-a\-la\-ku\-lás folytonossá válik. A fázisátalakulás során tanulmányoztam az optimális gráf szerkezetét is, ahol vizsgálatokból kitűnt, hogy a gráf a skálamentes struktúráját egészen a kritikus hőmérsékletig megtartja, fölötte viszont ez teljesen szétesik.

\item[$\mathcal{II}/c:$] Ezen kívül igyekeztem meghatározni - maximális rendezetlenség mellett - a rendszer kritikus paramétereit. Iteratív számolások segítségével sikerült megállapítani a kritikus mágnesezettségi exponenst, valamint lokalizálni a fázisátalakulási pontot két egymástól független módszer segítségével. Eltérő méretek esetén számolt véges-méret kritikus hő\-mér\-sék\-le\-tek eloszlása a mérések alapján két különböző exponenssel skálázódott, amiket megállapítva a sikeresen skáláztam össze a különböző méretekhez tartozó eloszlásgörbéket.

\item[$\mathcal{II}/d:$] A kritikus mágnesezettségi exponenst sikerült visszakapnom az átlagos klaszterméretre való két-pont illesztés segítségével. Az így kapott értékek kompatibilisek a fent említett exponensek értékeivel. 
 \end{description}

A harmadik tanulmányozott probléma szintén a nagy-$q$ állapotú Potts-modelhez kapcsolódott. Ebben az esetben azon kritikus klaszterek sűrűségét vizsgáltam, melyek egy előre meghatározott módon tapadnak ki egy véges átmérőjű csík geometria egyes határfelületein \cite{Karsai2008}. Miután ez a rendszer konform-invarianciát mutat a kritikus pontban, ezért hasonló viselkedést vártunk, mint ahogy egy másik konform-invariáns rendszerben, a kritikus perkoláció esetén korábban egzaktul meg lett adva. A vizsgálódások során a fenti kombinatorikus optimalizációs algoritmust használva a sűrűség profilok átlagát hasonlítottam össze az elméletből származó analitikus görbékkel.

\begin{description}
\item[$\mathcal{III}/a:$] A vizsgálatok során a véletlen kötéseket bimodális eloszlás alapján választottam ki. A rendezetlenség erőssége erősen befolyásolta a kritikus klaszterek felbontási hosszát, ezért első lépésként a rendezetlenség megfelelő intenzitását kellet megállapítani úgy, hogy a felbontási hossz megfelelően kicsi legyen, de a rendszer elég messze van a perkolációs limittől.

\item[$\mathcal{III}/b:$] Ezt követően azon kritikus klaszterek sűrűségét tanulmányoztam, amelyek a csík\-ge\-o\-met\-ri\-a mindkét felületét érintették. Kiváló egybeesés mutatkozott az elméletből várt és a numerikusan számolt eredmények között a különböző lineáris méretek esetén. Ezen túl a határfelületek közelében a tömbi-, és felületi mág\-ne\-se\-zett\-sé\-gi viselkedésből származtatott exponens érvényességét is ellenőriztem. A megfigyelt viselkedés jól illeszkedett az exponens alapján várt algebrai alakra. Végül egy korrekciós tag hatását vizsgáltam meg, mely a felületek közelében megfelelően csökkentette a mért és elvárt elméleti eredmények közti eltérést.

\item[$\mathcal{III}/c:$] Ezek után azon klaszterek sűrűségét elemeztem, melyek legalább az egyik határfelületen kitapadtak. Ez a sűrűség profil megegyezik a rendszer rendparaméter profiljával rögzített-szabad határfeltételek mellett. Ezt a profilt a véges felbontási hosszra visszavezethető felületi hatások erősen perturbálták a rögzített határ mentén, azonban a szabad határon a sűrűségprofil jól megközelítette a jósolt analitikus görbét. A rögzített határhoz közel csak extrapolációval sikerült megbecsülnöm a profil alakját, de a termodinamikai limitben asszimptotikusan jó viselkedést találtam. 

\item[$\mathcal{III}/d:$] Az utolsóként vizsgált sűrűségprofil azon klaszterekhez tartozott, amelyek bármelyik határfelületet érintik, így vagy mindkét oldalon vagy legalább az egyiken kitapadtak. Ez a sűrűségprofil a mindkét végén rögzített rendparaméter profillal ekvivalens. Ezt a görbét mindkét végén erősen befolyásolta a véges felbontási hossz, így egy másik, de ugyanazon renparaméter profilhoz tartozó sűrűséget vezettem be az elméleti eredmények ellenőrzésére. Azonosan elvégezve a számolást kritikus perkolációra is, mindkét esetben az eredmények jól illeszkedtek a konform-analítikus görbékre.
\end{description}
 
A negyedik probléma az anti-ferromágneses Ising-modell dinamikai viselkedésének vizsgálata volt háromszögrácson, zéró hőmérsékleten, külső mágneses tér hiányában \cite{Karsai2009}. Ez a geometriailag frusztrált modell egyensúlyban érdekes viselkedést mutat, amely egy igen intenzíven vizsgált terület volt az elmúlt évtizedekben. Azonban a nemegyensúlyi viselkedésével kapcsolatban ellentmondó vélemények jelentek meg a szakirodalomban. Két eltérő nézőpont közül az egyik szerint a dinamika során a korreláció az általános dinamikai exponenssel jellemezhető diffúzív növekedést mutat, melyhez logaritmikus korrekciók járulnak. Ezzel szemben a másik állítás szerint egy szubdiffúzív viselkedés jellemzi a rendszert, amit effektív exponensekkel írhatunk csak le. A vizsgálódásaim célja az volt, hogy a dinamikai viselkedés egy új típusú vizsgálatával független bizonyítékokat találjak valamelyik magyarázathoz. Ezen kívül a nemegyensúlyi tartomány dinamikájának részletesebb vizsgálatát tűztem ki magam elé.

\begin{description}
\item[$\mathcal{IV}/a:$] A nemegyensúlyi viselkedés vizsgálatára egy új mennyiséget definiáltam, a $t_r$ nemegyensúlyi relaxációs időt, ami tisztán a rendszer dinamikájától függ, és alkalmas a $z$ dinamikai exponens közvetlen meghatározására. Ezen relaxációs idő eloszlását vizsgálva különböző méretű rendszerekre, meggyőző bizonyítékokat találtam a diffúzív viselkedést leíró $z=2$ exponens mellett, amihez egy logaritmikus korrekció adódott – hasonlóképpen, mint az azonos kritikus viselkedést mutató XY modellben, valamint a teljesen frusztrált Ising-modell esetén.

\item[$\mathcal{IV}/b:$] Más hasonló modelleken elvégzett vizsgálati módszerek alapján tanulmányoztam az autokorrelációs függvényt különböző várakozási idők mellett, miután a rendszer relaxálódott egy egyensúlyi alapállapotba. Az ilyen típusú dinamikát jellemző idő-transzlációs viselkedésen túl sikerült lokalizálni az egyensúlyi dinamikai exponenst, amely szintén az előző pontban említett értéket adta vissza.

\item[$\mathcal{IV}/c:$] Végül a nemegyensúlyi autokorrelációs függvény skálázását tanulmányoztam a kü\-lön\-bö\-ző várakozási idők esetében, két különböző skálázási változót használva. Az első, logaritmikus korrekcióktól mentes változóval történt behelyettesítés nem vezetett semmilyen releváns egybeeséshez a skálázás során, ebből arra következtethettem, hogy a rendszer valójában nem ezzel a változóval skálázódik. Ezzel szemben a logaritmikus korrekciókat tartalmazó skálázási változó as\-szimp\-to\-ti\-kus\-an jó összeskálázást mutatott, mely szintén a korábbi megállapításainkat támasztotta alá, miszerint a rendszert a nemegyensúlyi fázisban egy diffúzív dinamika jellemzi, additív logaritmikus korrekciókkal.
\end{description}


%% file: SummaryFr.tex
\selectlanguage{french}
\chapter{Résumé}


Durant ma thèse, j’ai étudié le comportement coopératif dans différents types de systèmes complexes à $N$ corps  sur des structures régulières et sur des réseaux sans échelle de longueur. Je me suis intéressé à leur comportement critique statique et dynamique près de leur transition de phase. J’ai effectué des calculs de haute performance en utilisant des méthodes numériques récentes, afin de caractériser, à travers des moyennes numériques, les propriétés critiques de ces systèmes. Après une courte introduction au sujet, je passerai en revue brièvement les principaux résultats que j’ai obtenus dans quatre sujets.
\vspace{.1in}

Le phénomène coopératif se manifeste dans différents domaines de la science, où les entités, qui interagissent, essaient de satisfaire certaines conditions optimales. De tel comportement sont observables en sociologie, économie, biologie ou physique etc.…, où les agents interagissants sont définis différemment, mais où leur comportement présente des attributs universels. En ce sens la complexité du système peut être définie par ses caractéristiques interconnectées résultant des propriétés de ses parties individuelles. De telles parties peuvent aussi bien être celles qui gouvernent les interactions que les règles dynamiques contrôlant l’évolution dans le temps, ou encore le contexte qui peut également influencer le comportement du système. Ce genre d’effets combinés mène à un système complexe avec un comportement coopératif non trivial et soulève de nombreuses intéressantes questions qui sont le moteur d’études avancées.

Les systèmes corrélés peuvent être examinés efficacement dans le cadre de la physique statistique. Le sujet de ce domaine est d’étudier les systèmes qui dépendent de variables aléatoires, et décrit leur comportement obtenu d’un grand nombre d’observations utilisant la terminologie de la physique. Un sujet intéressant de cette branche est celui des transitions de phase qui a été intensément étudié depuis le début du XX\textsuperscript{ème} siècle. Les transitions de phase apparaissent dans de nombreuses parties de la Nature où la compétition entre différents processus, contrôlée par un paramètre extérieur, détermine l’état du système. Les exemples les plus connus en physique sont la vaporisation, les transitions paramagnétique/ferromagnétique, mais des comportements similaires peuvent être observés dans des systèmes modèles tels que dans différents modèles de spin, dans des problèmes épidémiques étendus ou dans la percolation critique.

La première approche pertinente fut la \textit{théorie de champ moyen} qui donne une description phénoménologique des transitions de phase, du phénomène critique et est en mesure de décrire une grande gamme des systèmes modèles. Cela a tout d’abord été défini par Pierre Weiss en 1907 pour des systèmes ferromagnétiques où il a déterminé que les spins interagissent entre eux grâce à un champ moléculaire, proportionnel à une aimantation moyenne \cite{Weiss1907}. Cette méthode d’approximation du champ moléculaire néglige les interactions entre les particules individuelles en les remplacent par un champ moyen effectif, ce qui permet de simplifier la solution du problème. Dans le même temps l’inconvénient de cette méthode provient également du champ moléculaire moyen, puisqu’il néglige toute sorte de fluctuations dans le système. Donc l’approximation de champ moyen est valide seulement dans les systèmes à grande dimension ou en cas de modèle à faible fluctuations.

L’une des plus simples et plus élégantes spéculations de champ moyen, concernant la forme possible générale d’une potentielle thermodynamique proche du point critique, a été introduite par Lev. D. Landau en 1936 \cite{Landau1936,Landau1980}. La \textit{théorie de Landau} permet une reproduction phénoménologique des transitions de phase continues basée sur la symétrie du paramètre d’ordre. Il a démontré que l’énergie libre est une fonction analytique et qu’elle peut s’étendre à des séries de puissance du paramètre d’ordre où les seuls termes qui contribuent sont ceux compatibles avec la symétrie du système.
\vspace{.1in}

L’observation du fait que la longueur de la corrélation diverge au seuil de la transition de phase et que les fluctuations dans le système sont les mêmes tout a toute l’échelle amène à la reconnaissance d’un comportement invariant d'échelle des systèmes critiques. La première approche mathématique d’un tel phénomène a été faite par Léo Kadanoff dans les années 60 \cite{RevModPhys.39.395,Kadanoff1966}. Quelques années plus tard au début des années 70, basé sur ces résultats, Kenneth G. Wilson, a engagé un débat sur ce sujet, dans ses célèbres écrits \cite{Wilson1971,Wilson1971b}. Il a introduit la \textit{théorie du groupe de renormalisation} qui a produit les prédictions à propos du comportement critique en accord avec les résultats expérimentaux et a donné la possibilité de classifier les systèmes critiques en classe d’universalité, du fait de leur comportement singulier. Ses recherches ont été récompensées par le prix Nobel en 1982.

Au début des années 70 une autre théorie émerge et complète les connaissances concernant le comportement universel autour des phénomènes critique. La \textit{théorie conforme} a d’abord été proposée par Polyakov \cite{Polyakov1972}, qui a tiré bénéfice du fait que le groupe des applications conformes est équivalent au groupe des fonctions complexes analytiques en deux dimensions \cite{Belavin1984,Henkel1999}. En exploitant cette propriété et l'invariance d'échelle, la fonction de partition du système critique est calculable ce qui amène à un calcul exact des exposants critiques. Dans le cas de transformations conformes, la fonction corrélation du paramètre d’ordre peut être également calculée ce qui permet de déduire le profile du paramètre d’ordre le long des bords du système. 
\vspace{.1in}

En première approximation, le système examiné est généralement supposé homogène, cette condition simplifie son étude et sa description physique. Toutefois dans la nature, une substance est souvent caractérisée par un certain degré d’inhomogénéité, qui peut perturber son comportement critique \cite{Ziman1979}. De fréquents exemples établis de cette anormalité sont  les défauts et les impuretés dans les cristaux. Dans une description théorique ce genre de caractéristique est introduit par le concept de désordre qui est défini par des valeurs aléatoires distribuées suivant une certaine loi, caractéristique du \textit{modèle désordonné} examiné. Ces propriétés aléatoires sont capables aussi bien de modifier les caractéristiques critiques d’un système que de changer l’ordre de la transition de phase ou de modifier le point critique et les exposants, ce qui range alors le modèle dans  une nouvelle classe d’universalité.

Toutefois, l’inhomogénéité peut aussi provenir également de la structure géométrique du matériel de base. Au-delà de la physique du solide, où les cristaux ont une géométrie régulière, de nombreux milieux auto-organisés dans la nature peuvent former des structures aléatoires. Deux mathématiciens hongrois, P\'al Erd\H{o}s et Alfréd Rényi ont défini le 1\textsuperscript{er} \textit{modèle graphe aléatoire} connu dans leurs papiers pionniers au début des années 60, ce qui s’est avéré être l’origine de la nouvelle science des réseaux \cite{Erdos1959,Erdos1960,Erdos1961}. Toutefois, comme les informations des réseaux du monde réel devinrent accessibles grâce à l’émergence de vastes bases de données, une attention particulière aux  principes organisateurs et implicites a suggéré une structure plus complexe. Au début des années 90 Barab\'asi, Albert et Jeong, après avoir examiné la structure du {\guillemotleft}World Wide Web{\guillemotright}, ont trouvé une décroissance algébrique de la distribution des degrés. Cette observation les a fait réaliser la véritable règle qui gouverne l’évolution de ce genre de \textit{réseau complexe} \cite{Barabasi1999}. Ils ont défini un modèle de réseau dynamique croissant, le \textit{modèle Barab\'asi-Albert}, où les sites sont connectés de façon homogène mais suivent une règle d’attachement préférentiel conduisant à un \textit{graphe sans échelle} avec une loi de puissance de la distribution des degrés. Ce modèle donna une très bonne approximation des réseaux complexes réels et est  devenu populaire après que l’on ait trouvé que beaucoup de systèmes complexes en science présente la même structure \cite{albert-2002-74}. Ce comportement, observé dans plusieurs disciplines, qui suggère des règles universelles derrière les réseaux auto-organisés porte cette discipline en première ligne de la science jusqu’à aujourd’hui.
\vspace{.1in}

Analyser un système complexe à $N$ corps était difficile autrefois, puisque généralement ces systèmes ont beaucoup de degrés de liberté et que ces derniers peuvent être dans de nombreux états possibles. Toutefois grâce à l’amélioration des moyens de calculs de nouvelles possibilités sont apparues et des méthodes numériques de haute précision ont été développées pour examiner les modèles pertinents. Cela a été à la base de la nouvelle discipline physique numérique, qui ensuite devint le 3\textsuperscript{ème} pilier de la science après les expériences et la théorie. En utilisant les calculs numériques, la description de système à $N$ corps devint accessible ce qui a amené à la renaissance de la physique statistique. Cette relation induit l’évaluation des méthodes Monte-Carlo qui ont fourni des techniques spéciales de calculs pour les simulations de physique statistiques.

L’une des 1ères, et des plus utilisées, méthodes, qui soit capable de simuler des systèmes spins interagissant, est l’\textit{algorithme Metropolis}, introduit en 1953 \cite{metropolis:1087}. Cet algorithme de retournement à un spin fait évoluer le système, à travers une chaine Markov où le changement d’énergie du système dépend des configurations locales. Toutefois, d’autres algorithmes, celui de \textit{Swendsen-Wang} et celui de \textit{Wolff} \cite{PhysRevLett.58.86,PhysRevLett.62.361}, fournissent des méthodes plus efficaces où au lieu de retourner un seul spin, des domaines complets sont retournés en une étape et conduise plus vite le système vers l'équilibre.

Au-delà de cela, de nombreux autres genres de méthodes statistiques ont été développés afin de calculer certaines fonctions thermodynamiques en utilisant des considérations mathématiques. Un récent algorithme qui utilise l’optimisation combinatoire, est capable de calculer exactement l’énergie libre pour des modèles où la fonction énergie libre est une fonction submodulaire \cite{dAuriac2002,dAuriac2004}. Cette méthode itérative, qui fournit des résultats en un temps fortement polynomial, a été appliquée fréquemment au cours de mon travail.
\vspace{.1in}

Ma motivation lors de mon doctorat fut d’examiner le comportement coopératif dans des systèmes complexes en utilisant les méthodes précédemment mentionnées, de la physique statistique et de l'informatique. Le but de mon travail fut d’étudier le comportement critique des systèmes à $N$ corps durant leurs transitions de phase et de décrire de façon analytique leurs caractéristiques universelles, au moyen de calculs numériques. Afin d’y arriver j’ai effectué des études dans quatre sujets différents qui sont présentés dans la dissertation de la manière suivante:

Après une brève introduction, j’ai résumé les points capitaux en relation avec les résultats théoriques. J’ai brièvement abordé le sujet des transitions de phase et des phénomènes critiques, de même que la théorie des classes d’universalité et des exposants critiques. Ensuite j’ai introduit les modèles statistiques important qui sont examinés plus tard dans la thèse et j’ai donné une petite description des modèles désordonnés. Dans le chapitre suivant, j’ai tout d’abord mis en avant les définitions de la théorie des graphes dont j’ai eu besoin pour introduire les structures géométriques appliquées et j’ai passé en revue les principales propriétés des réseaux régulières et j’ai défini les conditions de bord généralement utilisées. J’ai terminé ce chapitre avec une petite introduction sur les réseaux complexes. Le chapitre suivant contient les méthodes numériques appliquées que j’ai utilisées au cours des études numériques. J’ai écrit quelques mots sur les méthodes de Monte-Carlo et j’ai introduit l'algorithme d’optimisation combinatoire utilisé, et ses justifications mathématiques. Pour terminer j’ai décrit mes propres techniques pour générer des réseaux sans échelle.
\vspace{.1in}

Suite à cette introduction théorique les résultats scientifiques ont été présentés de la manière suivante:
\vspace{.1in}

Le 1\textsuperscript{er} sujet auquel je me suis intéressé est une étude des transitions de phase hors équilibre dans les réseaux sans échelle de longueur, où la distribution des connectivités était ajustée, de telle façon qu’une transition de phase puisse être réalisée même dans les réseaux réalistes ayant un degré exposant $\gamma<3$ \cite{Karsai2006}. Le système hors équilibre étudié  était le {\guillemotleft}contact process{\guillemotright} qui est un modèle de réaction-diffusion appartenant à la classe d’universalité de la percolation dirigé.

\begin{description}
\item[$\mathcal{I}/ a:$] Tout d’abord j’ai donné la solution de champ moyen dynamique du modèle dans les réseaux sans échelle et j’ai localisé les trois régimes critiques connus auparavant. Le 1\textsuperscript{er} régime est celui où le comportement de champ moyen est conventionnel. Ensuite cette solution devient non conventionnelle puisque les propriétés critiques sont dépendantes de $\gamma$. Dans le 3\textsuperscript{ème} régime, le système est toujours dans une phase active.

\item[$\mathcal{I}/ b:$] J’ai également fait quelques considérations théoriques afin de généraliser les récents résultats de théorie des champs à propos des échelles de taille finie, qui sont censées être valide au-dessus de la  dimension critique supérieure, autrement dit dans le régime de champ moyen conventionnel. J’ai introduit de manière simple le volume du réseau dans les fonctions échelles que j’ai dérivées dans les deux cas,  où l’infection est initié soit par un site typique connecté, soit par le vertex le plus connecté du réseau.

\item[$\mathcal{I}/ c:$] J’ai exécuté des simulations numériques de haute performance afin de stimuler le processus contact dans le régime de champ moyen conventionnel. J’ai localisé le seuil de transition de phase et j’ai calculé les exposants de taille finie pour le site connecté maximal et typique que j’ai trouvé en accord avec les prédictions de théorie des champs. J’ai également déterminé l’exposant du volume corrélé qui s’est avéré égal dans les deux cas et compatible avec les predictions théoriques.  Finalement j’ai analysé le {\guillemotleft}scaling{\guillemotright} dynamique du système au point critique pour les deux cas mentionné ci-dessus. A l'aide d’extrapolations j’ai montré que  les exposants reliés aux deux cas ci-dessus (site typique ou maximalement connecté) étaient différents mais compatible avec les prédictions de champ  moyen et de loi d'échelle en taille finie.
\end{description}
\vspace{.1in}

Le deuxième problème que j’ai étudié fut le modèle de Potts aléatoire  ferromagnétique avec de grandes valeurs de $q$ sur des réseaux évolutifs sans échelle \cite{Karsai2007}. Ce problème est équivalent à un problème de coopération optimale, où les agents essaient de trouver une situation optimale, où les bénéfices de coopération de paire (ici les couplages de Potts) et la somme totale du support, qui est la même pour tous les projets (introduite ici comme la température), sont maximisés. Une transition de phase apparaît dans le système entre un état où tous les agents sont corrélés, et un état désordonné à haute température. J’ai examiné ce modèle en utilisant un algorithme d’optimisation combinatoire sur les réseaux de Barabasi-Albert sans échelle de longueur avec des couplages homogènes et aussi avec  des couplages pondérés par des variables aléatoires indépendantes, suivant une distribution quasi-continue avec différents intensité  de désordre.

\begin{description}
\item[$\mathcal{II}/ a:$] Tout d’abord j’ai donné la solution exacte du système pour une large classe de réseaux évoluant avec des couplages homogènes. La transition de phase s’est avérée être du premier ordre et le point critique a été déterminé par simples considérations théoriques.

\item[$\mathcal{II}/ b:$] Grâce à des calculs numériques j’ai examiné la courbe d’aimantation pour différentes intensités du désordre et j’ai trouvé la transition de phase prédite théoriquement dans le cas pur. Toutefois la transition de phase devient du deuxième ordre pour n’importe quel désordre non nul.  J’ai examiné la structure de l'ensemble optimal  et j’ai trouvé qu'elle était modifiée lorsque la température variait.

\item[$\mathcal{II}/ c:$] J’ai étudié également les propriétés critiques du système et j’ai calculé la distribution des températures critiques de taille finie pour différentes tailles en cas de désordre maximal. J’ai déduit par des calculs itératifs  l’exposant critique d’aimantation et j’ai localisé la température critique en utilisant deux méthodes indépendantes. La loi d'échelle en taille fine pour la distribution des températures de transition est caractérisée par deux exposants distincts que j’ai déterminés et utilisés pour vérifier que  ces courbes des distributions de températures critiques se ramenaient toutes à une courbe unique si l'on utilisait les variables réduites.

\item[$\mathcal{II}/ d:$] J’ai également déduit l’exposant critique d’aimantation par un ajustement à deux points en utilisant la taille moyenne de l'amas critique. J’ai obtenu des valeurs compatibles avec les résultats précédents en tenant compte des arrondis de calculs.
\end{description}
\vspace{.1in}

Le troisième problème examiné fut en rapport également avec le modèle de Potts ferromagnétique aléatoire à grand nombre d'états \cite{Karsai2008}. J’ai examiné la densité critique des amas qui touchent l'un ou l'autre des bords dans une géométrie rectangulaire. Conformément à une prédiction de la théorie conforme je me suis attendu au même comportement que celui dérivé exactement pour la percolation critique dans des bandes infinies \cite{Simmons2007}. J’ai calculé des moyennes à l'aide de l’algorithme d’optimisation combinatoire  mentionné ci-dessus et j’ai comparé les moyennes numériques aux courbes théoriques attendues.

\begin{description}
\item[$\mathcal{III}/ a:$] Au cours de cette étude j’ai utilisé une forme bimodale pour la distribution des couplages aléatoires, dont l’intensité controle la longueur de brisure des amas critiques. J’ai choisi une valeur appropriée de l'intensité du désordre pour la  longueur de brisure des amas critiques soit suffisamment courte mais en évitant la limite de percolation.

\item[$\mathcal{III}/ b:$] Tout d’abord j’ai examiné les densités critiques des amas qui touchent les deux longueurs du réseau rectangulaire. J’ai trouvé une bonne concordance entre les valeurs prédites par les théories conformes et les moyennes calculées pour différentes tailles. J’ai vérifié la validité de la combinaison de l'exposant d’aimantation {\guillemotleft}bulk{\guillemotright} et de l'exposant d'aimantation de surface prédite par les lois d’échelle, par l’étude du comportement de la densité près des bords et j’ai trouvé une concordance raisonnable. J’ai également appliqué une correction près des bords afin d’obtenir une meilleure adéquation entre les courbes prédites et calculées.

\item[$\mathcal{III}/ c:$] En second, j’ai considéré la densité des amas qui touche l’un des bords de la bande. Cette densité est analogue au profile du paramètre d’ordre dans le système avec des conditions de bords fixe-libre. Ce profile près des bords fixes est fortement perturbé par les effets de surface, qui sont dus à la longueur de brisure des amas. Toutefois aux bords libres les courbes de densité s’approche de la fonction prédite par la théorie conforme. Près du bord fixe, j’ai estimé le comportement asymptotique  de la courbe à échelle en extrapolant et j'ai obtenu des valeurs en accord avec les résultats des théories conformes.

\item[$\mathcal{III}/ d:$] Finalement j’ai considéré la densité des points des amas qui touchent un des bords ou les deux. Cette densité est analogue au profile du paramètre d’ordre avec des conditions de bord fixe. Ce profile de densité s’est trouvé fortement perturbé par la longueur de brisure des amas aux deux bords, c’est pourquoi j’ai étudié une autre densité définie uniquement pour les amas traversant l'échantillon.  On suppose que la même forme d’échelle s'applique aux deux cas. J’ai également accompli le même calcul pour le problème de la  percolation et j’ai trouvé que les résultats numériques et analytiques pour ce type de profil de densité sont en concordance dans les deux cas.
\end{description}
\vspace{.1in}

Le dernier problème que j’ai étudié fut le modèle antiferromagnétique d’Ising bidimensionnel sur réseau triangulaire à température zéro en l’absence de champ extérieur \cite{Karsai2009}. Ce modèle a été intensément étudié au cours des deux dernières décennies, dans la mesure où il montre les caractéristiques exotiques à l'équilibre due à la frustration géométrique. Cependant des explications contradictoires ont été publiées dans la littérature à propos du comportement dynamique en hors équilibre, suivant qu’il était caractérisé par une croissance diffusive avec correction logarithmique \cite{Moore1999} ou par une dynamique sous  diffusives avec des exposants effectifs \cite{PhysRevE.68.066127,Kim2007}. Mon but fut de trouver des preuves indépendantes pour l’une des explications et d’examiner le comportement dynamique dans le régime de vieillissement.

\begin{description}
\item[$\mathcal{IV}/ a:$] Pour étudier le comportement en hors équilibre du système j’ai introduit une nouvelle quantité de temps de relaxation hors-équilibre $t_r$, qui dépend uniquement de la dynamique et qui est capable de déterminer directement la vraie valeur de l’exposant dynamique $z$. L’analyse de la distribution fournit une preuve forte que  les dynamiques dans le régime de vieillissement  sont régies par un exposant $z=2$ avec correction logarithmique, tel que cela a été trouvé aussi dans des problèmes reliés telsque le modèle XY bidimensionnel et le modèle d’Ising complètement frustré.

\item[$\mathcal{IV}/ b:$] D’après des méthodes récemment publiées \cite{Walter2008}, j’ai également étudié numériquement la fonction d’auto corrélation a deux temps dans le régime d'équilibre avec différents temps d’attente, après que le système soit relaxé dans un état d'équilibre. J’ai ainsi trouvé le comportement attendu d'invariance par  translation de la fonction d’auto corrélation et j’ai retrouvé  même valeur de l’exposant dynamique que  ci-dessus, mais à travers une combinaison d’exposants.

\item[$\mathcal{IV}/ c:$] Finalement j’ai considéré le comportement d’échelle de la fonction d’auto corrélation après différents temps d’attente dans le régime de vieillissement où j’ai utilisé les deux types de variables d’échelle introduits plus haut. La variable d'échelle sans correction logarithmique n’a pas montré de courbe universelle, ce qui suggéra que ce choix ne convient pas. Cependant la variable logarithmiquement corrigée a bien présenté une courbe universelle ce qui prouva la validité des corrections logarithmiques lors des dynamiques hors équilibrées en accord avec les considérations précédentes.
\end{description}